\documentclass[graybox,natbib]{svmult}
\bibpunct{(}{)}{;}{a}{}{,} 

\usepackage{helvet}         
\usepackage{courier}        
\usepackage{type1cm}        

\usepackage{graphicx}        
\usepackage{multicol}        
\usepackage[bottom]{footmisc}
\usepackage[normalem]{ulem}	
\usepackage{soul}   
\usepackage{textgreek}
\usepackage{subfig}
\usepackage{amsfonts}
\usepackage{amsmath}
\usepackage{booktabs}
\usepackage[utf8]{inputenc}
\usepackage[linesnumbered,ruled,vlined]{algorithm2e}
\usepackage{
algorithm2e,
amsmath,
amssymb,
array,
appendix,
bm,
bbm,
colortbl,
epsf,
enumitem,
float,
graphicx,
listings,
mathtools,
mathrsfs,
pifont,
setspace,
textcomp,
tikz,
titletoc,
units,
url,
ulem,
comment,
cite
}

\begin{document}
\title*{Bias and Coverage Properties of the WENDy-IRLS Algorithm}
\titlerunning{Bias and Coverage and Properties of WENDy} 
\author{Abhi Chawla, David M.~Bortz, Vanja Dukic\thanks{corresponding author}}
\authorrunning{Chawla et al.} 
\institute{Abhi Chawla, David M.~Bortz, Vanja Dukic \at Department of Applied Mathematics, University of Colorado, Boulder, CO, 80309-0526, USA \email{vanja.dukic@colorado.edu}
}
%
%
\maketitle
\abstract{The Weak form Estimation of Nonlinear Dynamics (WENDy) method is a recently proposed class of parameter estimation algorithms that exhibits notable noise robustness and computational efficiency. This work examines the coverage and bias properties of the original WENDy-IRLS algorithm's parameter and state estimators in the context of the following differential equations: Logistic, Lotka-Volterra, FitzHugh-Nagumo, Hindmarsh-Rose, and a Protein Transduction Benchmark. The estimators' performance was studied in simulated data examples,  under four different noise distributions (normal, log-normal, additive censored normal, and additive truncated normal), and a wide range of  noise, reaching  levels  much higher than previously tested for this algorithm.}
\section*{Keywords}
Weak form Estimation of Non-linear Dynamics, Coverage, Bias, FitzHugh-Nagumo, Hindmarsh-Rose, and Protein Transduction Benchmark.
\section{Introduction}
Weak form Scientific Machine Learning (WSciML) is a recently and actively developed framework for data-driven modeling, inference, and scientific discovery. 
The Weak form Sparse Identification of Nonlinear Dynamics (WSINDy) method learns governing mathematical models directly from data \citep{MessengerBortz2021JComputPhys,MessengerBortz2021MultiscaleModelSimul}, while the Weak form Estimation of Nonlinear Dynamics (WENDy) method efficiently estimates parameters when the model is known; WENDy-IRLS \citep{BortzMessengerDukic2023BullMathBiol} utilizes iteratively reweighted least squares algorithm,   and WENDy-MLE \citep{RummelMessengerBeckerEtAl2025arXiv250208881} utilizes maximum likelihood. Several works demonstrate that the weak form methods have a high tolerance for noise \citep{BortzMessengerDukic2023BullMathBiol,MessengerBortz2021JComputPhys,MessengerBortz2021MultiscaleModelSimul,MessengerBurbyBortz2024SciRep,MinorMessengerDukicEtAl2025JGeophysResMachLearnComput,RummelMessengerBeckerEtAl2025arXiv250208881,TranHeMessengerEtAl2024ComputMethodsApplMechEng}, and one publication investigates the asymptotic properties of WSINDy in the limit of continuum data \citep{MessengerBortz2024IMAJNumerAnal}. Another publication demonstrates how this approach, along with parametric bootstrapping, could be used for state uncertainty quantification \citep{MessengerDwyerDukic2024JRSocInterface}. However, while there has been a systematic investigation of the statistical properties of the maximum-likelihood extension of the weak-form WENDy method \citep{RummelMessengerBeckerEtAl2025arXiv250208881}, there has been no systematic investigation of the statistical properties of the original weak-form WENDy-IRLS method \citep{BortzMessengerDukic2023BullMathBiol}. This work studies the bias and coverage properties of the original WENDy-IRLS algorithm (which, for simplicity, is referred to as the WENDy algorithm for the rest of this paper) \citep{BortzMessengerDukic2023BullMathBiol} applied to five benchmark ordinary differential equation systems arising in the modeling of biological and physical phenomena.

\subsection{Output Error-based Non-linear Least Squares}
Dynamical systems are often used to model biological phenomena, such as population growth, ecological predator-prey relationships, and neuron behavior. These relationships are frequently represented as a system of ordinary differential equations (ODEs) of the form:
\begin{align}
    \dfrac{d\mathbf{u}}{dt} = \sum_{j=1}^{J} w_j\mathbf{f}_j(\mathbf{u}), \\  \mathbf{u}(t_0) = \mathbf{u}_0 \in \mathbb{R}^d \nonumber
\end{align}
Here, $\mathbf{u}(\cdot): \mathbb{R} \rightarrow \mathbb{R}^d$ represents the solution to a system with $d$ states, or dependent variables, which depend on one independent variable $t$. The independent variable $t$ usually denotes time; indeed, for the remainder of this work, $t$ will be referred to as time. The $J$ dimensional vector $\mathbf{w} = [w_1, w_2, \cdots, w_J]^T$ contains the \emph{parameters} of the system that are multiplied by the \emph{features} $\mathbf{f}_j$. These features $\mathbf{f}_j$ are, or can in principle be, functions of all the states $\mathbf{u}$. 

An essential component of the modeling process involves estimating the parameters $\mathbf{w}$ from the observed data $\mathbf{U} \in \mathbb{R}^{(M+1) \times d}$ collected at $M+1$ time points. Accurate parameter estimation not only enhances the model’s predictive capabilities but also provides insights into the underlying theory governing the system dynamics. However, this task can be challenging, since many ODE systems used in biological and physical modeling are often non-linear and do not offer closed-form solutions. 

A widely used approach for parameter estimation is the \textit{Output Error-based Non-linear Least Squares} (OE-NLS) method. This method begins with an initial parameter guess, $\mathbf{w}_0$, then numerically solves the ODE system and evaluates a discrepancy between the computed state variables and the observed data (typically, this discrepancy is evaluated via the loss function called the "least squares" or the $L_2$-norm). A nonlinear optimization algorithm is then employed to adjust the parameters and minimize this loss iteratively. This process is repeated until a parameter set is obtained that minimizes the discrepancy between the model and the data.

It is easy to see that the OE-NLS algorithm is inefficient as it requires resolving the ODE numerically for every new set of parameter values at each iteration. If the ODE is chaotic (highly sensitive to initial values), then repeatedly solving the ODE may result in large errors. Lastly, OE-NLS methods are not very robust to increasing levels of noise, limiting their utility for many datasets and frequently requiring informative priors in the Bayesian framework  (see \citep{McGoffMukherjeePillai2015StatistSurv, DukicLopesPolson2012JAmStatAssoc, ElderdDukicDwyer2006ProcNatlAcadSciUSA, NardiniBortz2019InverseProbl}). WENDy aims to offer a better alternative by eliminating the step of numerically solving the ODEs. The advantages of WENDy stem from reformulating the problem in the weak form, thus providing a framework that is more robust to high levels of noise. 

\subsection{Weak-form Estimation of Non-linear Dynamics}

The WENDy algorithm \citep{BortzMessengerDukic2023BullMathBiol} is a parameter estimation method in which data is substituted into the model equation and the norm of a (weak form) residual is minimized. Methods based on EE residuals were originally developed in the aerospace engineering literature \citep{Greenberg1951NACATN2340}. Despite their computational efficiency, widespread adoption was hindered by poor performance with noisy data. To address the noise issue, several weak form EE methods have been proposed over the years (dating back to the 1950s \citep{Shinbrot1954NACATN3288,Shinbrot1957TransAmSocMechEng} and 1960s \citep{LoebCahen1963Automatisme,LoebCahen1965IEEETransAutomControl}), but they never received as much attention as OE methods due to their sensitivity to the choice of test function. The WENDy algorithm maintains computational efficiency while simultaneously offering a mathematically guided strategy for test function selection.

To implement the weak-form approach,  the ODE system is first expressed in a matrix form
\begin{align} \label{matrix form}
    \dfrac{d\mathbf{u}}{dt} = \Theta(\mathbf{u})\mathbf{W},
\end{align}
where the row vector of $J$ features is $\Theta(\mathbf{u}) \equiv [f_1(\mathbf{u})\;\; f_2(\mathbf{u})\;\;\cdots\;\;f_J(\mathbf{u})], $ and the parameter matrix is defined to have the parameter of the $i^{th}$ ODE and the $j^{th}$ feature in the system: $ \left[\mathbf{W}\right]_{ji} \equiv w_{ji}$ where $1 \leq j \leq J$ and $1 \leq i \leq d.$ 

Next,  (\ref{matrix form}) is multiplied by a test function, $\phi (t) \in C^{\infty}_c [0, T]$, which is infinitely differentiable and compactly supported on the given interval such that, $\phi (0) = \phi (T) = 0$. Both sides are then integrated, and  integration by parts is used to reduce the order of the derivatives on $\mathbf{u}$, to obtain:
\begin{align}
    \phi(t)\mathbf{u}(t)\bigr\vert^T_0 - \int_0^T \dot{\phi}\mathbf{u}dt = \int_0^T \phi\Theta(\mathbf{u})\mathbf{W}dt \label{IBP}
    \\ 
    \Rightarrow  - \int_0^T \dot{\phi}\mathbf{u}dt = \int_0^T \phi\Theta(\mathbf{u})\mathbf{W}dt .
\end{align}

Suppose  data have been collected at equally spaced $M+1$ time-points, $\{t_m\}_{m=0}^{M}$, with $\Delta t = t_m - t_{m-1} $ for each of the $d$ states in the system represented by the matrix, $\mathbf{U} \in \mathbb{R}^{(M+1)\times d}$. The reduced order of derivative in the integral equation~\ref{IBP} allows for direct estimation of the integrals using data $\mathbf{U}$ and quadrature methods. This leaves 
\begin{align}
    -\dot{\Phi} \mathbf{Q}\mathbf{U} \approx \Phi \mathbf{Q}\Theta(\mathbf{U})\mathbf{W}
\end{align}
where $\Phi$ and $\dot{\Phi}$ are the evaluations of the test function and its derivative at time points $\{t_m\}$, and $\mathbf{Q} = diag(\frac{\Delta t}{2},\Delta t,\ldots, \Delta t, \frac{\Delta t}{2} )$ is the quadrature matrix for the trapezoidal rule. 

Suppose $K$ test functions, centered at distinct time-points in the interval $[0, T]$, have been used: 
\begin{align}
    \Phi = \begin{bmatrix}
    \phi_1(t_0) & \cdots & \phi_1(t_M) \\
    \phi_2(t_0) & \cdots & \phi_2(t_M) \\
    \vdots  & \ddots & \vdots \\
    \phi_K(t_0) & \cdots & \phi_K(t_M)
    \end{bmatrix},
   \nonumber
\end{align}
 such that the matrix, $ \mathbf{G} \equiv \Phi \mathbf{Q}\Theta(\mathbf{U}) \in \mathbb{R}^{K \times J}$, is full rank. Then the following weak-form linear least squares problem becomes
\begin{align} \label{LLS}
    \min \limits_{\mathbf{W}} \Vert \textsf{vec}(\mathbf{GW} - \mathbf{B})\Vert_2 ^2,
\end{align}
where $\textsf{vec}$ is the column-major vectorization of the matrices, and $\mathbf{B} \equiv -\dot{\Phi} \mathbf{Q}\mathbf{U} \in \mathbb{R}^{K\times d}$. This problem can now be solved with the normal equations,
\begin{align}
    \mathbf{W} = (\mathbf{G}^T\mathbf{G})^{-1}\mathbf{G}^T\mathbf{B}
\end{align}

The weak-form, therefore, allows one to eliminate the need to forward-solve the system. One can further improve this approach by realizing that the problem posed in (\ref{LLS}) is an Errors-In-Variables problem, since both $\mathbf{G}$ and $\mathbf{B}$ depend on $\mathbf{U}$, and therefore, contain errors. Thus, Iteratively Reweighted Least Squares (IRLS) is utilized, where the covariance matrix of the residuals, $\mathbf{C}^{(n)}$, is calculated at the $n$-th iteration, and the parameters are recalculated using
\begin{align}
    \mathbf{W}^{(n+1)} = (\mathbf{G}^T(\mathbf{C}^{(n)})^{-1}\mathbf{G})^{-1}\mathbf{G}^T(\mathbf{C}^{(n)})^{-1}\mathbf{B}
\end{align}
until some convergence criterion is reached. For more information regarding the details of calculating the covariance matrix, $\mathbf{C}^{(n)}$, and the specifics of the test-functions choice and their radii, see \citep{BortzMessengerDukic2023BullMathBiol}.

\subsection{Coverage}  

Assessing coverage is a crucial concept in statistical inference. Coverage is defined as the probability that a parameter's confidence interval contains the true parameter value. For normally distributed estimators, like WENDy, the true parameter, $w^*_i$, is covered by the 95\% confidence interval if,
\begin{align}
    w_i - 1.96 \cdot \widehat{\sigma}_{w_i} \leq w^*_i \leq w_i + 1.96 \cdot \widehat{\sigma}_{w_i},
\end{align}
where $1.96$ is the 97.25-th quantile of the standard normal distribution,  $w_i$ is the parameter estimate, and $\widehat{\sigma}_{w_i}$ is the estimated standard error of the parameter estimator. The key question of the coverage assessment is whether the actual coverage of WENDy confidence intervals equals their nominal coverage; in other words, whether the fraction of $C\%$ confidence intervals that contain the true parameter values is close to $C\%$, for every $C$. 
In other words, coverage assessment will help assess how often WENDy confidence intervals can capture true parameter values in comparison with the expected nominal coverage of those confidence intervals.

Confidence interval calculations require covariance and standard errors. Obtaining the parameter covariance matrix with OE-NLS techniques, however, is often difficult (see \citet{McGoffMukherjeePillai2015StatistSurv}). However, with WENDy, the covariance matrix of the parameters, $\mathbf{S}$,$\widehat{\sigma}_{w_i}$, can be directly estimated based on the covariance matrix of residuals, $\widehat{\mathbf{C}}$, as follows
\begin{align}
    \mathbf{S} \equiv \widehat{\sigma} ^2 ((\mathbf{G}^T\mathbf{G})^{-1} \mathbf{G}^T)\;\widehat{\mathbf{C}}\;(\mathbf{G}(\mathbf{G}^T\mathbf{G})^{-1})
    \\
    \Rightarrow
 (\widehat{\sigma}_{w_1}, \ldots, \widehat{\sigma}_{w_J}) \equiv \sqrt{diag(\mathbf{S})}, \;\;\;\;\;\;\;\; \nonumber
\end{align}
where, $\widehat{\sigma} ^2$ is the estimate for the measurement variance. The measurement variance is estimated by convolving each compartment of the data U with a high-order filter ${\bf{f}}$ and taking the Frobenius norm of the resulting convolved data matrix ${\bf{f}}*{\bf{U}}$ (see \citet{BortzMessengerDukic2023BullMathBiol} for more details). 

\subsection{Bias}  

One of the most commonly utilized forms of assessing estimator performance in statistical research is bias.  Bias is  defined as the expectation of the discrepancy between the true parameter, $\mathbf{w}^*$, and the parameter estimator,  $\mathbf{w}$:
\begin{align*}
    \text{Bias} = E({\mathbf{w}^* - \mathbf{w}}),
\end{align*}
Similarly, relative bias is defined as:
\begin{align*}
    \text{Relative Bias} = \dfrac{E(\mathbf{w}^* - \mathbf{w})}{\mathbf{w}^*},
\end{align*}
where the division is carried out element-wise. Quantifying the bias will provide a direct assessment of how accurate WENDy estimators are, as well as of how WENDy handles different parameters that modulate different order terms in a model.

\subsection{Outline}

The rest of this paper is organized as follows: Section 2 provides details about the ODE systems and data used, as well as the coverage and bias assessment methods under different normal and log-normal noise scenarios. Section 3 provides detailed assessment results for each of the systems. The paper ends with a discussion in Section 4, summarizing the main findings and future work.


\section{Methods}

\subsection{Benchmark Models}
This section presents the five benchmark models that will be used throughout the paper: logistic growth, Lotka-Volterra, FitzHugh-Nagumo, Hindmarsh-Rose, and Protein Transduction Benchmark (PTB). These models have also been studied in \citet{BortzMessengerDukic2023BullMathBiol}. This section also describes the methodological experiments conducted to assess the WENDy bias and coverage in each model.

\subsubsection{Logistic Growth Model}
The Logistic model is used to describe the growth of a state variable with a saturation level. The canonical Logistic equation is represented as 
\begin{equation} \label{1.1.1}
    \frac{dP}{dt} = kP(1 - \dfrac{P}{L}), 
\end{equation}
where $k$ is the growth rate of the state variable, $P$, and $L$ is the saturation level (also called the carrying capacity) imposed on the system by some spatial or resource constraint. Although this is a fairly simple model and has a closed-form solution, it is valuable as a proof of concept to test the coverage and bias of WENDy.  Note that the following equivalent version of the logistic growth model is used in order to accommodate the form required by the WENDy algorithm:
\begin{equation} \label{1.1.2}
    \frac{du}{dt} = w_1 u + w_2 u^2.
\end{equation}
Thus, the parameters to be estimated are $w_1$, $w_2$, and the state variable is $u$. 

\[\sigma_{NR} : = \frac{\mathbb{E}[(n-n^\star)^2]}{\|n^\star\|_{RMS}^2} = \mathbb{E}[(\varepsilon - 1)^2] = e^{\sigma^2}(e^{\sigma^2}-1) + (e^{\sigma^2/2}-1)^2.\]

\begin{figure} 
    \centering
    \includegraphics[width=0.9\linewidth]{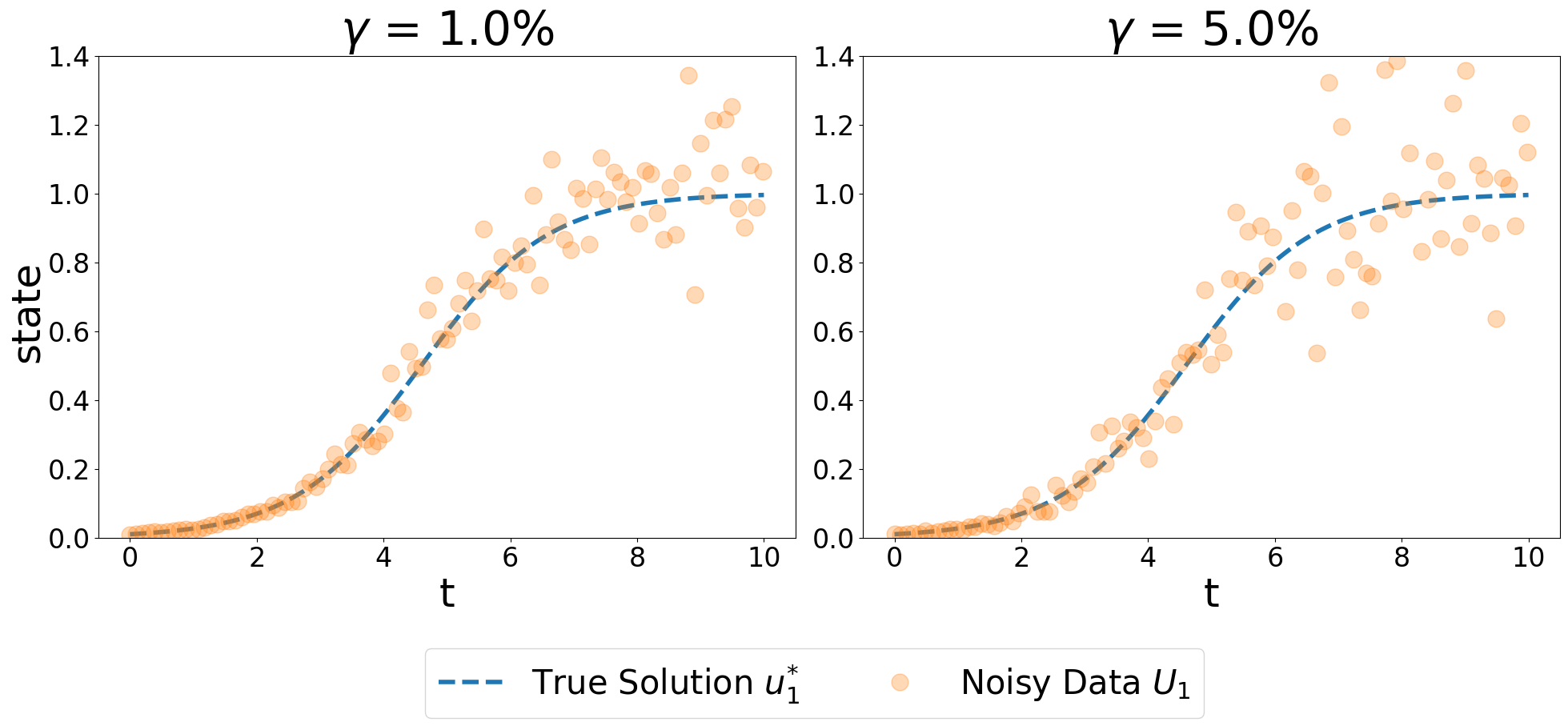}
    \caption{Logistic growth curve with 103 data points, and multiplicative log-normally distributed noise: left 1\% noise; right: 5\% noise. The following initial conditions and true parameter values were used: $u_0 = 0.01$; $\mathbf{w}^* = (1, -1)$}
    \label{fig:LogisticGrowth}
\end{figure}

To illustrate this system, Figure~\ref{fig:LogisticGrowth}
 displays a simulated logistic growth curve model with 103 data points and varying levels of multiplicative log-normal noise used to preserve non-negativity of the system states. As one can see, even at 5\%, the variability of the observations gets quite pronounced over time. Descriptions of the state and noise generation are provided in more detail in later sections.
 
\subsubsection{Lotka-Volterra Model}
The Lotka-Volterra model is a system of non-linear equations that models the dynamics of competing state variables. It is often used to model predator-prey populations in ecological applications \citep{Lotka1978TheGoldenAgeofTheoreticalEcology1923-1940}.

\begin{equation}
    \begin{cases}
    
    \dfrac{dx}{dt} = \alpha x - \beta xy 
    \\[10pt]
    \dfrac{dy}{dt} = -\gamma y + \delta xy
    \end{cases}
\end{equation}
Here, $x$ represents a prey, and $y$ represents the predator whose only food source is $x$. The parameters, $\alpha$ and $\gamma$ are the growth rates of $x$ and $y$ respectively; and $\beta$ and $\delta$ modulate the interactions between the two populations. The following equivalent model is used instead to accommodate the required form by WENDy: 
\begin{equation} 
    \begin{cases}
    
    \dfrac{du_1}{dt} = w_1 u_1 + w_2 u_1 u_2 
    \\[10pt]
    \dfrac{du_2}{dt} = w_3 u_2 + w_4 u_1 u_2
    \end{cases}
\end{equation}
The parameters to be estimated are $w_1, w_2, w_3, w_4$ and the states are $u_1, u_2$. 

\begin{figure} 
    \centering
    \includegraphics[width=0.9\linewidth]{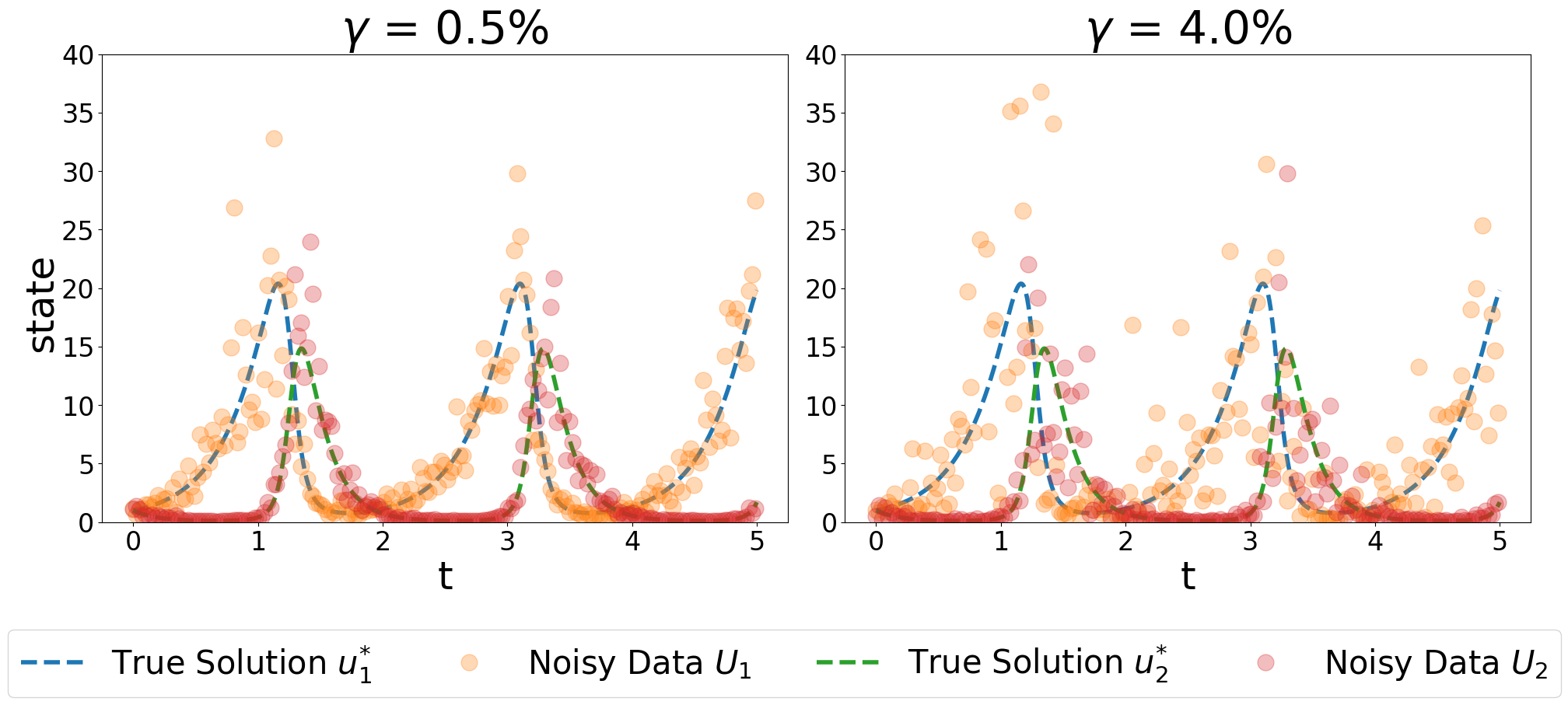}
    \caption{Lotka-Volterra model with varying levels of multiplicative log-normal noise. The following initial conditions and parameter values where used: $\mathbf{u}_0 = (1, 1)$; $\mathbf{w}^* = (3, -1, -6, 1)$}
    \label{fig:LV}
\end{figure}

To illustrate this system, Figure~\ref{fig:LV}
 displays a simulated Lotka-Volterra  model with 205 data points and varying levels of multiplicative log-normal noise used to preserve non-negativity of the system states. 

\subsubsection{FitzHugh-Nagumo Model}
The FitzHugh-Nagumo (FHN) model simulates an excitable system like a neuron with a non-linear system of ODEs \citep{FitzHugh1961BiophysicalJournal}. The FHN model is represented canonically with
\begin{equation} \label{1.1.3}
    \begin{cases}
    \dfrac{dv}{dt} =  v - v^3 -  w 
    \\[10pt]
    \dfrac{dw}{dt} = \frac{1}{\epsilon}(v - bw + a)
    \end{cases}
\end{equation}
where, when modeling action potentials, $v$ corresponds to the membrane potential while $w$ corresponds to the recovery variable that acts as a blocking mechanism for the membrane potential. The parameter, $\epsilon$, is the time-scaling factor for the blocking mechanism, the parameter $b$ is the strength of the blocking mechanism, and the parameter $a$ is the threshold voltage for the neuron spiking. The following equivalent equation was used to accommodate the  required form for WENDy:
\begin{equation}
    \begin{cases}
    \dfrac{du_1}{dt} =  w_1 u_1 + w_2 u_1^3 +  w_3 u_2
    \\[10pt]
    \dfrac{du_2}{dt} = w_4 u_1 + w_5(1) + w_6 u_2
    \end{cases}
\end{equation}
where the parameters to be estimated are $w_1, w_2, w_3, w_4, w_5, w_6$ and the states are $u_1$ and $u_2$. 

\begin{figure} 
    \centering
    \includegraphics[width=0.9\linewidth]{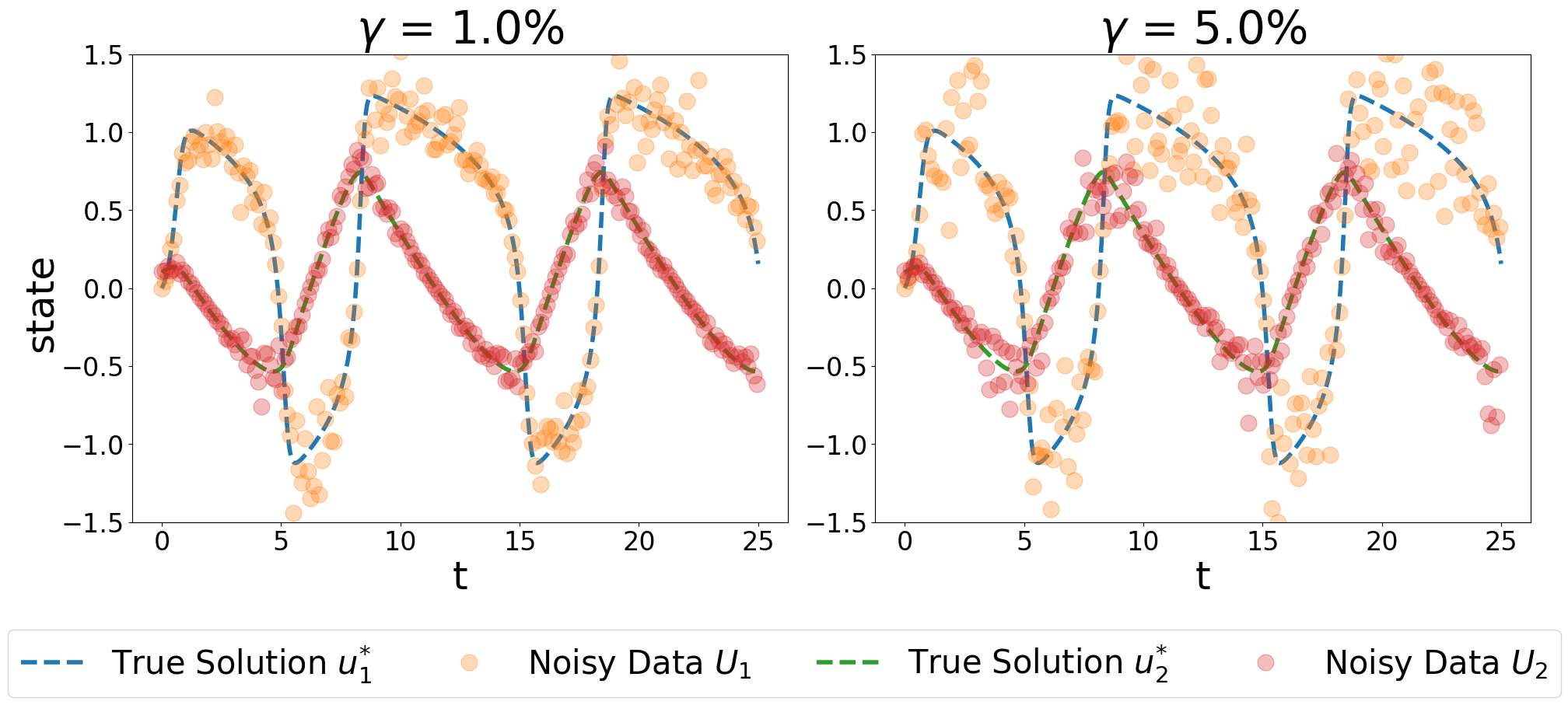}
    \caption{FitzHugh-Nagumo model with varying levels of log-normal noise. The following initial conditions and parameter values where used: $\mathbf{u}_0 = (0, 0.1)$; $\mathbf{w}^* = (3, -3, 3, -1/3, 17/150, 1/15)$}
    \label{fig:FHN}
\end{figure}

To illustrate this system, Figure~\ref{fig:FHN}
 displays a simulated FHN model with 205 data points and varying levels of additive normal noise.  Note that the states can be negative in this system, and thus, additive normally distributed noise can be appropriate. 

\subsubsection{Hindmarsh-Rose Model}
The Hindmarsh-Rose (HMR) model is used to model the spiking and bursting activity observed experimentally in neurons \citep{HindmarshRose1984ProcRSocLondBBiolSci}. The canonical model is a system of 3 ODEs,
\begin{align}
    \begin{cases}
        \dfrac{dx}{dt} =  y - a x^3 +  b x^2 - z
    \\[10pt]
        \dfrac{dy}{dt} = c - d x^2 - y
    \\[10pt]
        \dfrac{dz}{dt} = r(s(x-x_R)-z)
    \end{cases}
\end{align}
where, similar to the FHN model, when modeling neural activity, $x$ represents the membrane potential, $y$ represents the transport of sodium and potassium ions through the fast ion channels, and $z$ represents the adaptation current, which allows for the bursting behavior of the model. The parameters $a$ and $b$ affect the sharpness of the spiking and threshold potential for spiking, respectively; $c$ and $d$ both control the strength of the blocking mechanism of membrane potential; and $r$ is the time-scaling factor for $z$, and $s$ is the strength of the adaptation. Lastly, $x_R$ represents the threshold potential for bursting. The following equivalent form was used in the WENDy trials:
\begin{align}
    \begin{cases}
        \dfrac{du_1}{dt} =  w_1 u_2 + w_2  u_1^3 +  w_3 u_1^2 + w_4 u_3
    \\[10pt]
        \dfrac{du_2}{dt} = w_5 (1) + w_6 u_1^2 + w_7 u_2
    \\[10pt]
        \dfrac{du_3}{dt} = w_8 u_1 + w_9 (1) + w_{10} u_3
    \end{cases}
\end{align}
where there are 10 parameters to be estimated and $u_1, u_2$, and $u_3$ are states. 

\begin{figure} 
    \centering
    \includegraphics[width=0.9\linewidth]{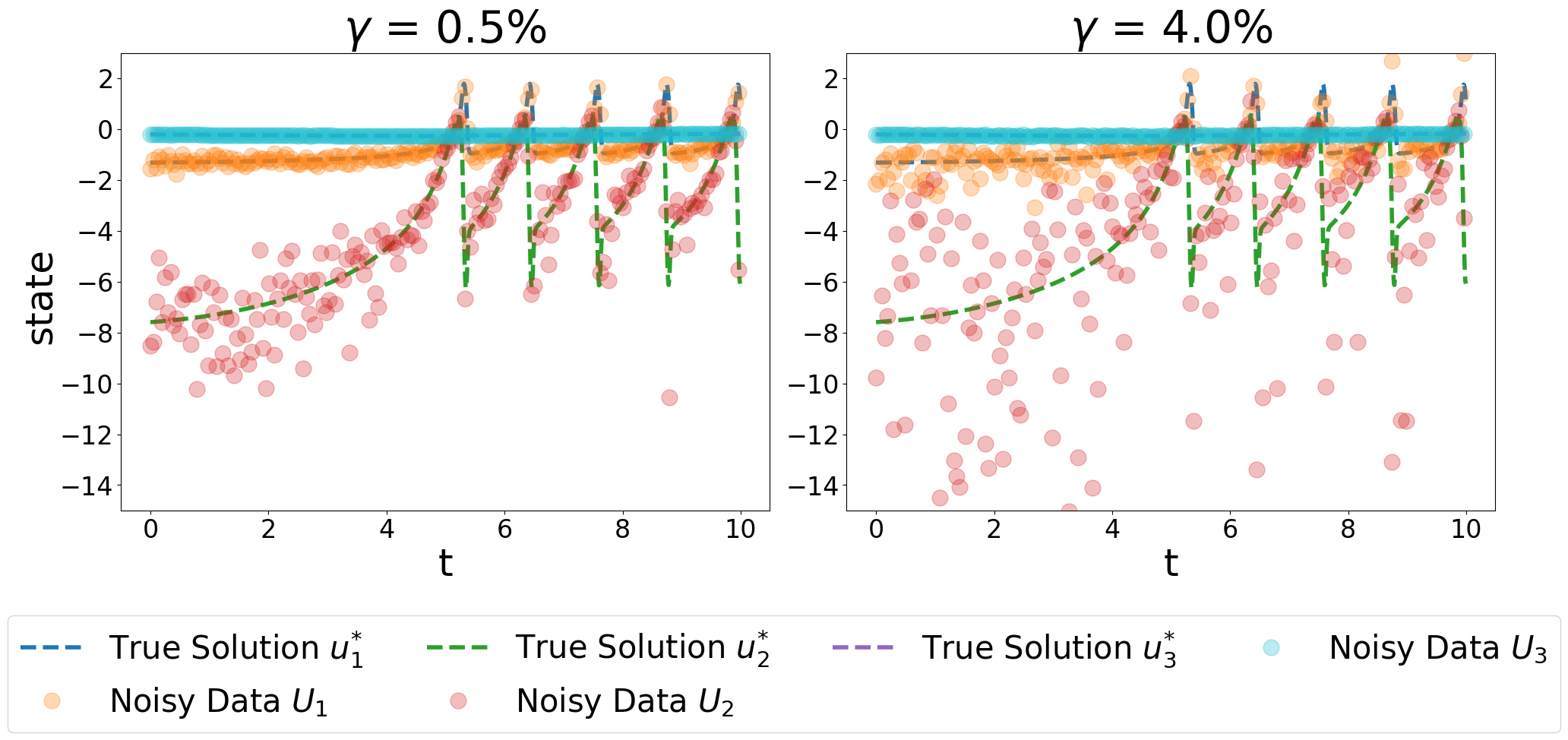}
    \caption{Hindmarsh-Rose model with varying levels of log-normal noise. The following initial conditions and parameter values were used: $\mathbf{u}_0 = (-1.31, -7.6, -0.2)$; $\mathbf{w}^* = (10, -10, 30, -10, 10, -50, -10, 0.04, 0.0319, -0.01)$}
    \label{fig:HMR}
\end{figure}

To illustrate this system, Figure~\ref{fig:HMR}
 displays a simulated HMR model with 205 data points and varying levels of additive normal noise. Note that the states can be negative in this system, and thus, additive normally distributed noise can be appropriate. 

\subsubsection{Protein Transduction Benchmark Model} 
The Protein Transduction Benchmark (PTB) Model is a five-compartment system used to model protein signaling transduction pathways in a signal transduction cascade utilized in \citet{SchoeberlEichler-JonssonGillesEtAl2002NatBiotechnol}.

The model is canonically represented as
\begin{align*}
    \begin{cases}
        \dfrac{dS}{dt} =  -k_1 S -k_2 S\cdot R + k_3C_{RS}
    \\[10pt]
        \dfrac{d\Tilde{S}}{dt} = k_1 S
    \\[10pt]
        \dfrac{dR}{dt} =  -k_2 S\cdot R + k_3C_{RS} + \dfrac{VR_{pp}}{K_m + R_{pp}}
     \\[10pt]
     \dfrac{dC_{RS}}{dt} = k_2 S\cdot R - k_3C_{RS} - k_4 C_{RS}
    \\[10pt]
     \dfrac{dR_{pp}}{dt} = k_4 C_{RS} - \dfrac{VR_{pp}}{K_m + R_{pp}}
    \end{cases}
\end{align*}
where $R$ is a protein, undergoing phosphorylation into $R_{pp}$, catalyzed by an enzyme, $S$. The state, $C_{RS}$, represents the activated enzyme-protein complex, and $\Tilde{S}$ represents the degraded enzyme $S$. The parameters $k_1, k_2, k_3$, and $k_4$ are proportionality constants; $V$ is the limiting rate approached by the system when the protein $R$ reaches its saturation concentration for the enzyme $S$; and $K_m$ is the Michaelis constant, which is half the protein's saturation concentration. 

The following equivalent form was used for the WENDy trials
\begin{align}
    \begin{cases}
        \dfrac{du_1}{dt} =  w_1u_1 +w_2u_1u_3 + w_3u_4
    \\[10pt]
        \dfrac{du_2}{dt} = w_4 u_1
    \\[10pt]
        \dfrac{du_3}{dt} =  w_5 u_1 u_3 + w_6 u_4 + w_7\dfrac{u_5}{0.3 + u_5}
     \\[10pt]
     \dfrac{du_4}{dt} = w_8u_1u_3 + w_9 u_4
    \\[10pt]
     \dfrac{du_5}{dt} = w_{10}u_4 + w_{11}\dfrac{u_5}{0.3 + u_5}
    \end{cases}
\end{align}
where there are 11 parameters to be estimated and $u_1$ through $u_5$ are states. Note that terms with the same features but different constants were combined into one term (see fourth ODE in system). 

\begin{figure} 
    \centering
    \includegraphics[width=0.9\linewidth]{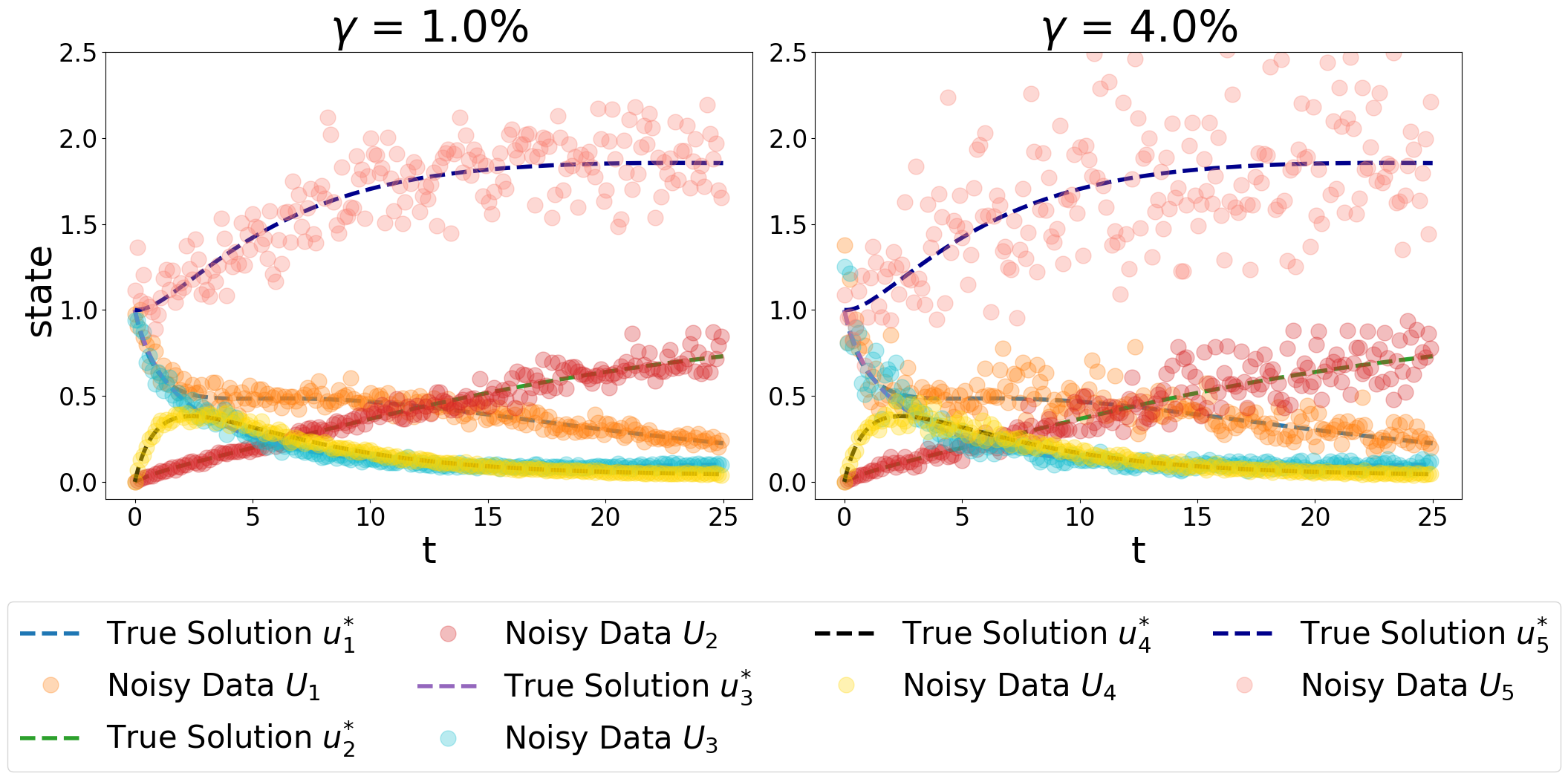}
    \caption{Protein Transduction Benchmark Model with varying levels of log-normal noise. The following initial conditions and parameter values were used: $\mathbf{u}_0 = (1, 0, 1, 0, 1)$; $\mathbf{w}^* = (-0.07, -0.6, 0.35, 0.07, -0.6, 0.05, 0.17, 0.6, -0.35, 0.3, -0.017)$}
    \label{fig:PTB}
\end{figure}

To illustrate this system, Figure~\ref{fig:PTB}
 displays a simulated PTB model with 205 data points and varying levels of additive normal noise. Note that the states can be negative in this system, and thus, the additive normally distributed noise is appropriate.

\subsection{Data Generation} 
To evaluate the coverage properties and bias of WENDy estimators,  the algorithm was applied to synthetically generated data designed to mimic the real data closely. This involved varying the distribution of the noise, noise levels, and temporal resolution (i.e., number of data points) during the data generation process.

Starting from the true dynamics, noise was added using one of four distributions: additive normal, multiplicative log-normal (MLN), additive truncated normal (ATN), or additive censored normal (ACN). Additive normal noise was utilized for all models, but was used more so as a benchmark for Logistic, LV, and PTB models since negative state values are not physically meaningful for such models. For the FHN and HMR models, only MLN and additive Normal noise were utilized since negative state values are physically meaningful. 

The data-generating process was thus carried out as follows. For every time point $j = 1, ..., M+1$ and every state variable  $u_{ij}$ (where $i=1,...,d$), the data are generated:
\begin{itemize}
    \item Additive normal normal noise:  
    $${u}_{ij} = {u}_{ij}^* + {z}_{ij}$$ where ${z}_{ij} \sim N(0,\sigma_i)$ and the standard deviation 
    $\sigma_i \equiv (\max \mathbf{U}_i - \min \mathbf{U}_i) \cdot \gamma$
    \item Additive truncated normal noise:  
    $${u}_{ij} = {u}_{ij}^* + {z}_{ij}$$ where ${z}_{ij} \sim TN(0,\sigma_i)$ and the standard deviation 
    $\sigma_i \equiv (\max \mathbf{U}_i - \min \mathbf{U}_i) \cdot \gamma$
    \item Multiplicative log-normal noise: 
    $${u}_{ij} = {u}_{ij}^* {z}_{ij}$$ where $\log(z_{ij}) \sim N(0,\sigma_i)$ and the standard deviation, $\sigma_i$, is obtained by solving the equation: $
    \exp(\sigma_i^2 - 1)*\exp(\sigma_i^2)= (\max\mathbf{U}_i - \min\mathbf{U}_i) \cdot \gamma$
\end{itemize}
\noindent Note that $\sigma_i$ is used to generically denote the standard deviation of the respective distribution for the $i$th state (or the logarithm of the state) in the model, and $\gamma \in (0, 1]$ is a percent of the range of that state. Increasing values of $\gamma$ were used to make datasets with increasing levels of noise. In the remainder of the paper, the noise level is referred to by its $\gamma$ value; for instance, a $10\%$ noise level corresponds to a standard deviation with $\gamma = 0.1$. Note that this interpretation of noise level is different from the one used in \citep{BortzMessengerDukic2023BullMathBiol}, where they utilized a noise ratio which was approximated by:
\begin{align}
    \sigma_{NR} \approx \frac{\lVert \mathbf{u}^* - \mathbf{U} \rVert_{\text{rms}}}{\lVert \mathbf{U} \rVert_{\text{rms}}}
\end{align}
\begin{algorithm}
\caption{Data Generation}
\label{alg:noise_gen}
\SetKwInOut{Input}{Input} 
\SetKwInOut{Output}{Output} 
\Input{True states $\{\mathbf{u}^*\}$, standard deviation $\{\boldsymbol{\sigma}\}$, noise algorithm (0 = additive normal, 1 = ACN, 2 = ACN, and 3 = MLN)}
\Output{Noisy data $\{\mathbf{U}\}$} 
\BlankLine
\If{noise algorithm $=$  3}{
    \tcp{Multiplicative log-normal noise}
    $\mathbf{N} \leftarrow$ GenerateNormalNoise(mean $= 0$, stdDev $= \boldsymbol{\sigma}$, dimension $=$ $\dim\mathbf{u}^*)$\\
    $\mathbf{U} \gets \exp(\mathbf{N})\otimes \mathbf{u}^*$ \tcp{element-wise}
    \Return $\mathbf{U}$}
\Else{
    \If{noise algorithm $<$2}{
    \tcp{Additive normal noise}    
    $\mathbf{N} \leftarrow$ GenerateNormalNoise(mean $= 0$, stdDev $= \boldsymbol{\sigma}$, dimension $=$ $\dim\mathbf{u}^*)$\\
    $\mathbf{U} \gets \mathbf{N} + \mathbf{u}^*$\\
    \If{noise algorithm == 1}{
    \tcp{Additive censored normal noise}    
    $\mathbf{U} \gets \max(0, \mathbf{U})$ \tcp{element-wise}}
    \Return $\mathbf{U}$}
    \Else{
    \tcp{Additive truncated normal noise}    
    $\mathbf{N} \leftarrow$ GenerateTruncatedNormalNoise(mean $= 0$, stdDev $= \boldsymbol{\sigma}$, lb = $-\mathbf{u}^*$, dimension $=$ $\dim\mathbf{u}^*)$\\
    $\mathbf{U} \gets \mathbf{N} + \mathbf{u}^*$  \\
     \Return $\mathbf{U}$}}
\end{algorithm}
\subsection{Coverage and Bias}
Coverage and bias of parameter estimators were assessed across datasets with varying noise levels and resolutions. For each noise or resolution level, 1,000 datasets were generated and fed into the WENDy algorithm. 

For each dataset, the WENDy estimates of the parameters and the corresponding standard errors were obtained from the final iteration of the IRLS algorithm and used to construct 95\% confidence intervals. A parameter was considered successfully covered if its true value was included within the corresponding confidence interval. Coverage was then taken as the proportion of confidence intervals that included the true parameter values across the 1,000 datasets. 

For the bias calculation, bias for each parameter estimator was computed as the difference between the mean of 1000 estimates and the true parameter value for that parameter. 

While varying noise, noise levels were increased until either the coverage of one of the parameters dropped below 50\% or  a noise level of 90\% was reached, which was the maximum allowed. The 50\% coverage threshold is used as a way to decipher the point at which WENDy estimators essentially become unreliable for inference, indicating that the uncertainty in parameter estimates has grown large enough to compromise their practical interpretability.
While varying data resolution,  datasets at multiple resolution levels were simulated to assess how WENDy estimators responded to changes in the number and density of available data points. Lastly, 100 sample WENDy-estimated solutions curves were found and plotted using parametric bootstrap from the asymptotic distribution of the WENDy estimator, as in \citep{MessengerDwyerDukic2024JRSocInterface}.          

\begin{algorithm} 
\SetKwInOut{Input}{Input}
\SetKwInOut{Output}{Output}

\Input{True parameter values $\{\mathbf{w}^*\}$, estimated values $\{\mathbf{w}\}$, and standard errors $\{\boldsymbol{\widehat{\sigma}}_{\mathbf{w}}\}$}
\Output{Array $C$ where $C_i = 0$ if $w^*_i$ is outside the confidence interval, and $C_i = 1$ if $w^*_i$ is within it}

Initialize $C$ as an array of zeros with the same length as $\mathbf{w}^*$

\For{$i = 1$ \KwTo length$(C)$}{
    \If{$w_i - 1.96 \cdot \widehat{\sigma}_{w_i} \leq w^*_i \leq w_i + 1.96 \cdot \widehat{\sigma}_{w_i}$}{
        $C_i \gets 1$
    }
}
\Return $C$

\caption{Calculate Coverage}
\end{algorithm}

\section{Results}

This section discusses the coverage and bias of WENDy estimators with perturbations in the distribution and level of noise in the data for the five benchmark models. WENDy parameter estimators have been shown to be asymptotically normally distributed and unbiased under additive Gaussian noise. The work below investigates how deviations from this assumption of additive Normal noise impact the parameter and state estimator performance, specifically in terms of their bias and variance. Also investigated are the coverage properties of the parameter confidence intervals.

\subsection{Logistic Growth Curve Model}

\subsubsection{Varying Noise Type and Level}

\subsubsection*{Additive Normal Noise}
\begin{figure} 
    \centering
    \begin{tabular}{c}
{\includegraphics[width=0.8\linewidth]{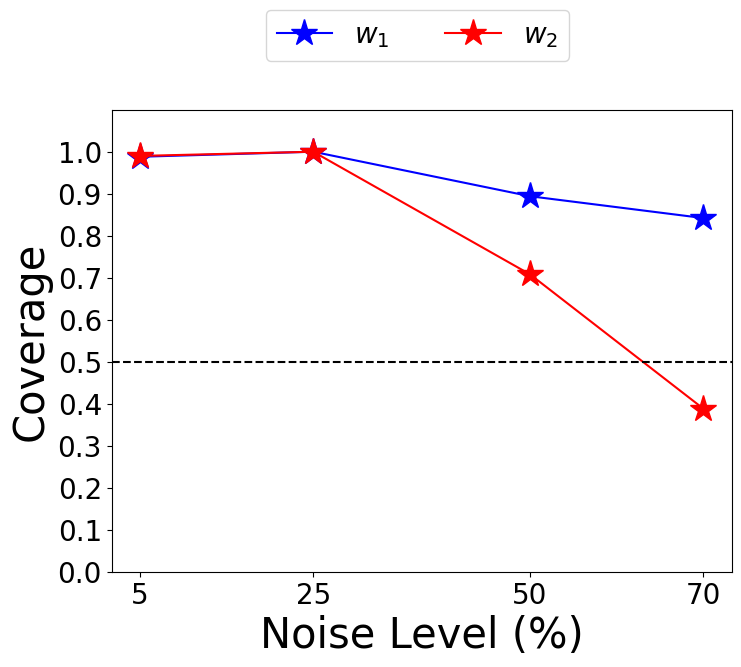}}
    \\
     \text{(a)}
     \\
     \includegraphics[width=1\linewidth]{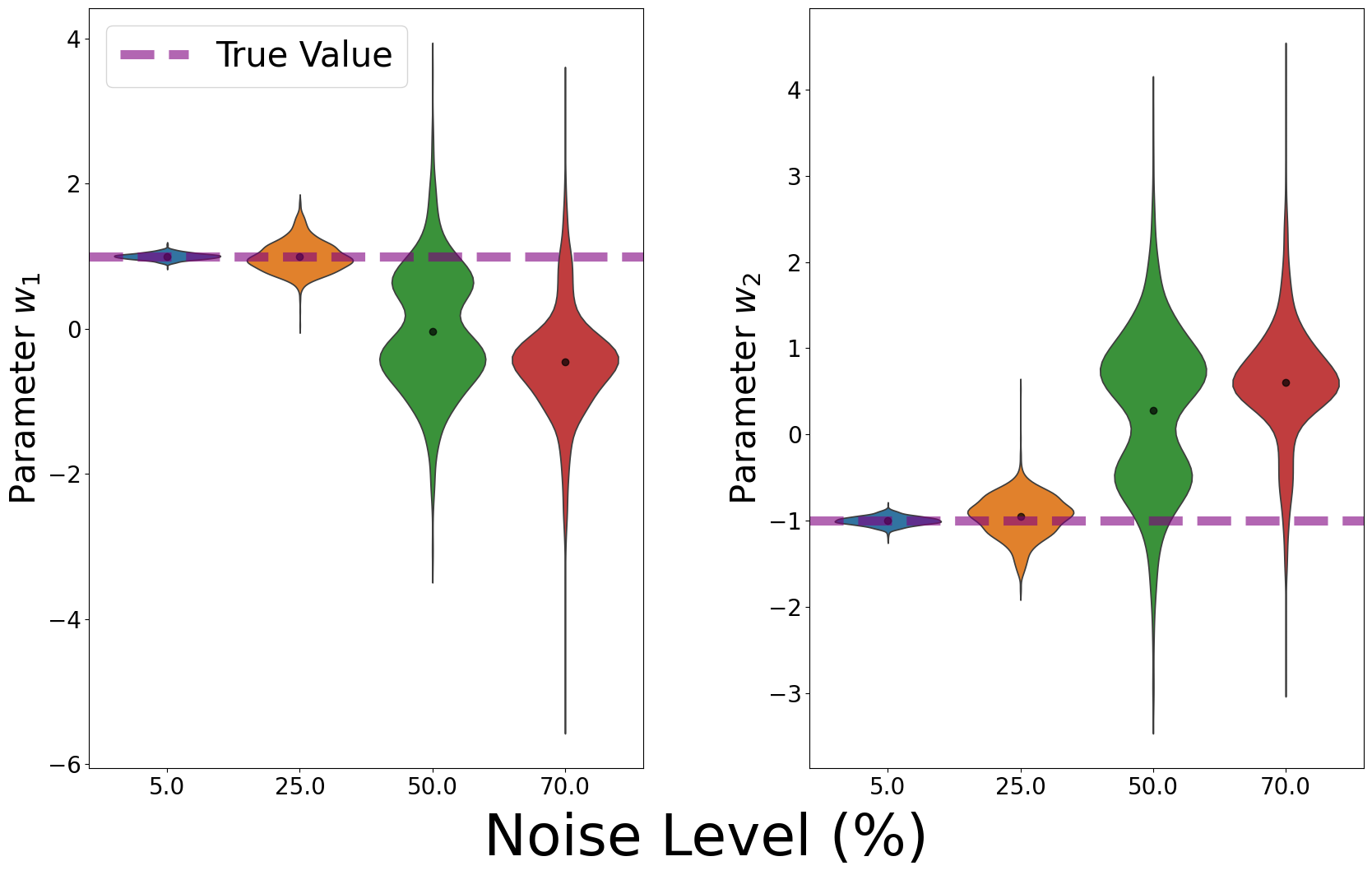}
    \\
      \text{(b)}
    \end{tabular}
     \caption{Logistic model parameter estimation performance with increasing normal noise (1000 datasets per level, 103 data points each). (a) coverage across four noise levels. (b) violin plots of parameter estimates, with the dashed red line indicating the true parameter values.}
    \label{fig:CovBiasLogisticN}
\end{figure}

As shown in Figure~\ref{fig:CovBiasLogisticN}(a), the coverage of the 95\% confidence intervals for both parameters remained above nominal for 5\% and 25\% noise. As noise increased to 70\%, the coverage of the $w_2$ parameter decreased to below 50\%. The $w_1$ parameter decreased slightly as noise increased to 70\% but remained above 80\%. The worse coverage of the $w_2$ parameter compared to the $w_1$ parameter is seen consistently with the Logistic model with all types of noise. This may be caused by the fact that the $w_2$ parameter modulates the quadratic term, whereas the $w_1$ parameter modulates the linear term; hence, the identifiability is more difficult for the $w_2$ parameter due to the nonlinearity.

\begin{figure} 
    \centering
    \begin{tabular}{c}
    
    \includegraphics[width=1\linewidth]{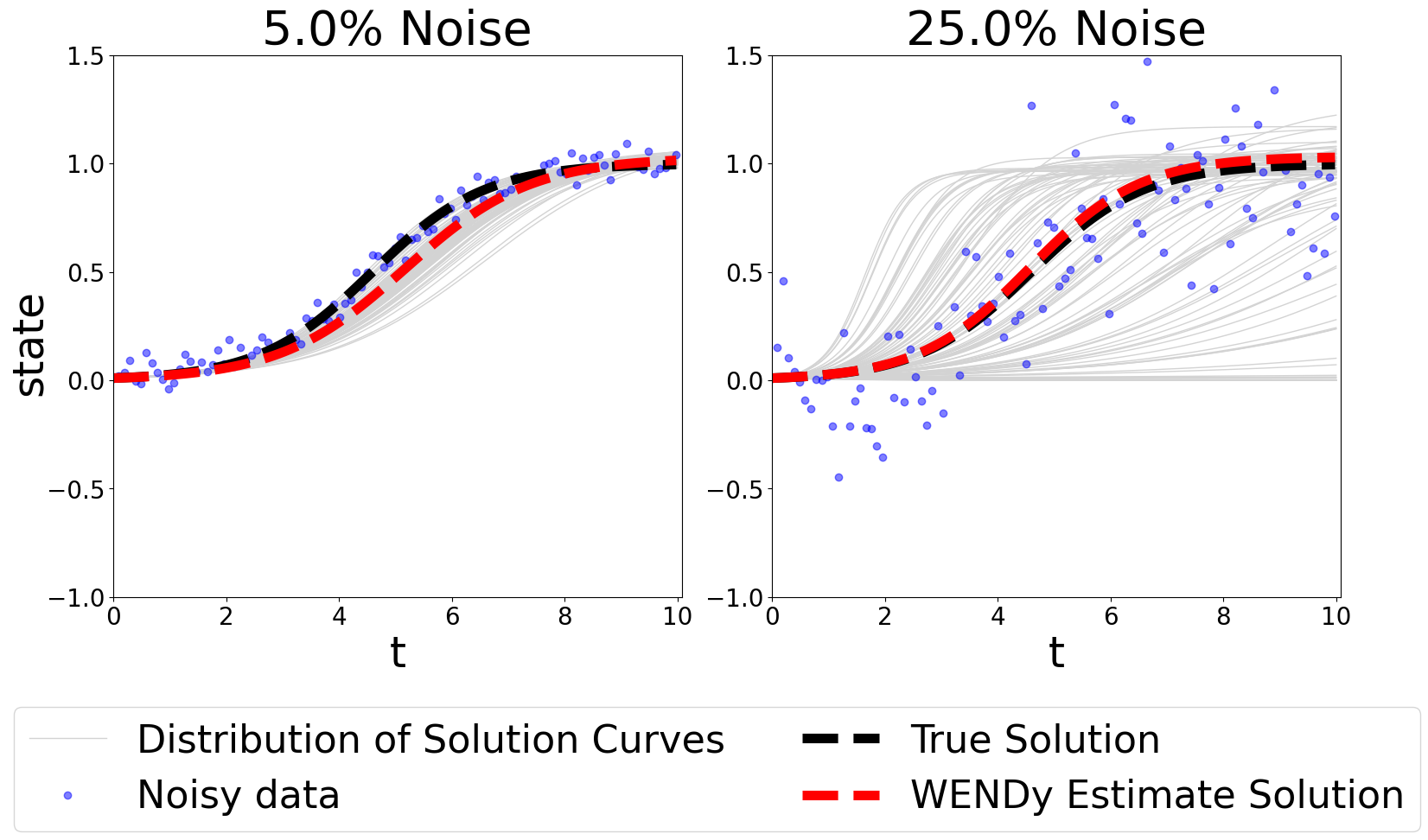} 
    \\
    \text{(a)}
    \\
    \includegraphics[width=1\linewidth]{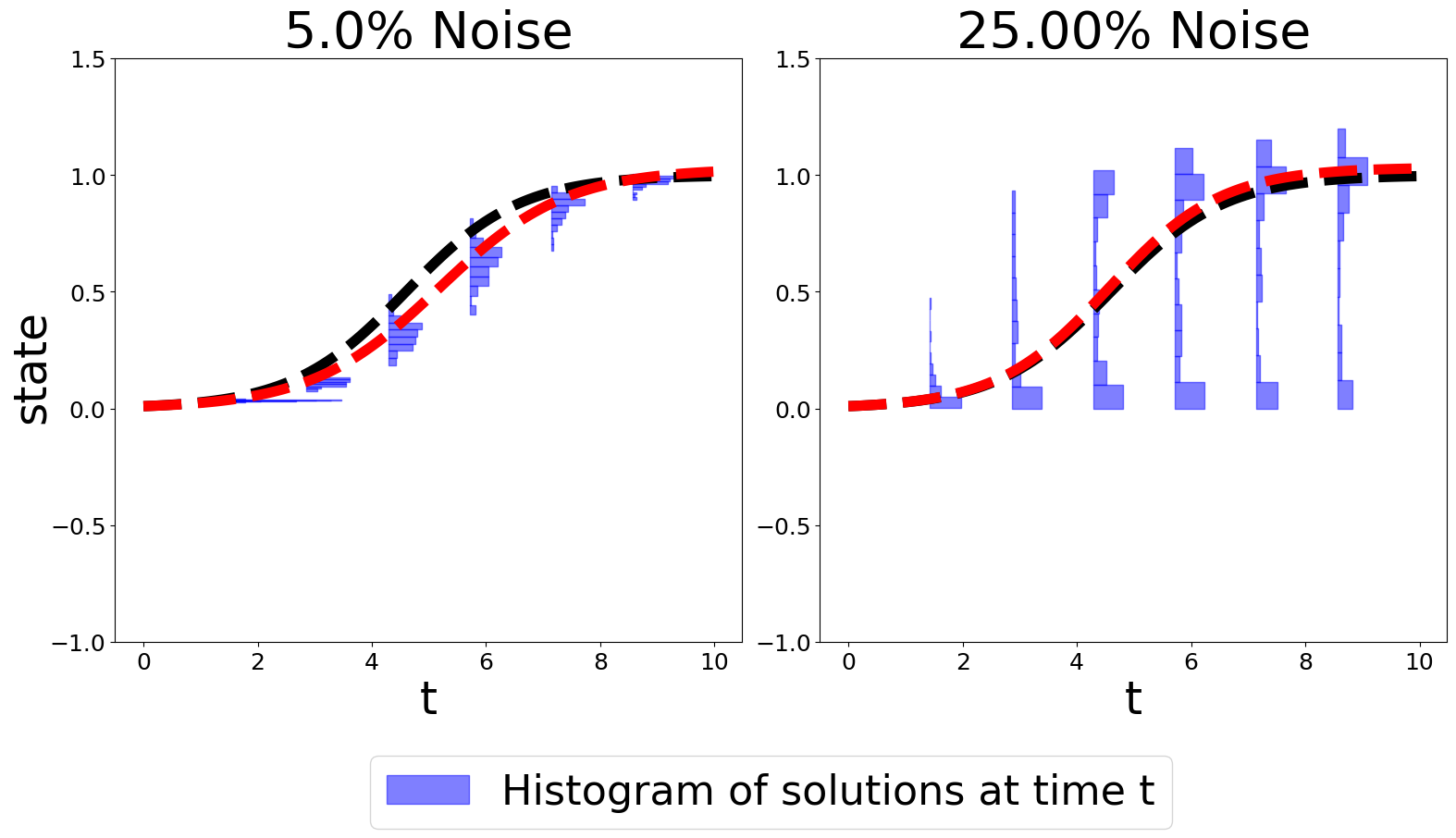}
    \\
    \text{(b)}
    \end{tabular}
     \caption{Top row (a): Logistic model parameter estimation example and uncertainty quantification on two datasets: one dataset with low uncertainty and high coverage (left) and one dataset with high uncertainty and low coverage (right). The light gray curves are used to illustrate the uncertainty around the WENDy solutions; they are obtained via parametric bootstrap, as a sample of WENDy solutions corresponding to a random sample of 1000 parameters from their estimated asymptotic estimator distribution.  Bottom row (b): WENDy solution and histograms of state distributions across specific points in time for the datasets in (a).}
    \label{fig:SamplePlotLogisticN}
\end{figure}

 As seen in Figure~\ref{fig:CovBiasLogisticN}(b), the bias in the parameters increased. The violin plot at 50\% noise showed some bimodality for both parameters. 
As seen in Figure~\ref{fig:SamplePlotLogisticN}(a) and (b), the distribution of solution states (at selected time points) is skewed for but unimodal for lower noise levels and bimodal for higher noise levels. This is likely caused by the states being too low or negative during the early time-points leading to a smaller growth rate, since the growth rate heavily depends on the initial points in Logistic growth.

\subsubsection*{Additive Censored Normal Noise}

\begin{figure} 
    \centering
    \begin{tabular}{c}
{\includegraphics[width=0.8\linewidth]{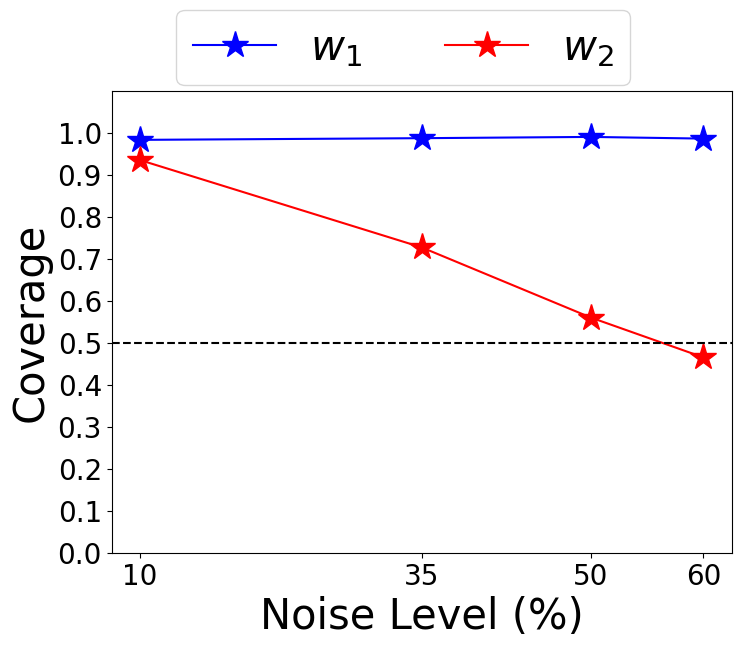}}
    \\
    \text{(a)}
    \\
    \includegraphics[width=1\linewidth]{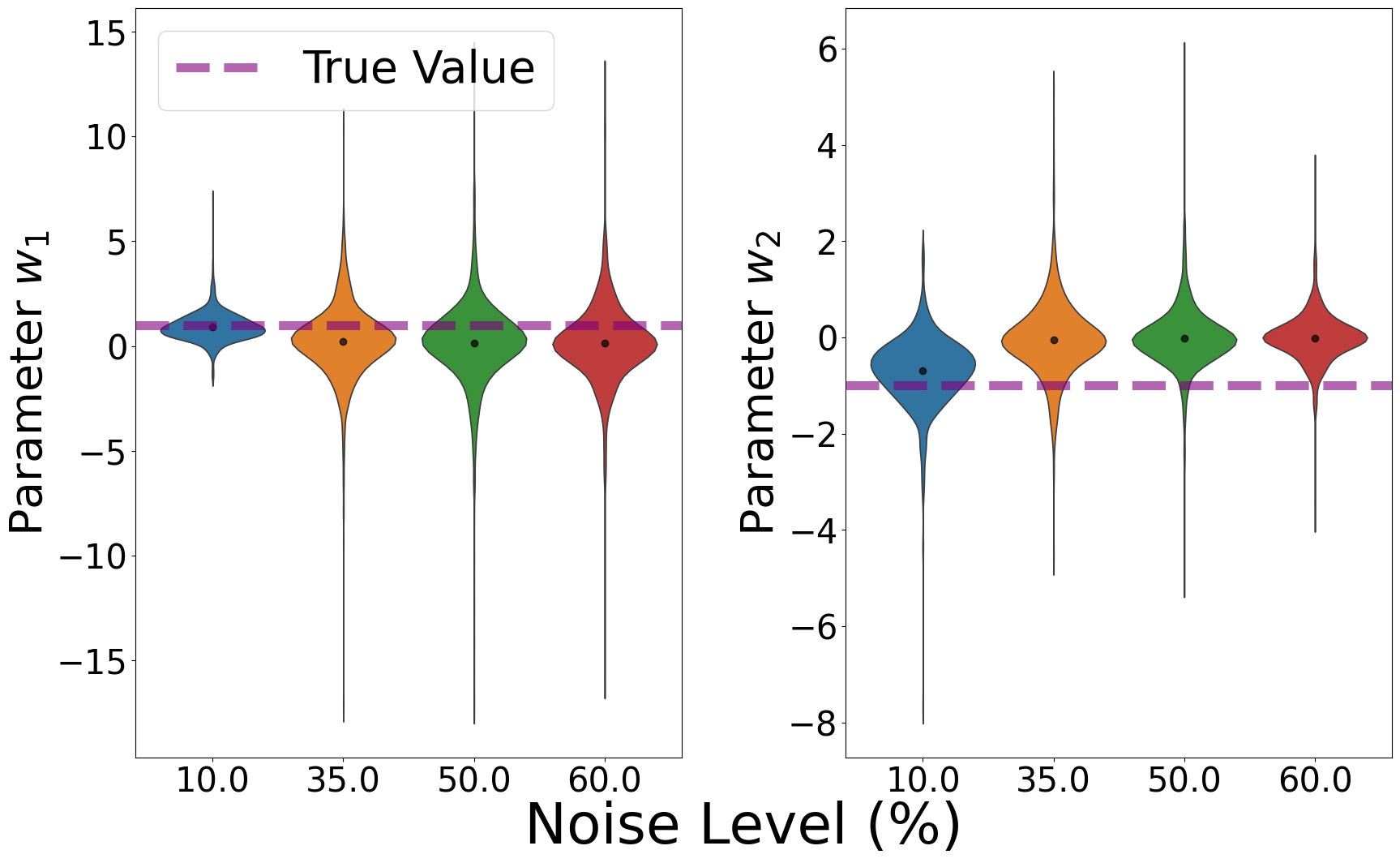}
    \\
      \text{(b)} \;\;\;\;\;\;\;\;\;\;\;\;\;\;\;\;\;\;\;\;\;\;\;\;\;\;\;\;\;\;\;
    \end{tabular}
     \caption{Logistic model parameter estimation performance with increasing ACN noise (1000 datasets per level, 103 data points each). (a) coverage across four noise levels. (b) violin plots of parameter estimates, with the dashed red line indicating the true parameter values.}
    \label{fig:CovBiasLogisticCN}
\end{figure}

As shown in Figure~\ref{fig:CovBiasLogisticCN}(a), the coverage of the 95\% confidence intervals for the $w_1$ parameter remained slightly above nominal 95\% level for all noise levels from 10\% to 60\%. However, the coverage for the $w_2$ parameter kept decreasing with increasing levels of ACN noise, from the nominal 95\% coverage at 10\% noise, to coverage below 50\%  at  60\% noise.
As WENDy parameter estimators have been shown to be asymptotically Normally distributed under additive Gaussian noise \citep{BortzMessengerDukic2023BullMathBiol}, the coverage performance under  ACN noise is thus comparable to the coverage performance under additive Normal noise for the $w_1$ parameter. However, the coverage becomes increasingly inferior to additive Normal noise coverage for the $w_2$ parameter as the noise level increases. 


WENDy parameter estimators have also been shown to be asymptotically unbiased \citep{BortzMessengerDukic2023BullMathBiol} under additive Gaussian noise. However, as Figure~\ref{fig:CovBiasLogisticCN}(b) shows,  the bias for both parameters under  ACN noise increased slightly as the level of noise increased, but plateaued at the higher noise levels from 30\% through 60\%. This is expected, given that the censored normal density errors only differ substantially from the normal density errors  when the states are low in magnitude, which happens only in the early part of the logistic growth curve.

\begin{figure}
\clearpage
\centering
    \begin{tabular}{c}
    
    \includegraphics[width=1\linewidth]{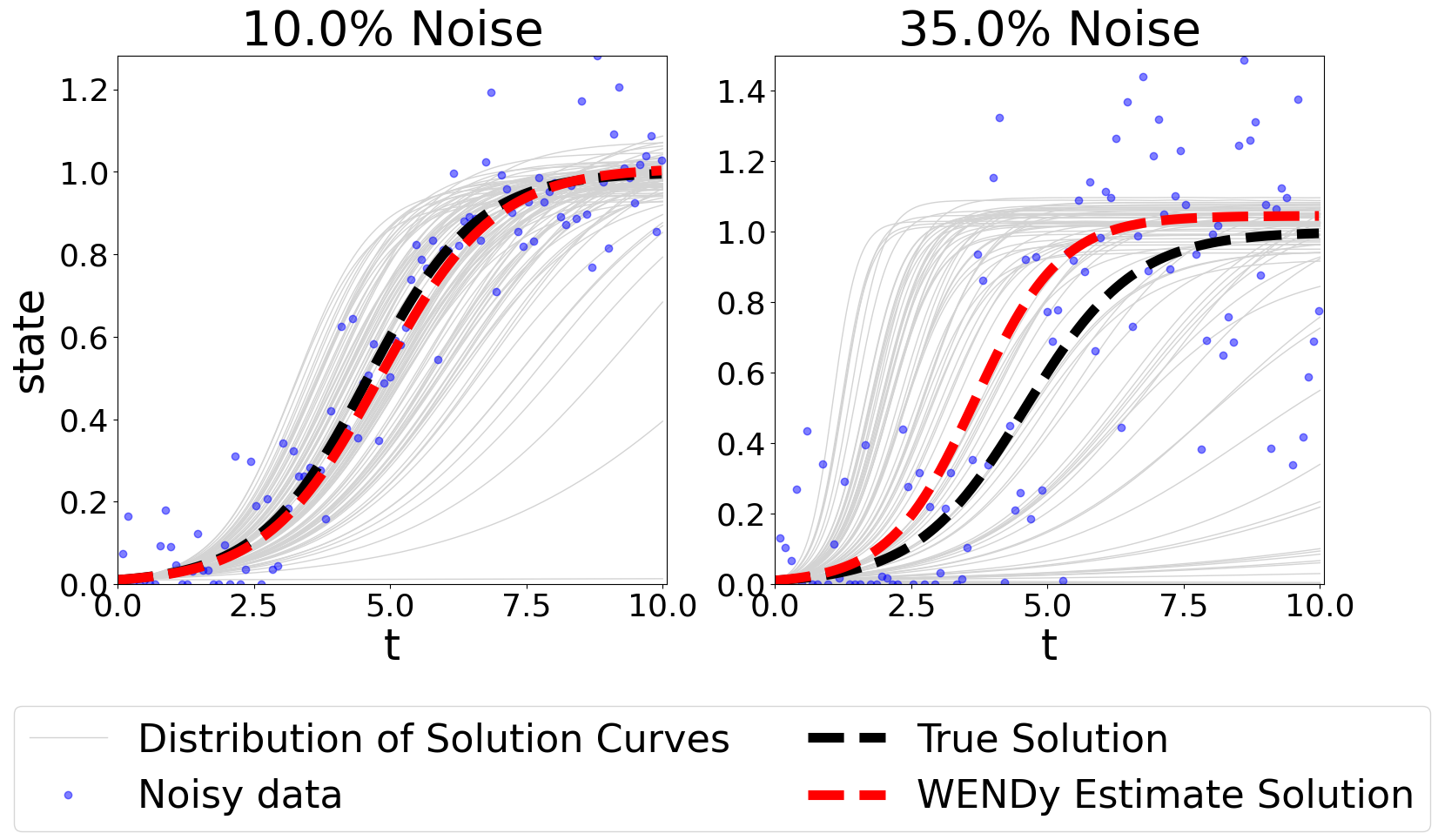} 
    \\
    \text{(a)}
    \\
    \includegraphics[width=1\linewidth]{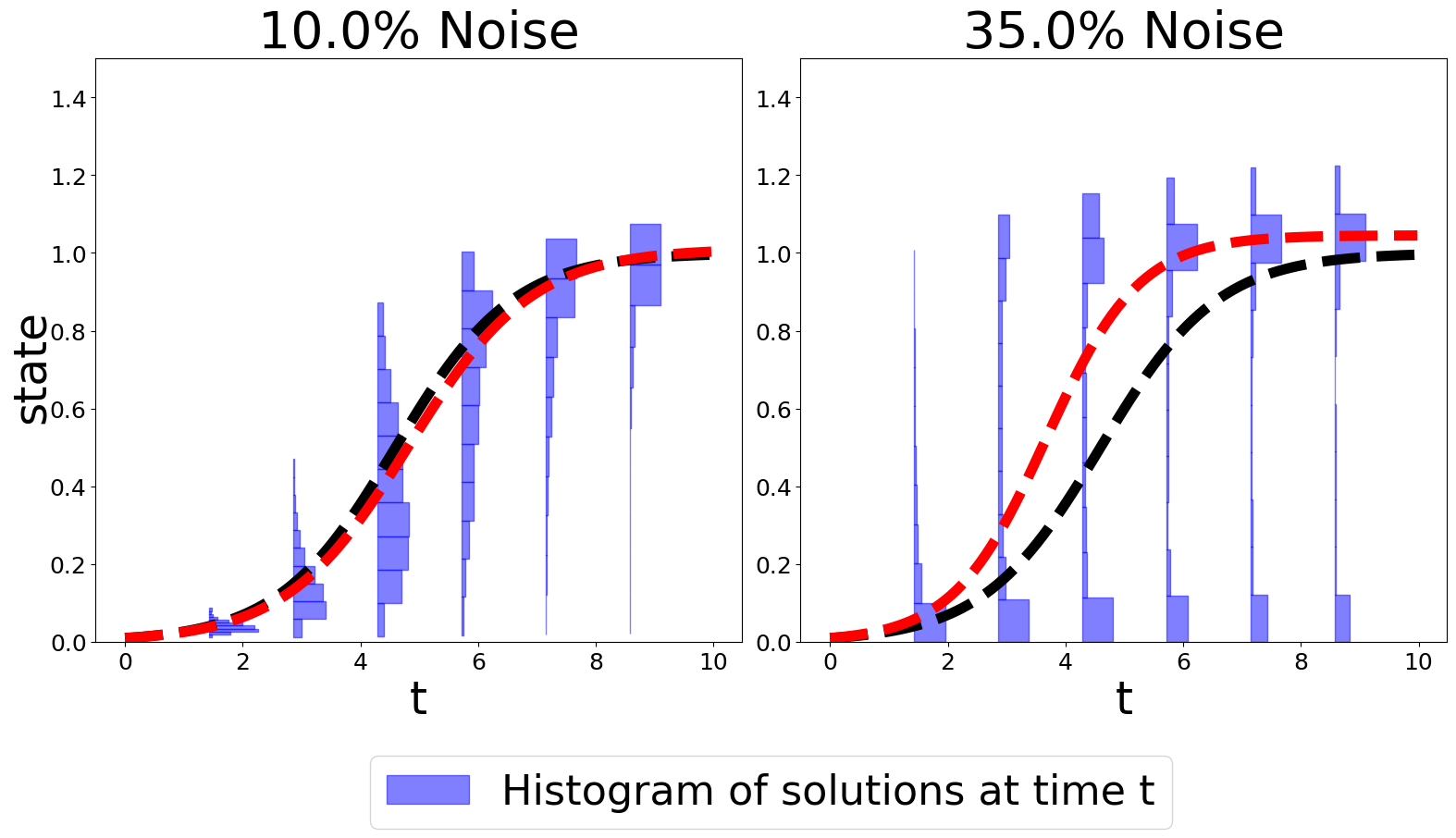}
    \\
    \text{(b)}
    \end{tabular}
     \caption{Top row (a): Logistic model parameter estimation example and uncertainty quantification on two datasets: one dataset with low uncertainty and high coverage (left) and one dataset with high uncertainty and low coverage (right). The light gray curves are used to illustrate the uncertainty around the WENDy solutions; they are obtained via parametric bootstrap, as a sample of WENDy solutions corresponding to a random sample of 1000 parameters from their estimated asymptotic estimator distribution.  Bottom row (b): WENDy solution and histograms of state distributions across specific points in time for the datasets in (a).}
    \label{fig:SamplePlotLogisticCN}
\end{figure}

As Figures~\ref{fig:SamplePlotLogisticCN}(a) and (b) show, the distribution of states (values of solution curves at different time points) for lower noise levels is skewed but unimodal, as expected under the ACN structure since ACN is a non-symmetric noise distribution and the states are derived as non-linear functions of that noise. At higher noise levels, the distribution of states becomes bimodal at time points corresponding to states with higher magnitude, but remains unimodal at time points corresponding to states with lower magnitude. This bimodality results from a sizable fraction of states being zero due the high probability because of large values of noise.

\subsubsection*{Multiplicative Log-Normal Noise}

\begin{figure} 
    \centering
    \begin{tabular}{c}
{\includegraphics[width=0.8\linewidth]{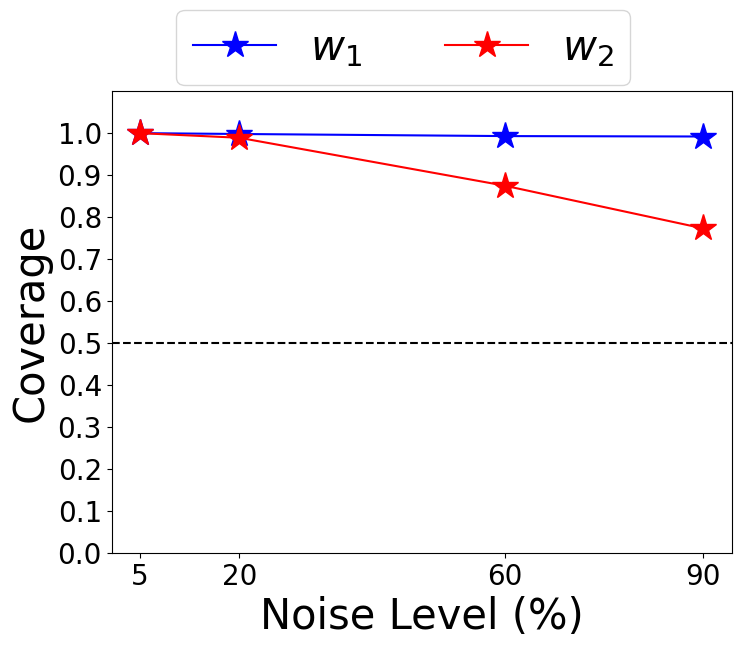}}
    \\
    \text{a}
    \\
\includegraphics[width=1\linewidth]{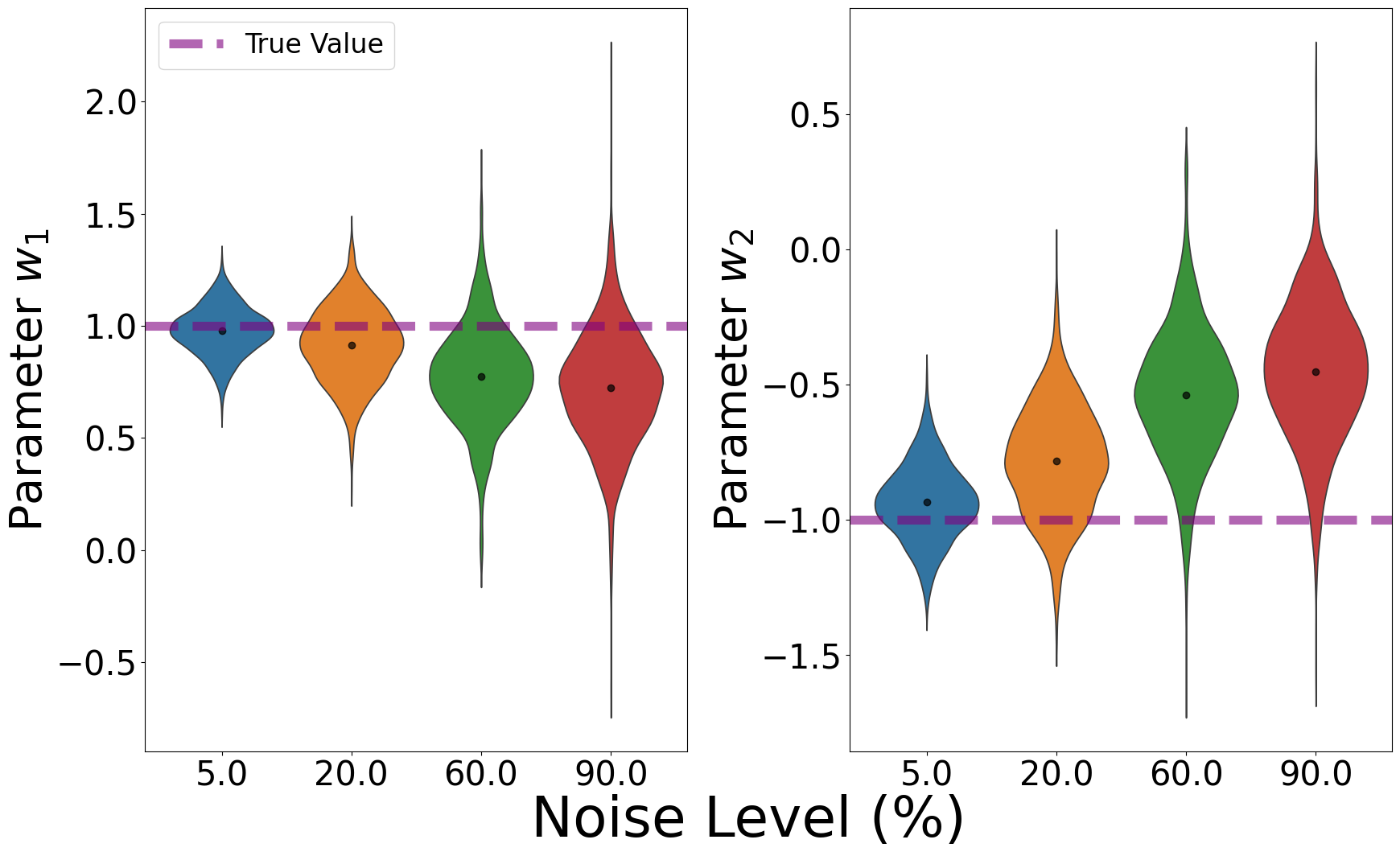}
    \\
      \text{(b)}
    \end{tabular}
     \caption{Logistic model parameter estimation performance with increasing MLN noise (1000 datasets per level, 103 data points each). (a) coverage across four noise levels. (b) violin plots of parameter estimates, with the dashed red line indicating the true parameter values.}
    \label{fig:CovBiasLogisticLN}
\end{figure}

As shown in Figure~\ref{fig:CovBiasLogisticLN}(a), the coverage of the 95\% confidence intervals for the $w_1$ parameter remained slightly above the nominal 95\% level for all noise levels from 5\% to 90\%. In contrast, the coverage of the 95\% confidence intervals for the $w_2$ parameter consistently decreased as noise increased, though it never dropped below 50\%. As seen in Figure~\ref{fig:CovBiasLogisticLN}(b), both the bias and variance of the parameter estimators increased consistently with higher noise levels. The violin plots visually depict the excellent coverage of MLN noise estimators since the line depicting the true parameter values falls within the area of the violin plots even with 90\% noise. 

\begin{figure} 
    \centering
    \begin{tabular}{c}
    
    \includegraphics[width=1\linewidth]{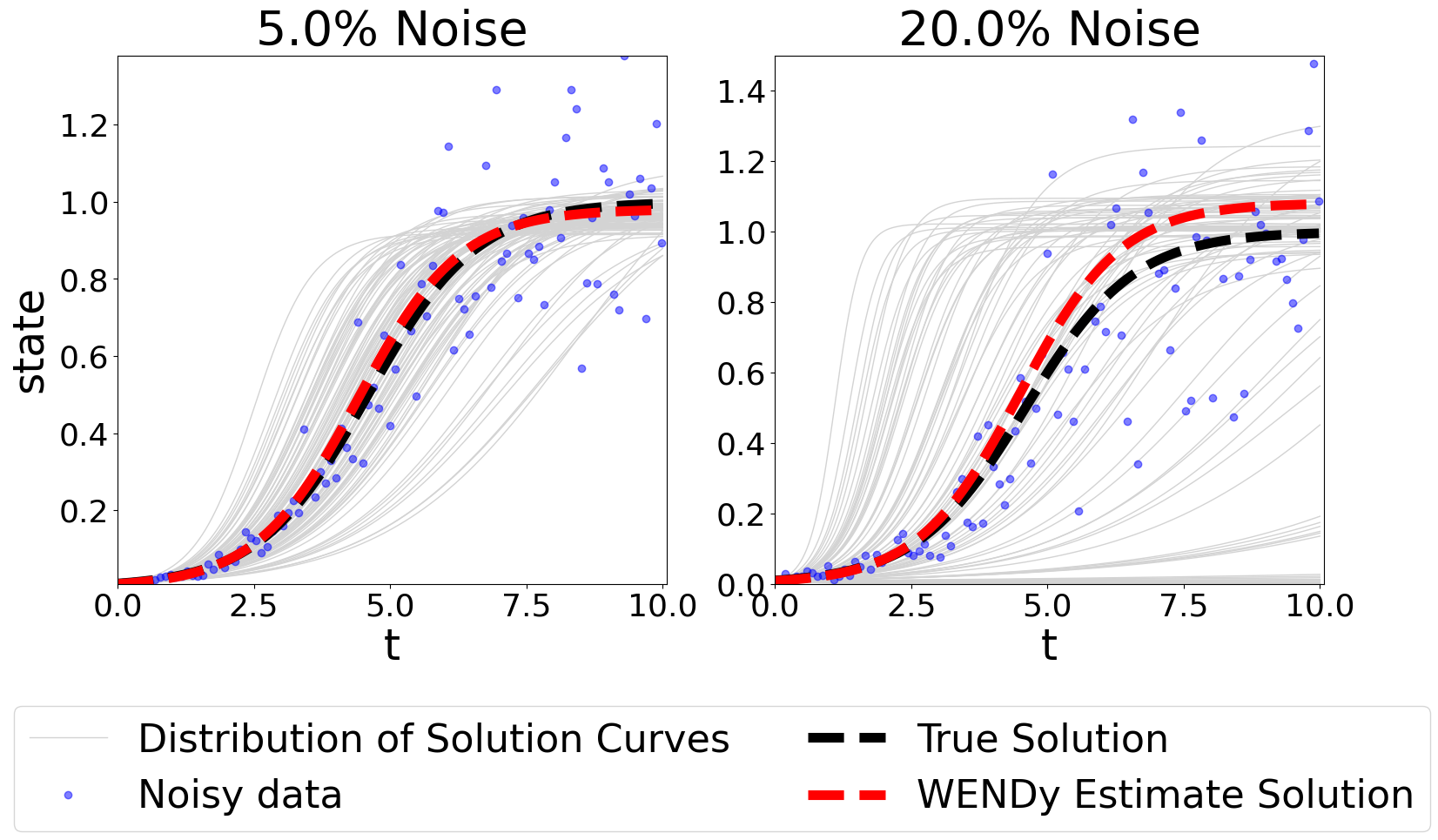} 
    \\
    \text{(a)}
    \\
    \includegraphics[width=1\linewidth]{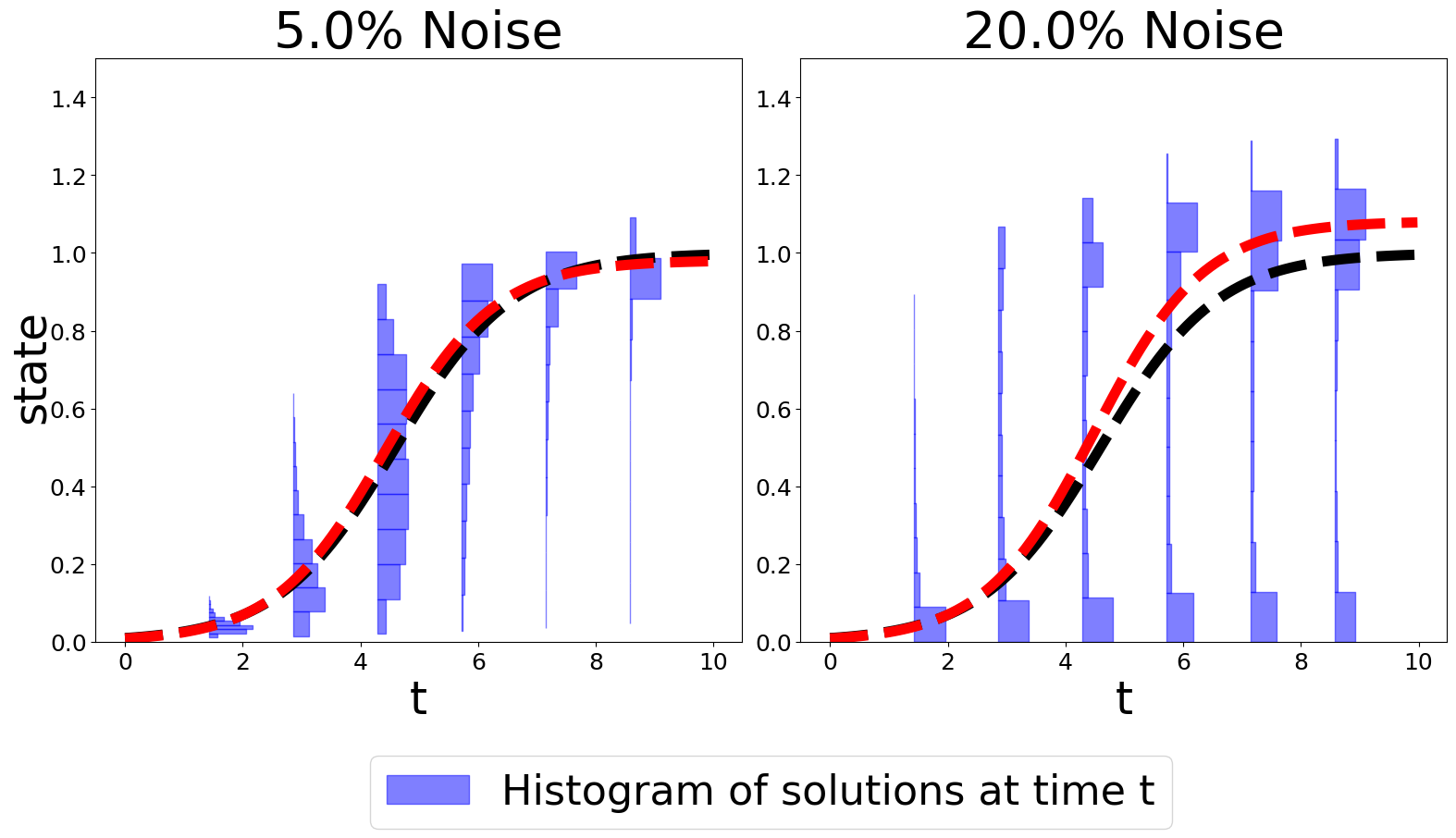}
    \\
    \text{(b)}
    \end{tabular}
     \caption{Top row (a): Logistic model parameter estimation example and uncertainty quantification on two datasets: one dataset with low uncertainty and high coverage (left) and one dataset with high uncertainty and low coverage (right). The light gray curves are used to illustrate the uncertainty around the WENDy solutions; they are obtained via parametric bootstrap, as a sample of WENDy solutions corresponding to a random sample of 1000 parameters from their estimated asymptotic estimator distribution.  Bottom row (b): WENDy solution and histograms of state distributions across specific points in time for the datasets in (a).}
    \label{fig:SamplePlotLogisticLN}
\end{figure}

As Figures~\ref{fig:SamplePlotLogisticLN}(a) and (b) show, the distribution of solution states (at selected time points) at lower noise levels is skewed but remains unimodal. At higher noise levels, however, the distributions become bimodal at time points corresponding to states with larger magnitudes, while remaining unimodal at time points with smaller magnitudes. Since MLN noise structure is heteroskedastic, the lower magnitude state values don't get distorted much, leading to unimodal histograms, while higher magnitude state values are subject to much more distortion, which can explain the bimodality at those state values.

\subsubsection*{Additive Truncated Normal Noise}
\begin{figure}
    \centering
    \begin{tabular}{c}
{\includegraphics[width=0.8\linewidth]{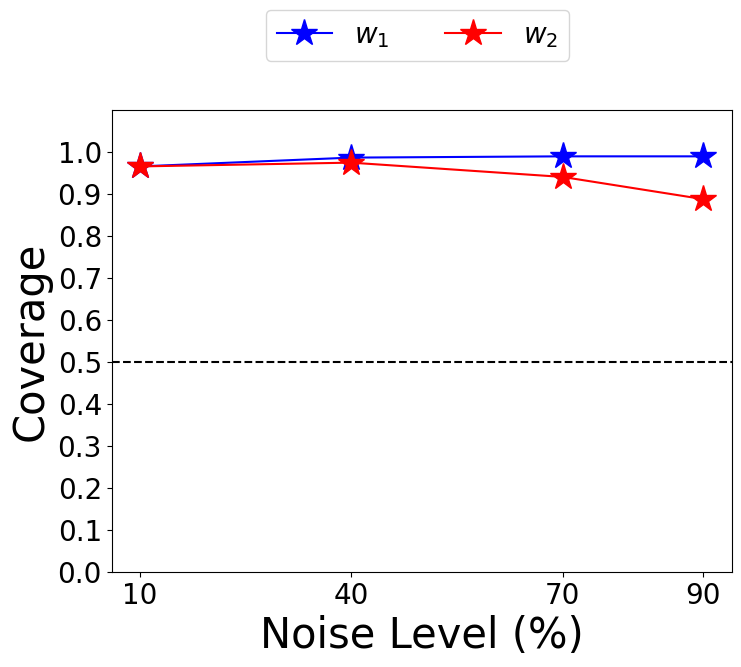}}
    \\
    \text{(a)}
    \\
    \includegraphics[width=1\linewidth]{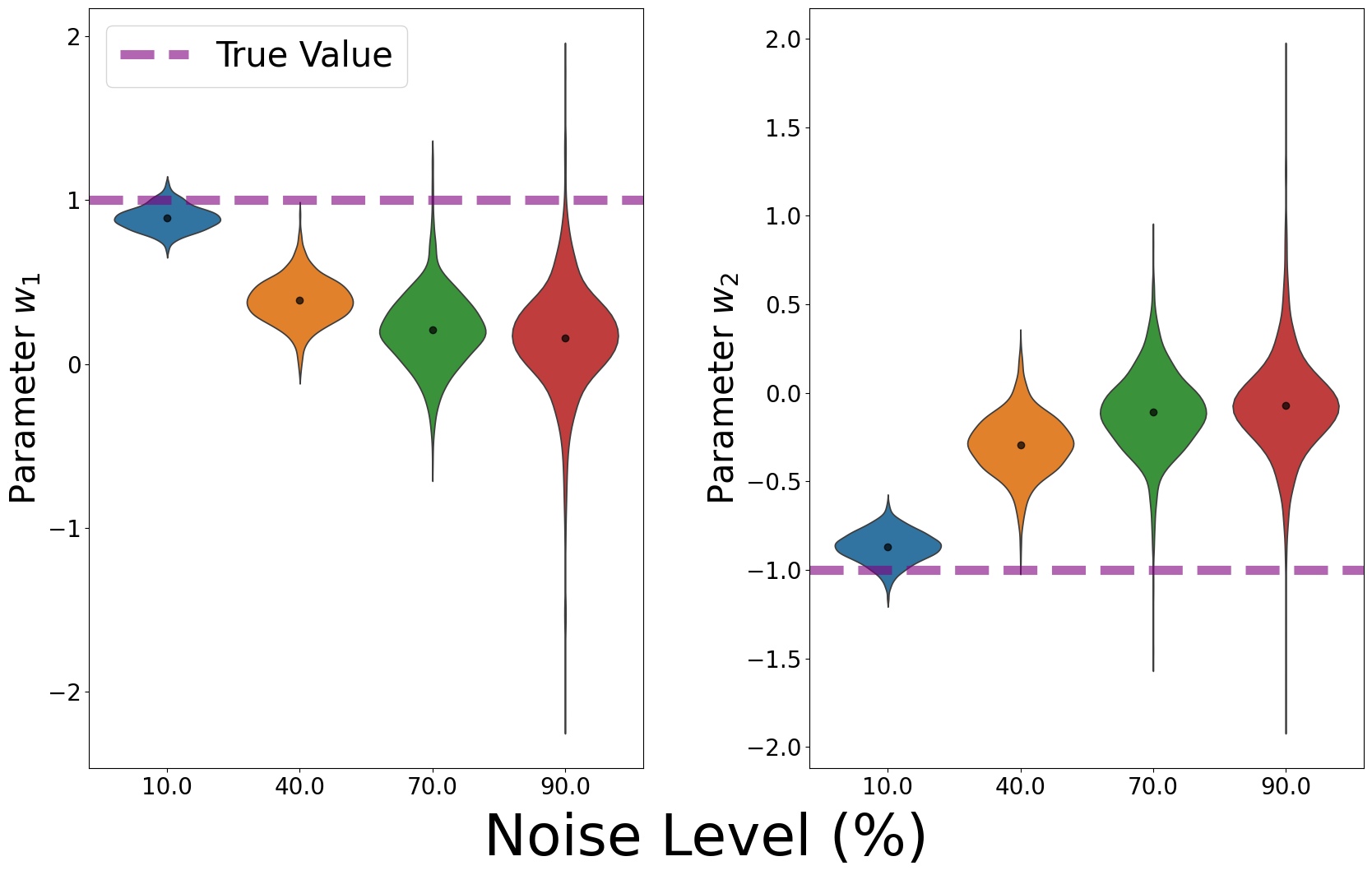}
    \\
      \text{(b)} 
    \end{tabular}
     \caption{Logistic model parameter estimator performance with increasing ATN noise levels (1000 datasets per noise level, 103 data points in each dataset): (a) coverage; (b) violin plots of parameter estimates, with the dashed red line indicating the true parameter values.}
    \label{fig:CovBiasLogisticTN}
\end{figure}

As shown in Figure~\ref{fig:CovBiasLogisticTN}(a), the coverage of the 95\% confidence intervals for both parameters remained close to the nominal 95\% level across all noise levels from 10\% to 90\%. As Figure~\ref{fig:CovBiasLogisticTN}(b) shows, the bias for both parameters increased modestly as the noise level rose, but then plateaued at the higher noise levels (70\%–90\%). The higher coverage can be attributed to the fact that WENDy assumes normally distributed errors, and although ATN noise structures are non-symmetric, it behaves like normal noise for higher magnitude state values while keeping state values positive, which allows for better identifiability of the growth rate.

\begin{figure} 
 \clearpage
 \centering
    \begin{tabular}{c}
    
    \includegraphics[width=1\linewidth]{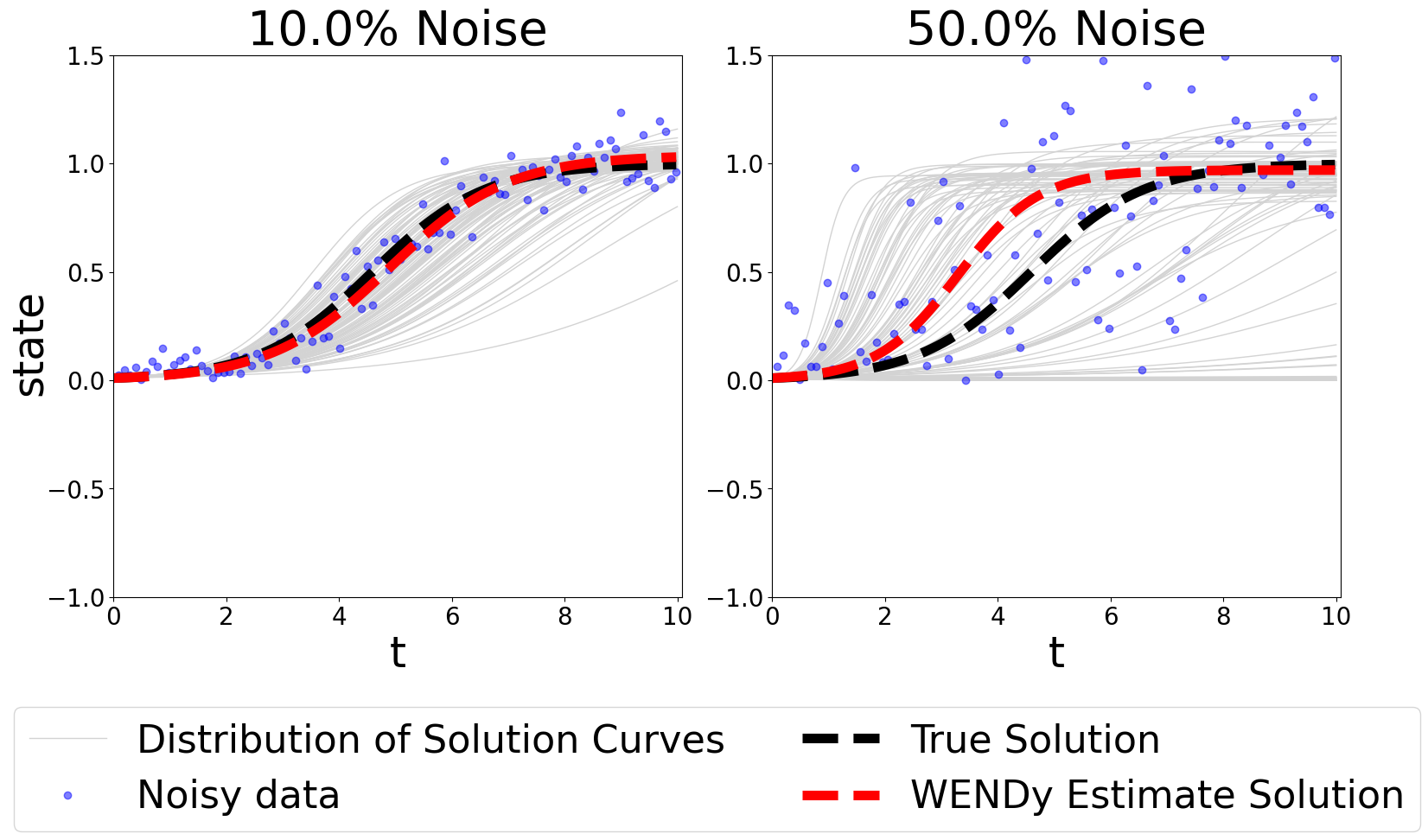} 
    \\
    \text{(a)}
    \\
    \includegraphics[width=1\linewidth]{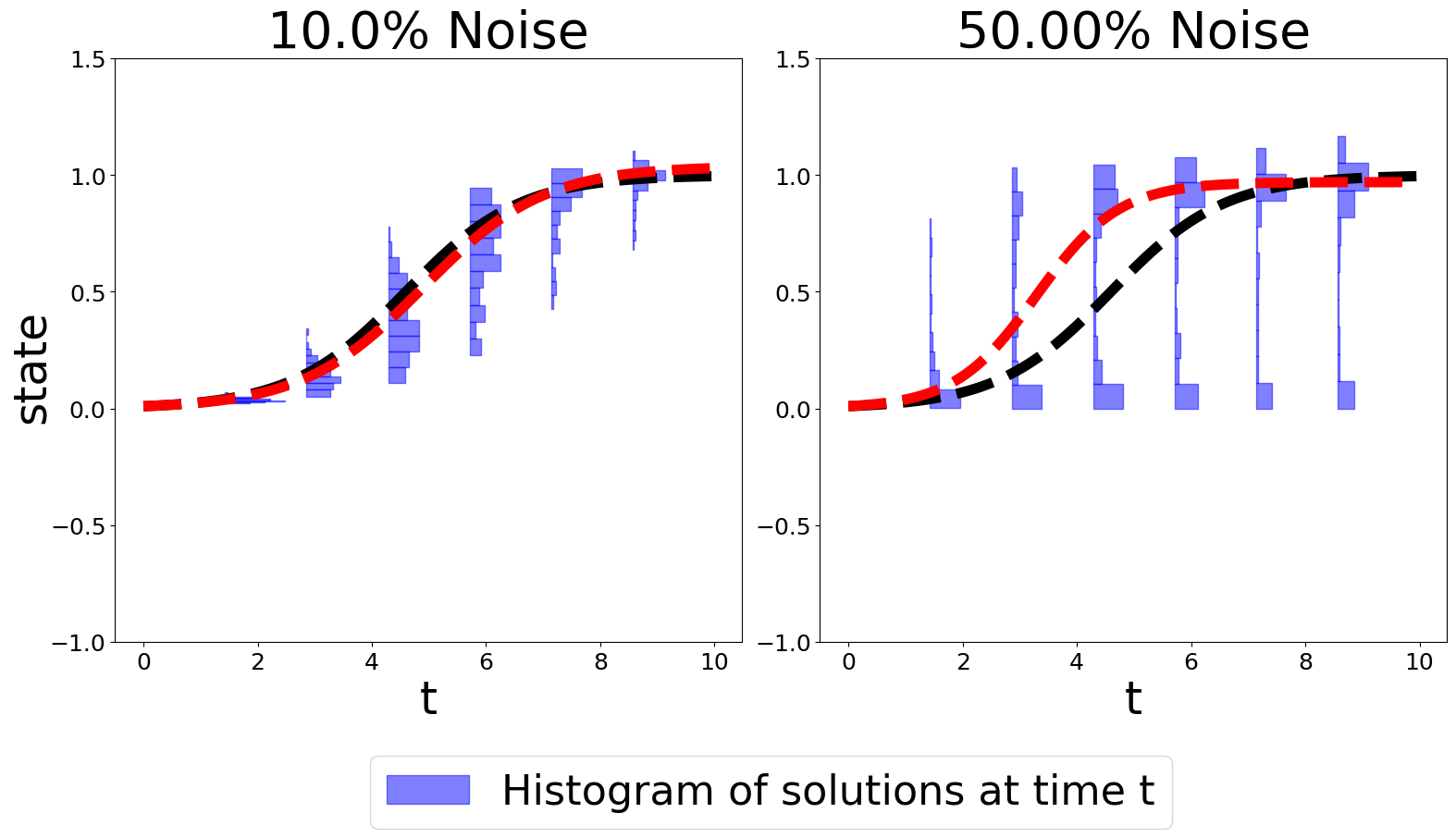}
    \\
    \text{(b)}
    \end{tabular}
     \caption{Top row (a): Logistic model parameter estimation example and uncertainty quantification on two datasets: one dataset with low uncertainty and high coverage (left) and one dataset with high uncertainty and low coverage (right). The light gray curves are used to illustrate the uncertainty around the WENDy solutions; they are obtained via parametric bootstrap, as a sample of WENDy solutions corresponding to a random sample of 1000 parameters from their estimated asymptotic estimator distribution.  Bottom row (b): WENDy solution and histograms of state distributions across specific points in time for the datasets in (a).}
    \label{fig:SamplePlotLogisticTN}
\end{figure}

As Figures~\ref{fig:SamplePlotLogisticTN}(a) and (b) show, the distribution of solution states (at selected time points) at lower noise levels is skewed but unimodal. At higher noise levels, however, the distribution becomes bimodal at time points corresponding to states with larger magnitudes, while remaining unimodal at time points with smaller magnitudes. This bimodality arises because the ATN noise affects the symmetry of the errors more strongly at early time-points, where states are farther from equilibrium, leading to clusters of solution trajectories at distinct magnitudes.

\begin{figure} 
    \centering
    \includegraphics[width=0.38\linewidth]{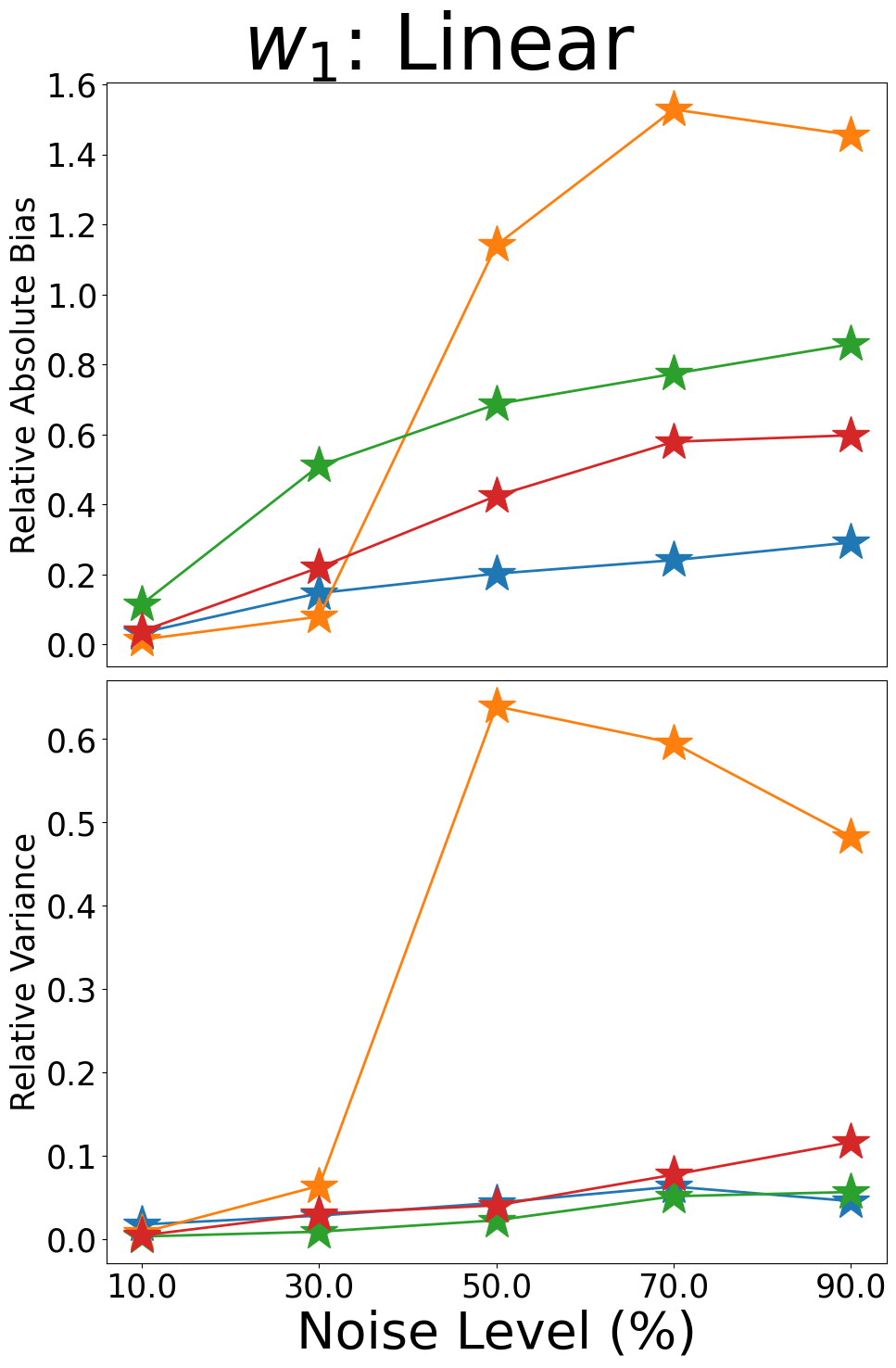}\;\;\;\;    
    \includegraphics[width=0.4\linewidth]{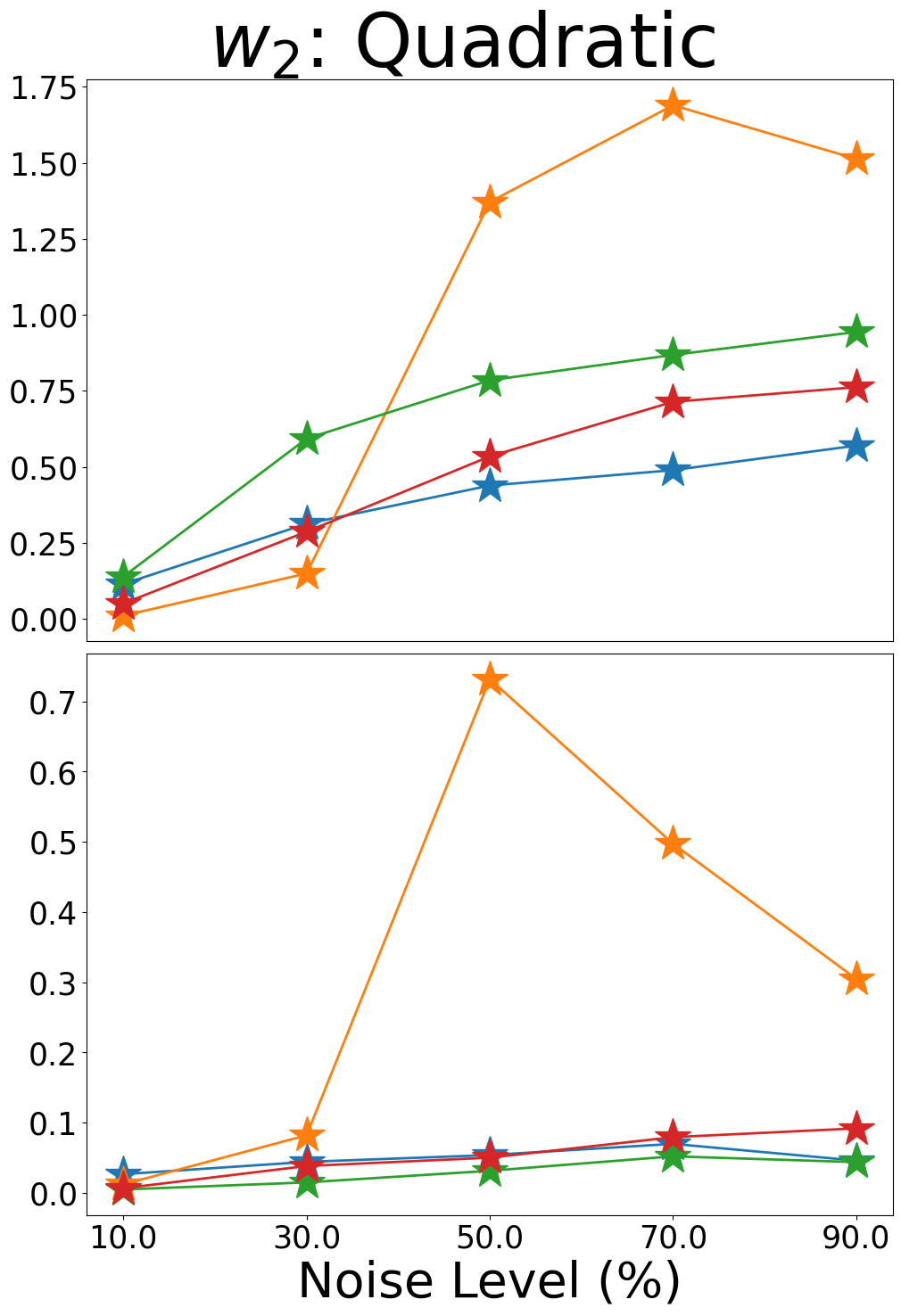}\;\;\;\;   \\ 
    \includegraphics[width=0.3\linewidth]{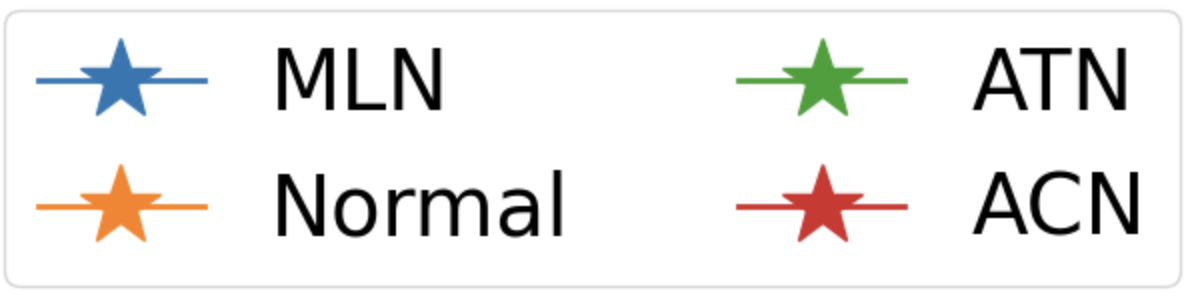}
    
    \caption{Relative bias magnitude (top row) and variance (bottom row) for WENDy estimators of Logistic model states across 100 datasets with increasing levels of additive normal, MLN, ACN, and ATN.}
    \label{fig:LogisticBiasVarNoise}
\end{figure}
As seen in Figure~\ref{fig:LogisticBiasVarNoise}
the relative bias present in WENDy estimators is for 10\% and 30\% noise is highest for ATN noise but becomes the highest for normal noise with 50\% noise and up for both parameters. The relative variance is highest for MLN noise at 10\% noise but then becomes the highest for normal noise from 30\% noise and up for both parameters. A likely explanation for the observed behavior is that since the logistic model only allows for positive states, it is likely that many states end up negative, which can throw off the growth rate of the model since negative state solutions lie across an equilibrium solution. The high coverage seen earlier of the normal noise WENDy estimators can be attributed the the high variance levels. estimators from datasets with MLN noise were also observed to have the smallest relative bias at high levels of noise, which would explain the sustained coverage seen earlier.
\subsubsection{Varying Data Resolution}
\subsubsection*{Additive Normal Noise}
\begin{figure}
    \centering
    \begin{tabular}{c}
{\includegraphics[width=0.8\linewidth]{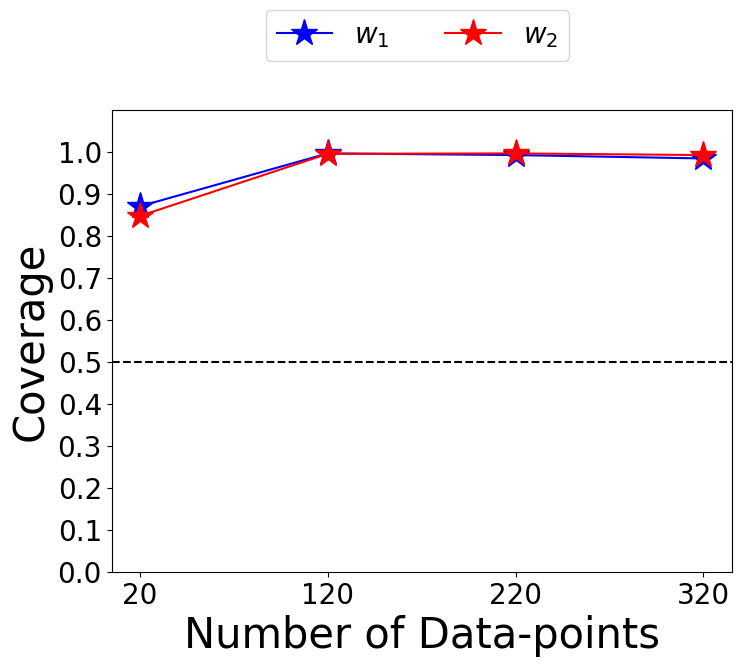}}
    \\
    \text{(a)}
    \\
    \includegraphics[width=1\linewidth]{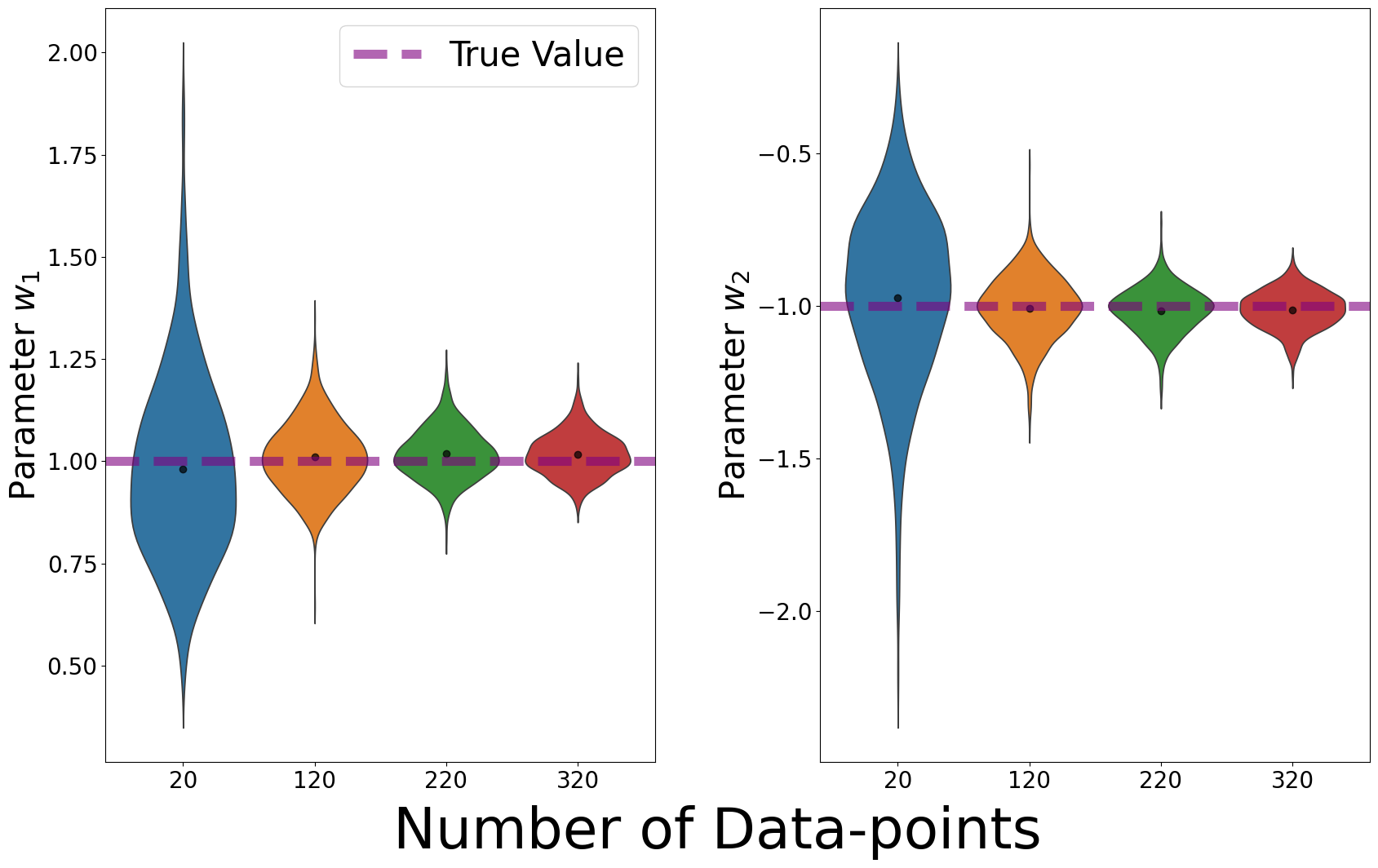}
    \\
      \text{(b)} 
    \end{tabular}
     \caption{Logistic model parameter estimation performance with increasing data resolution (1000 datasets per level, 5\% additive normal noise). (a) coverage across four noise levels. (b) violin plots of parameter estimates, with the dashed red line indicating the true parameter values.}
    \label{fig:CovBiasResLogisticN}
\end{figure}

As shown in Figure~\ref{fig:CovBiasResLogisticN}(a), the coverage of the 95\% confidence intervals for both parameters started just below nominal at 20 data-points and then rose to above nominal at 120 data-points, where it remained for the rest of the resolution levels. This is consistent with the fact that WENDy parameter estimators have been shown to be asymptotically Normally distributed under additive Gaussian noise \citep{BortzMessengerDukic2023BullMathBiol}.


As Figure~\ref{fig:CovBiasResLogisticN}(b) shows,  the bias for both parameters decreased slightly with increasing levels of resolution, while the variance decreased greatly. Again, this is consistent with the fact that WENDy parameter estimators have also been shown to be asymptotically unbiased \citep{BortzMessengerDukic2023BullMathBiol} under additive Gaussian noise.

\begin{figure}
\clearpage
    \centering
    \begin{tabular}{c}
    
    \includegraphics[width=1\linewidth]{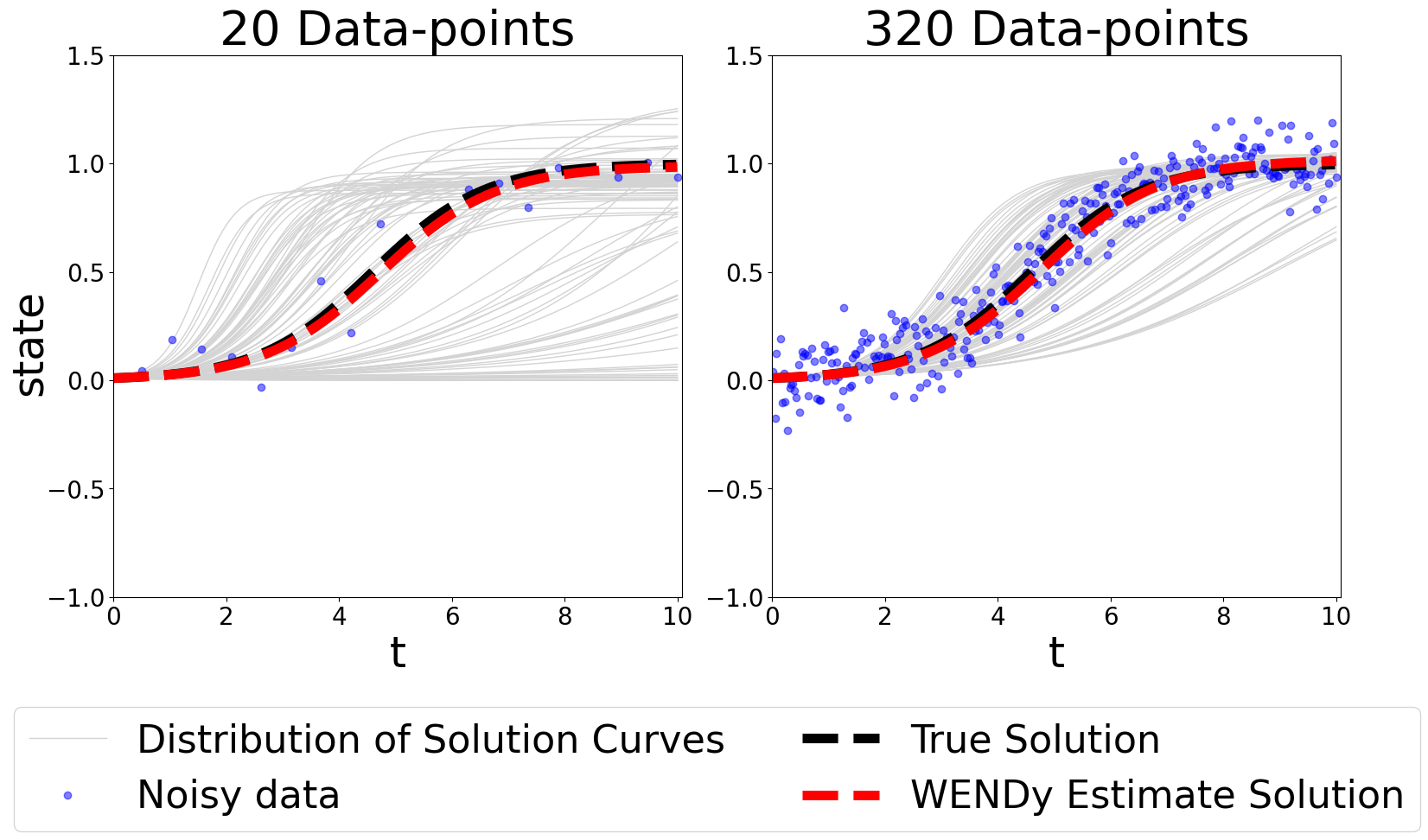} 
    \\
    \text{(a)}
    \\
    \includegraphics[width=1\linewidth]{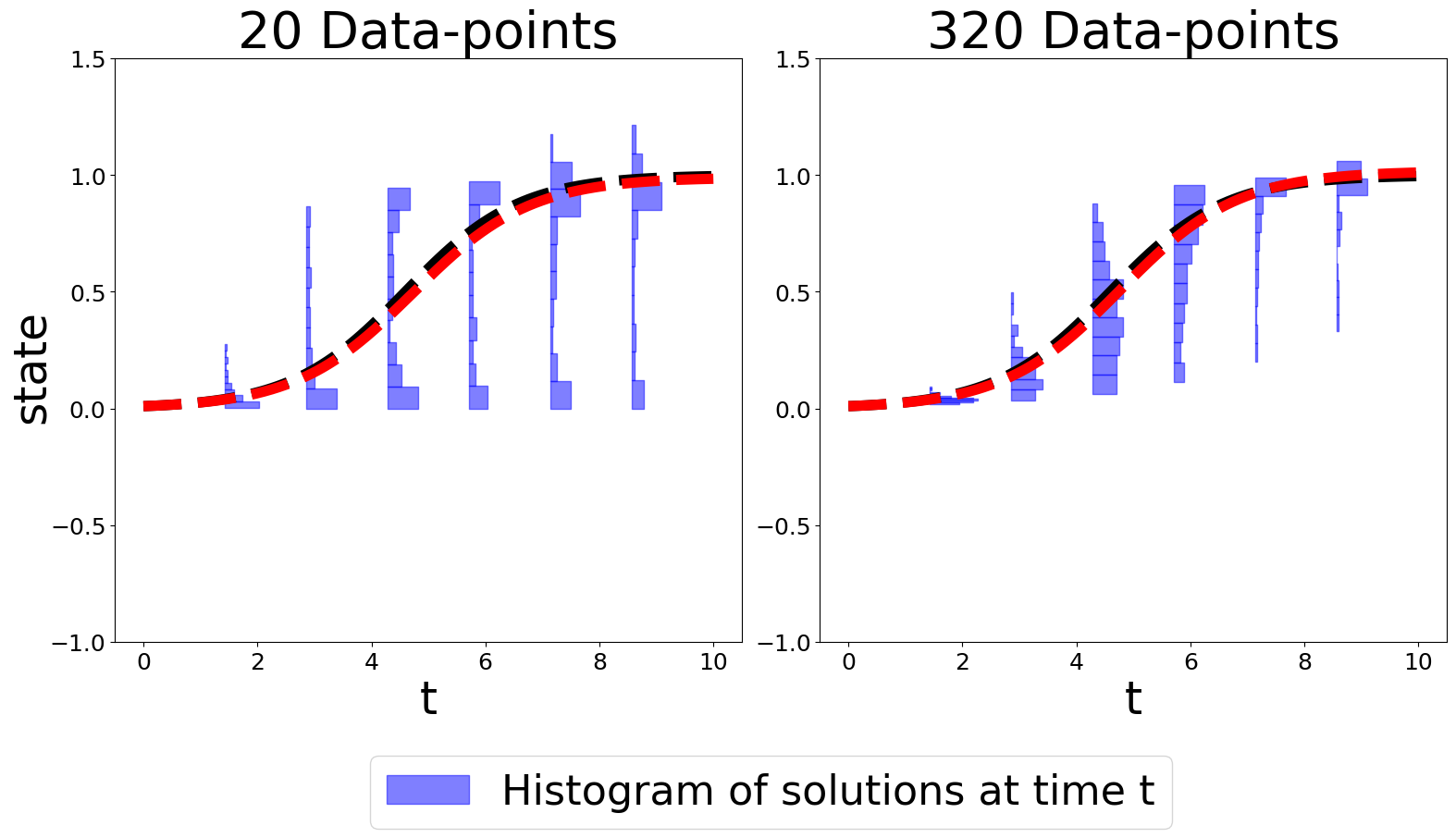}
    \\
    \text{(b)}
    \end{tabular}
     \caption{Top row (a): Logistic model parameter estimation example and uncertainty quantification on two datasets: one dataset with low (left) and one dataset with high data resolution (right). The light gray curves are used to illustrate the uncertainty around the WENDy solutions; they are obtained via parametric bootstrap, as a sample of WENDy solutions corresponding to a random sample of 1000 parameters from their estimated asymptotic estimator distribution.  Bottom row (b): WENDy solution and histograms of state distributions across specific points in time for the datasets in (a).}
    \label{fig:SamplePlotResLogisticN}
\end{figure}

As Figure~\ref{fig:SamplePlotResLogisticN}(a) and (b) show, the distribution of solution states (at selected time points) for lower and higher resolution levels is bimodal for higher magnitude data points and unimodal but skewed at low magnitude data points. At higher resolution levels, the bimodality goes away and the histograms become narrower. This is expected as more points reveal more features of the dynamics, allowing for better identification of parameters.


\subsubsection*{Additive Censored Normal Noise}
\begin{figure}
    \centering
    \begin{tabular}{c}
{\includegraphics[width=0.8\linewidth]{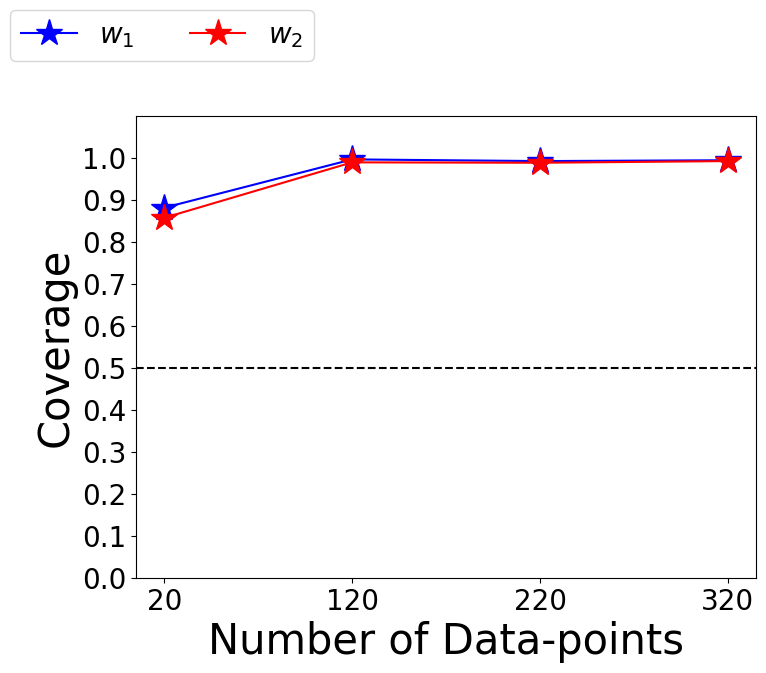}}
    \\
    \text{(a)}
    \\
    \includegraphics[width=1\linewidth]{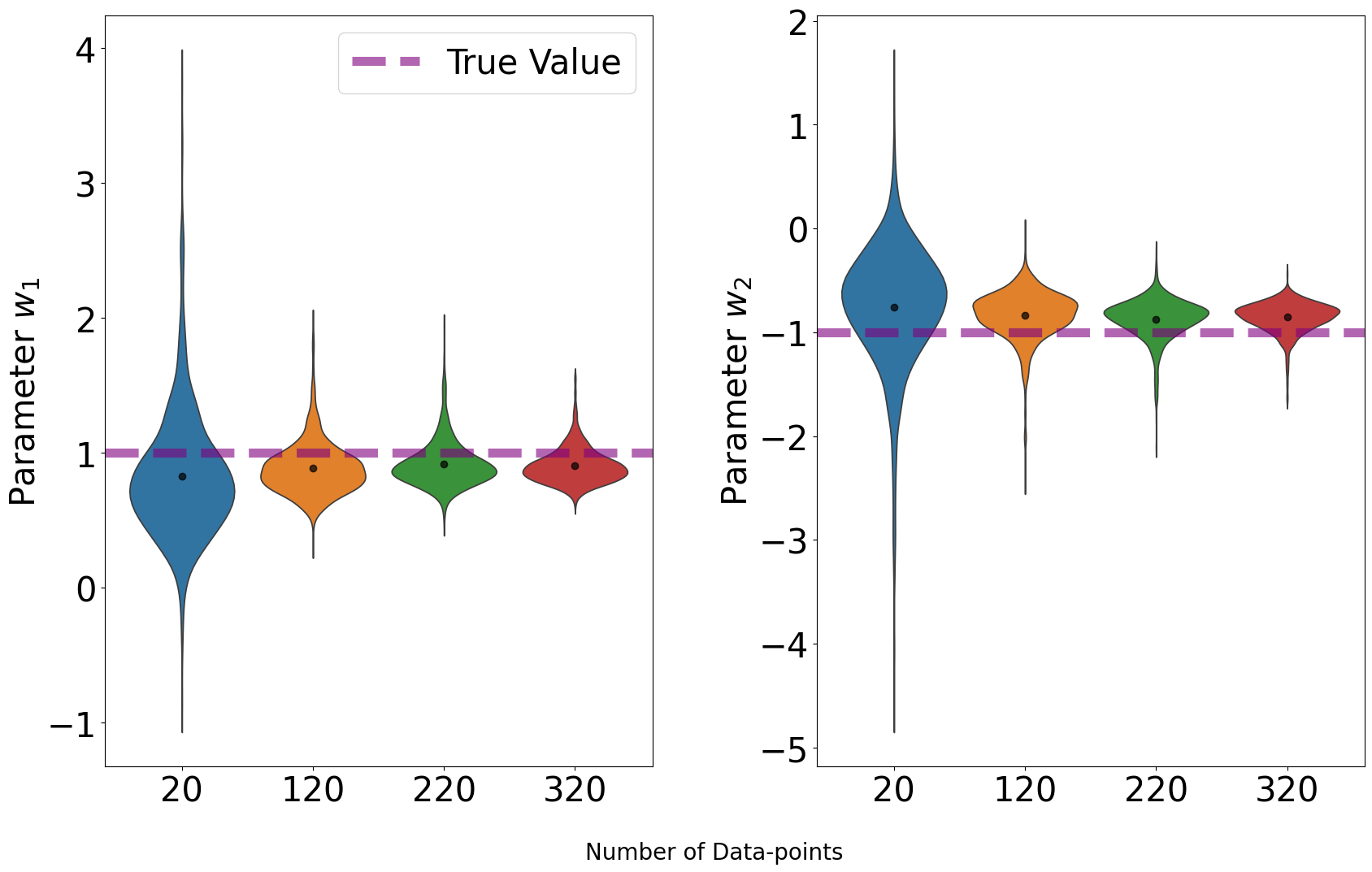}
    \\
      \text{(b)} 
    \end{tabular}
     \caption{Logistic model parameter estimation performance with increasing data resolution (1000 datasets per level, 10\% ACN). (a) coverage across four noise levels. (b) violin plots of parameter estimates, with the dashed red line indicating the true parameter values.}
    \label{fig:CovBiasResLogisticCN}
\end{figure}

\begin{figure}
\clearpage
\centering
    \begin{tabular}{c}
    
    \includegraphics[width=1\linewidth]{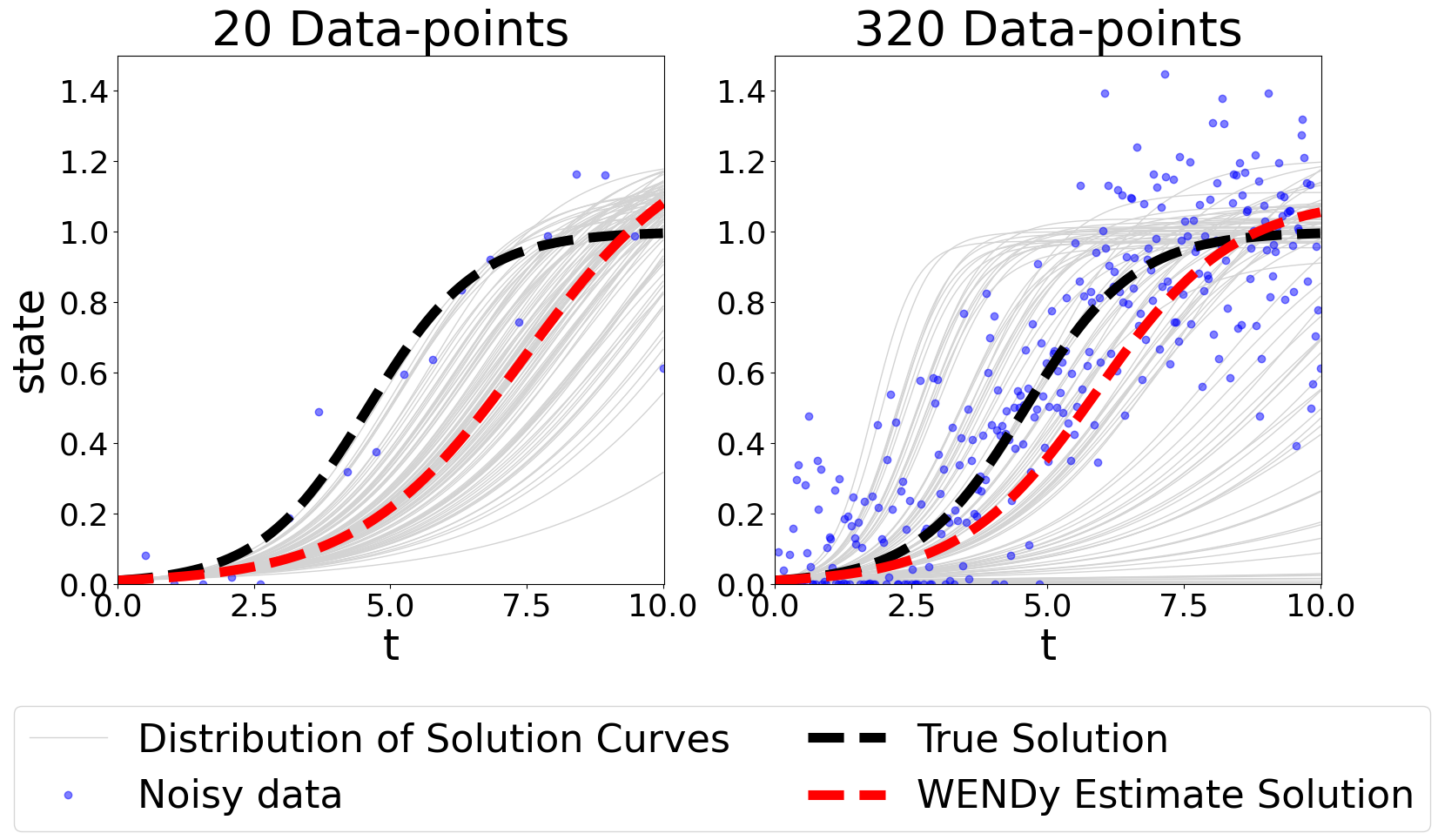} 
    \\
    \text{(a)}
    \\
    \includegraphics[width=1\linewidth]{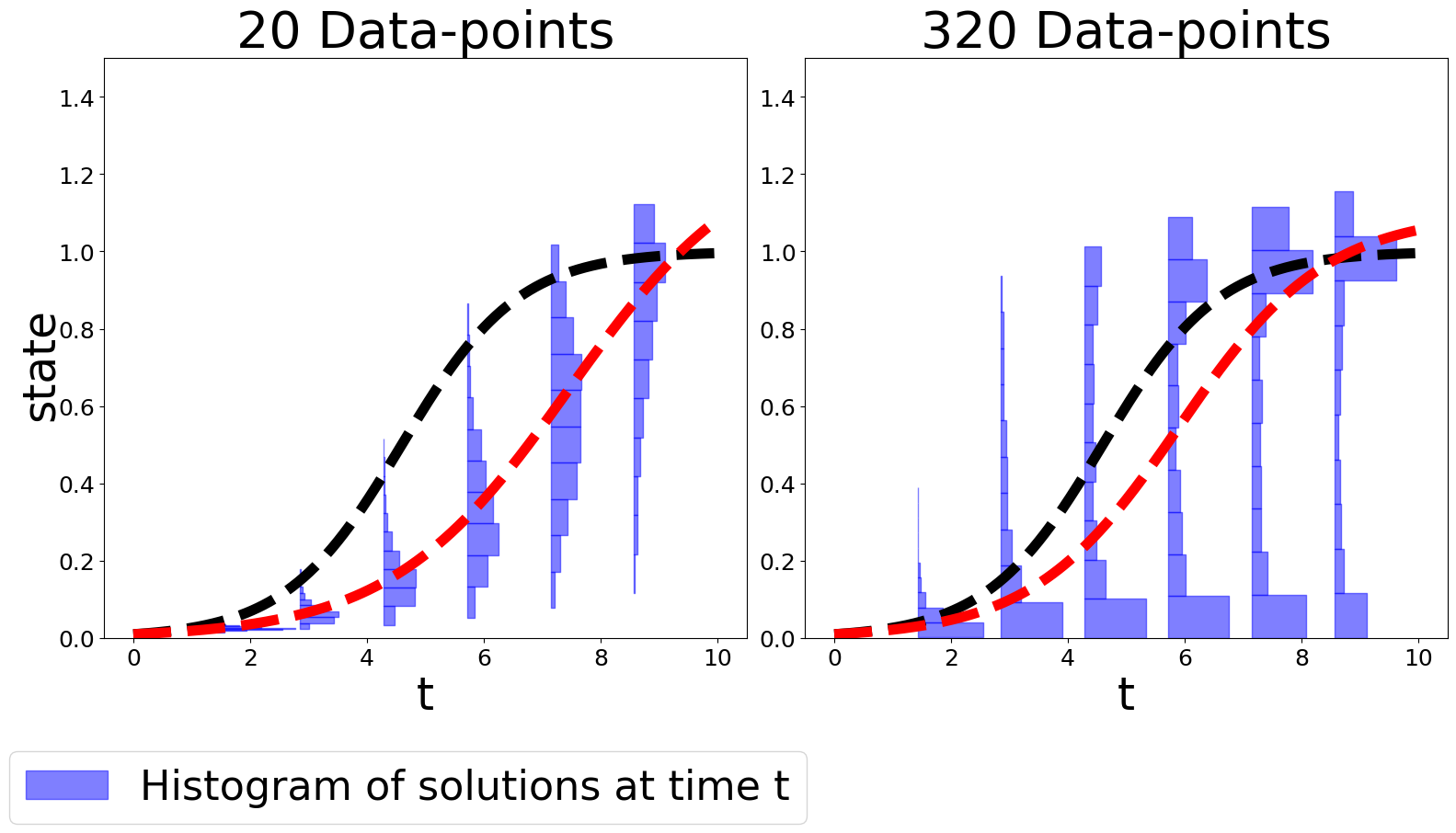}
    \\
    \text{(b)}
    \end{tabular}
     \caption{Top row (a): Logistic model parameter estimation example and uncertainty quantification on two datasets: one dataset with low (left) and one dataset with high data resolution (right). The light gray curves are used to illustrate the uncertainty around the WENDy solutions; they are obtained via parametric bootstrap, as a sample of WENDy solutions corresponding to a random sample of 1000 parameters from their estimated asymptotic estimator distribution.  Bottom row (b): WENDy solution and histograms of state distributions across specific points in time for the datasets in (a).}
    \label{fig:SamplePlotResLogisticCN}
\end{figure}

As shown in Figure~\ref{fig:CovBiasResLogisticCN}(a), the coverage of the 95\% confidence intervals for both $w_1$ and $w_2$ was slightly below nominal at low resolution (20 data points), but increased rapidly with more data. By 120 data points, coverage for both parameters was essentially nominal and remained stable as resolution increased further. As Figure~\ref{fig:CovBiasResLogisticCN}(b) shows, bias for both parameters decreased slightly with increasing resolution, while variance dropped substantially, as expected. 
As Figures~\ref{fig:SamplePlotResLogisticCN}(a) and (b) illustrate, the distribution of solution states (at selected time points) at higher resolutions is bimodal at higher-magnitude time points, while at lower resolutions the distribution becomes skewed and unimodal. This might be due to the fact that the higher the resolution, the greater portion of data points will be censored to 0, which might skew certain solution curves to lower state values, hence, producing the bimodal histograms.

\subsubsection*{Multiplicative Log-Normal Noise}
\begin{figure}
    \centering
    \begin{tabular}{c}
{\includegraphics[width=0.8\linewidth]{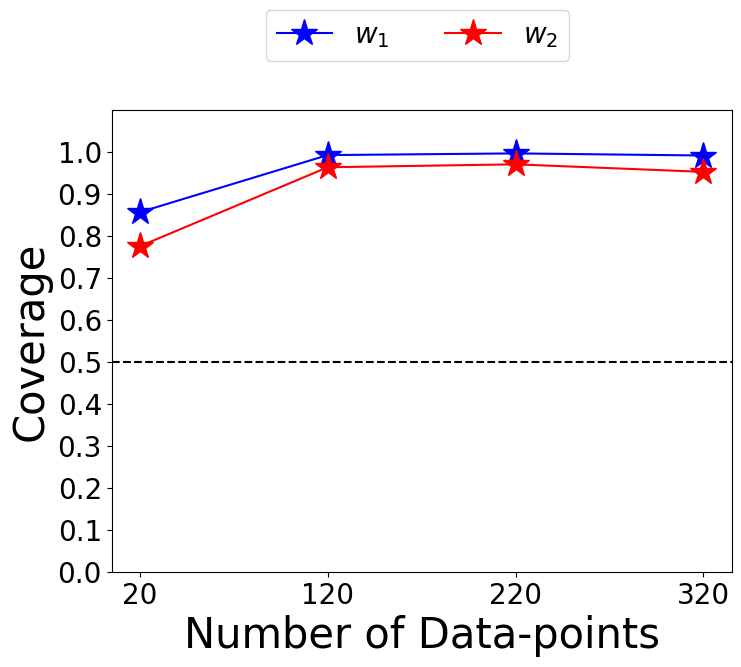}}
    \\
    \text{(a)}
    \\
    \includegraphics[width=1\linewidth]{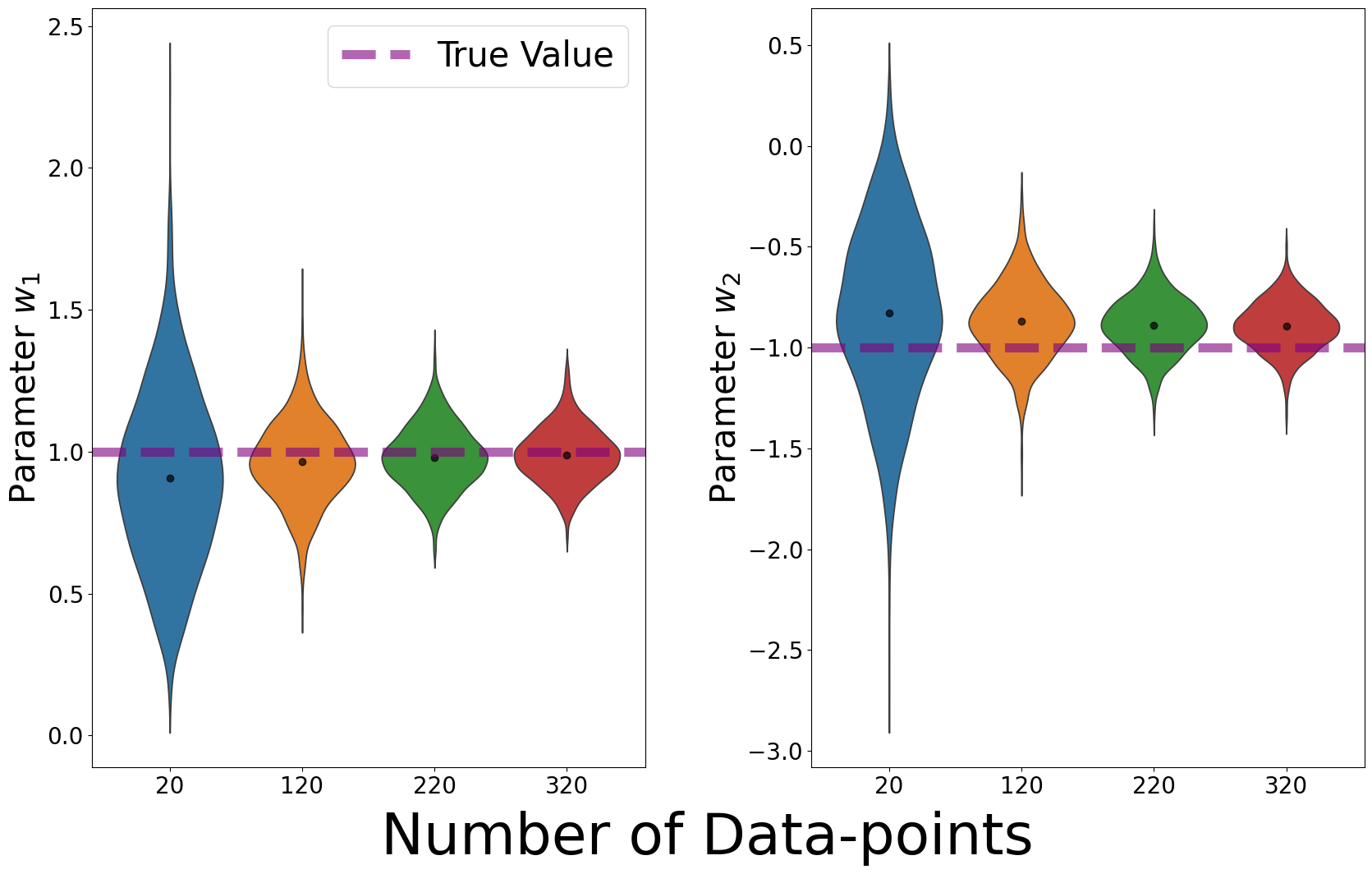}
    \\
      \text{(b)} 
    \end{tabular}
     \caption{Logistic model parameter estimation performance with increasing data resolution (1000 datasets per level, 10\% MLN noise). (a) coverage across four noise levels. (b) violin plots of parameter estimates, with the dashed red line indicating the true parameter values.}
    \label{fig:CovBiasResLogisticLN}
\end{figure}

As shown in Figure~\ref{fig:CovBiasResLogisticLN}(a), the coverage of the 95\% confidence intervals for both the parameters was below nominal at 20 data-points, but then rose to around nominal at 120 data-points and stayed there for increasing resolution. The $w_2$ parameter had slightly worse coverage, which might be caused by the fact that it modulates the non-linear term in the ODE, making it tougher to estimate.

As Figure~\ref{fig:CovBiasResLogisticLN}(b) shows,  the bias for the $w_1$ parameter decreased significantly for increasing levels of resolution, while the bias for the $w_2$ parameter decreased slightly, which would explain it having worse coverage than $w_1$.

\begin{figure}
    \centering
    \begin{tabular}{c}
    
    \includegraphics[width=1\linewidth]{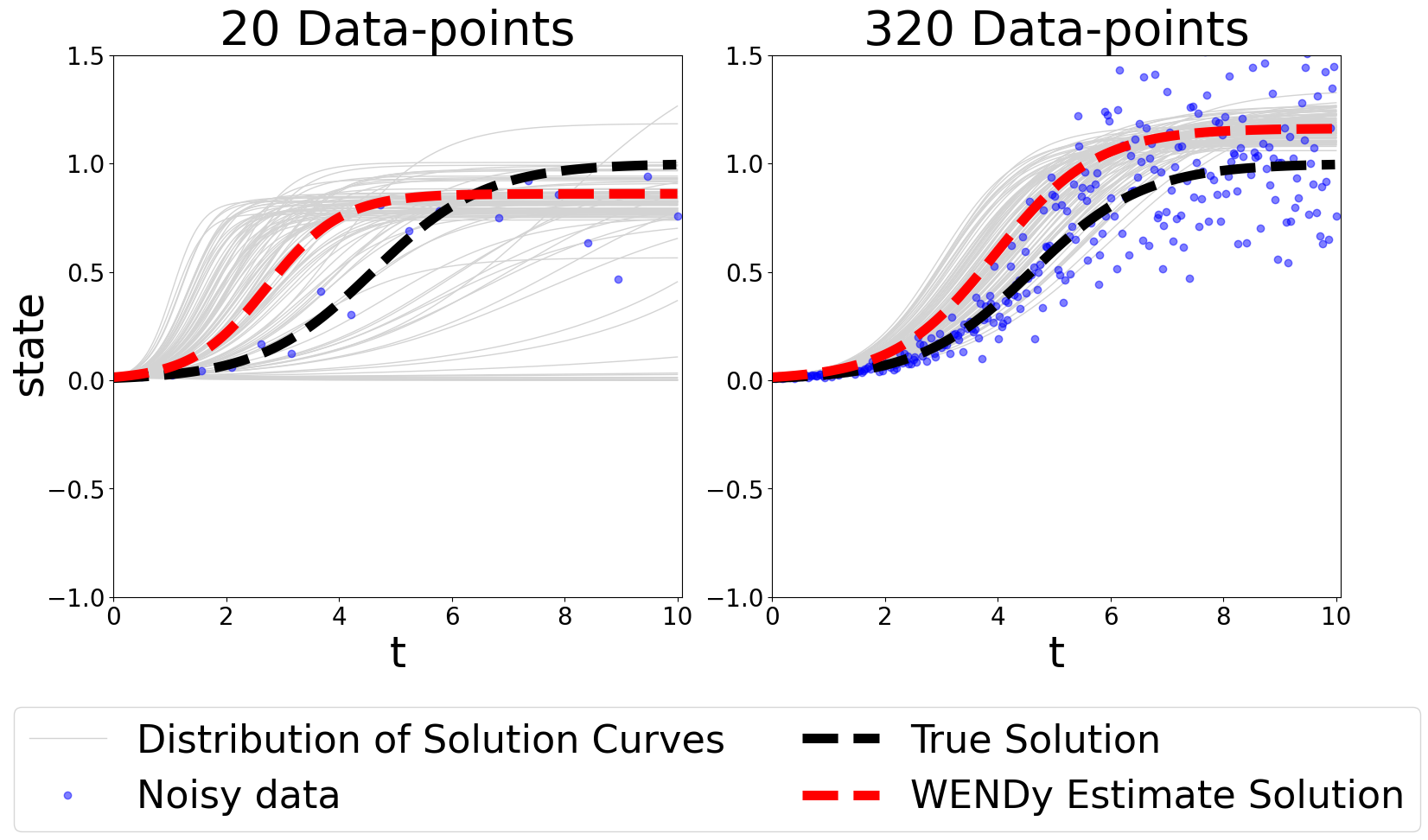} 
    \\
    \text{(a)}
    \\
    \includegraphics[width=1\linewidth]{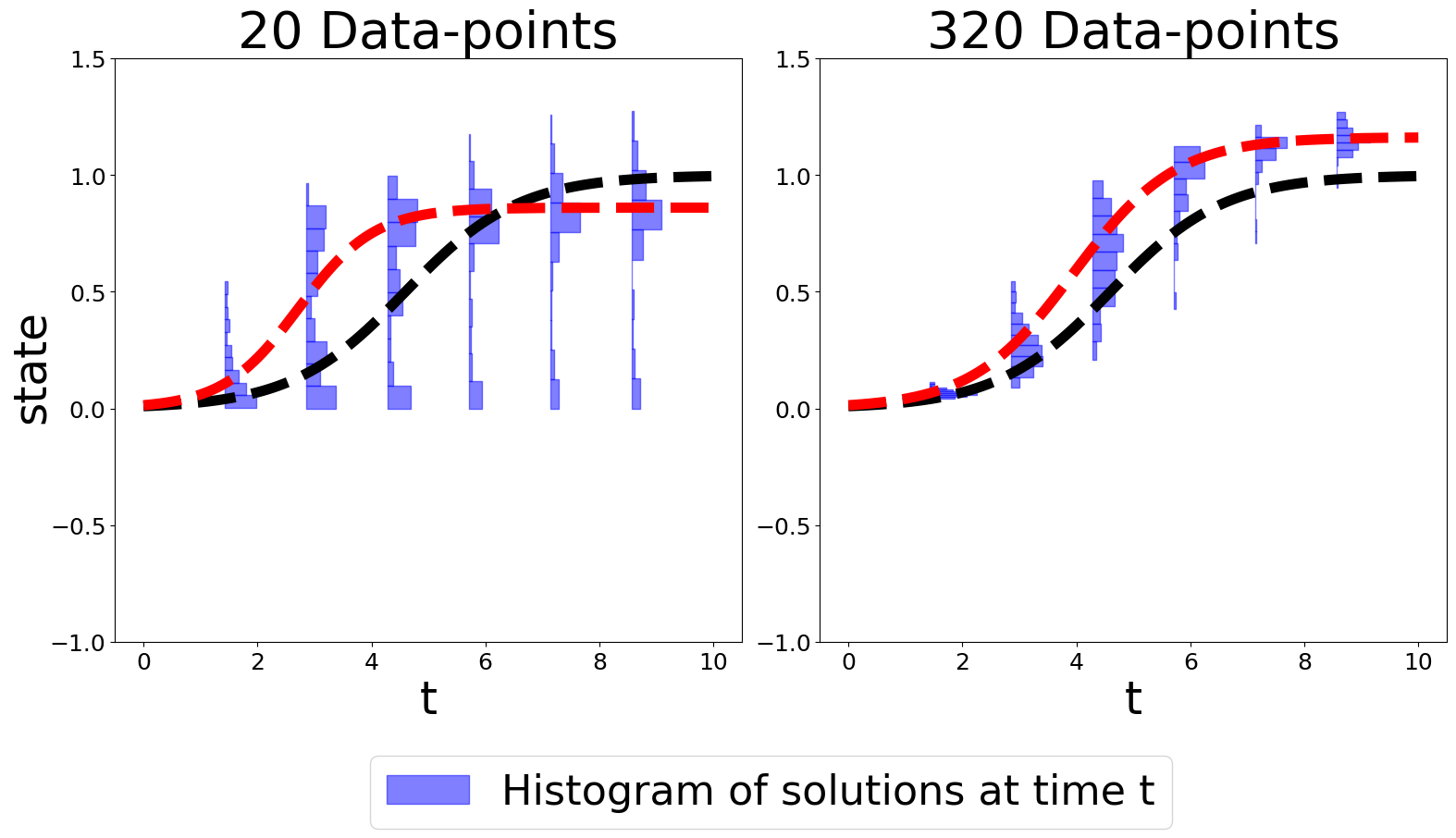}
    \\
    \text{(b)}
    \end{tabular}
     \caption{Top row (a): Logistic model parameter estimation example and uncertainty quantification on two datasets: one dataset with low (left) and one dataset with high data resolution (right). The light gray curves are used to illustrate the uncertainty around the WENDy solutions; they are obtained via parametric bootstrap, as a sample of WENDy solutions corresponding to a random sample of 1000 parameters from their estimated asymptotic estimator distribution.  Bottom row (b): WENDy solution and histograms of state distributions across specific points in time for the datasets in (a).}
    \label{fig:SamplePlotResLogisticLN}
\end{figure}

As Figure~\ref{fig:SamplePlotResLogisticLN}(a) and (b) show, the distribution of solution states (at selected data-points) is bimodal at certain time points for lower resolution. While for higher resolution, it remains unimodal and somewhat skewed.

\subsubsection*{Additive Truncated Normal Noise}
\begin{figure} 
    \centering
    \begin{tabular}{c}
{\includegraphics[width=0.8\linewidth]{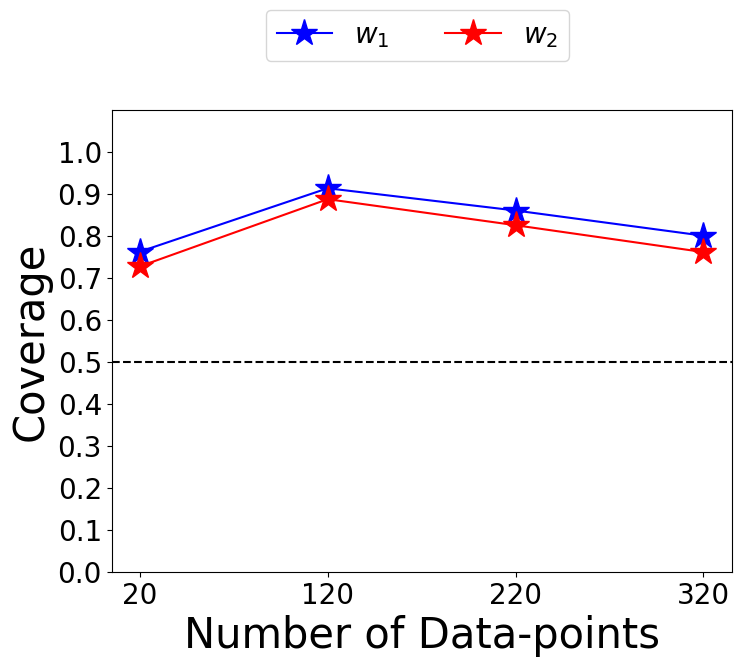}}
    \\
    \text{(a)} 
     \\
     \includegraphics[width=1\linewidth]{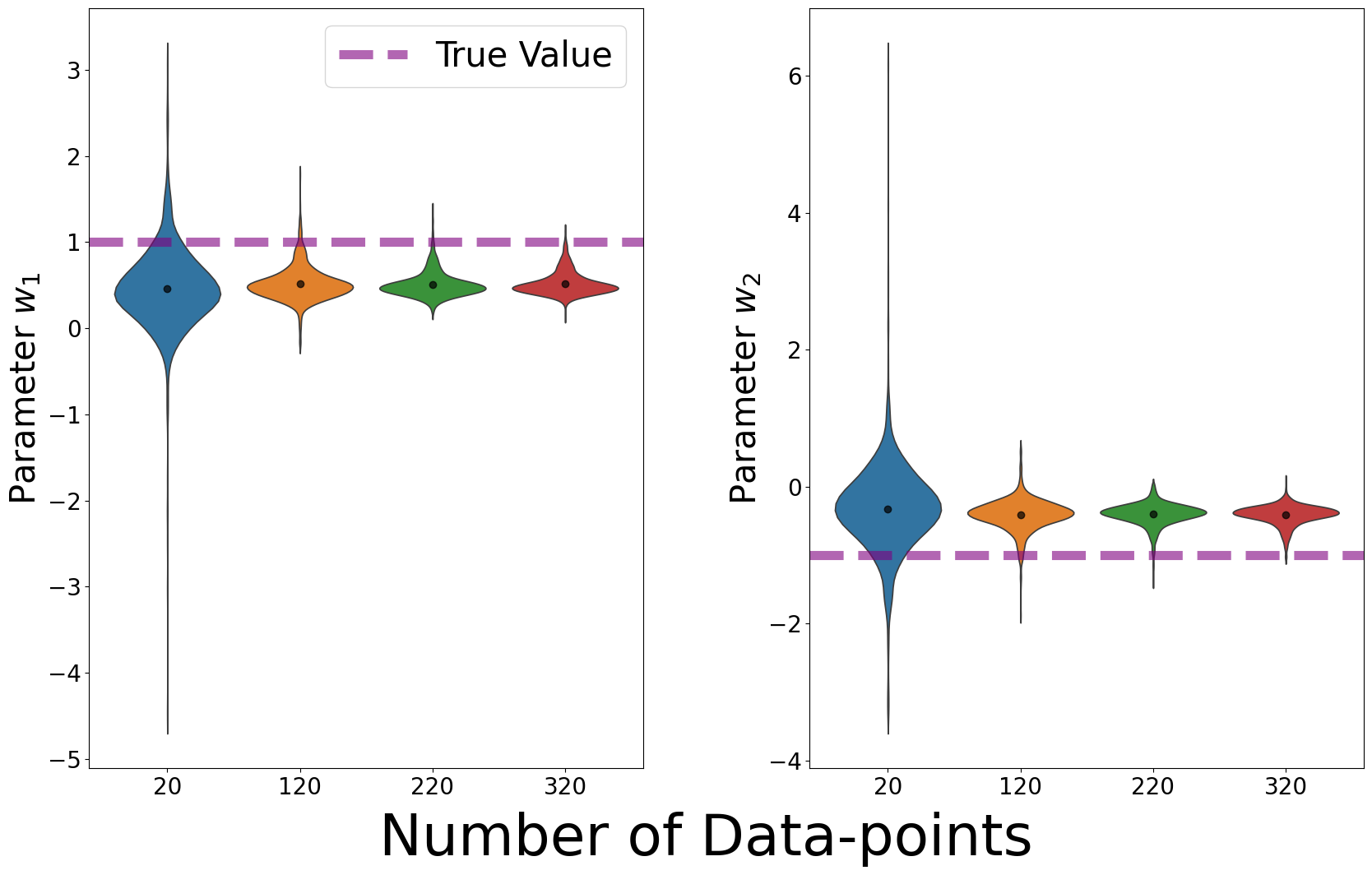}
    \\
      \text{(b)} 
    \end{tabular}
     \caption{Logistic model parameter estimation performance with increasing data resolution (1000 datasets per level, 10\% ATN noise). (a) coverage across four noise levels. (b) violin plots of parameter estimates, with the dashed red line indicating the true parameter values.}
    \label{fig:CovBiasResLogisticTN}
\end{figure}

As shown in Figure~\ref{fig:CovBiasResLogisticTN}(a), the coverage of the 95\% confidence intervals for both parameters was between 70\%  and 80\% at 20 data-points, but increased to just below nominal 95\% at 120 data-points, before dropping again to around 80\% as resolution further increased to 320 points. As Figure~\ref{fig:CovBiasResLogisticTN}(b) shows,  the bias for both parameters was stable for increasing levels of resolution, while the variance dropped considerably. This again explains the drop in coverage, as for the corresponding average parameter estimates, the standard error is much smaller for higher resolution levels; hence, it is likely that the true parameters are covered by the 95\% confidence intervals.

\begin{figure} 
    \centering
    \begin{tabular}{c}
    
    \includegraphics[width=1\linewidth]{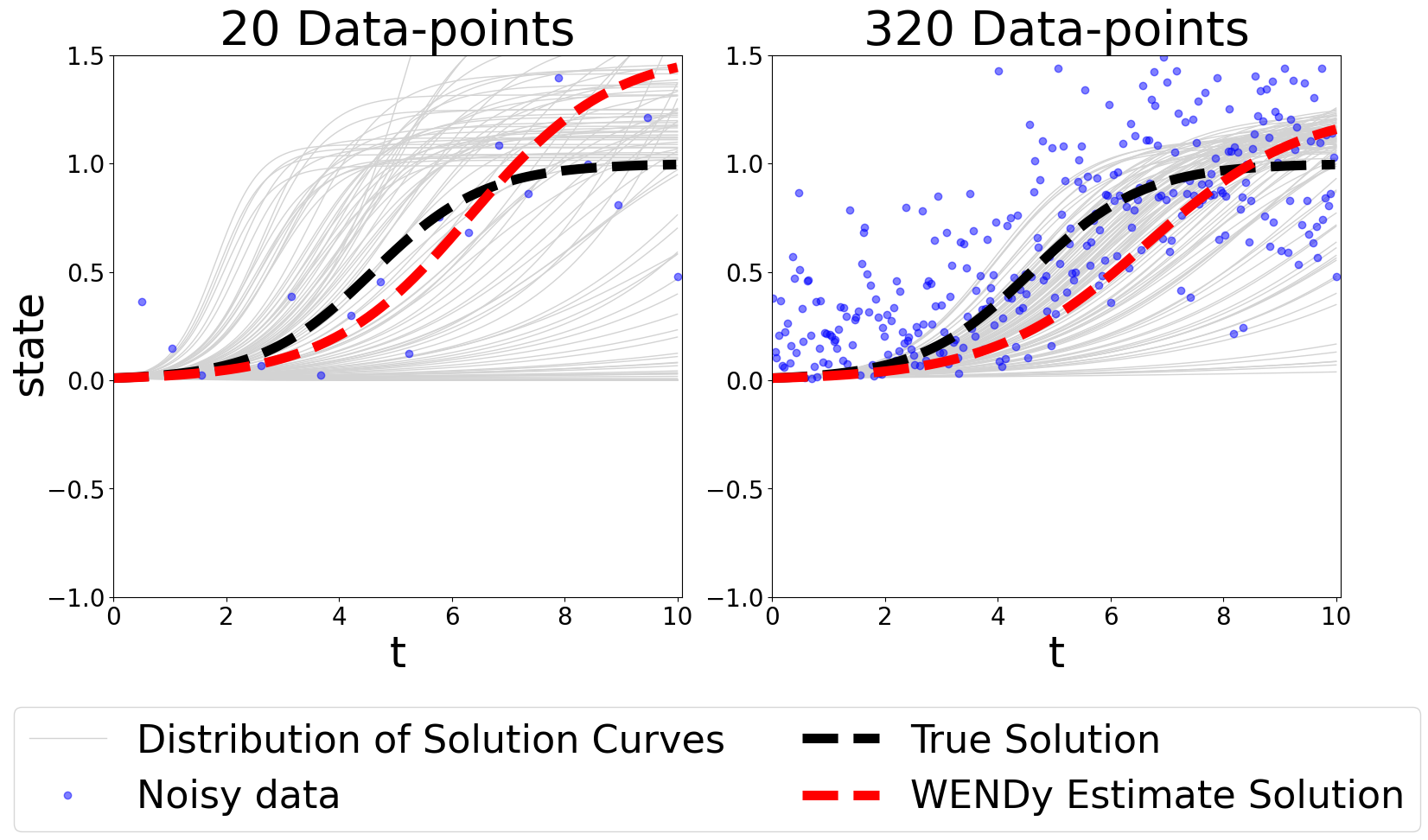} 
    \\
    \text{(a)}
    \\
    \includegraphics[width=1\linewidth]{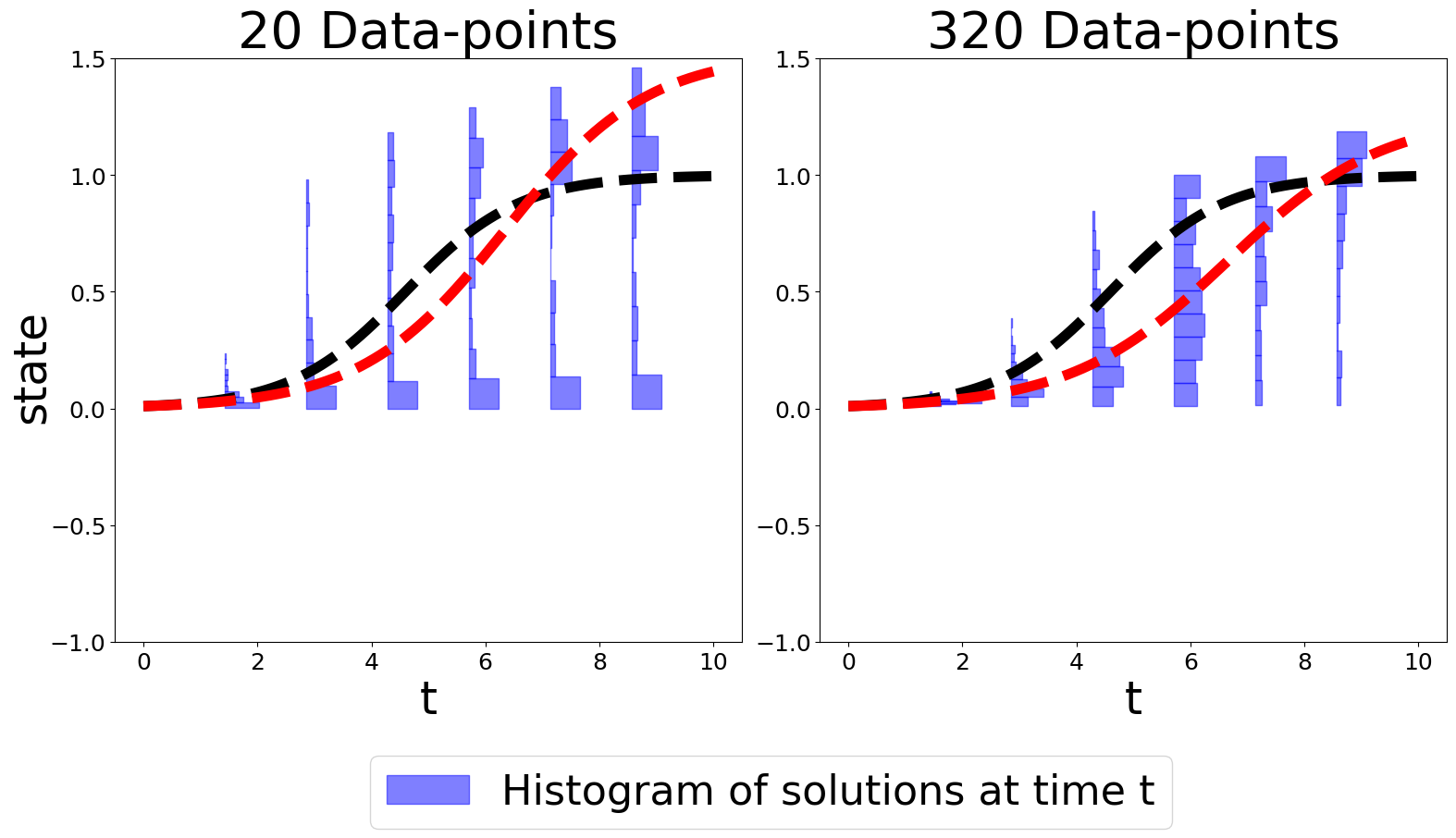}
    \\
    \text{(b)}
    \end{tabular}
     \caption{Top row (a): Logistic model parameter estimation example and uncertainty quantification on two datasets: one dataset with low (left) and one dataset high dataset (right). The light gray curves are used to illustrate the uncertainty around the WENDy solutions; they are obtained via parametric bootstrap, as a sample of WENDy solutions corresponding to a random sample of 1000 parameters from their estimated asymptotic estimator distribution.  Bottom row (b): WENDy solution and histograms of state distributions across specific points in time for the datasets in (a).}
    \label{fig:SamplePlotResLogisticTN}
\end{figure}

As Figure~\ref{fig:SamplePlotResLogisticTN}(a) and (b) show, the distribution of solution states (at selected time-points) for lower resolution levels is bimodal for higher magnitude data-points, while for higher resolution levels, the distribution is unimodal but skewed.


\subsection{Lotka-Volterra}
\subsubsection{Varying Noise Level}
\subsubsection*{Additive Normal Noise}
\begin{figure} 
    \centering
    \begin{tabular}{c}
{\includegraphics[width=0.8\linewidth]{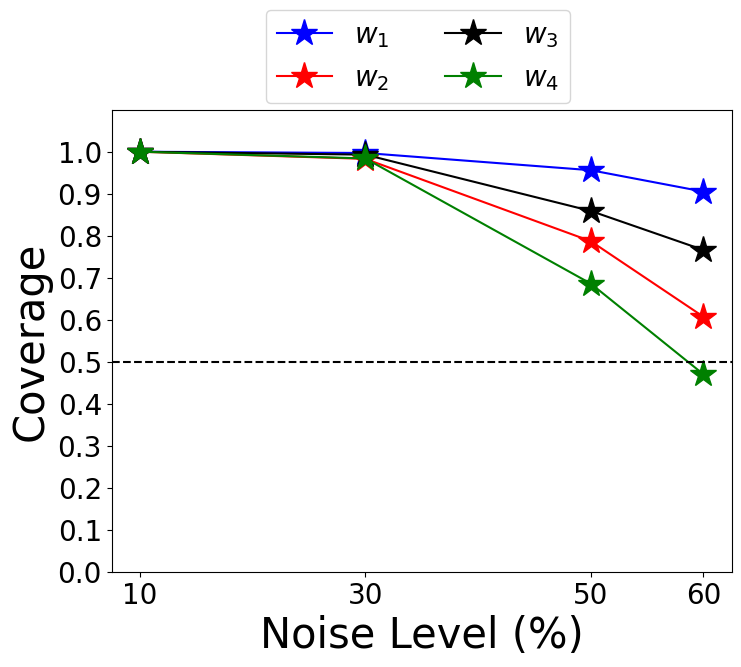}}
    \\
    \text{(a)}
    \\
    \includegraphics[width=1\linewidth]{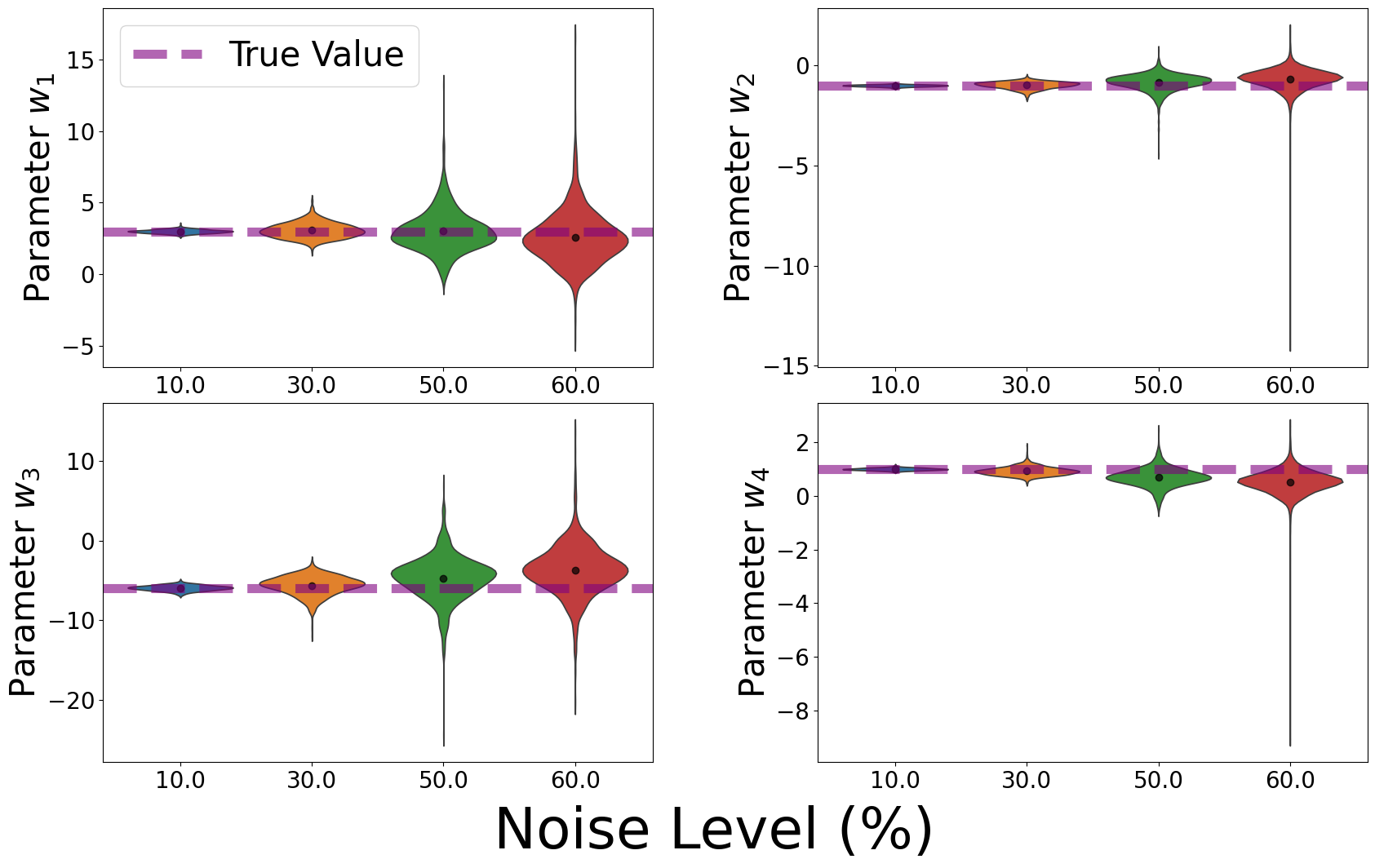}
    \\
      \text{(b)} 
    \end{tabular}
     \caption{Lotka-Volterra model parameter estimation performance with increasing additive normal noise (1000 datasets per level, 205 data points each). (a) coverage across four noise levels. (b) violin plots of parameter estimates, with the dashed red line indicating the true parameter values.}
    \label{fig:CovBiasLVN}
\end{figure}

As shown in Figure~\ref{fig:CovBiasLVN}(a), the coverage of the 95\% confidence intervals for all the parameters remained slightly above the nominal 95\% level at 10\% and 30\% noise before decreasing at 50\% and 60\% noise. The coverage for the $w_4$ parameter dropped below 50\% at 60\% noise. The $w_4$ parameter modulates the quadratic term in the LV model, and its coverage is the most sensitive to noise, as observed across all noise distributions. As shown in Figure~\ref{fig:CovBiasLVN}(b), the bias and variance for all parameters increased with higher noise levels, as expected.

\begin{figure} 
    \centering
    \begin{tabular}{c}
    
    \includegraphics[width=1\linewidth]{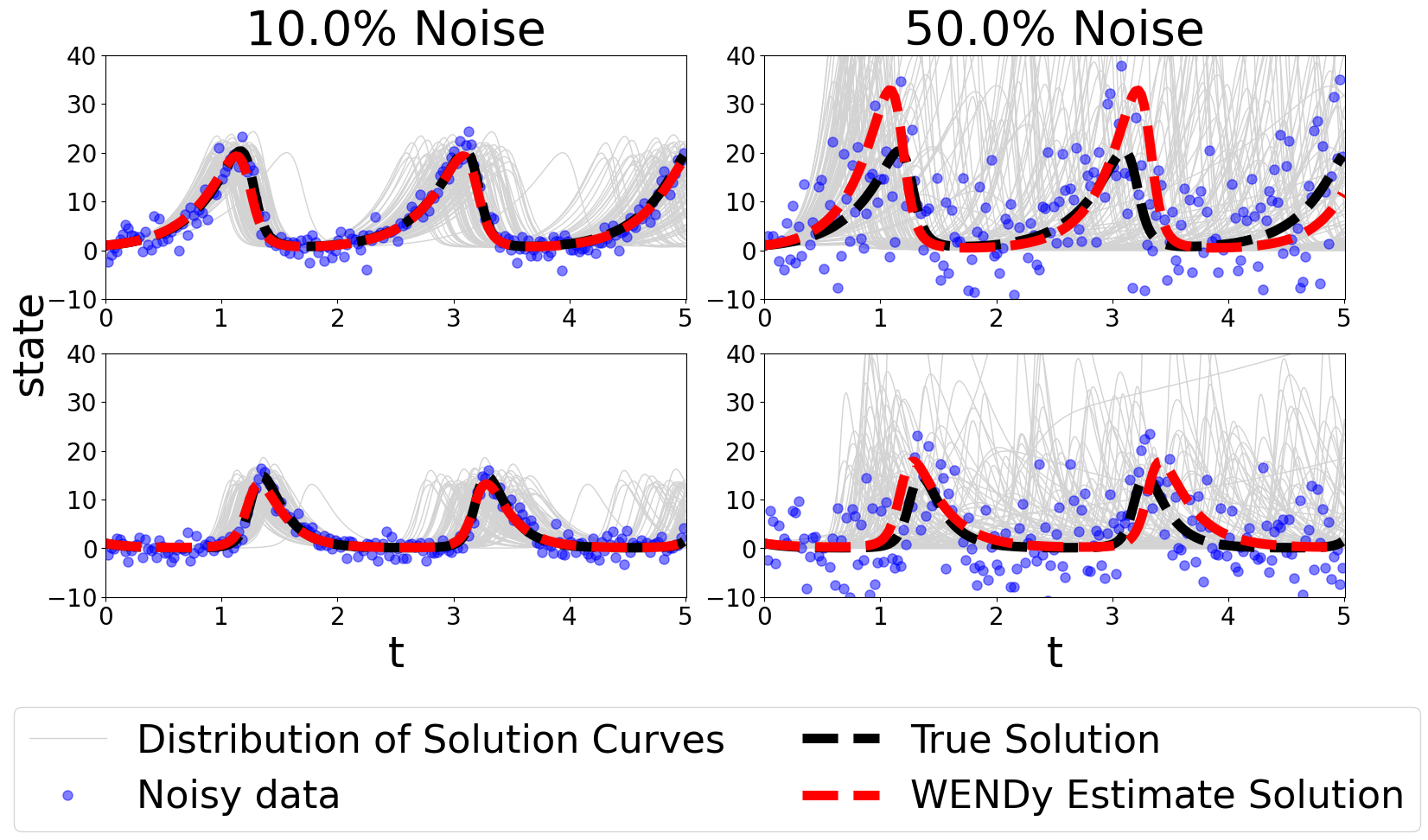} 
    \\
    \text{(a)}
    \\
    \includegraphics[width=1\linewidth]{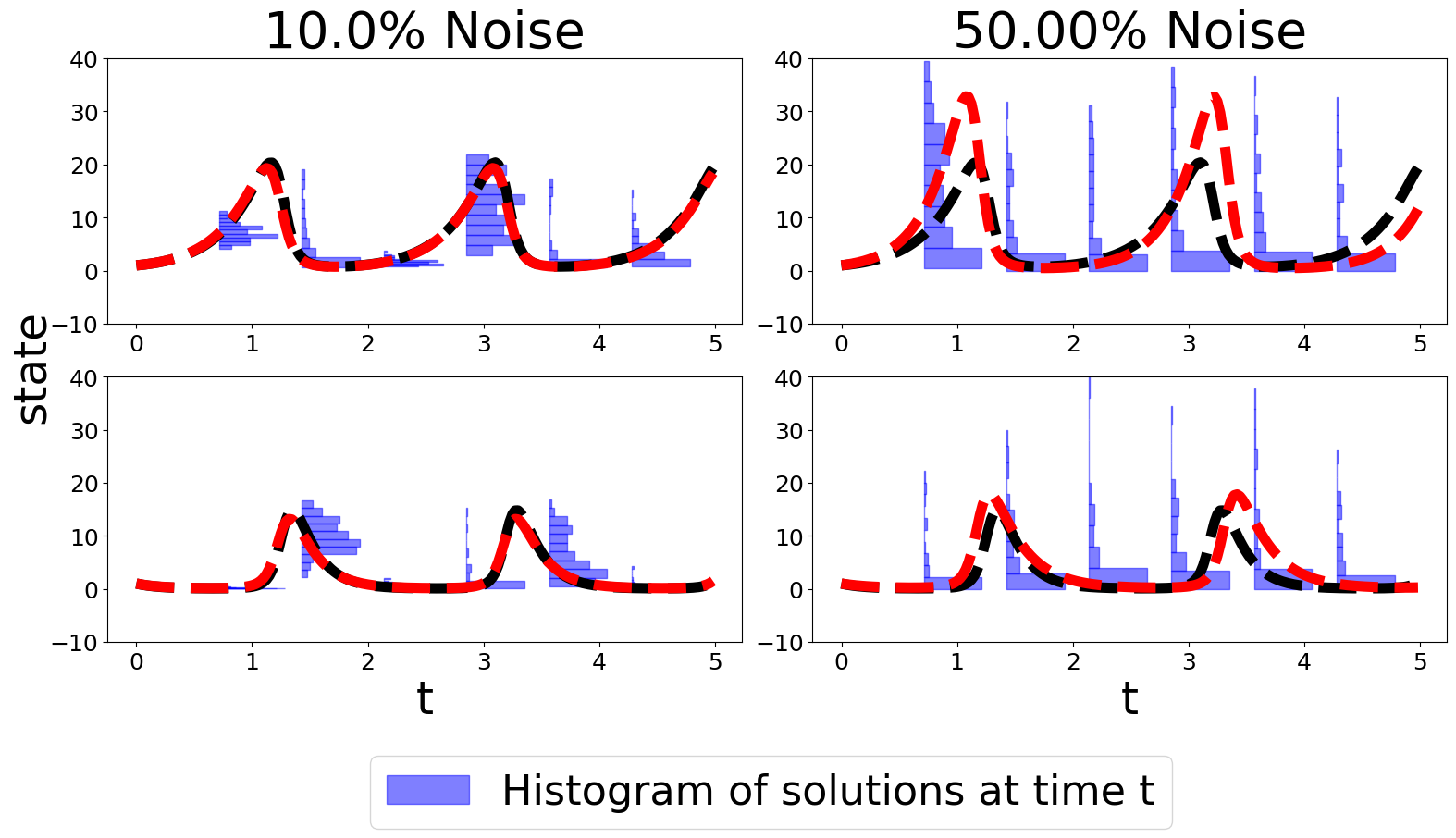}
    \\
    \text{(b)}
    \end{tabular}
     \caption{Top row (a): Lotka-Volterra model parameter estimation example and uncertainty quantification on two datasets: one dataset with low uncertainty and high coverage (left) and one dataset with high uncertainty and low coverage (right). The light gray curves are used to illustrate the uncertainty around the WENDy solutions; they are obtained via parametric bootstrap, as a sample of WENDy solutions corresponding to a random sample of 1000 parameters from their estimated asymptotic estimator distribution.  Bottom row (b): WENDy solution and histograms of state distributions across specific points in time for the datasets in (a).}
    \label{fig:SamplePlotLVN}
\end{figure}

As Figures~\ref{fig:SamplePlotLVN}(a) and (b) show, the distribution of states (values of solution curves at different time points) for higher noise levels is more skewed compared to lower noise levels. The high number of negative state values at higher noise levels likely makes parameter identifiability more difficult, since the negative state values do not contribute to the dynamics. This may explain why WENDy estimates produce a poor fit at higher noise levels.

\clearpage

\subsubsection*{Additive Censored Normal Noise}
\begin{figure}
    \centering
    \begin{tabular}{c}
{\includegraphics[width=0.8\linewidth]{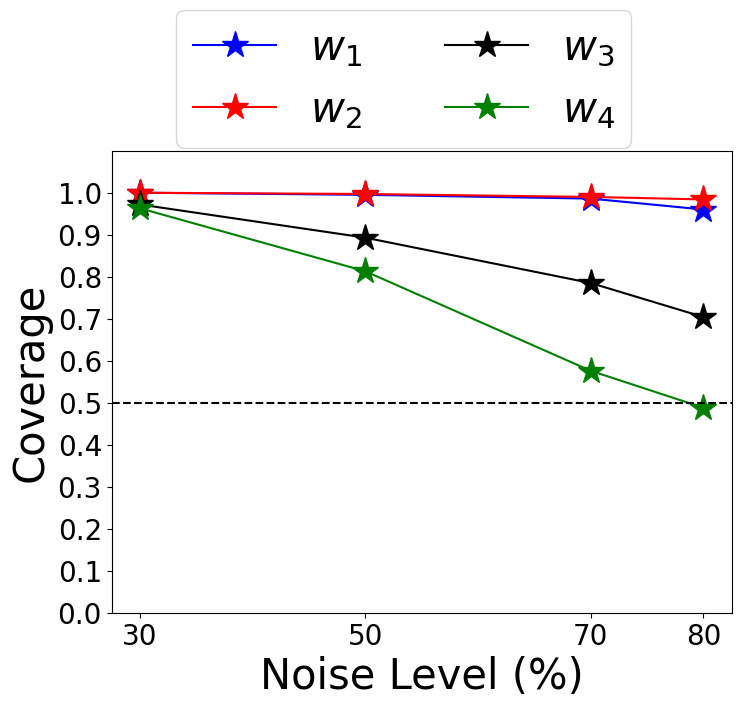}}
    \\
    \text{(a)}
\\    \includegraphics[width=1\linewidth]{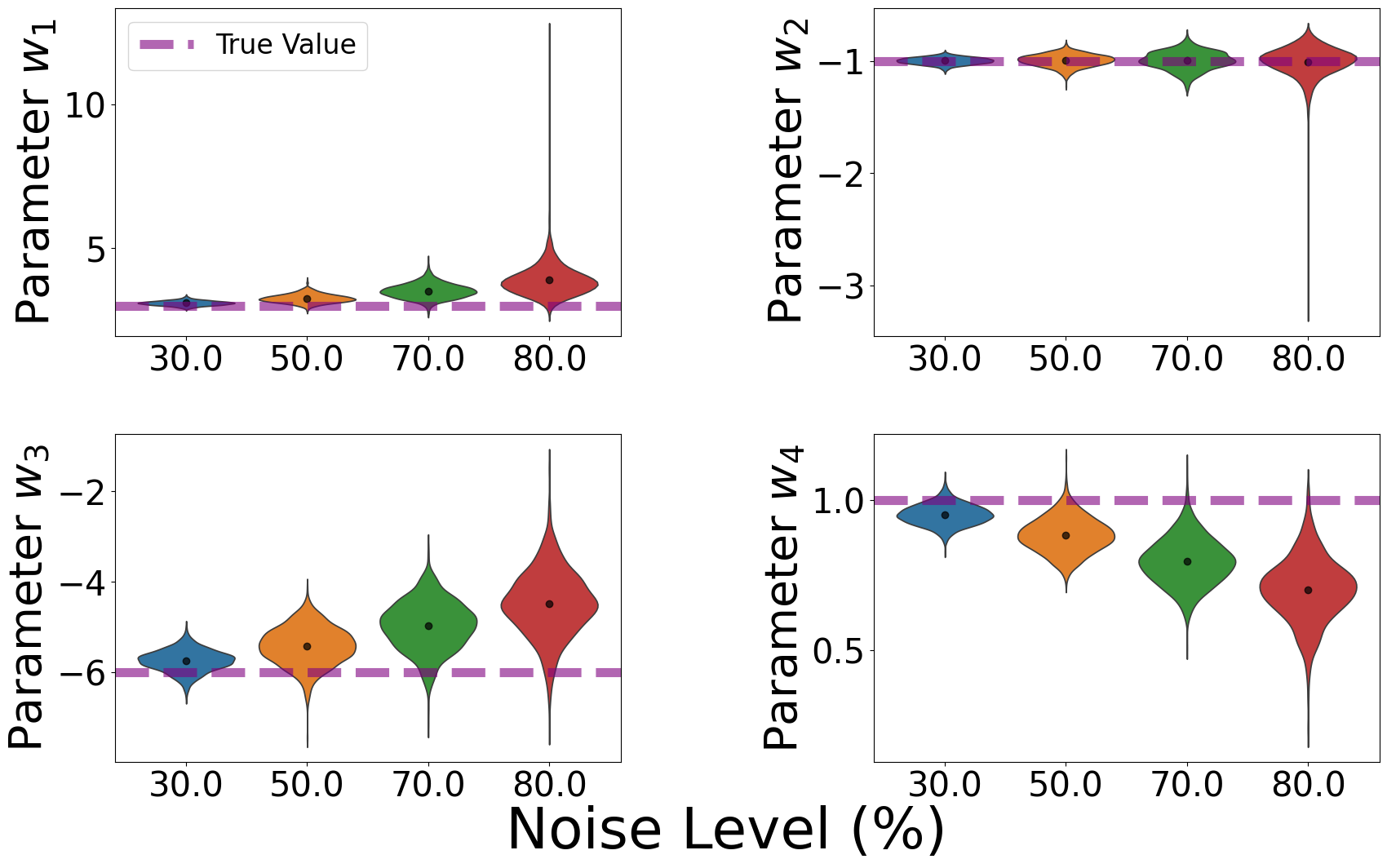}
    \\
      \text{(b)} 
    \end{tabular}
     \caption{Lotka-Volterra model parameter estimation performance with increasing ACN noise (1000 datasets per level, 205 data points each). (a) coverage across four noise levels. (b) violin plots of parameter estimates, with the dashed red line indicating the true parameter values.}
    \label{fig:CovBiasLVCN}
\end{figure}

As shown in Figure~\ref{fig:CovBiasLVCN}(a), the coverage of the 95\% confidence intervals for the $w_1$ and $w_2$ parameters remained slightly above the nominal 95\% level for all noise levels from 30\% to 80\%. However, the coverage of the 95\% confidence intervals for the $w_3$ and $w_4$ parameters consistently decreased with increasing noise levels, starting at the nominal 95\% coverage and dropping below 50\% for $w_4$ at 80\% noise. As shown in Figure~\ref{fig:CovBiasLVCN}(b), the bias for all parameters increased slightly with higher noise levels. Notably, the variance of the $w_3$ and $w_4$ parameters increased steadily across noise levels, whereas the variance of the $w_1$ and $w_2$ parameters rose sharply at 80\%, as indicated by the elongated tails of the violin plots. This difference in variance behavior helps explain the contrasting coverage results between the two parameter pairs at high noise levels.  

\begin{figure} 
    \centering
    \begin{tabular}{c}
    
    \includegraphics[width=1\linewidth]{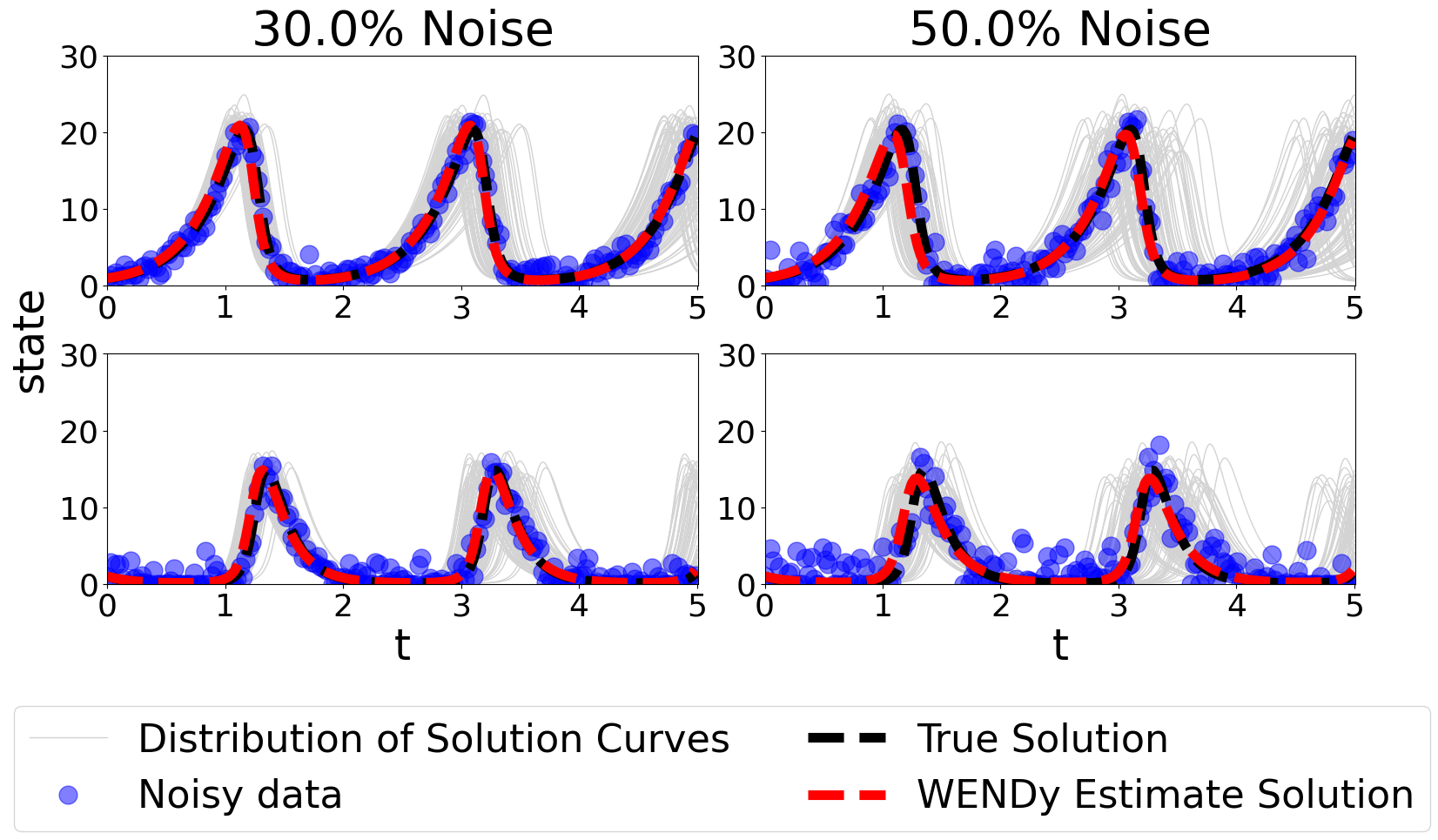} 
    \\
    \text{(a)}
    \\
    \includegraphics[width=1\linewidth]{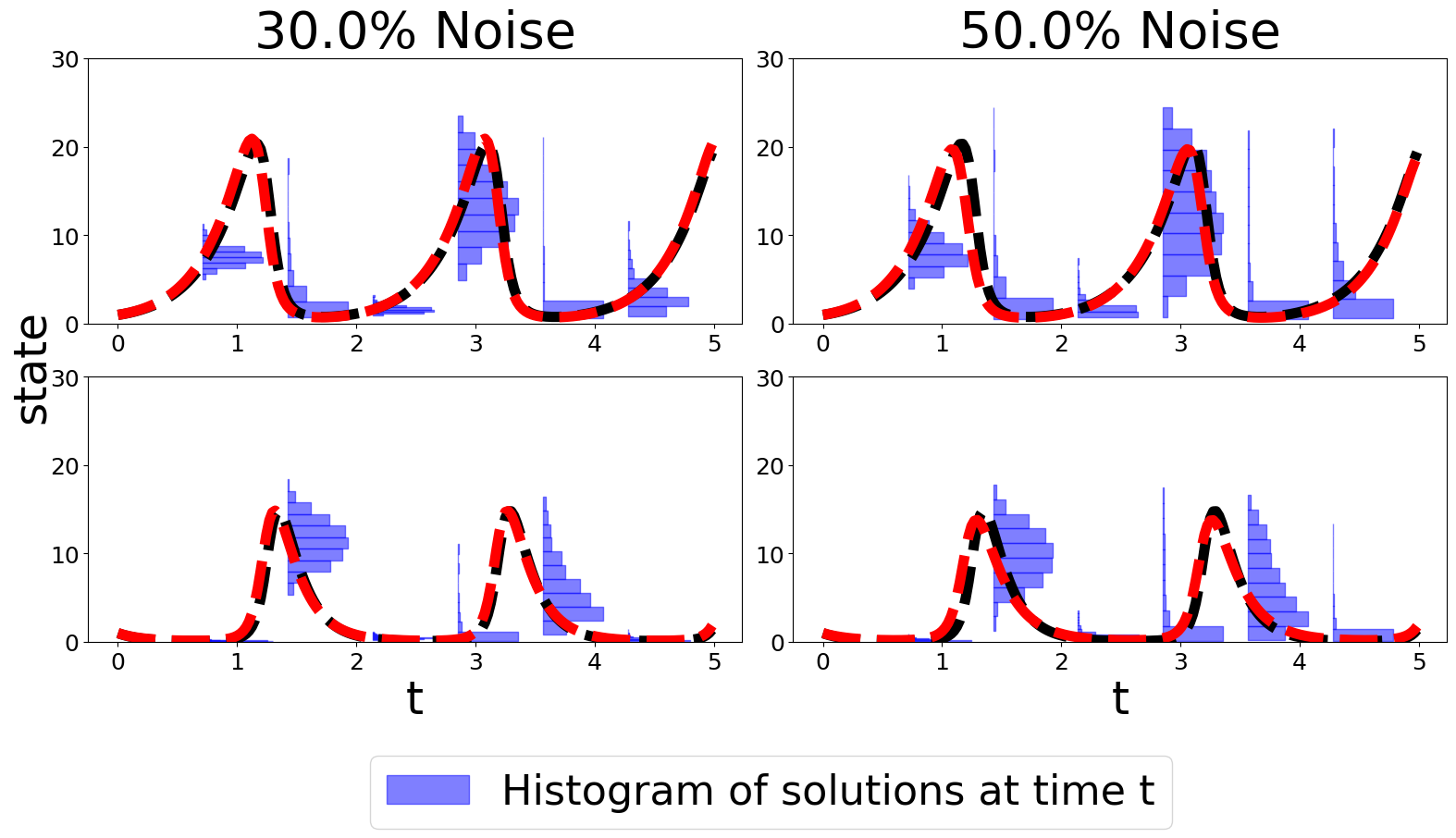}
    \\
    \text{(b)}
    \end{tabular}
     \caption{Top row (a): Lotka-Volterra model parameter estimation example and uncertainty quantification on two datasets: one dataset with low uncertainty and high coverage (left) and one dataset with high uncertainty and low coverage (right). The light gray curves are used to illustrate the uncertainty around the WENDy solutions; they are obtained via parametric bootstrap, as a sample of WENDy solutions corresponding to a random sample of 1000 parameters from their estimated asymptotic estimator distribution.  Bottom row (b): WENDy solution and histograms of state distributions across specific points in time for the datasets in (a).}
    \label{fig:SamplePlotLVCN}
\end{figure}

As Figures~\ref{fig:SamplePlotLVCN}(a) and (b) show, the distribution of states (values of solution curves at different time points) for lower and higher noise levels remained relatively similar, being more skewed at data points with low magnitude and showing greater variance at higher noise levels. Nevertheless, the fits for both noise levels were reasonable, since ACN noise at higher-magnitude points behaves like normal noise, making the peaks more identifiable and allowing for reasonable fits.

\clearpage
\subsubsection*{Multiplicative Log-Normal Noise}
\begin{figure} 
    \centering
    \begin{tabular}{c}
{\includegraphics[width=0.8\linewidth]{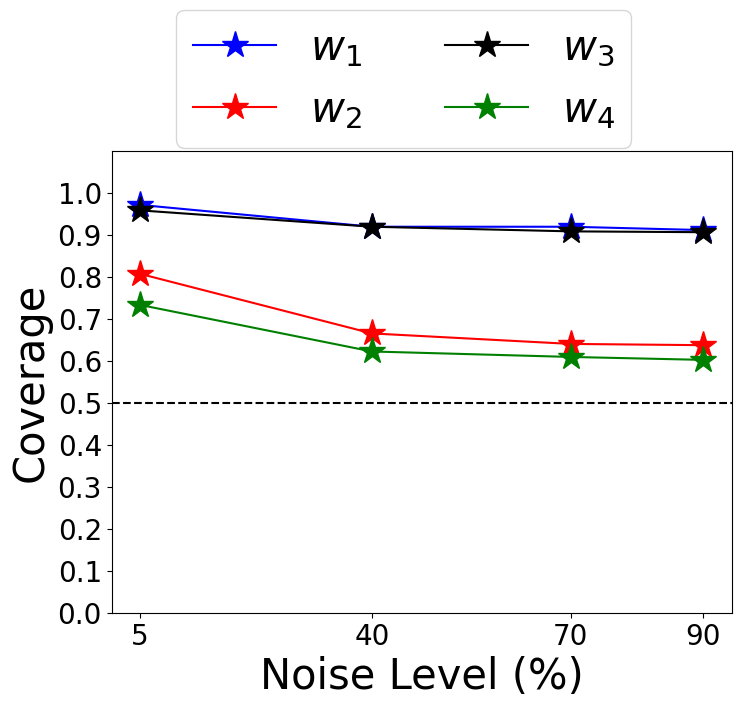}}
    \\
    \text{(a)}
    \\
    \includegraphics[width=1\linewidth]{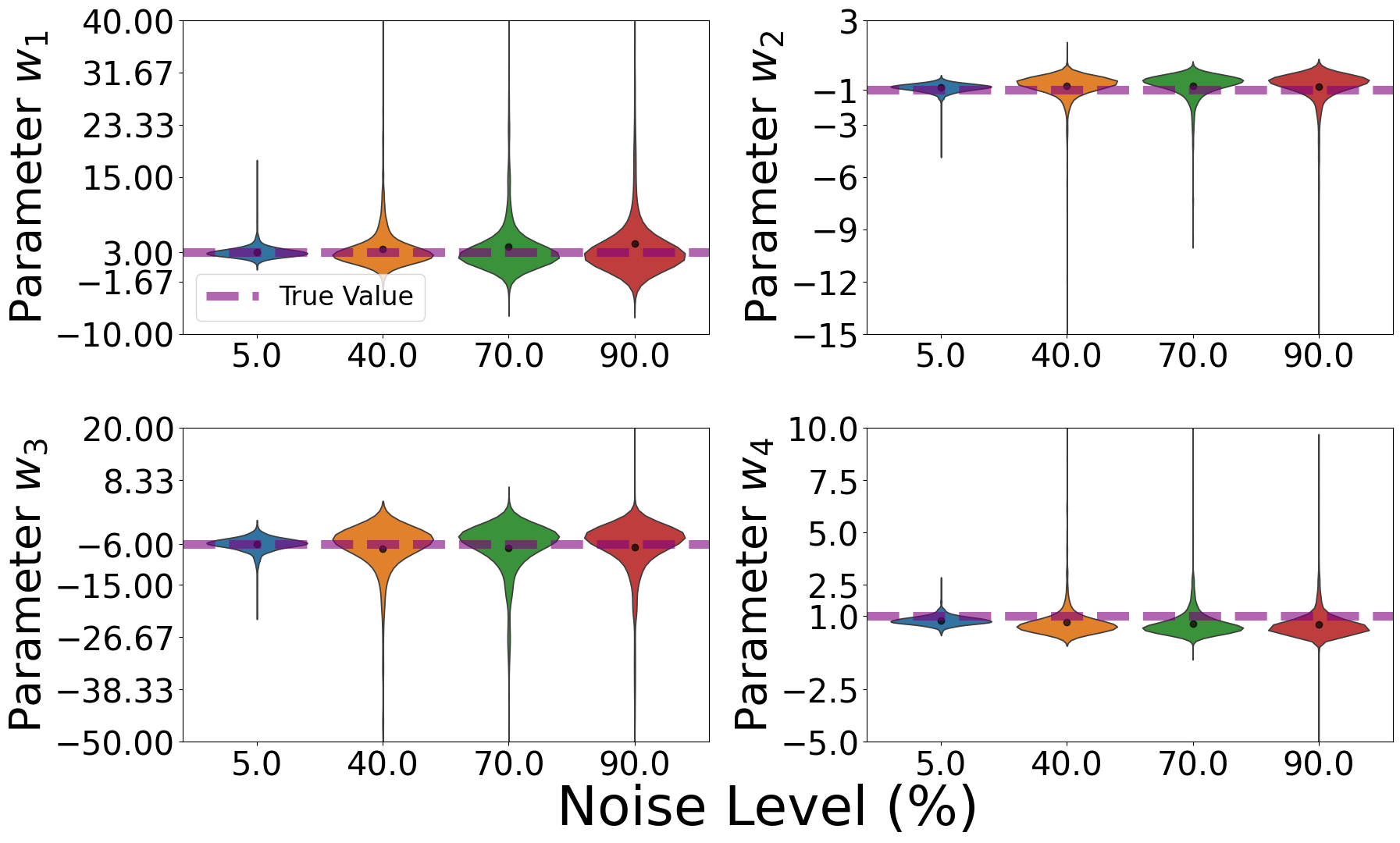}
    \\
      \text{(b)} 
    \end{tabular}
     \caption{Lotka-Volterra model parameter estimation performance with increasing MLN noise (1000 datasets per level, 205 data points each). (a) coverage across four noise levels. (b) violin plots of parameter estimates, with the dashed red line indicating the true parameter values.}
    \label{fig:CovBiasLVLN}
\end{figure}

As shown in Figure~\ref{fig:CovBiasLVLN}(a), the coverage of the 95\% confidence intervals for the $w_1$ and $w_2$ parameters remained near the nominal 95\% level for all noise levels from 5\% to 90\%. In contrast, the coverage of the 95\% confidence intervals for the $w_3$ and $w_4$ parameters decreased as noise increased from 5\% to 40\%, but then plateaued at higher noise levels. As shown in Figure~\ref{fig:CovBiasLVLN}(b), the bias for all parameters increased slightly with higher noise levels. The variance for all parameters increased significantly, especially compared to other noise distributions for the LV model. This explains the relatively steady coverage for all parameters.  

\begin{figure} 
    \centering
    \begin{tabular}{c}
    
    \includegraphics[width=1\linewidth]{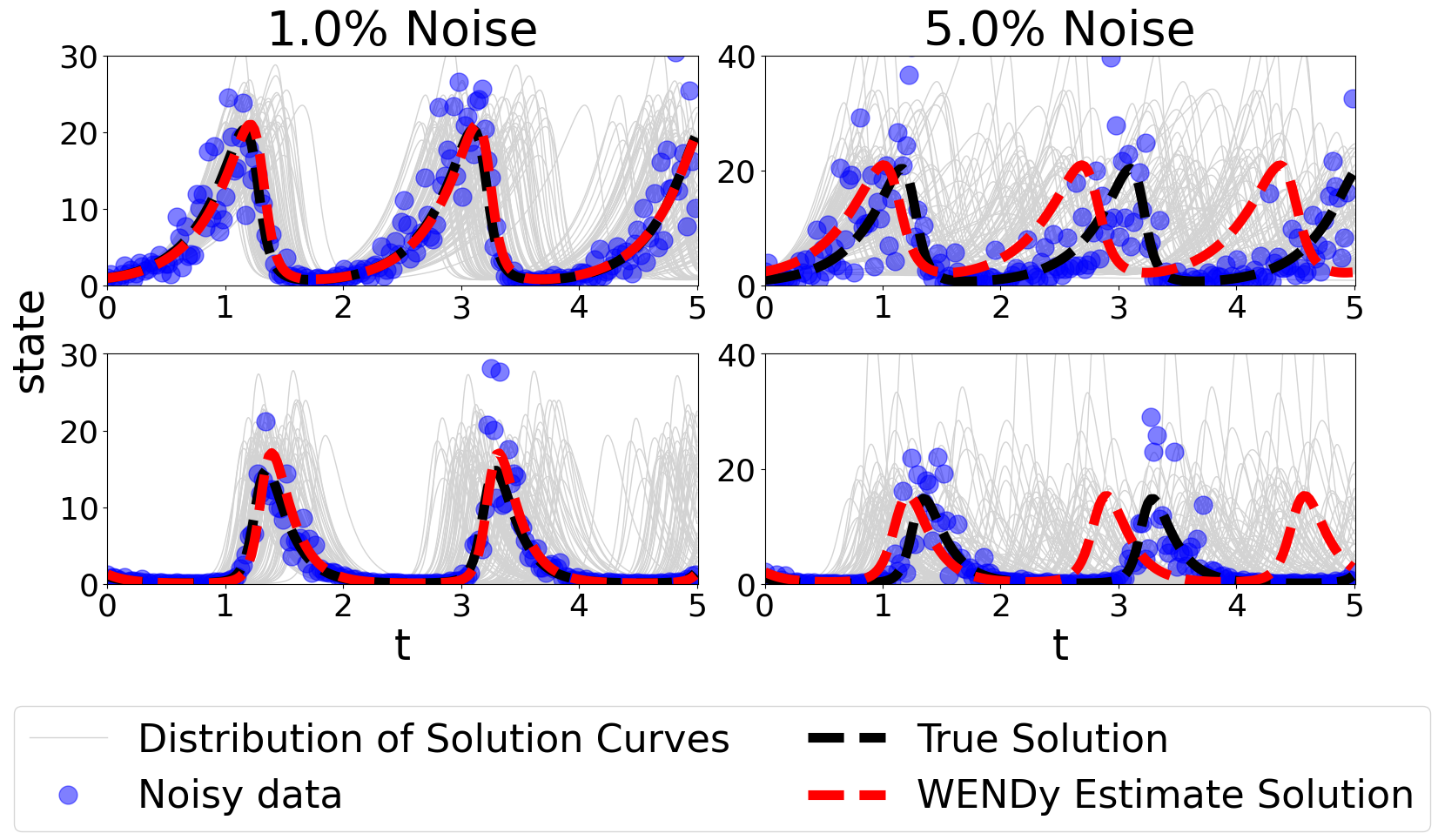} 
    \\
    \text{(a)}
    \\
    \includegraphics[width=1\linewidth]{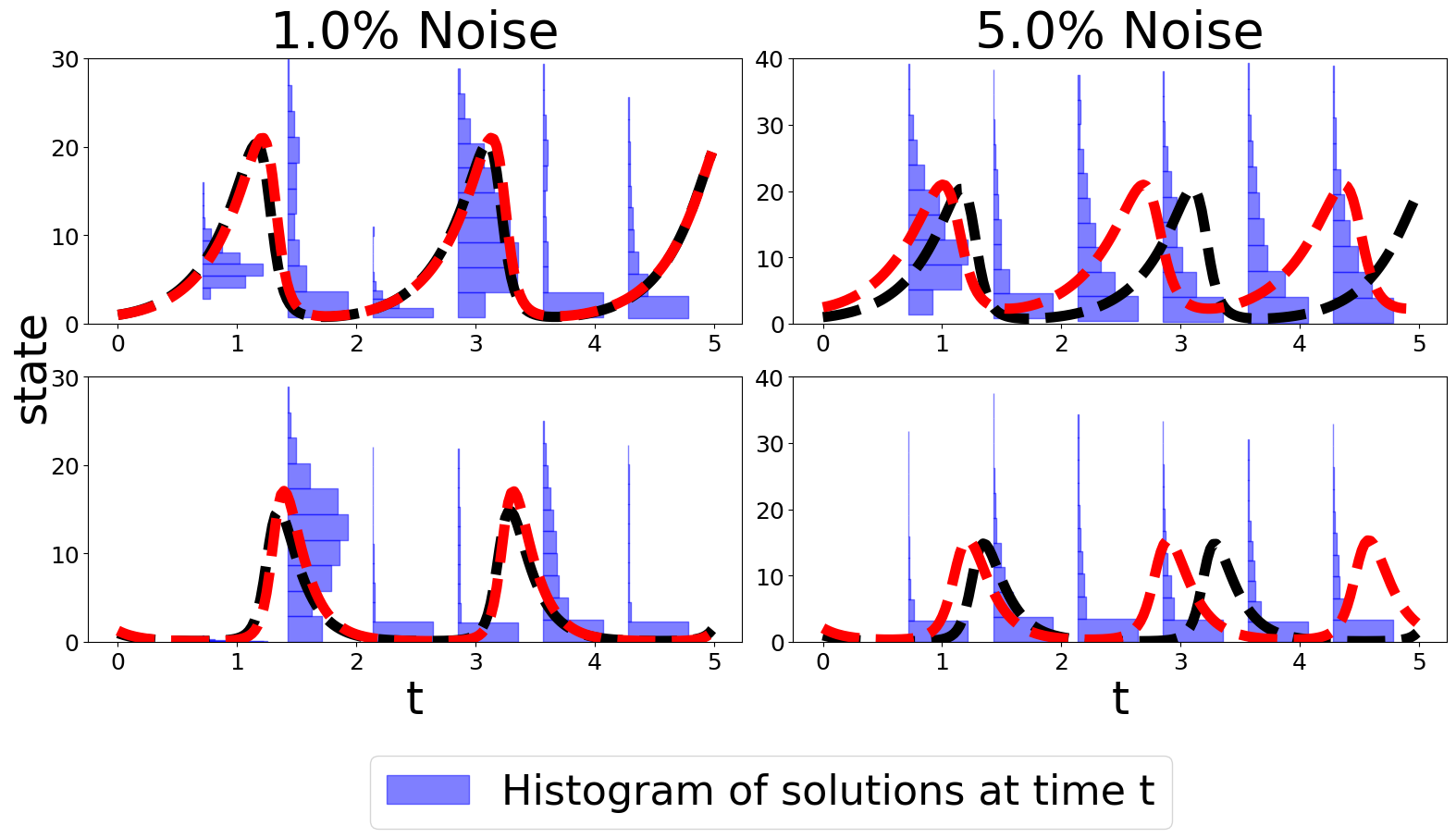}
    \\
    \text{(b)}
    \end{tabular}
     \caption{Top row (a): Lotka-Volterra model parameter estimation example and uncertainty quantification on two datasets: one dataset with low uncertainty and high coverage (left) and one dataset with high uncertainty and low coverage (right). The light gray curves are used to illustrate the uncertainty around the WENDy solutions; they are obtained via parametric bootstrap, as a sample of WENDy solutions corresponding to a random sample of 1000 parameters from their estimated asymptotic estimator distribution.  Bottom row (b): WENDy solution and histograms of state distributions across specific points in time for the datasets in (a).}
    \label{fig:SamplePlotLVLN}
\end{figure}

As illustrated in Figures~\ref{fig:SamplePlotLVLN}(a) and (b), the distribution of states (values of solution curves at different time points) became more skewed at higher noise levels, with noticeable phase shifts relative to the true solution. The heteroskedastic nature of MLN noise makes signal identification more difficult, which explains the phase shift of the estimated solutions. Noise levels above 5\% created substantial phase shifts and highly uncertain solution clouds; therefore, they were excluded here.

\clearpage
\subsubsection*{Additive Truncated Normal Noise}
\begin{figure} 
\clearpage
\centering
    \begin{tabular}{c}
{\includegraphics[width=0.8\linewidth]{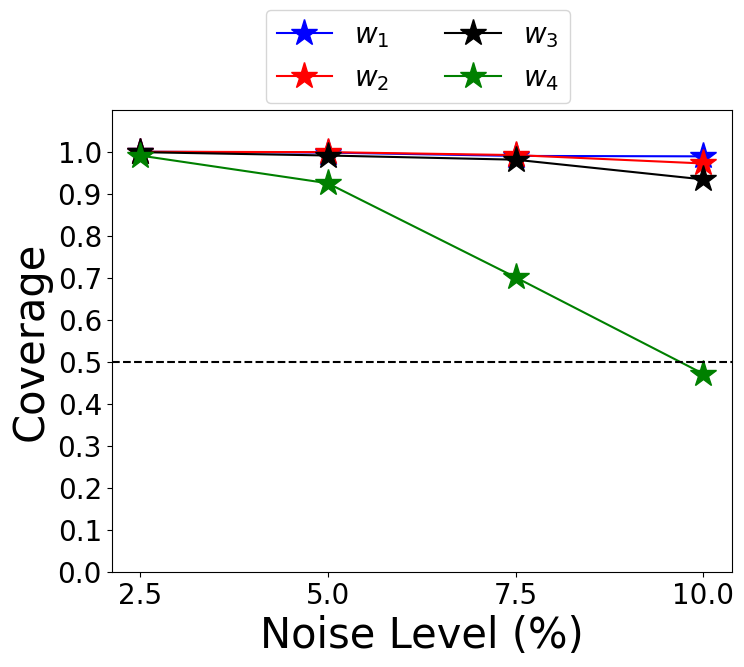}}
    \\
    \text{(a)}
    \\
    \includegraphics[width=1\linewidth]{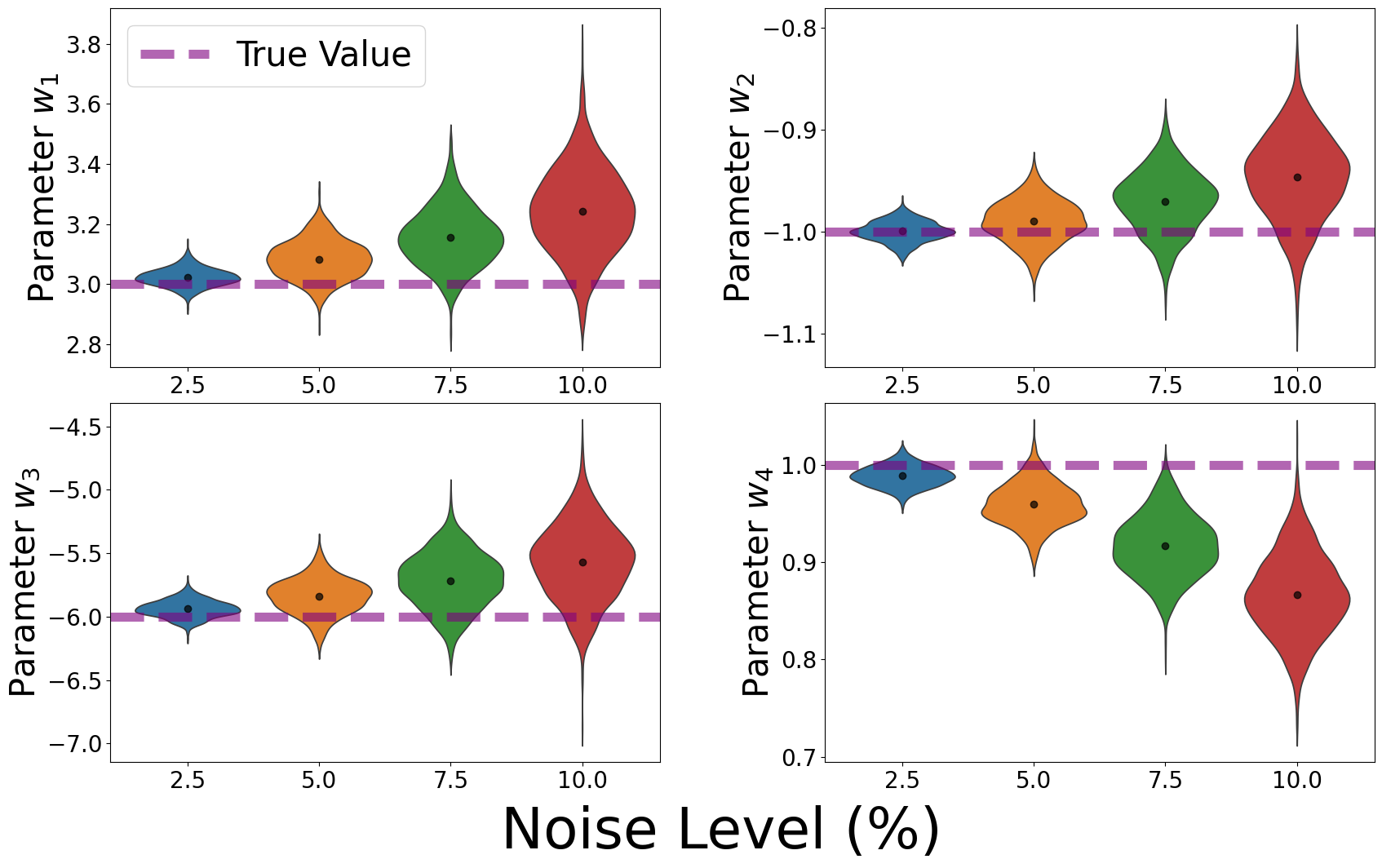}
    \\
      \text{(b)} 
    \end{tabular}
     \caption{Lotka-Volterra model parameter estimation performance with increasing ATN noise (1000 datasets per level, 205 data points each). (a) coverage across four noise levels. (b) violin plots of parameter estimates, with the dashed red line indicating the true parameter values.}
    \label{fig:CovBiasLVTN}
\end{figure}

As shown in Figure~\ref{fig:CovBiasLVTN}(a), the coverage of the 95\% confidence intervals for the $w_1$ and $w_2$ parameters remained slightly above the nominal 95\% level across all noise levels from 2.5\% to 10\%. The coverage for the $w_4$ parameter consistently decreased with increasing noise, starting at nominal coverage and dropping below 50\% at 10\% noise. As shown in Figure~\ref{fig:CovBiasLVTN}(b), the bias for all parameters increased modestly with higher noise levels. The $w_4$ parameter violin plot visually depicts the poor coverage of this parameter, as the violin distribution deviates from the true value.  

\begin{figure} 
    \centering
    \begin{tabular}{c}
    
    \includegraphics[width=1\linewidth]{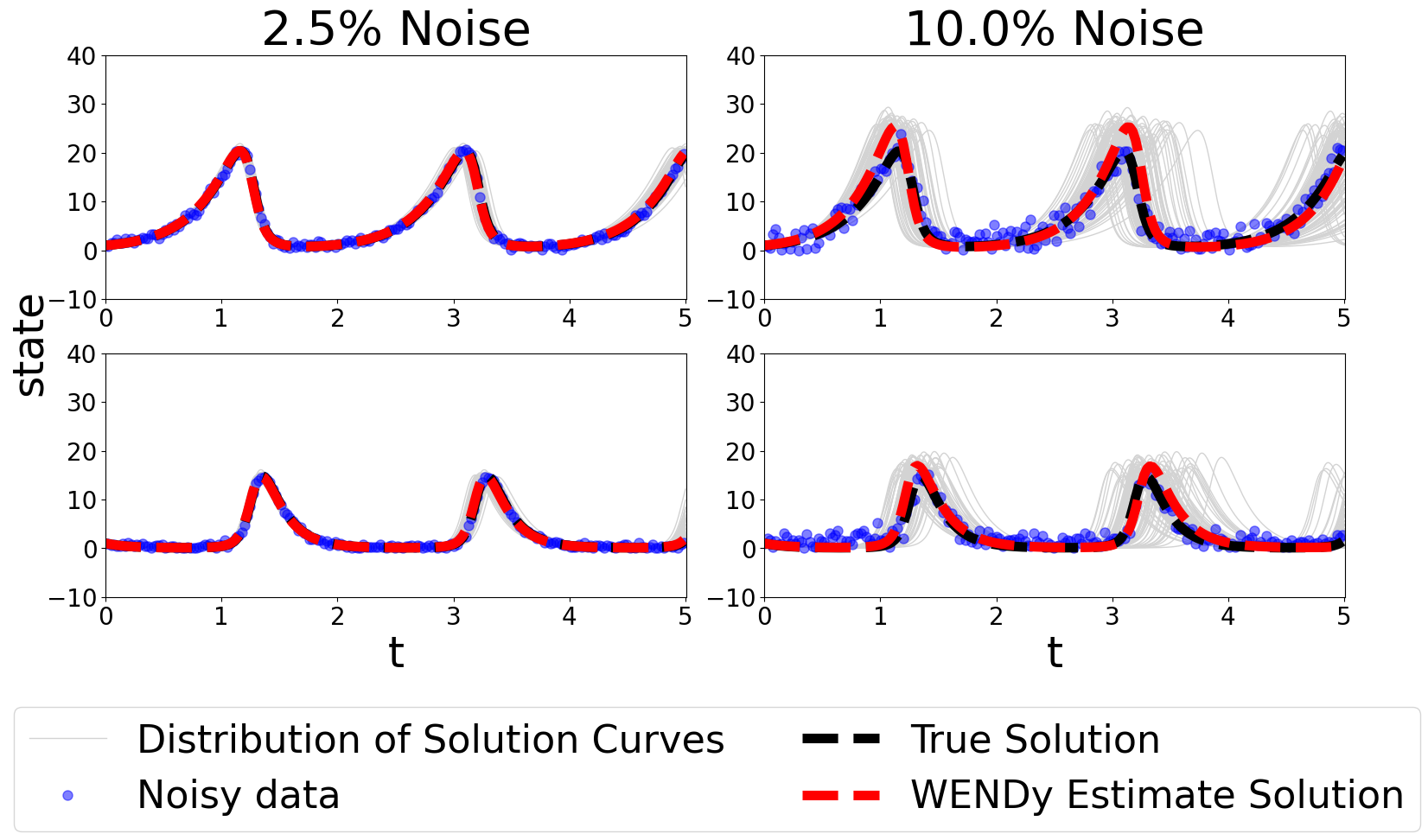} 
    \\
    \text{(a)}
    \\
    \includegraphics[width=1\linewidth]{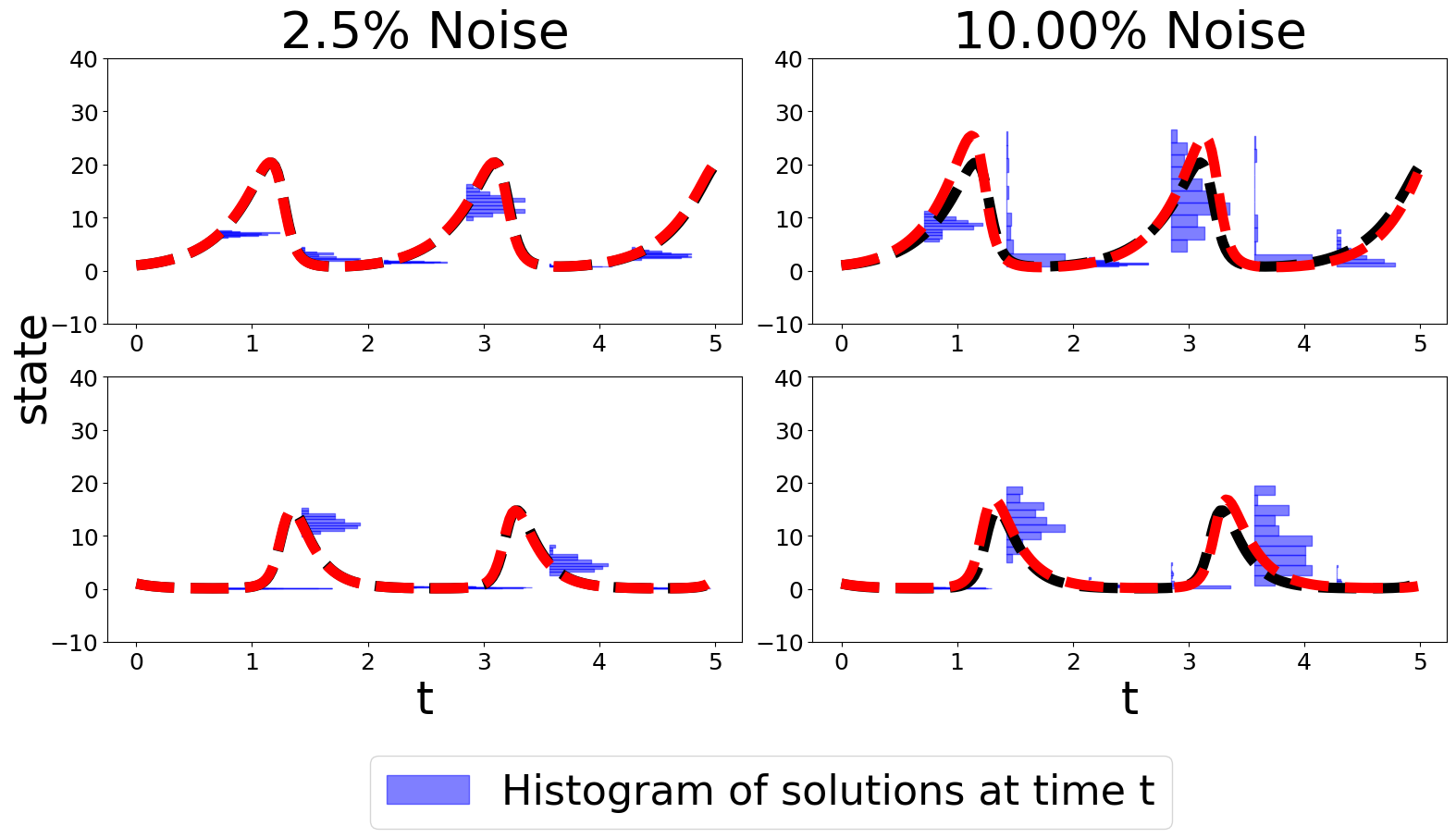}
    \\
    \text{(b)}
    \end{tabular}
     \caption{Top row (a): Lotka-Volterra model parameter estimation example and uncertainty quantification on two datasets: one dataset with low uncertainty and high coverage (left) and one dataset with high uncertainty and low coverage (right). The light gray curves are used to illustrate the uncertainty around the WENDy solutions; they are obtained via parametric bootstrap, as a sample of WENDy solutions corresponding to a random sample of 1000 parameters from their estimated asymptotic estimator distribution.  Bottom row (b): WENDy solution and histograms of state distributions across specific points in time for the datasets in (a).}
    \label{fig:SamplePlotLVTN}
\end{figure}

As illustrated in Figures~\ref{fig:SamplePlotLVTN}(a) and (b), the distribution of solution curves remained largely consistent across noise levels, with increased skew at lower-magnitude data points and greater variance at higher noise. Nevertheless, the fits for both noise levels were reasonable due to the relatively low noise magnitudes.

\begin{figure} 
    \centering
    \includegraphics[width=0.4\linewidth]{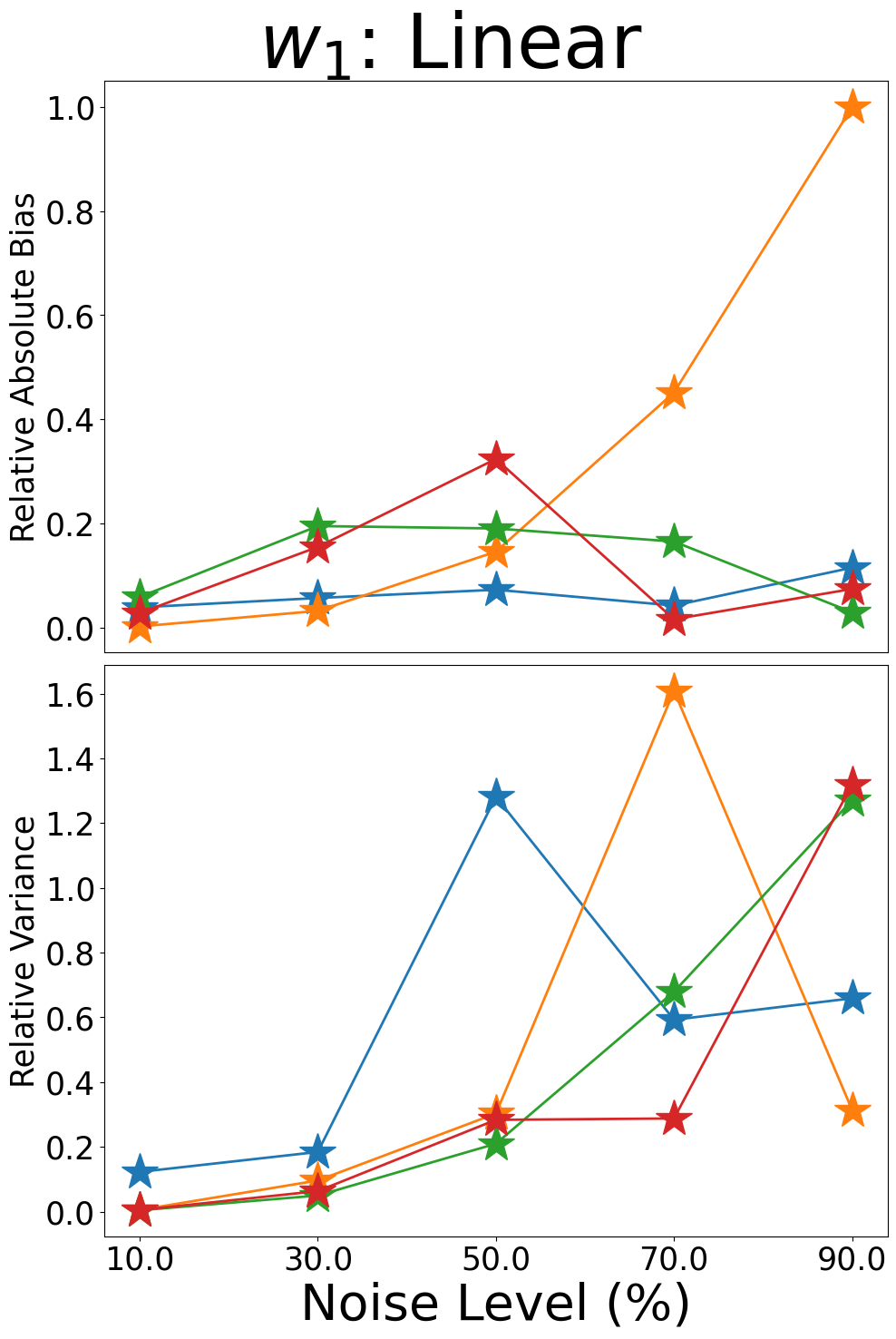}     \includegraphics[width=0.4\linewidth]{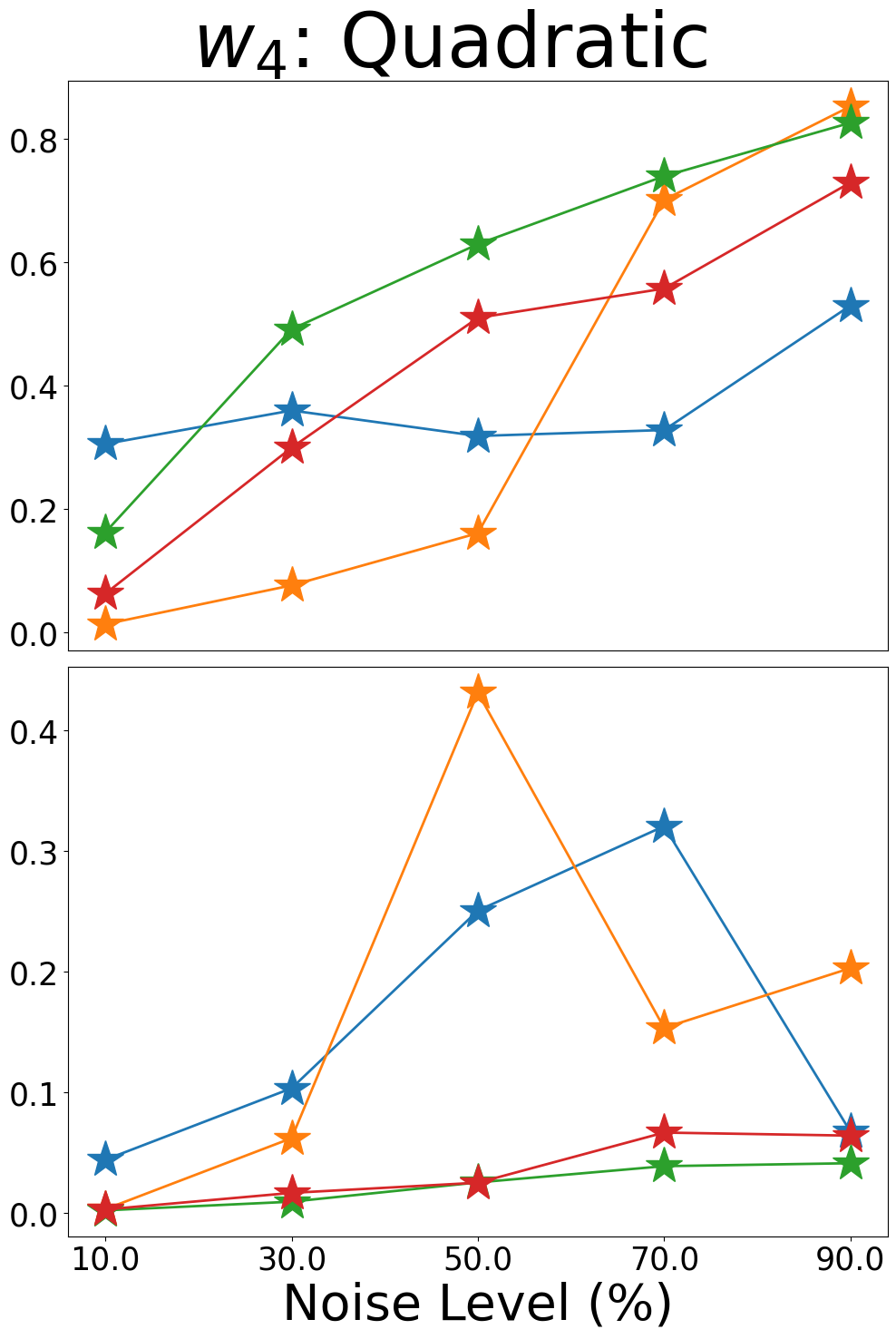} \\
    \includegraphics[width=0.3\linewidth]{NoiseDistrCompPlotsLogsitic/Legend4.png}
    \caption{Relative bias magnitude and variance for WENDy estimators of LV model across 100 datasets with additive normal, MLN, ACN, and ATN.}
    \label{fig:LVBiasVarNoise}
\end{figure}
As seen in Figure~\ref{fig:LVBiasVarNoise}, parameter estimators from datasets with additive noise had among the lowest bias for noise levels from 10\% to 50\%, while normal noise exhibited among the highest bias at higher noise levels. For noise levels below 90\%, it is clear that the $w_1$ parameter, which modulates the linear term in the LV model, had lower bias across almost all noise distributions compared to the $w_4$ parameter, which modulates the quadratic term in the LV model. Furthermore, at higher noise levels, the variance in the $w_4$ parameter estimators was also lower, which could explain its poorer coverage compared to the $w_1$ parameter.  

The noise distribution that produced the lowest bias was inconsistent across parameters; however, WENDy estimators from datasets with MLN noise consistently exhibited one of the lowest biases, which helps explain the stable coverage of WENDy estimators under MLN noise.

\subsubsection{Varying Data Resolution}
\subsubsection*{Additive Normal Noise}
\begin{figure} 
    \centering
    \begin{tabular}{c}
{\includegraphics[width=0.8\linewidth]{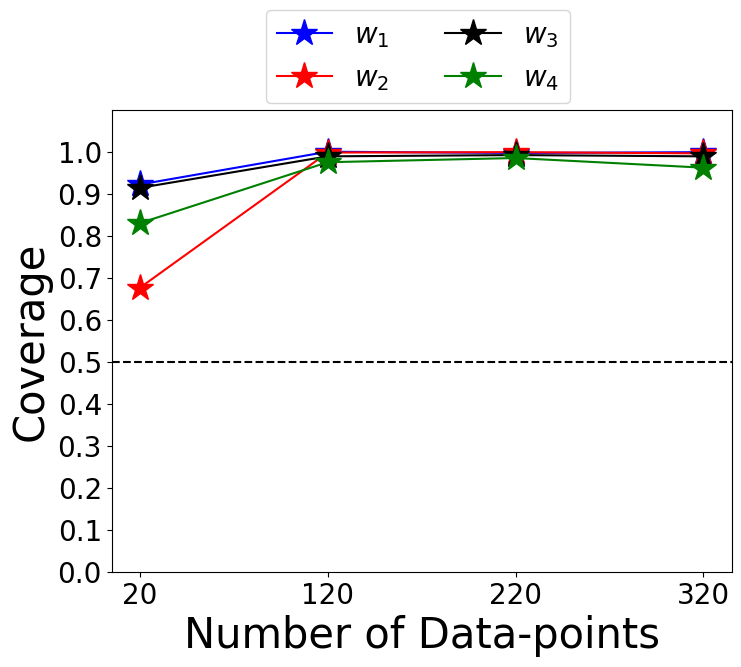}}
    \\
    \text{(a)}
    \\
    \includegraphics[width=1\linewidth]{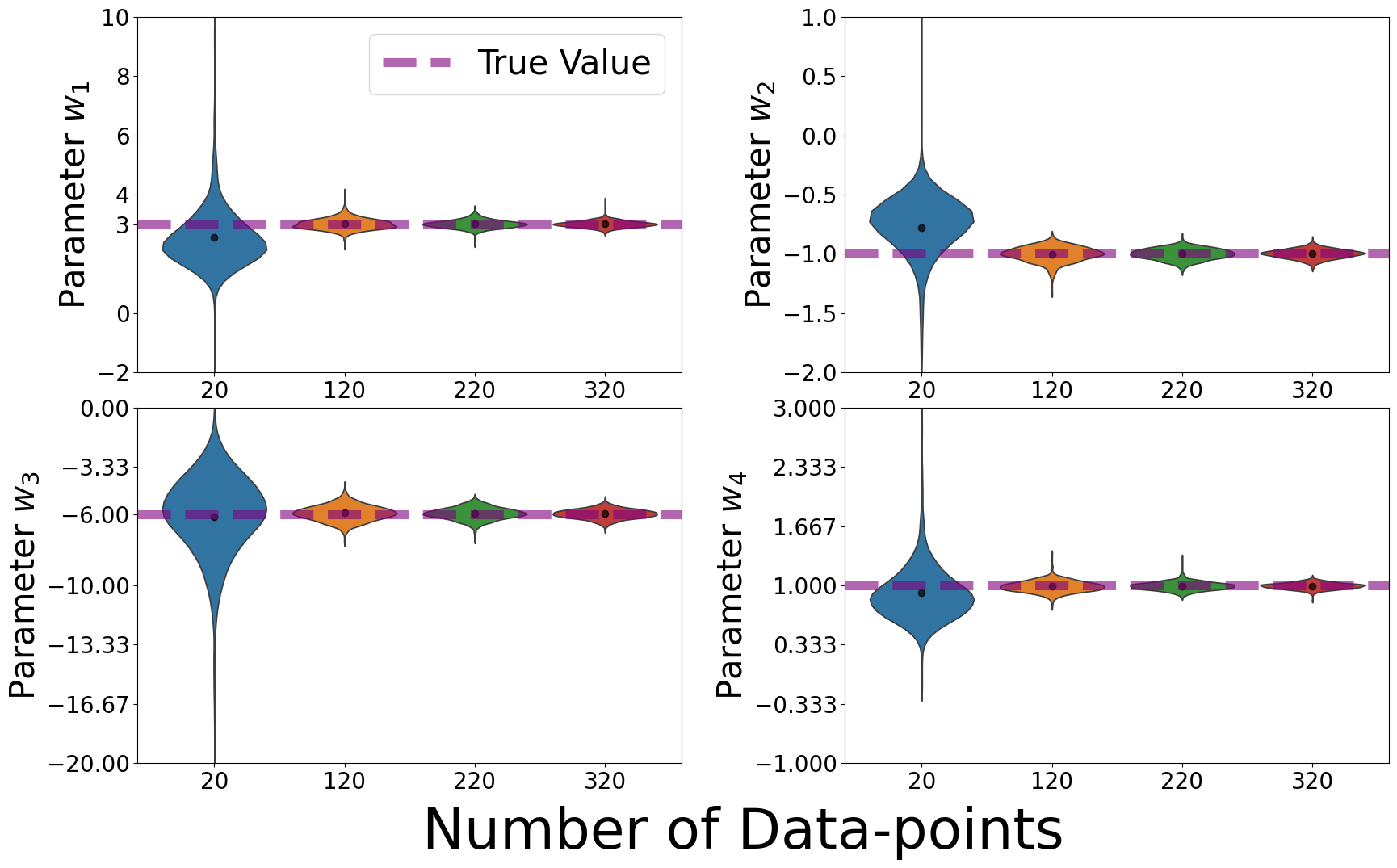}
    \\
      \text{(b)} \;\;\;\;\;\;\;\;\;\;\;\;\;\;\;\;\;\;\;\;\;\;\;\;\;\;\;\;\;\;\;
    \end{tabular}
     \caption{Lotka-Volterra model parameter estimation performance with increasing data resolution (1000 datasets per level, 30\% normal noise). (a) coverage across four noise levels. (b) violin plots of parameter estimates, with the dashed red line indicating the true parameter values.}
    \label{fig:CovBiasResLVN}
\end{figure}

As shown in Figure~\ref{fig:CovBiasResLVN}(a), the coverage of the 95\% confidence intervals for all the parameters was slightly below the nominal 95\% level at a resolution level of 20 data points and then rose to slightly above nominal at 120 data points. The $w_2$ parameter was the lowest at 20 data-points, indicating it was the most sensitive to resolution levels.


As Figure~\ref{fig:CovBiasResLVN}(b) shows, the $w_1$, $w_2$, and $w_4$ parameters' bias decreased significantly from 20 data-points to 120 data-points, whereas the $w_3$ parameter didn't show a significant decrease.

\begin{figure} 
    \centering
    \begin{tabular}{c}
    
    \includegraphics[width=1\linewidth]{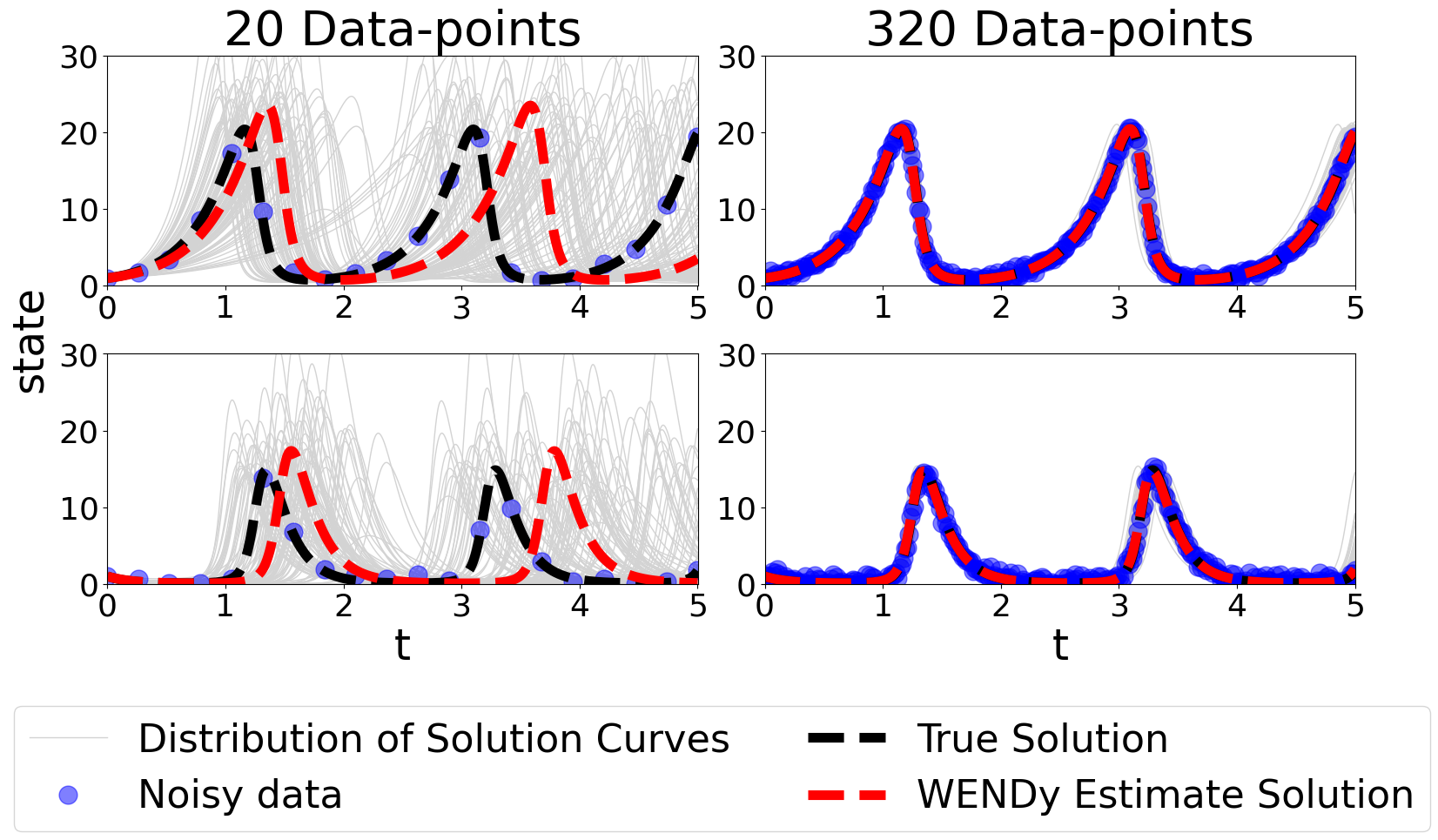} 
    \\
    \text{(a)}
    \\
    \includegraphics[width=1\linewidth]{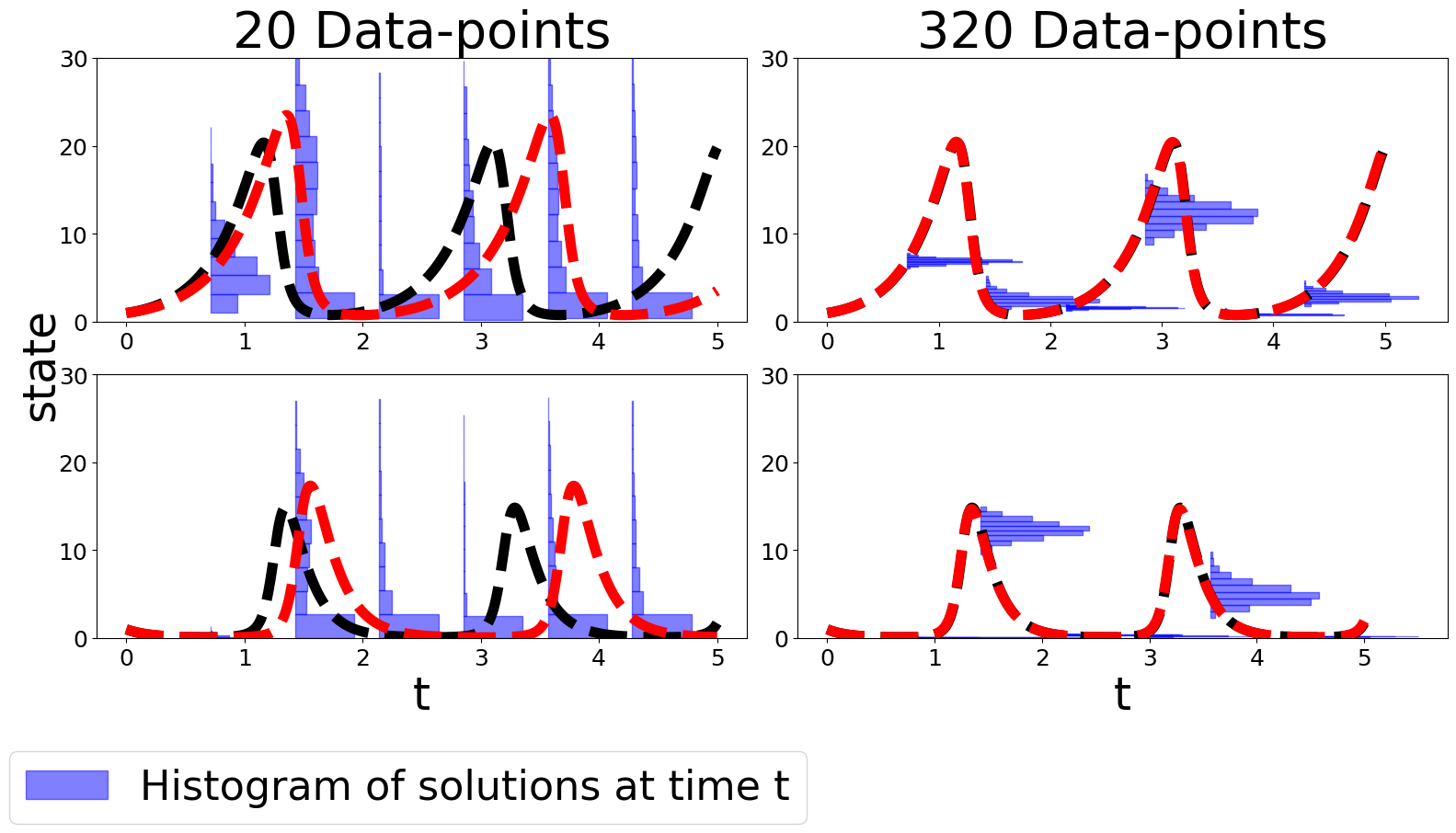}
    \\
    \text{(b)}
    \end{tabular}
     \caption{Top row (a): Lotka-Volterra model parameter estimation example and uncertainty quantification on two datasets: one dataset with low (left) and one dataset with high data resolution (right). The light gray curves are used to illustrate the uncertainty around the WENDy solutions; they are obtained via parametric bootstrap, as a sample of WENDy solutions corresponding to a random sample of 1000 parameters from their estimated asymptotic estimator distribution.  Bottom row (b): WENDy solution and histograms of state distributions across specific points in time for the datasets in (a).}
    \label{fig:SamplePlotResLVN}
\end{figure}

As Figure~\ref{fig:SamplePlotResLVN}(a) and (b) show, the distribution of states (at selected time-points) at lower resolution levels was more skewed toward lower state values and wider, whereas it became much more symmetric and narrower at higher resolution levels. WENDy estimators also exhibited greater phase shifts from the true solution at lower resolution levels compared to higher resolution levels. This is expected as lower-resolution data provide fewer time points to accurately capture the timing of population peaks and troughs, causing the estimated trajectories to be phase-shifted relative to the true dynamics when compared to higher resolutions.

\subsubsection*{Additive Censored Normal Noise}
\begin{figure} 
    \centering
    \begin{tabular}{c}
{\includegraphics[width=0.8\linewidth]{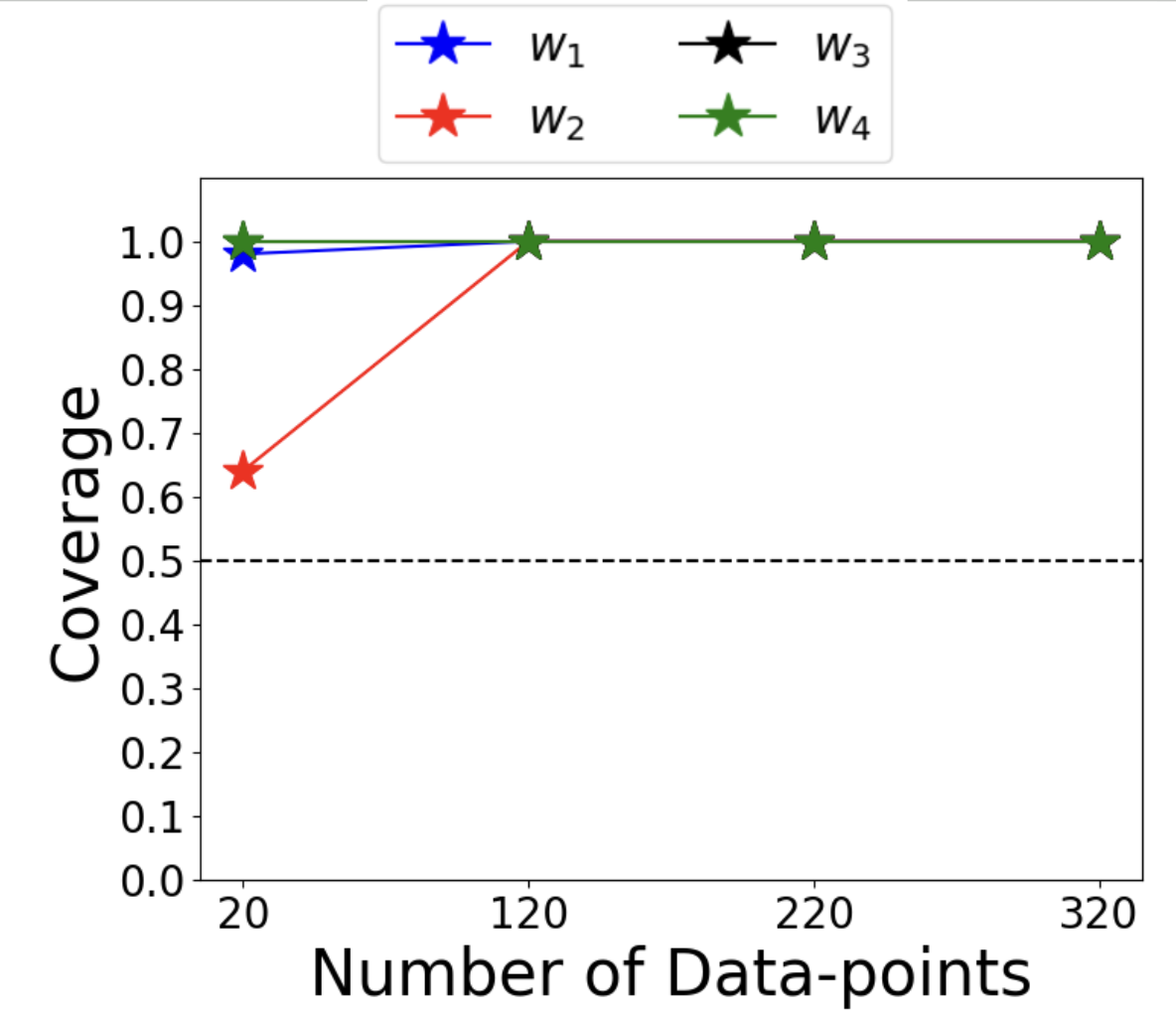}}
    \\
    \text{(a)}
    \\
    \includegraphics[width=1\linewidth]{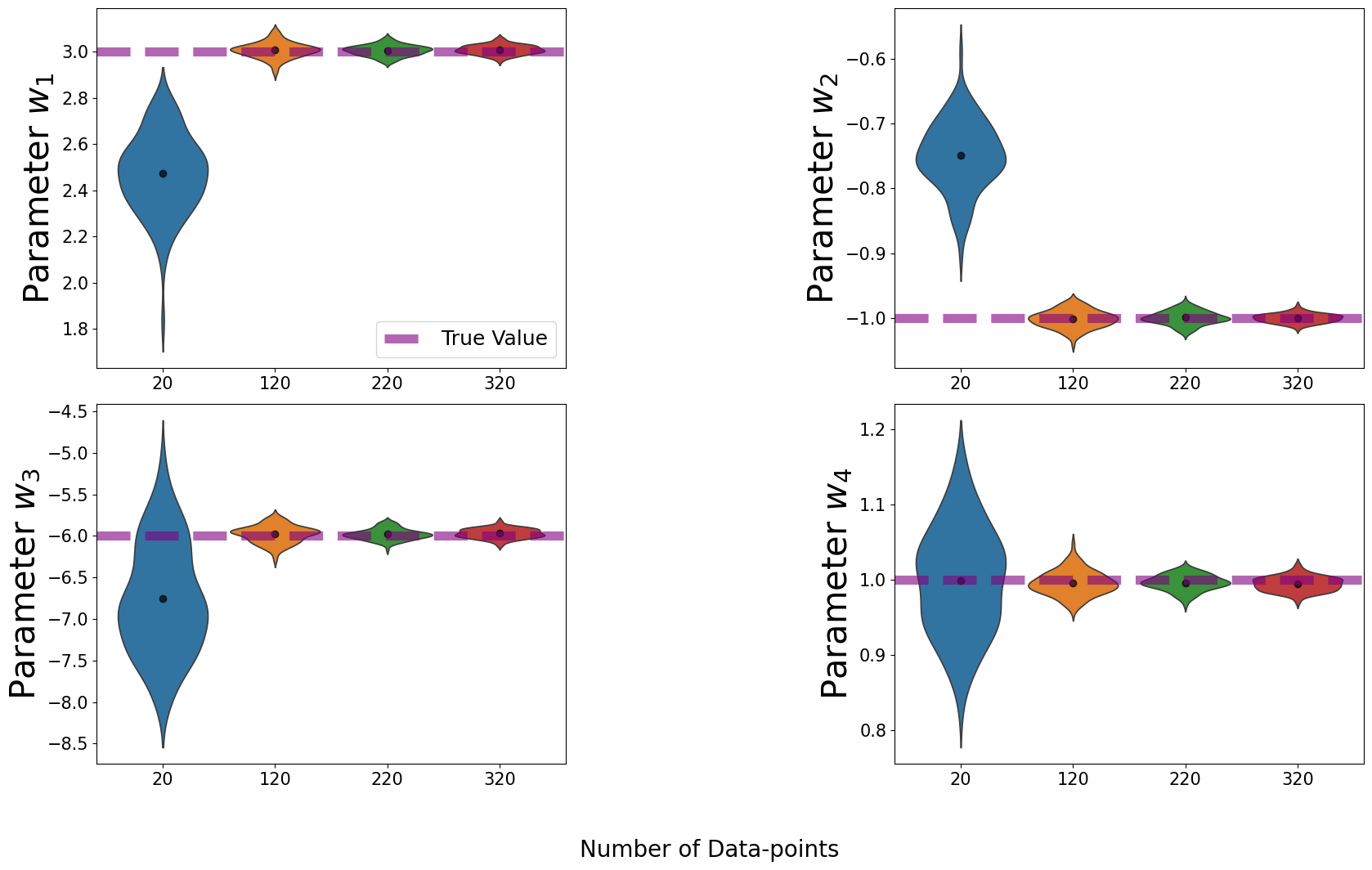}
    \\

    \text{(b)} \;\;\;\;\;\;\;\;\;\;\;\;\;\;\;\;\;\;\;\;\;\;\;\;\;\;\;\;\;\;\;
    \end{tabular}
     \caption{Lotka-Volterra model parameter estimation performance with increasing data resolution (1000 datasets per level, 30\% ACN noise). (a) coverage across four noise levels. (b) violin plots of parameter estimates, with the dashed red line indicating the true parameter values.}
    \label{fig:CovBiasResLVCN}
\end{figure}

As shown in Figure~\ref{fig:CovBiasResLVCN}(a), the coverage of the 95\% confidence intervals for the $w_1$, $w_3$, and $w_4$ parameters remained slightly above the nominal 95\% level across all resolution levels from 20 to 320 data points. In contrast, the coverage for the $w_2$ parameter only reached the nominal level at 120 data points and remained above nominal at higher resolution levels. This suggests that $w_2$ requires a sufficient number of data points to achieve reliable coverage, whereas the other parameters are relatively robust to changes in resolution. As shown in Figure~\ref{fig:CovBiasResLVCN}(b), the bias for all parameters decreased significantly as resolution increased from 20 to 120 data points, except for the $w_4$ parameter, whose bias remained steady.  

\begin{figure} 
    \centering
    \begin{tabular}{c}
    
    \includegraphics[width=1\linewidth]{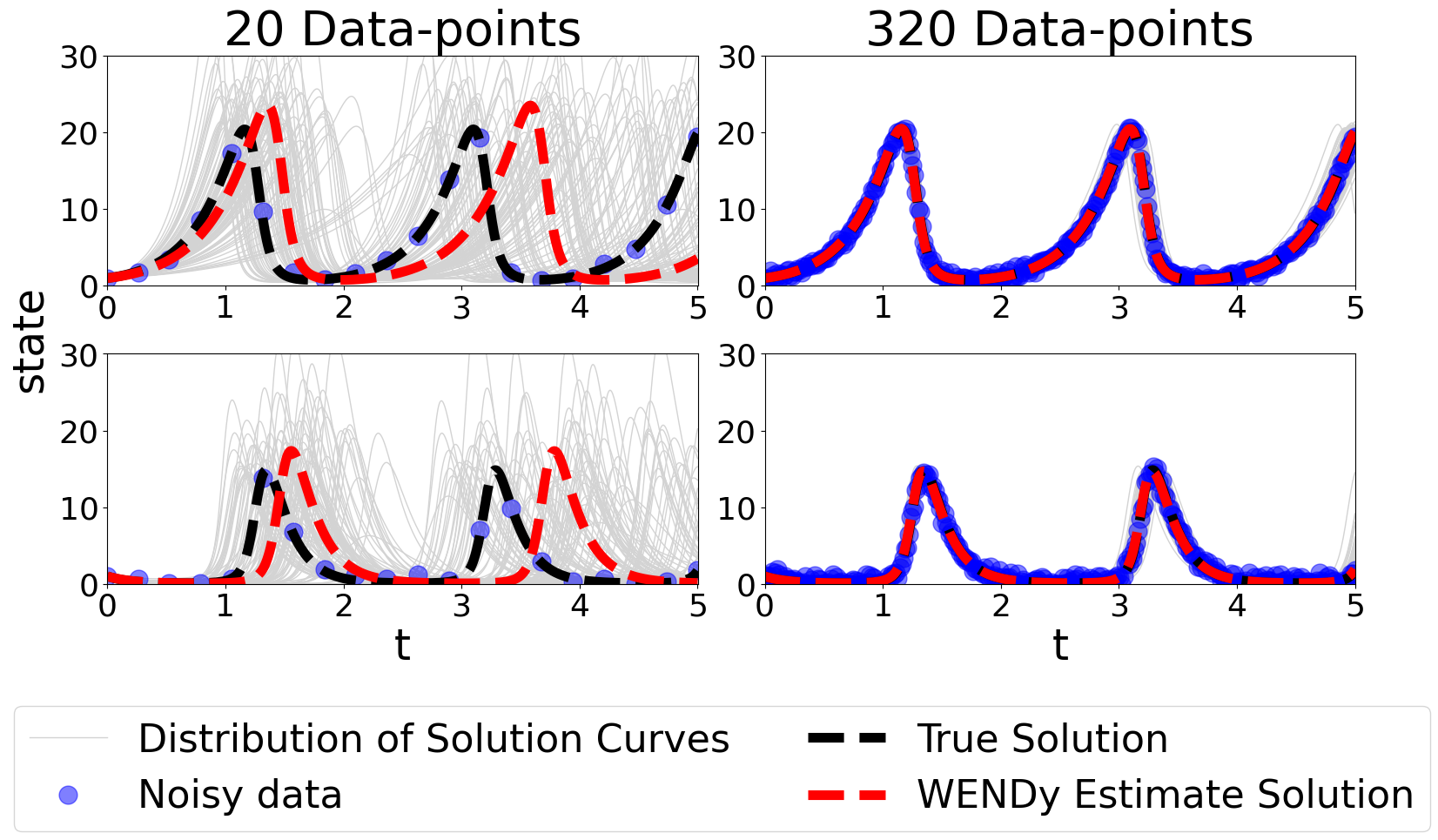} 
    \\
    \text{(a)}
    \\
    \includegraphics[width=1\linewidth]{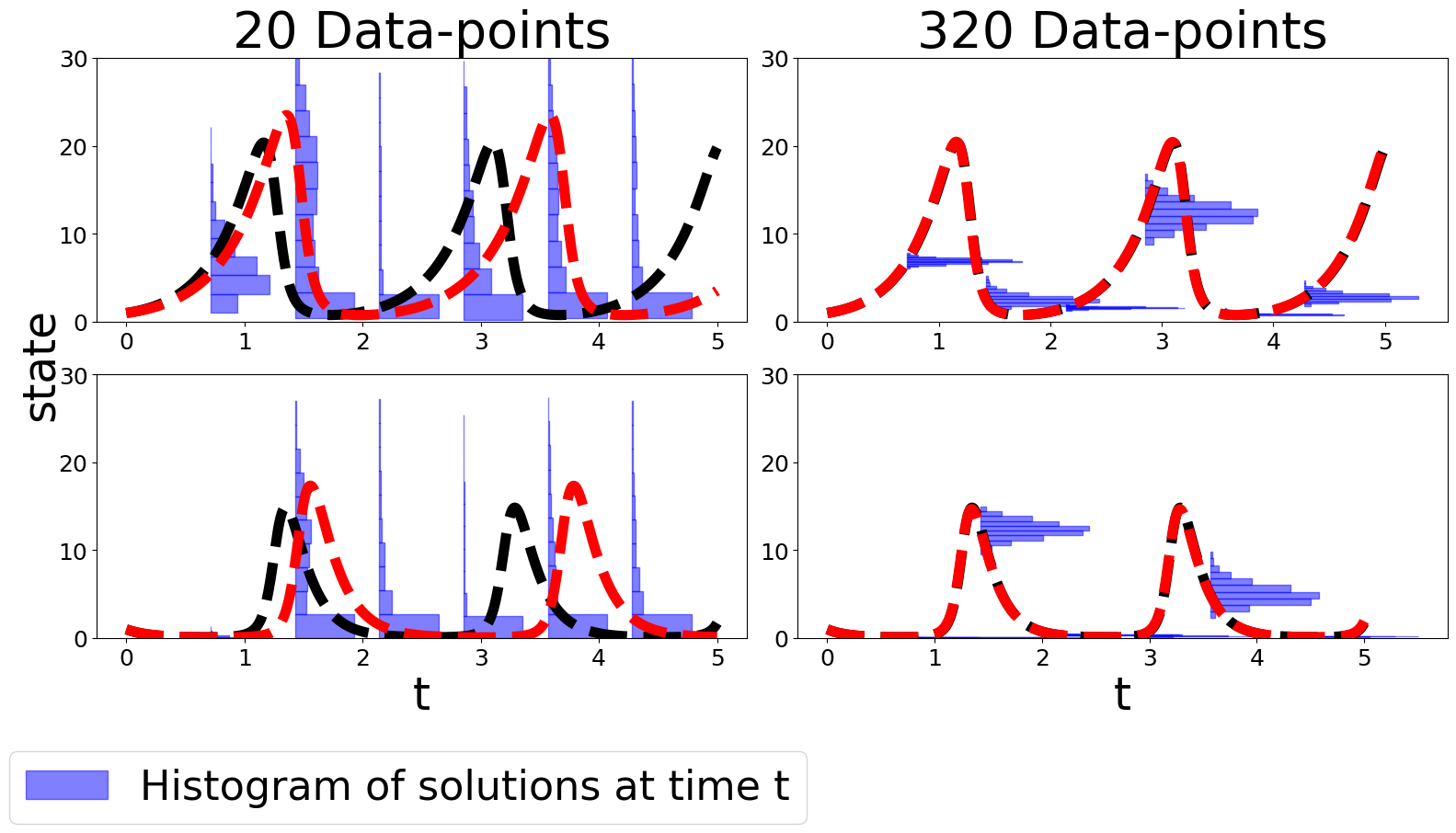}
    \\
    \text{(b)}
    \end{tabular}
     \caption{Top row (a): Lotka-Volterra model parameter estimation example and uncertainty quantification on two datasets: one dataset with low (left) and one dataset with high data resolution (right). The light gray curves are used to illustrate the uncertainty around the WENDy solutions; they are obtained via parametric bootstrap, as a sample of WENDy solutions corresponding to a random sample of 1000 parameters from their estimated asymptotic estimator distribution.  Bottom row (b): WENDy solution and histograms of state distributions across specific points in time for the datasets in (a).}
    \label{fig:SamplePlotResLVCN}
\end{figure}

As illustrated in Figures~\ref{fig:SamplePlotResLVCN}(a) and (b), the distribution of states (at selected time points) at lower resolution levels was more skewed toward lower state values and wider, whereas it became much more symmetric and narrower at higher resolution levels. WENDy estimators also exhibited greater phase shifts from the true solution at lower resolution levels compared to higher resolution levels. This is expected, as lower-resolution data provide fewer time points to accurately capture the timing of population peaks and troughs, causing the estimated trajectories to be phase-shifted relative to the true dynamics compared to higher resolutions.


\subsubsection*{Multiplicative Log-Normal Noise}
\begin{figure} 
    \centering
    \begin{tabular}{c}
{\includegraphics[width=0.8\linewidth]{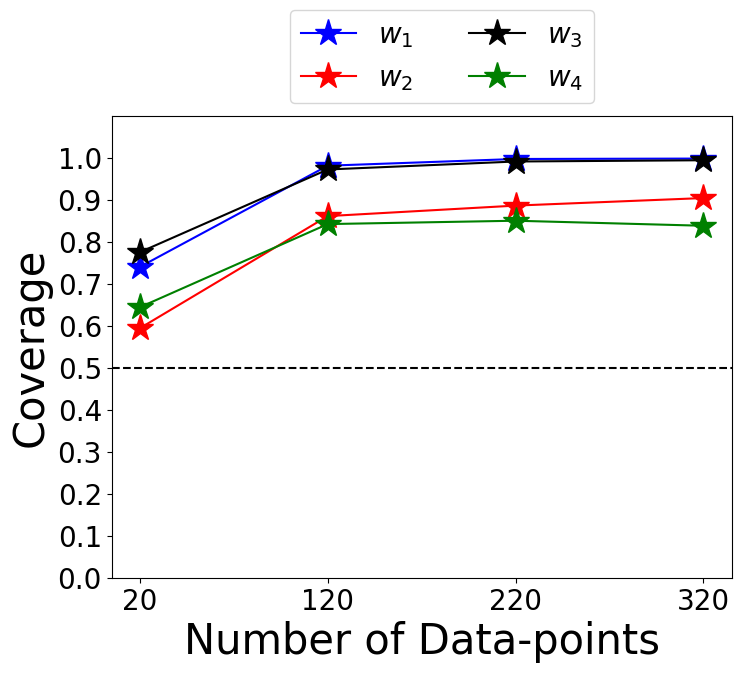}}
    \\
    \text{(a)}
    \\
    \includegraphics[width=1\linewidth]{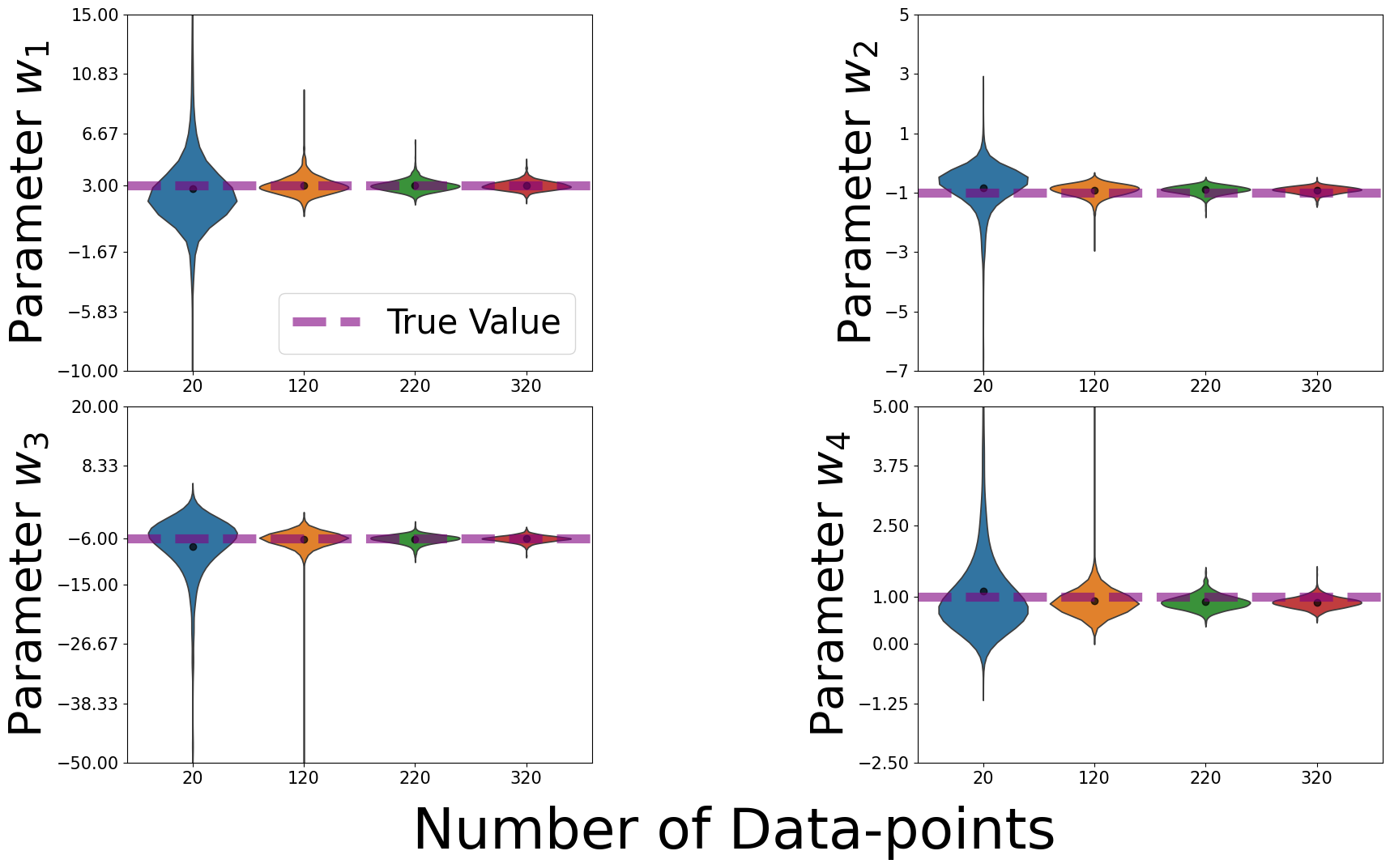}
    \\

      \text{(b)} 
    \end{tabular}
     \caption{Lotka-Volterra model parameter estimation performance with increasing data resolution (1000 datasets per level, 5\% MLN noise). (a) coverage across four noise levels. (b) violin plots of parameter estimates, with the dashed red line indicating the true parameter values.}
    \label{fig:CovBiasResLVLN}
\end{figure}

As shown in Figure~\ref{fig:CovBiasResLVLN}(a), the coverage of the 95\% confidence intervals for all parameters was below nominal at 20 data points. As the resolution increased, coverage for $w_1$ and $w_3$ rose slightly above nominal, while coverage for $w_2$ and $w_4$ stabilized around 80\%. This difference is expected, as $w_2$ and $w_4$ (corresponding to the $\beta$ and $\delta$ parameters) modulate the quadratic terms in the LV equations, making their dynamics more sensitive to low resolution and thus harder to accurately capture compared to the linear term parameters $w_1$ and $w_3$.


As Figure~\ref{fig:CovBiasResLVLN}(b) shows,  the bias for all parameters decreased with increasing resolution, but not nearly as much as with ACN noise. The decrease in variance was much more significant, as seen by the tails of the violin plots.

\begin{figure} 
    \centering
    \begin{tabular}{c}
    
    \includegraphics[width=1\linewidth]{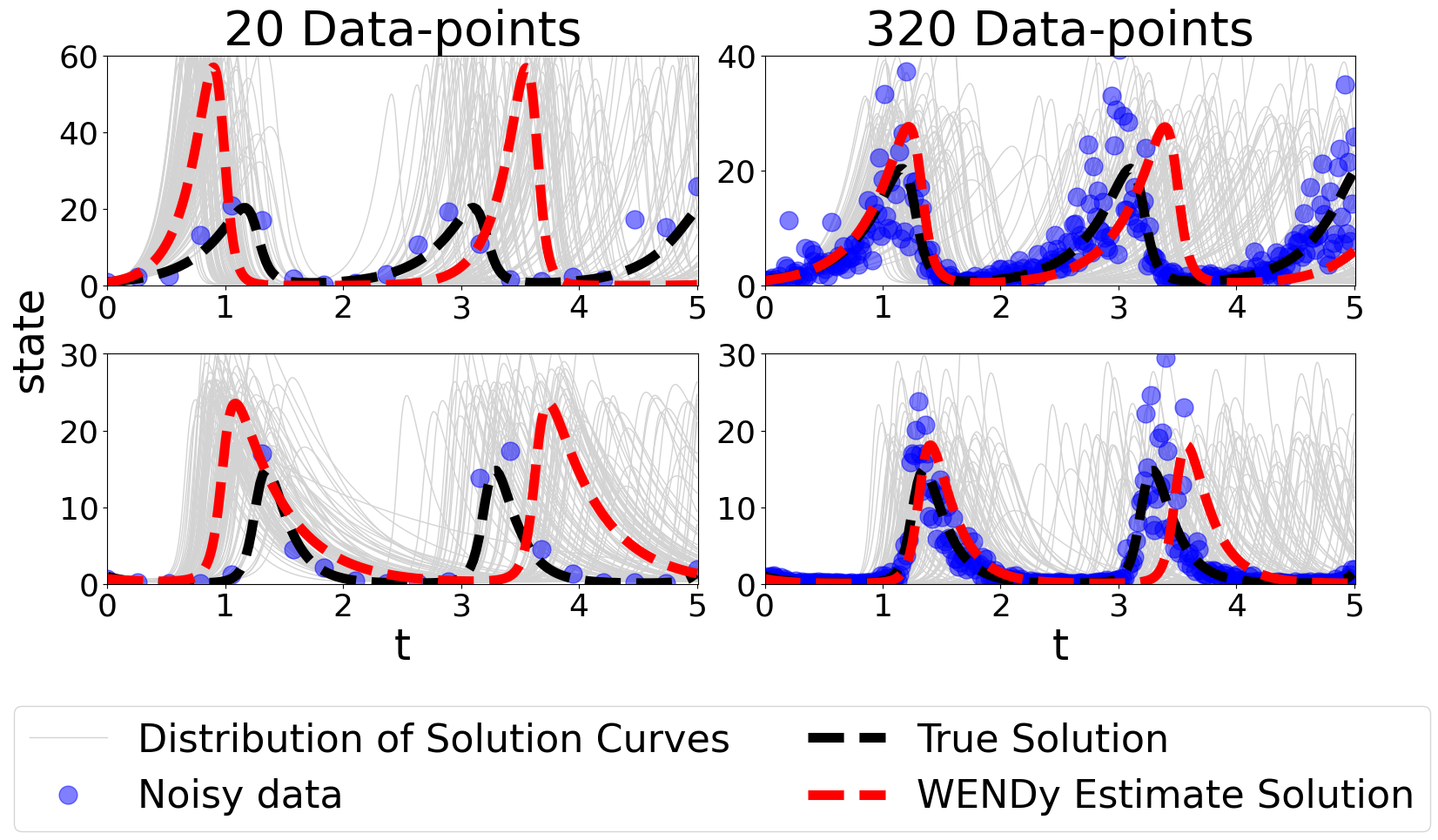} 
    \\
    \text{(a)}
    \\
    \includegraphics[width=1\linewidth]{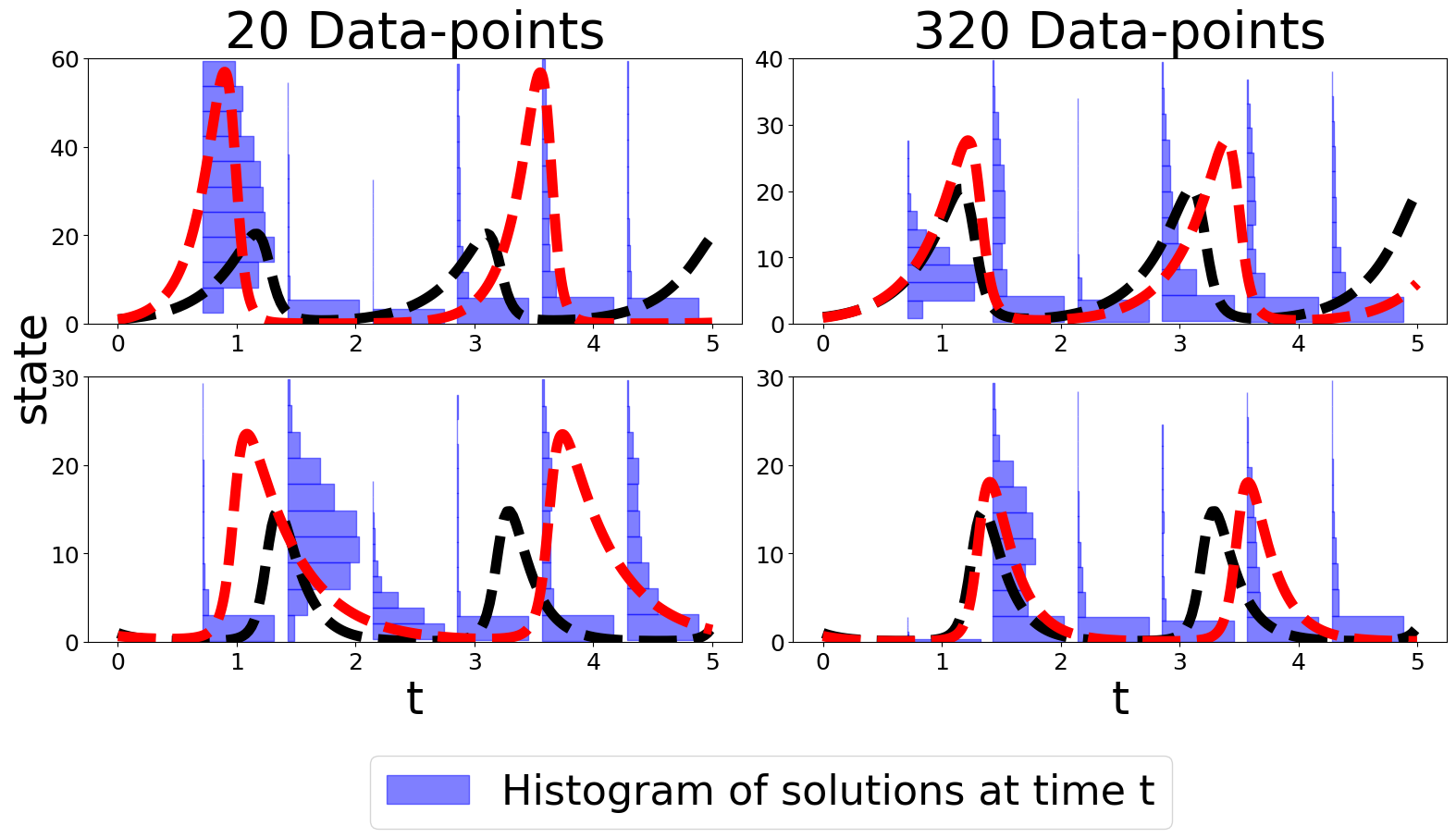}
    \\
    \text{(b)}
    \end{tabular}
     \caption{Top row (a): Logistic model parameter estimation example and uncertainty quantification on two datasets: one dataset with low (left) and one dataset with high data resolution (right). The light gray curves are used to illustrate the uncertainty around the WENDy solutions; they are obtained via parametric bootstrap, as a sample of WENDy solutions corresponding to a random sample of 1000 parameters from their estimated asymptotic estimator distribution.  Bottom row (b): WENDy solution and histograms of state distributions across specific points in time for the datasets in (a).}
    \label{fig:SamplePlotResLVLN}
\end{figure}

As Figure~\ref{fig:SamplePlotResLVLN}(a) and (b) show, the distribution of solution states (at selected time-points) for lower and higher resolution levels is skewed but mostly unimodal. The WENDy estimated solution struggles to capture the amplitude of the true curves at lower resolution levels compared to higher ones.

\subsubsection*{Additive Truncated Normal Noise}
\begin{figure} 
    \centering
    \begin{tabular}{c}
{\includegraphics[width=0.8\linewidth]{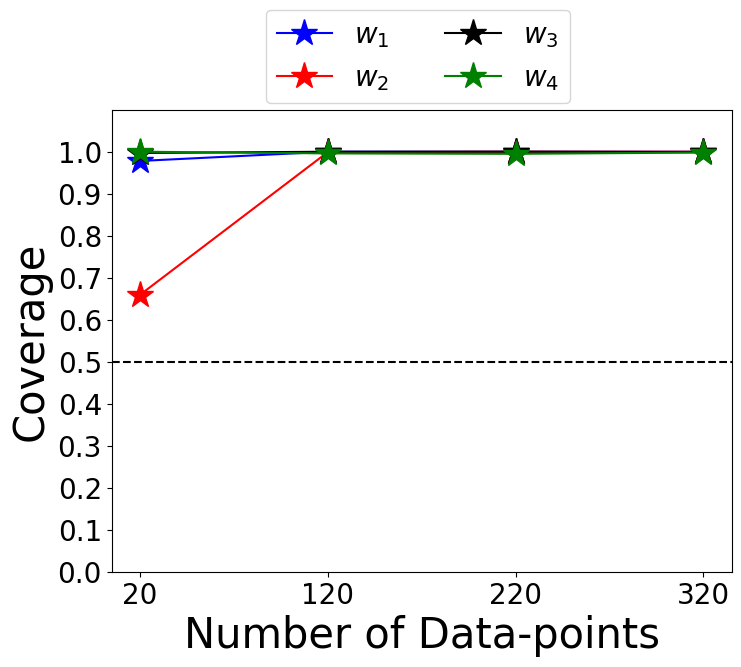}}
    \\
    \text{(a)}
    \\
    \includegraphics[width=1\linewidth]{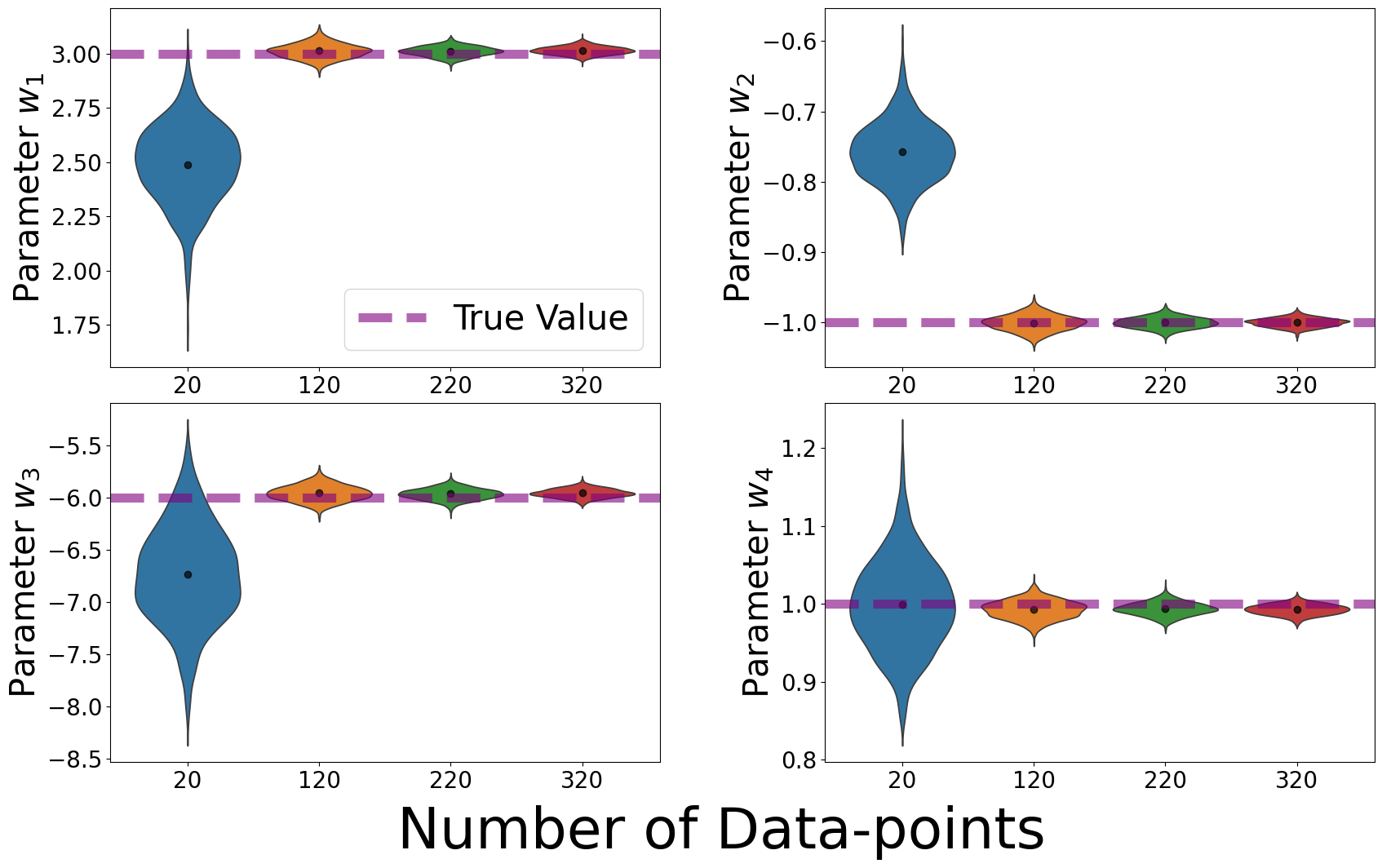}
    \\
     
      \text{(b)} 
    \end{tabular}
     \caption{Logistic model parameter estimation performance with increasing data resolution (1000 datasets per level, 90\% ATN noise). (a) coverage across four noise levels. (b) violin plots of parameter estimates, with the dashed red line indicating the true parameter values.}
    \label{fig:CovBiasResLVTN}
\end{figure}

As shown in Figure~\ref{fig:CovBiasResLVTN}(a), the coverage of the 95\% confidence intervals for the $w_1$, $w_3$, and $w_4$ parameters remained near nominal across all resolution levels. In contrast, coverage for the $w_2$ parameter was only around 60\% at 20 data points, increasing to nominal at 120 data points. Similar to ACN noise, the $w_2$ parameter was the most sensitive to low resolution and required more points to have good coverage.


As Figure~\ref{fig:CovBiasResLVTN}(b) shows,  the bias and variance for all parameters decreased with increasing resolution. The trend in the violin plots is similar to what was seen with ACN noise.

\begin{figure} 
    \centering
    \begin{tabular}{c}
    
    \includegraphics[width=1\linewidth]{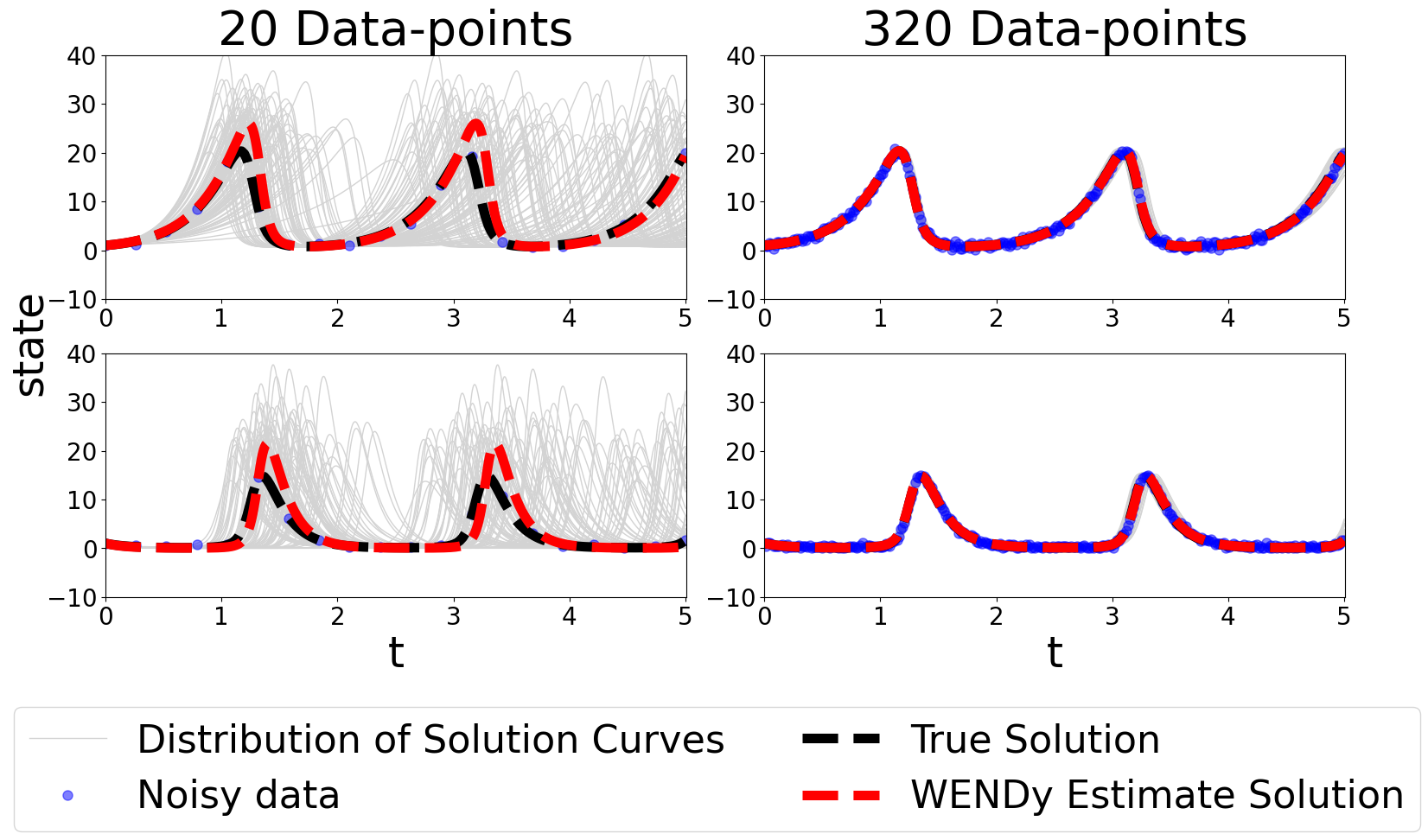} 
    \\
    \text{(a)}
    \\
    \includegraphics[width=1\linewidth]{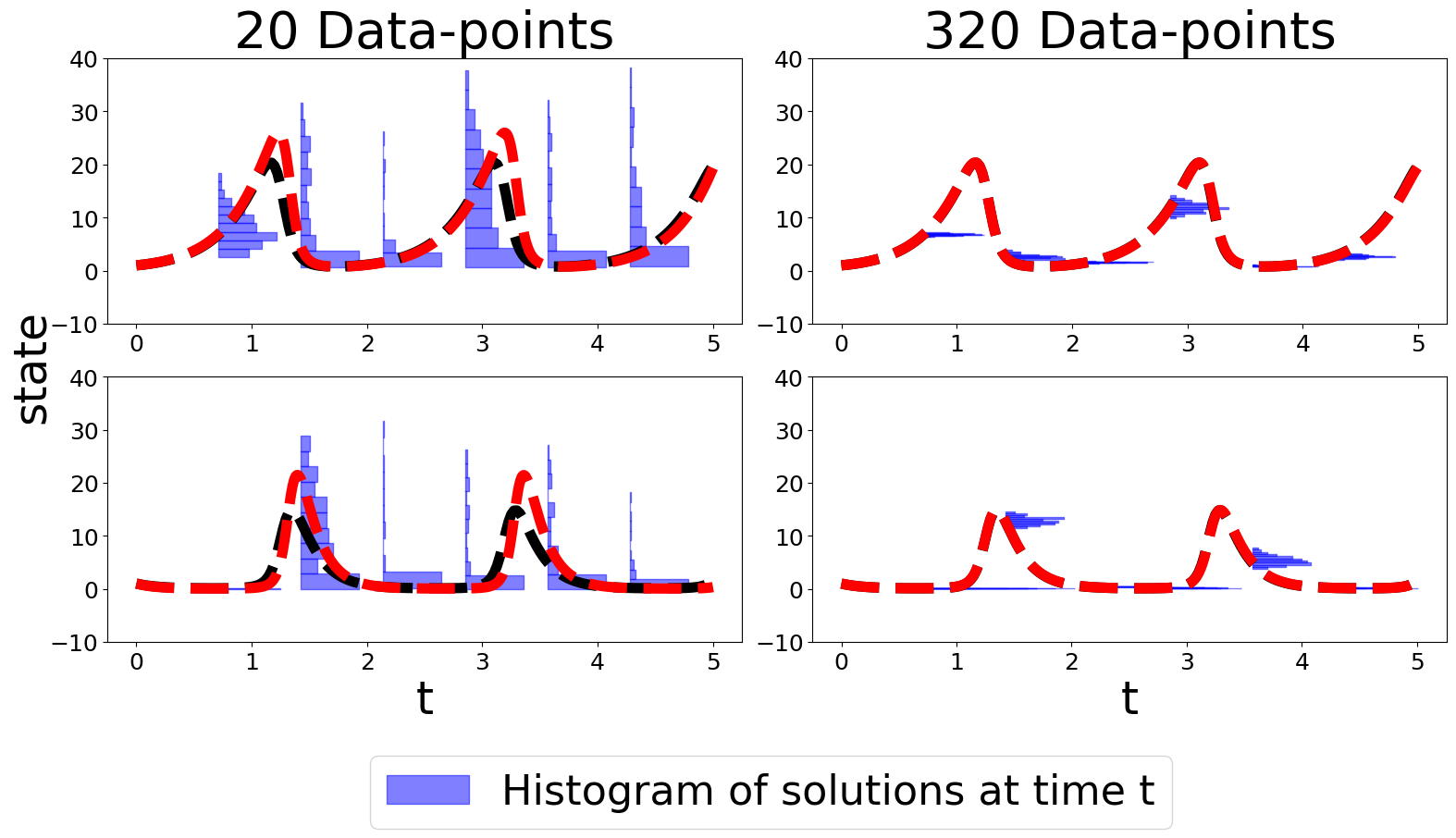}
    \\
    \text{(b)}
    \end{tabular}
     \caption{Top row (a): Logistic model parameter estimation example and uncertainty quantification on two datasets: one dataset with low (left) and one dataset high dataset (right). The light gray curves are used to illustrate the uncertainty around the WENDy solutions; they are obtained via parametric bootstrap, as a sample of WENDy solutions corresponding to a random sample of 1000 parameters from their estimated asymptotic estimator distribution.  Bottom row (b): WENDy solution and histograms of state distributions across specific points in time for the datasets in (a).}
    \label{fig:SamplePlotResLVTN}
\end{figure}

As Figure~\ref{fig:SamplePlotResLVTN}(a) and (b) show, the distribution of solution states (at selected time-points) for lower resolution levels is more skewed, whereas for high resolution levels is much more symmetric. The WENDy estimated solution is again much more phase-shifted at lower resolution levels compared to higher ones.

\clearpage
\subsection{FitzHugh-Nagumo}
\subsubsection{Varying Noise Level}
\subsubsection*{Additive Normal Noise}

\begin{figure} 
    \centering
    \begin{tabular}{c}
{\includegraphics[width=0.8\linewidth]{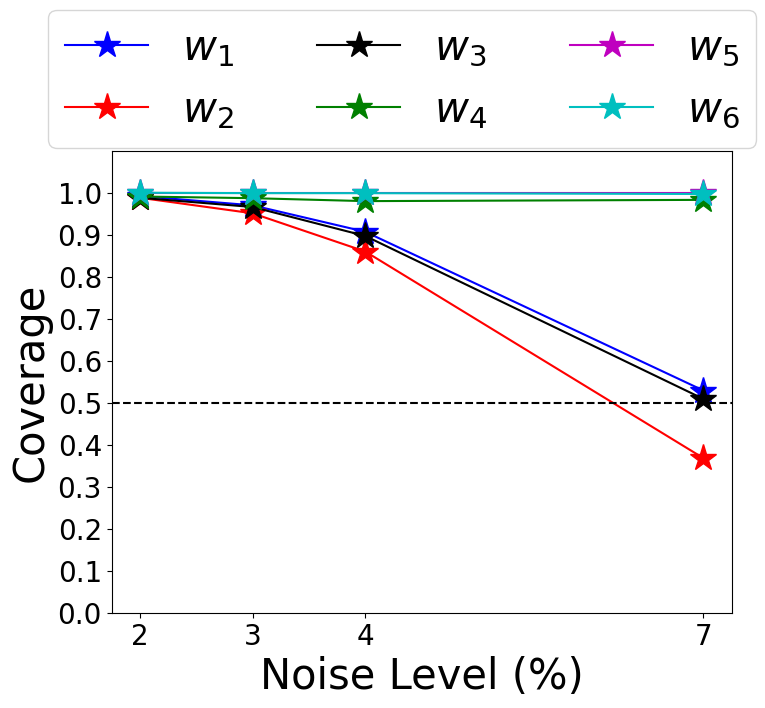}}
    \\
    \text{(a)}
    \\
    \includegraphics[width=1\linewidth]{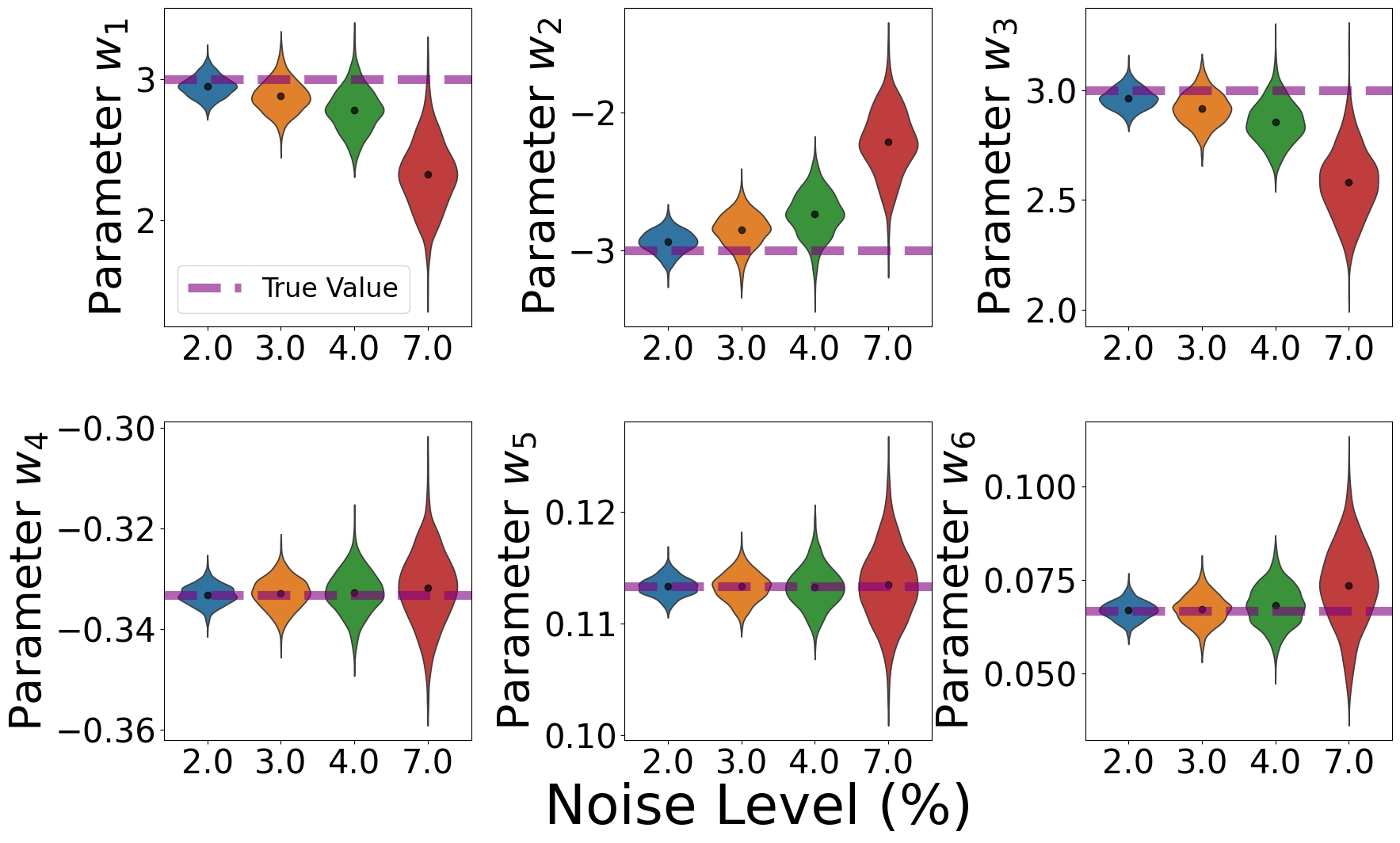}
    \\
      \text{(b)} 
    \end{tabular}
     \caption{FitzHugh-Nagumo model parameter estimation performance with increasing additive normal noise (1000 datasets per level, 205 data points each). (a) coverage across four noise levels. (b) violin plots of parameter estimates, with the dashed red line indicating the true parameter values.}
    \label{fig:CovBiasFHNN}
\end{figure}

As shown in Figure~\ref{fig:CovBiasFHNN}(a), the coverage of the 95\% confidence intervals for the $w_4$, $w_5$, and $w_6$ parameters remained slightly above the nominal 95\% level across all noise levels from 2\% to 7\%. In contrast, the coverage for the $w_1$, $w_2$, and $w_3$ parameters decreased as noise increased, dropping below 50\% at 7\% noise for the $w_2$ parameter. The FHN model contains many sharp spikes and is a stiff ODE; thus, at higher noise levels, more data points are required to accurately identify parameters.  

As shown in Figure~\ref{fig:CovBiasFHNN}(b), the bias in the $w_1$, $w_2$, and $w_3$ parameters increased substantially with higher noise, while the bias in the $w_4$, $w_5$, and $w_6$ parameters increased only modestly. This discrepancy explains the difference in coverage between the two sets of parameters.  

\begin{figure} 
    \centering
    \begin{tabular}{c}
    
    \includegraphics[width=1\linewidth]{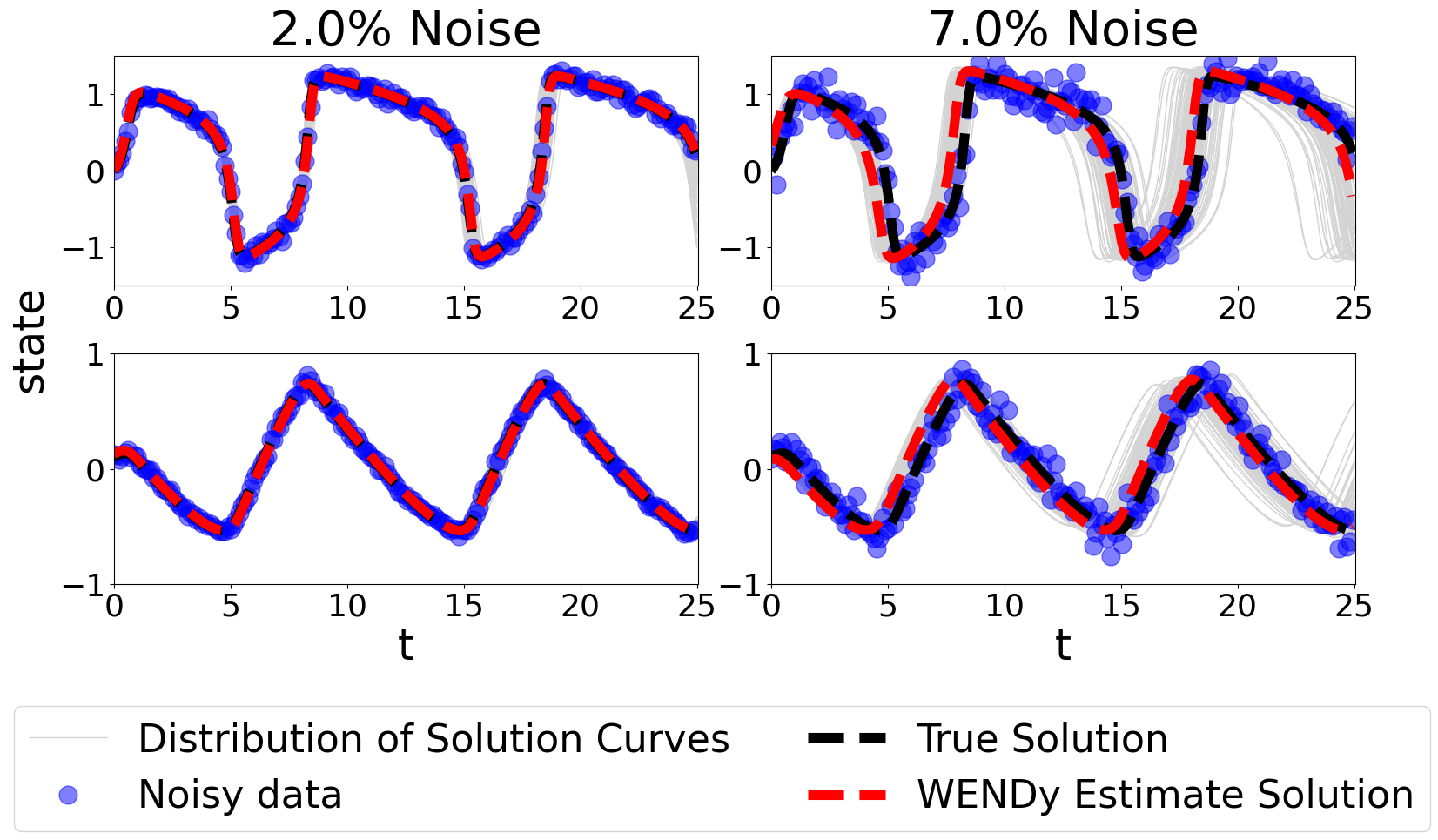} 
    \\
    \text{(a)}
    \\
    \includegraphics[width=1\linewidth]{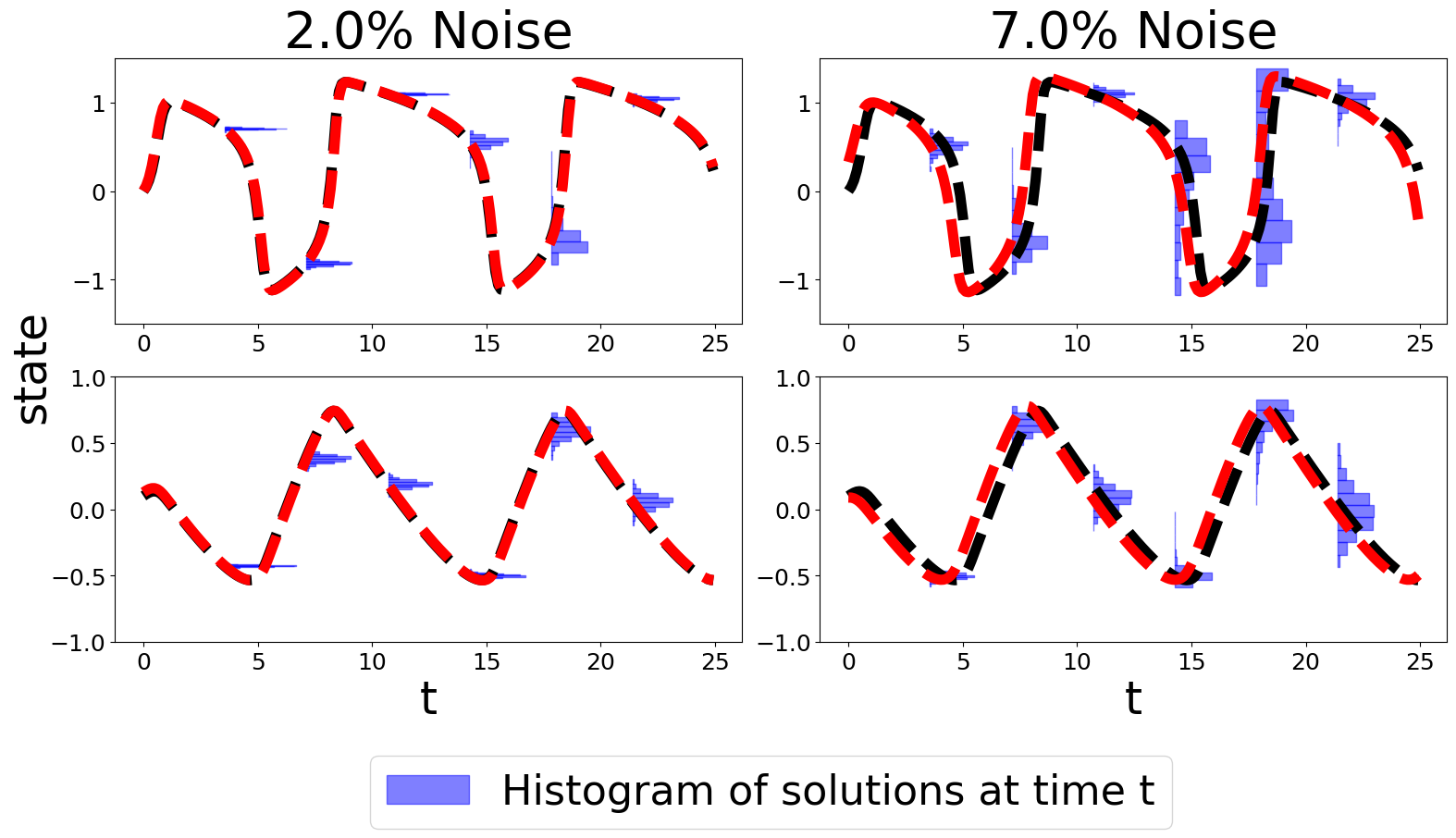}
    \\
    \text{(b)}
    \end{tabular}
     \caption{Top row (a): FitzHugh-Nagumo model parameter estimation example and uncertainty quantification on two datasets: one dataset with low uncertainty and high coverage (left) and one dataset with high uncertainty and low coverage (right). The light gray curves are used to illustrate the uncertainty around the WENDy solutions; they are obtained via parametric bootstrap, as a sample of WENDy solutions corresponding to a random sample of 1000 parameters from their estimated asymptotic estimator distribution.  Bottom row (b): WENDy solution and histograms of state distributions across specific points in time for the datasets in (a).}
    \label{fig:SamplePlotFHNTN}
\end{figure}

As illustrated in Figures~\ref{fig:SamplePlotFHNTN}(a) and (b), the distribution of solution curves at higher noise levels was more skewed and occasionally bimodal at certain time points compared to lower noise levels.

\clearpage
\subsubsection*{Multiplicative Log-Normal Noise}
\begin{figure} 
    \centering
    \begin{tabular}{c}
{\includegraphics[width=0.8\linewidth]{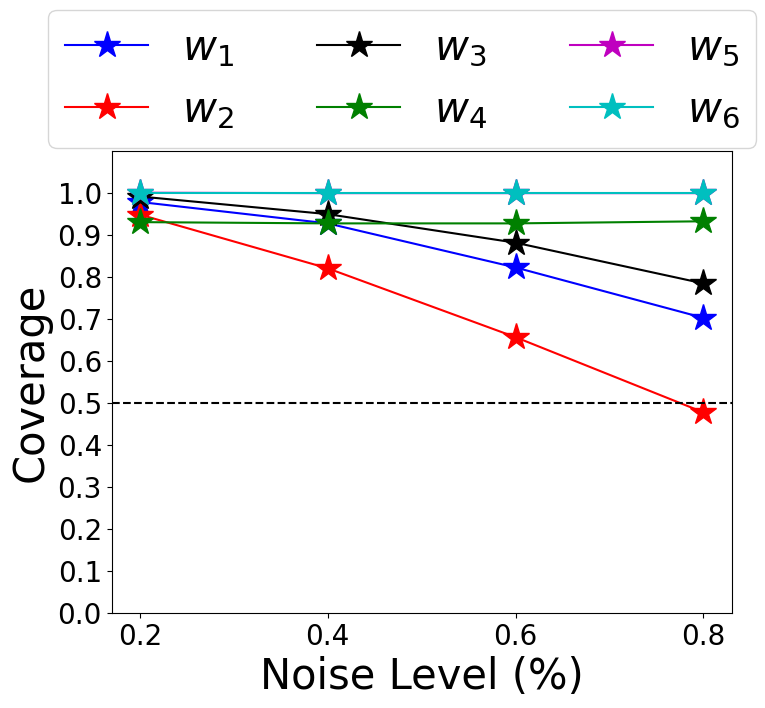}}
    \\
    \text{(a)}
    \\
    \includegraphics[width=1\linewidth]{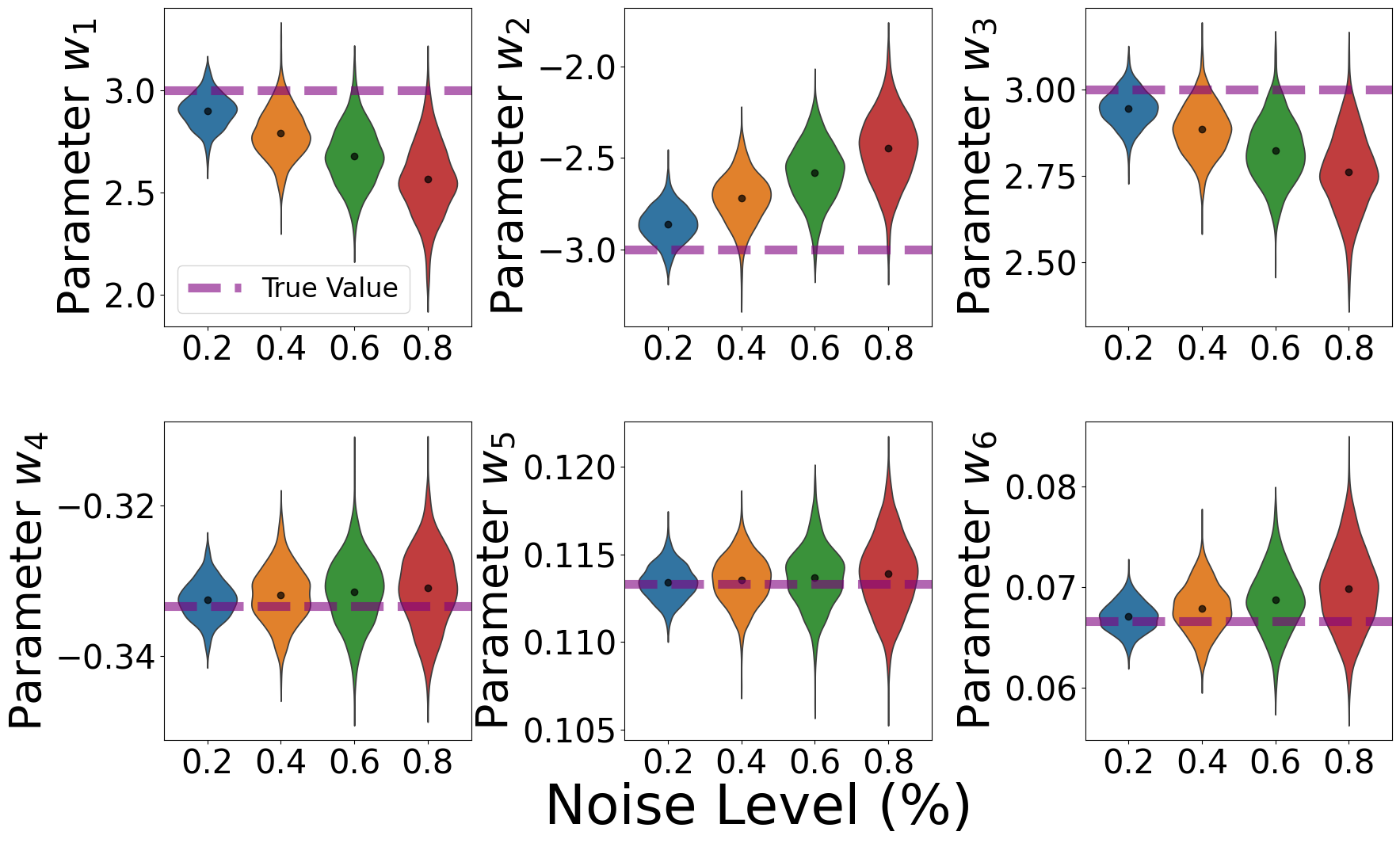}
    \\
     
      \text{(b)} 
    \end{tabular}
     \caption{FitzHugh-Nagumo model parameter estimation performance with increasing MLN noise (1000 datasets per level, 205 data points each). (a) coverage across four noise levels. (b) violin plots of parameter estimates, with the dashed red line indicating the true parameter values.}
    \label{fig:CovBiasFHNLN}
\end{figure}

As shown in Figure~\ref{fig:CovBiasFHNLN}(a), the coverage of the 95\% confidence intervals for the $w_4$, $w_5$, and $w_6$ parameters remained close to the nominal 95\% level across all noise levels from 0.2\% to 0.8\%. In contrast, the coverage for the $w_1$, $w_2$, and $w_3$ parameters decreased as noise increased, with the coverage for $w_2$ falling below 50\% at 0.8\% noise. Since MLN noise is heteroskedastic, regions of higher-magnitude states become noisier. Because the FHN model contains many spikes at higher-magnitude states and is stiff, WENDy estimator coverage is expected to deteriorate even at very low noise levels. In such cases, smoothing the data might substantially reduce noise around the spikes and improve coverage for WENDy estimators.  

As shown in Figure~\ref{fig:CovBiasFHNLN}(b), the bias in all parameters increased with noise. Similar to the normal noise case, the bias in $w_1$, $w_2$, and $w_3$ increased more sharply than in $w_4$, $w_5$, and $w_6$. This suggests that parameters in the $u_1$ state are more sensitive to noise than those in the $u_2$ state. This difference may arise from the cubic term in the ODE for the $u_1$ state, which makes parameter identifiability more difficult.  

\begin{figure} 
    \centering
    \begin{tabular}{c}
    
    \includegraphics[width=1\linewidth]{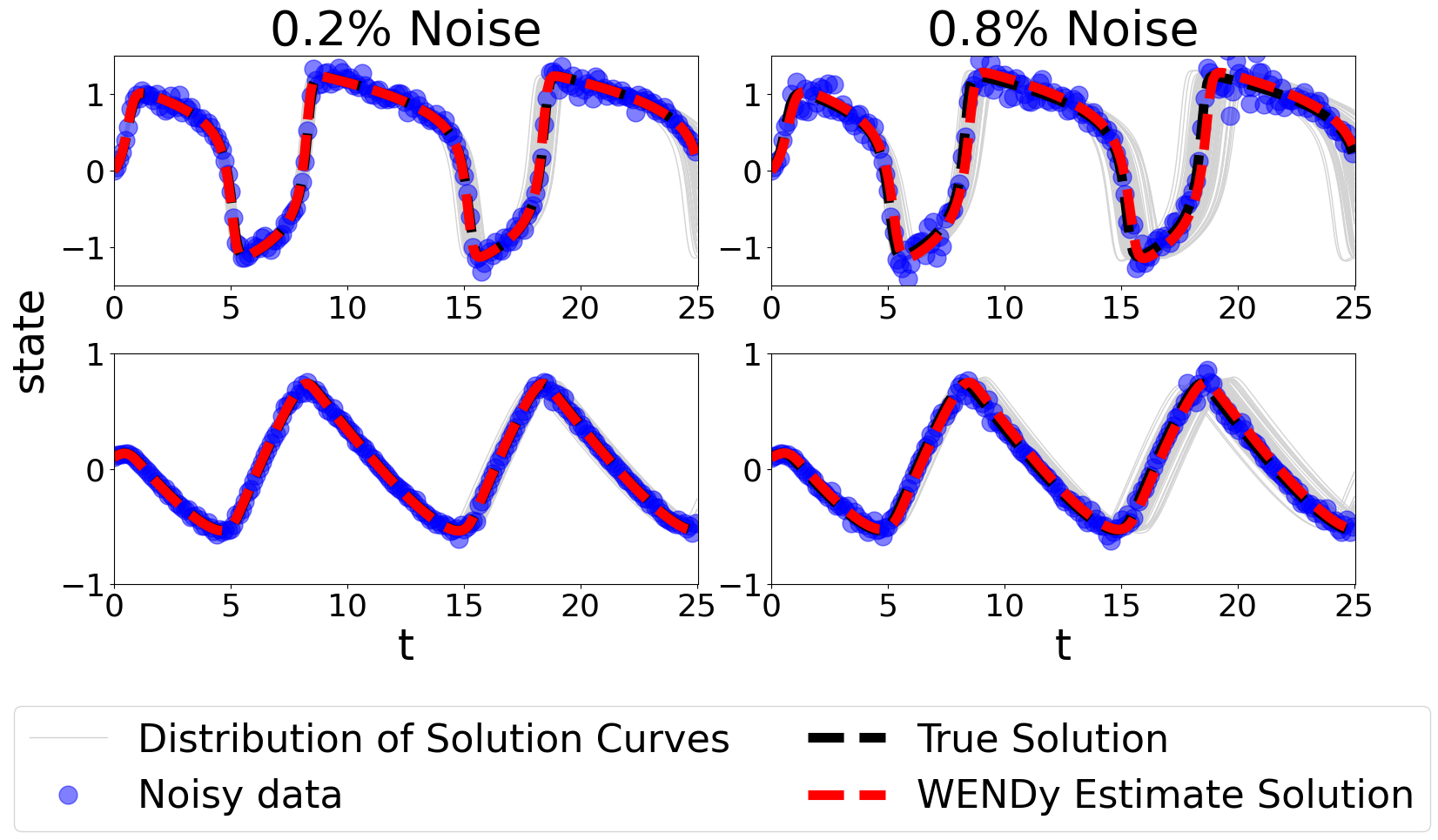} 
    \\
    \text{(a)}
    \\
    \includegraphics[width=1\linewidth]{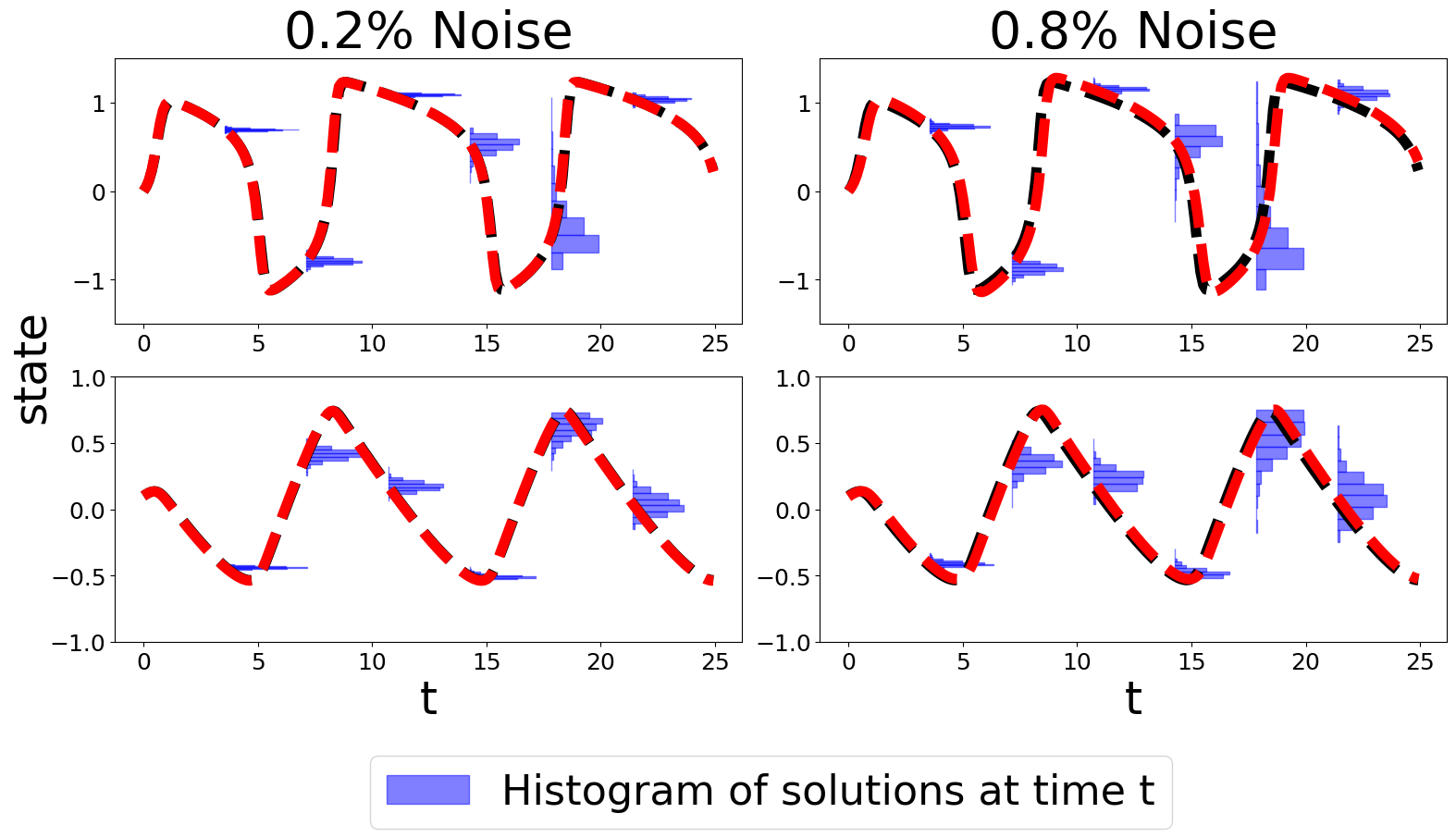}
    \\
    \text{(b)}
    \end{tabular}
     \caption{Top row (a): FitzHugh-Nagumo model parameter estimation example and uncertainty quantification on two datasets: one dataset with low uncertainty and high coverage (left) and one dataset with high uncertainty and low coverage (right). The light gray curves are used to illustrate the uncertainty around the WENDy solutions; they are obtained via parametric bootstrap, as a sample of WENDy solutions corresponding to a random sample of 1000 parameters from their estimated asymptotic estimator distribution.  Bottom row (b): WENDy solution and histograms of state distributions across specific points in time for the datasets in (a).}
    \label{fig:SamplePlotFHNLN}
\end{figure}

As illustrated in Figures~\ref{fig:SamplePlotFHNLN}(a) and (b), the distribution of solution curves at higher noise levels was wider compared to lower noise levels. Time points with higher-magnitude state values also exhibited more skewed solutions under both noise levels. Furthermore, the $u_1$ state histograms were wider than those of the $u_2$ state across both noise levels, consistent with the greater nonlinearity present in the $u_1$ state dynamics.

\clearpage

\begin{figure} 
    \centering
    \includegraphics[width=0.4\linewidth]{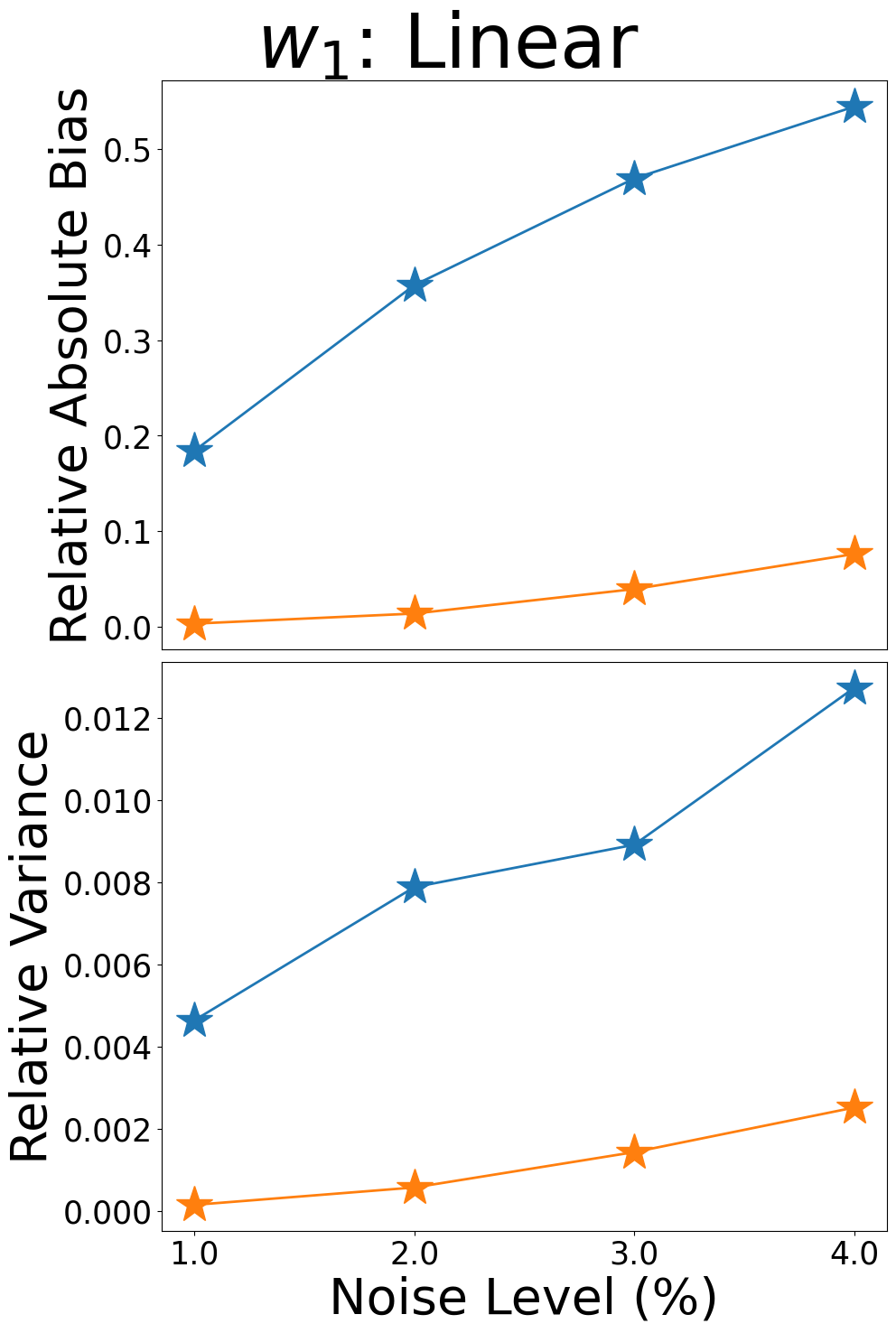}     \includegraphics[width=0.4\linewidth]{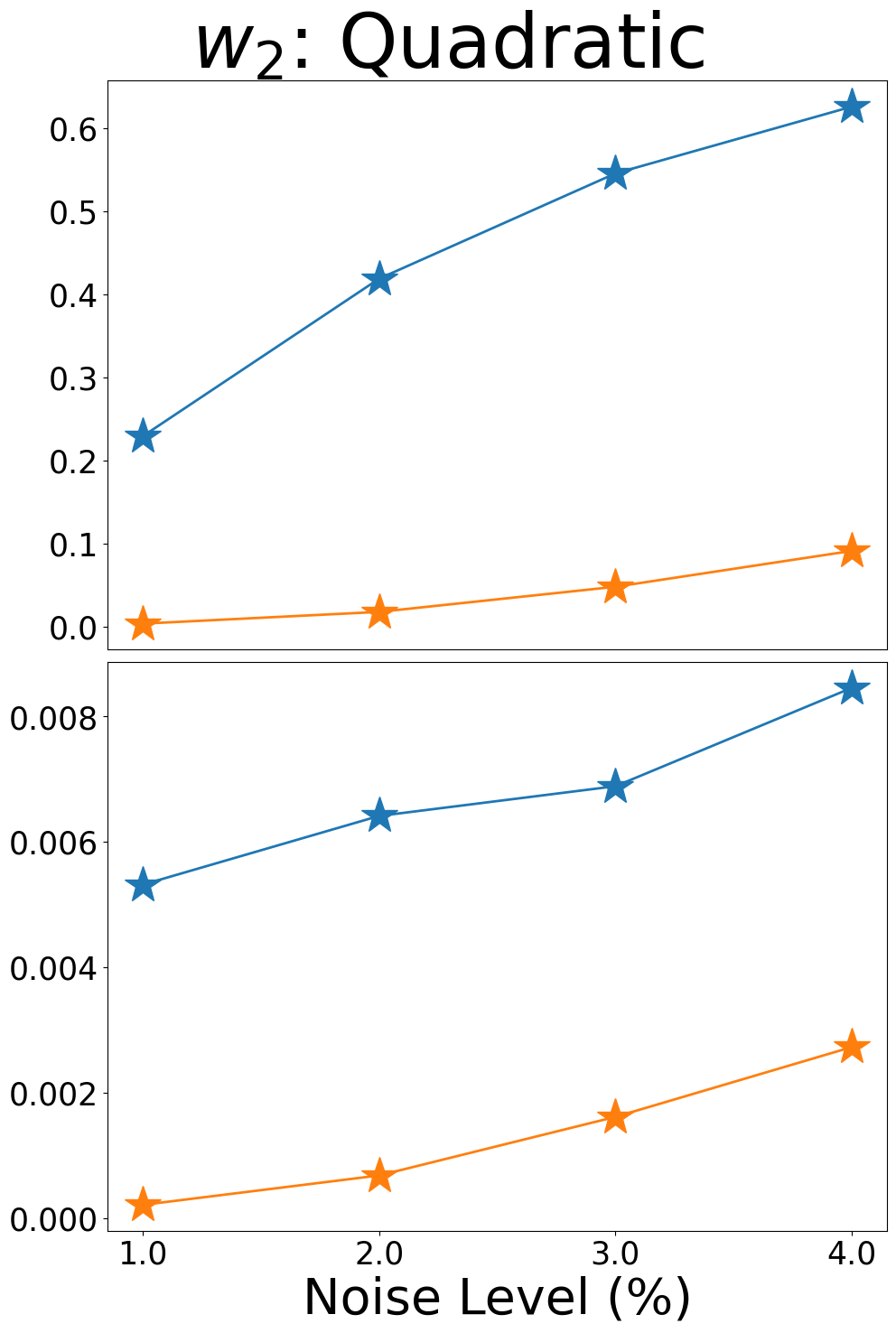}
    \includegraphics[width=0.4\linewidth]{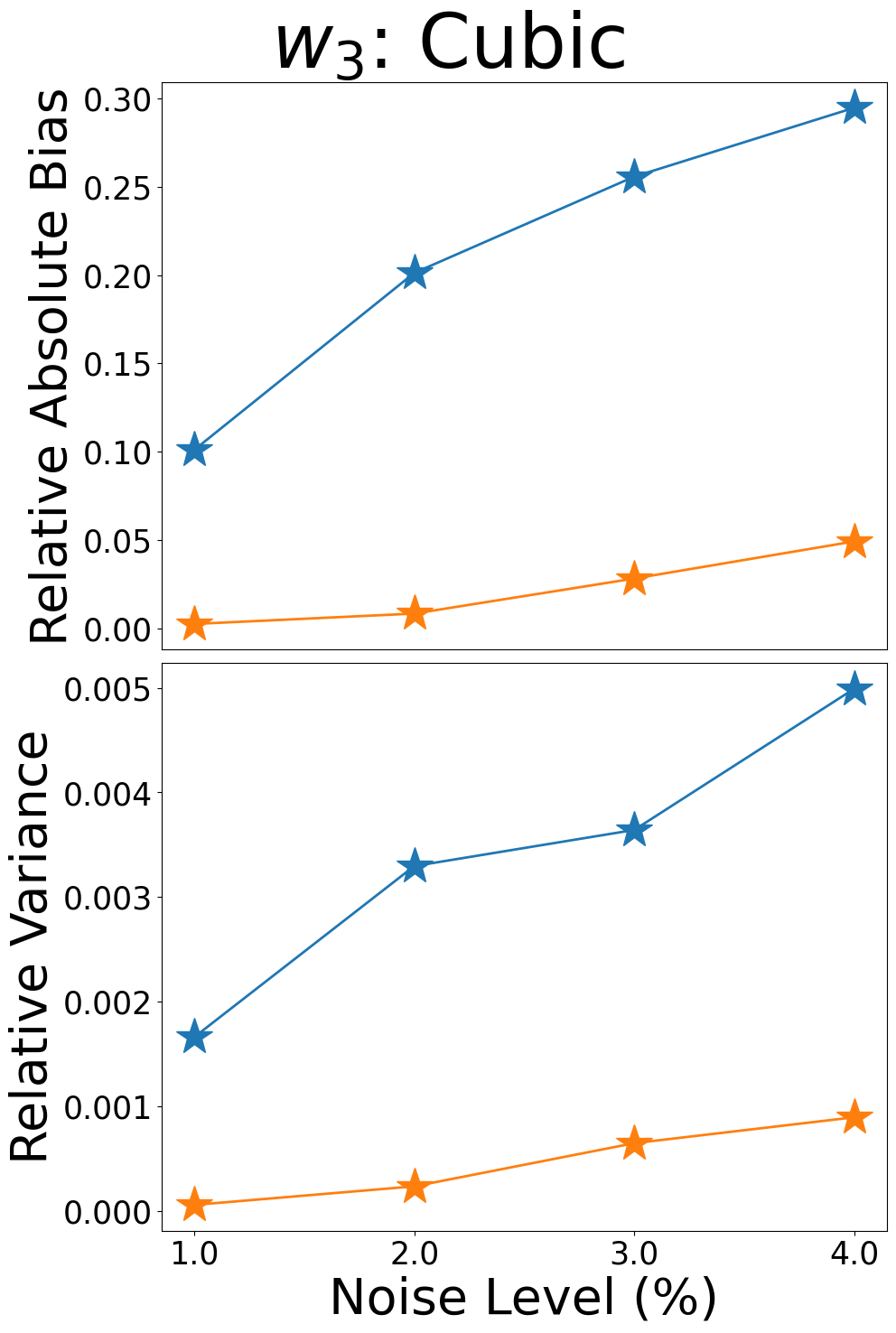}
    \includegraphics[width=0.4\linewidth]{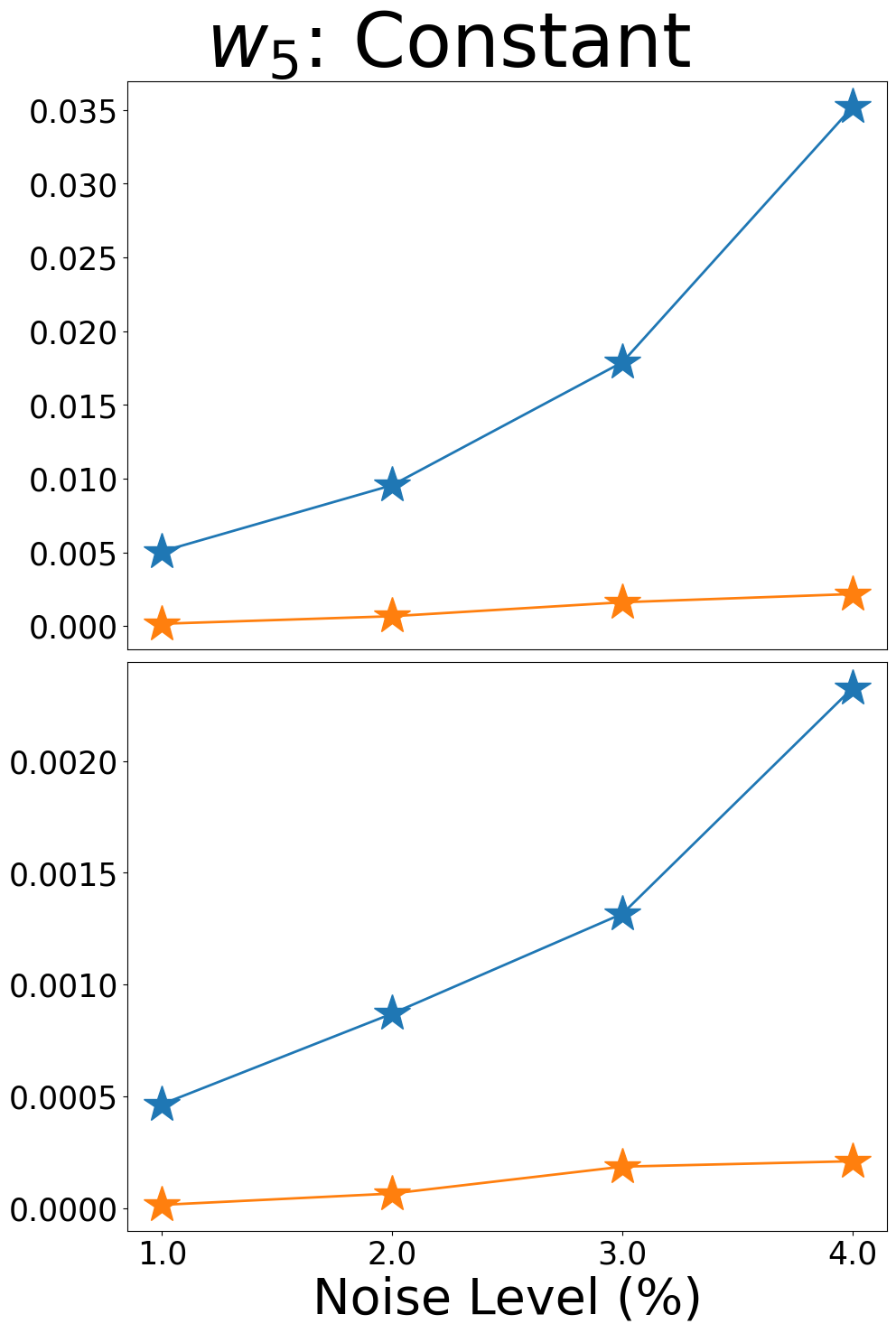}
    \\
    \includegraphics[width=0.3\linewidth]{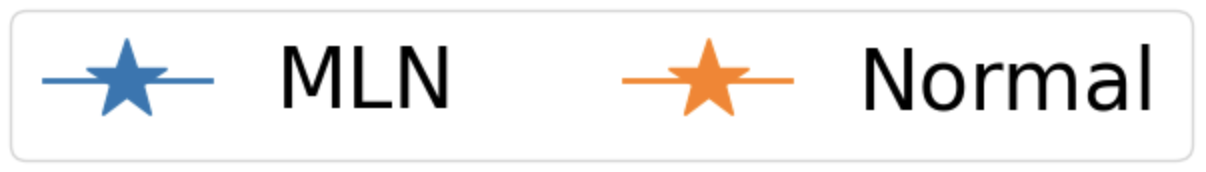}
    \caption{Relative bias magnitude and variance for WENDy estimators of LV model across 100 datasets with increasing levels of MLN and normal noise. Parameters modulating different order terms are selectively shown.}
    \label{fig:FHNBiasVarNoise}
\end{figure}
As seen in Figure~\ref{fig:FHNBiasVarNoise}, parameter estimators from datasets with log-normal noise exhibited substantially greater bias across all parameters compared to those with normal noise. In contrast, the variance of parameter estimators was higher under normal noise than under log-normal noise. This helps explain why coverage for MLN noise was much worse at very low noise levels, as the elevated bias in the parameter estimators dominated the results.

\subsubsection{Varying Data Resolution}
\subsubsection*{Additive Normal Noise}
\begin{figure} 
    \centering
    \begin{tabular}{c}
{\includegraphics[width=0.8\linewidth]{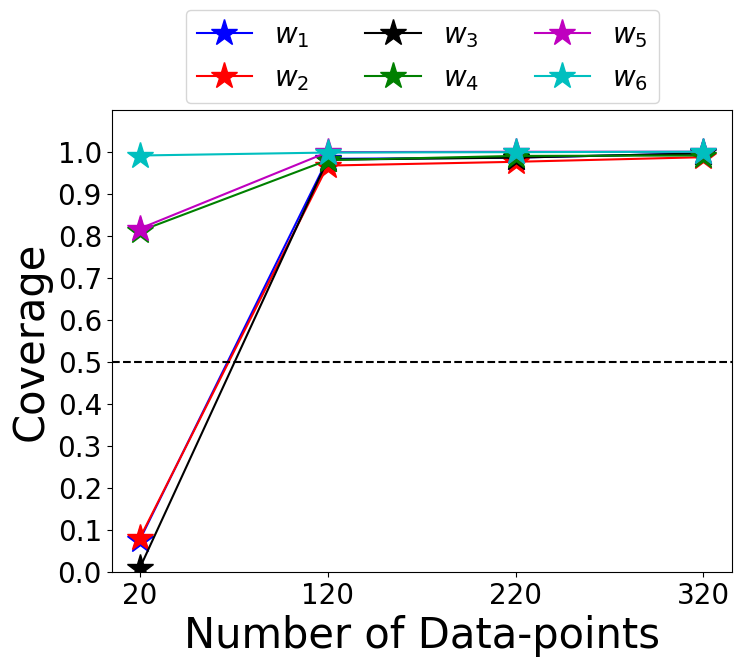}}
    \\
    \text{(a)}
    \\
\includegraphics[width=1\linewidth]{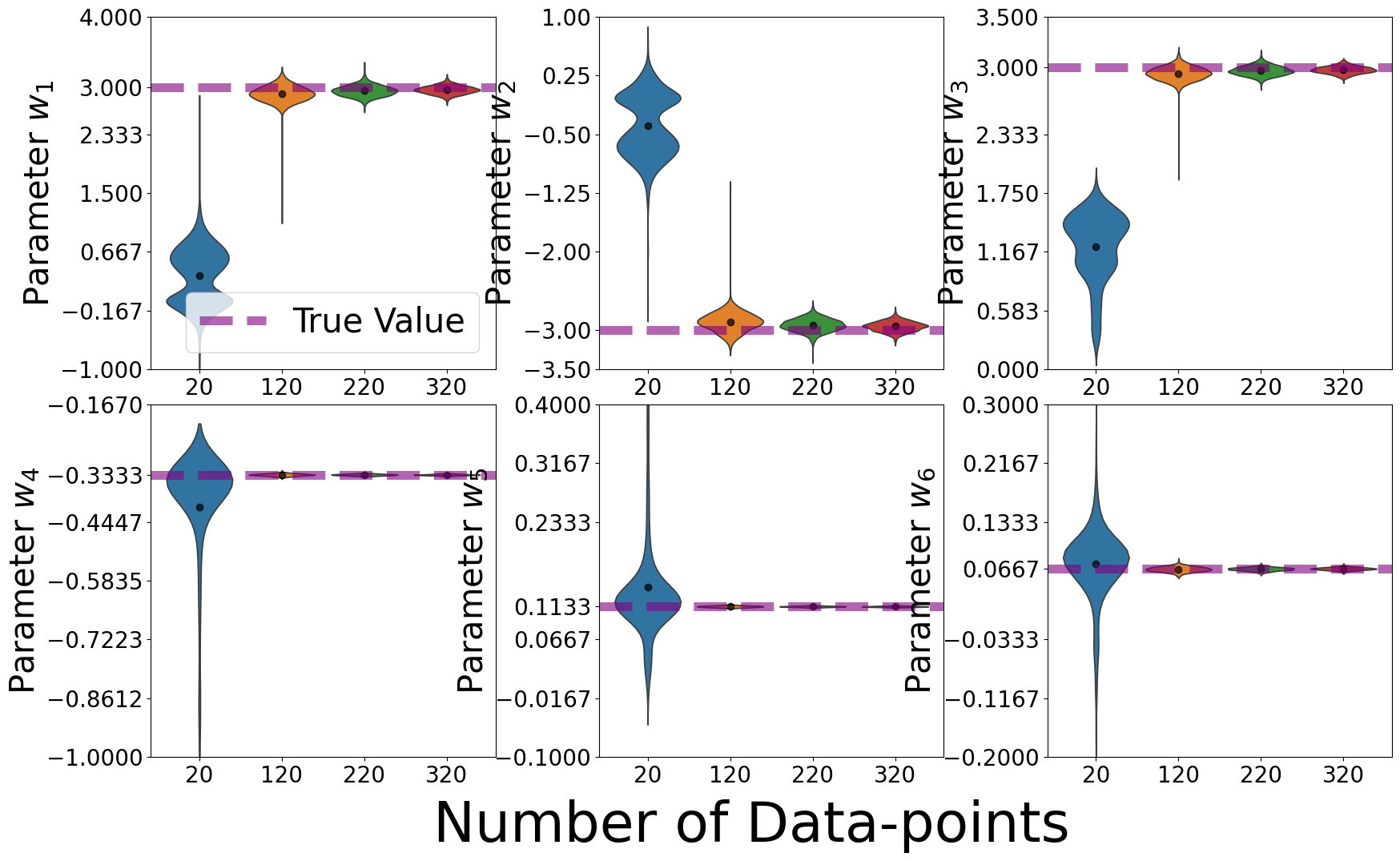}
    \\
      \text{(b)} 
    \end{tabular}
     \caption{FitzHugh-Nagumo model parameter estimation performance with increasing data resolution (1000 datasets per level, 2\% additive normal noise). (a) coverage across four noise levels. (b) violin plots of parameter estimates, with the dashed red line indicating the true parameter values.}
    \label{fig:CovBiasResFHNTN}
\end{figure}

As shown in Figure~\ref{fig:CovBiasResFHNTN}(a), the coverage of the 95\% confidence intervals for the $w_6$ parameter remained slightly above the nominal level as the resolution increased from 20 to 400 data points. The $w_1$, $w_2$, and $w_3$ parameters rose from below 50\% coverage at 20 data points to near nominal at 400 data points. The $w_4$ and $w_5$ parameters began around 80\% coverage and increased to above nominal with higher resolution. Once again, the coverage of the $u_1$-state parameters was much more sensitive to resolution, which can be attributed to the cubic term in the $u_1$ dynamics. This added nonlinearity makes parameter identifiability more difficult and requires more data points for reliable estimation. As shown in Figure~\ref{fig:CovBiasResFHNTN}(b), the bias and variance of all parameters decreased substantially as resolution increased. At 20 data points, the $w_1$, $w_2$, and $w_3$ parameters also exhibited signs of bimodality, which is likely due to the difficulty of identifying the phase and frequency of the dynamics under low-resolution data.  

\begin{figure} 
    \centering
    \begin{tabular}{c}
    
    \includegraphics[width=1\linewidth]{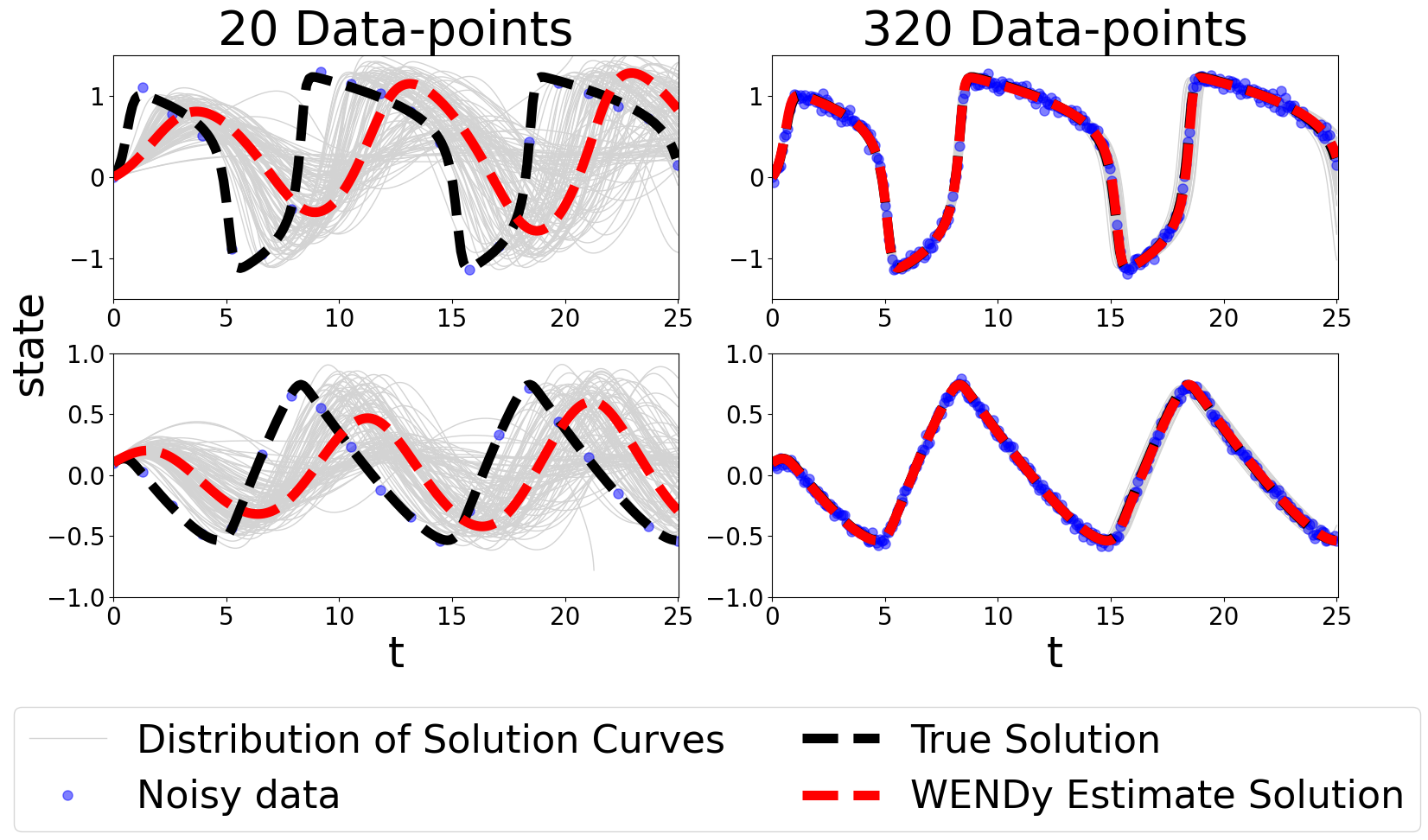} 
    \\
    \text{(a)}
    \\
    \includegraphics[width=1\linewidth]{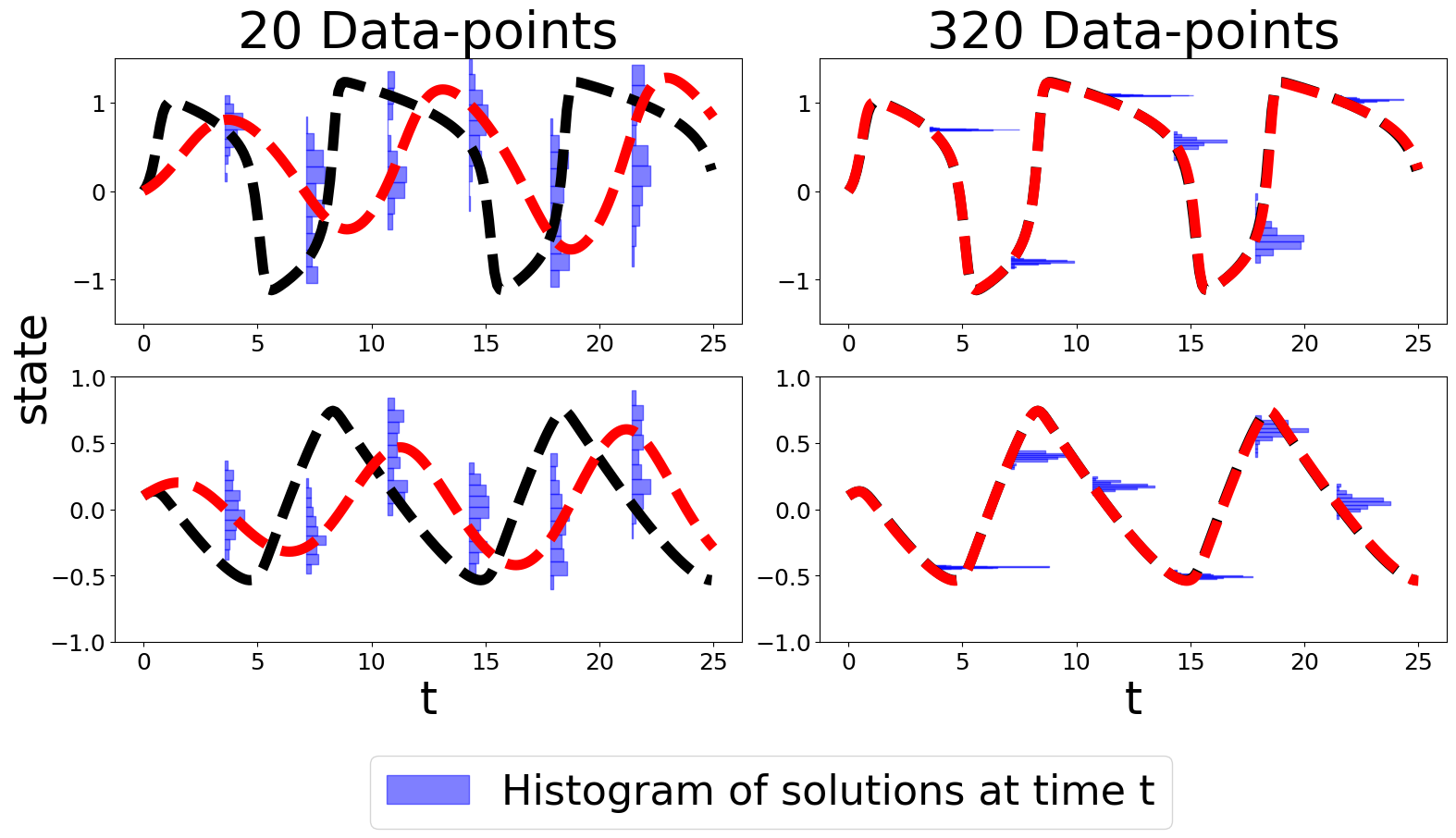}
    \\
    \text{(b)}
    \end{tabular}
     \caption{Top row (a): FitzHugh-Nagumo model parameter estimation example and uncertainty quantification on two datasets: one dataset with low resolution (left) and one dataset with resolution (right). The light gray curves are used to illustrate the uncertainty around the WENDy solutions; they are obtained via parametric bootstrap, as a sample of WENDy solutions corresponding to a random sample of 1000 parameters from their estimated asymptotic estimator distribution.  Bottom row (b): WENDy solution and histograms of state distributions across specific points in time for the datasets in (a).}
    \label{fig:SamplePlotResFHNTN}
\end{figure}

As shown in Figure~\ref{fig:SamplePlotResFHNTN}(a) and (b), the distribution of solution states (at selected time points) was wider at lower resolution levels compared to higher ones. Time points with higher-magnitude state values were also more skewed across both resolutions. At 20 data points, the fit struggled considerably to capture the steep growth and decay in the FHN model, despite the relatively low levels of noise.

\clearpage
\subsubsection*{Multiplicative Log-Normal Noise}
\begin{figure} 
    \centering
    \begin{tabular}{c}
{\includegraphics[width=0.8\linewidth]{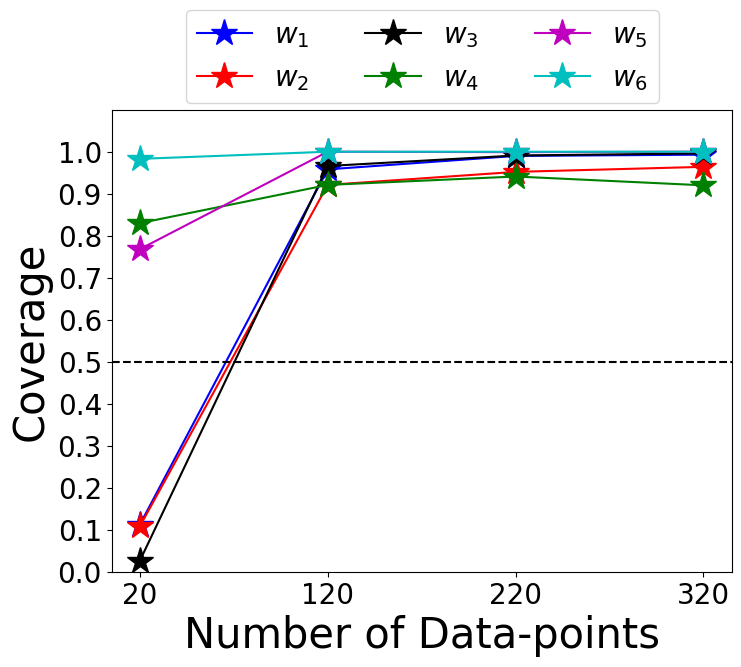}}
    \\
    \text{(a)}
    \\
    \includegraphics[width=1\linewidth]{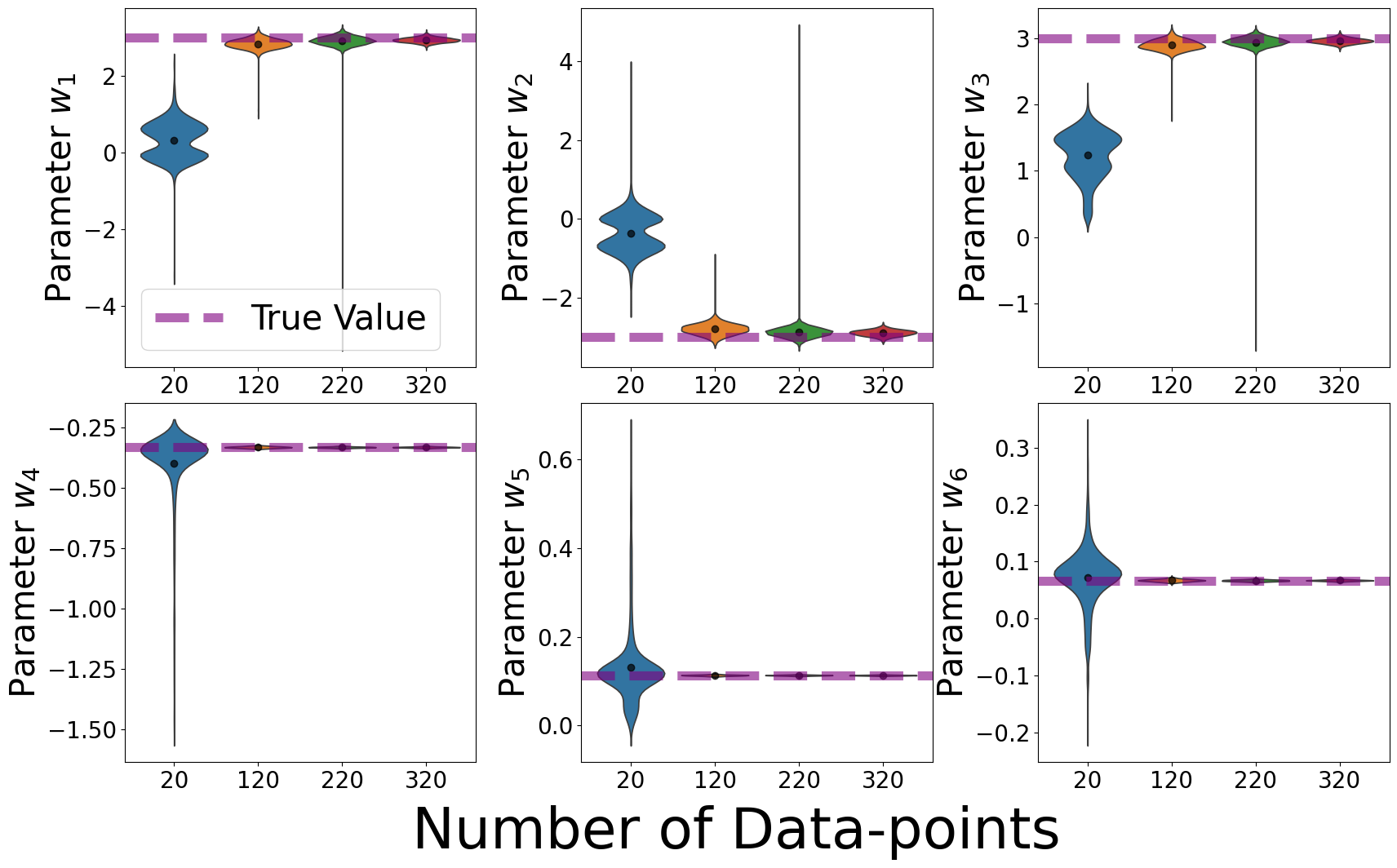}
    \\
     
      \text{(b)} 
    \end{tabular}
     \caption{FitzHugh-Nagumo model parameter estimation performance with increasing data resolution (1000 datasets per level, 2\% normal noise). (a) coverage across four noise levels. (b) violin plots of parameter estimates, with the dashed red line indicating the true parameter values.}
    \label{fig:CovBiasResFHNLN}
\end{figure}

As shown in Figure~\ref{fig:CovBiasResFHNLN}(a), similar to the additive normal noise case, the coverage of the 95\% confidence intervals for the $w_6$ parameter remained slightly above the nominal level as the resolution increased from 20 to 400 data points. The $w_1$, $w_2$, and $w_3$ parameters rose from below 50\% coverage at 20 data points to around nominal at 400 data points. The $w_4$ and $w_5$ parameters began near 80\% coverage before rising above nominal with increasing resolution. As shown in Figure~\ref{fig:CovBiasResFHNLN}(b), the bias and variance of all parameters decreased substantially as resolution increased. The bimodality in the violin plots of the $u_1$-state parameters was once again observed.  

\begin{figure} 
    \centering
    \begin{tabular}{c}
    
    \includegraphics[width=1\linewidth]{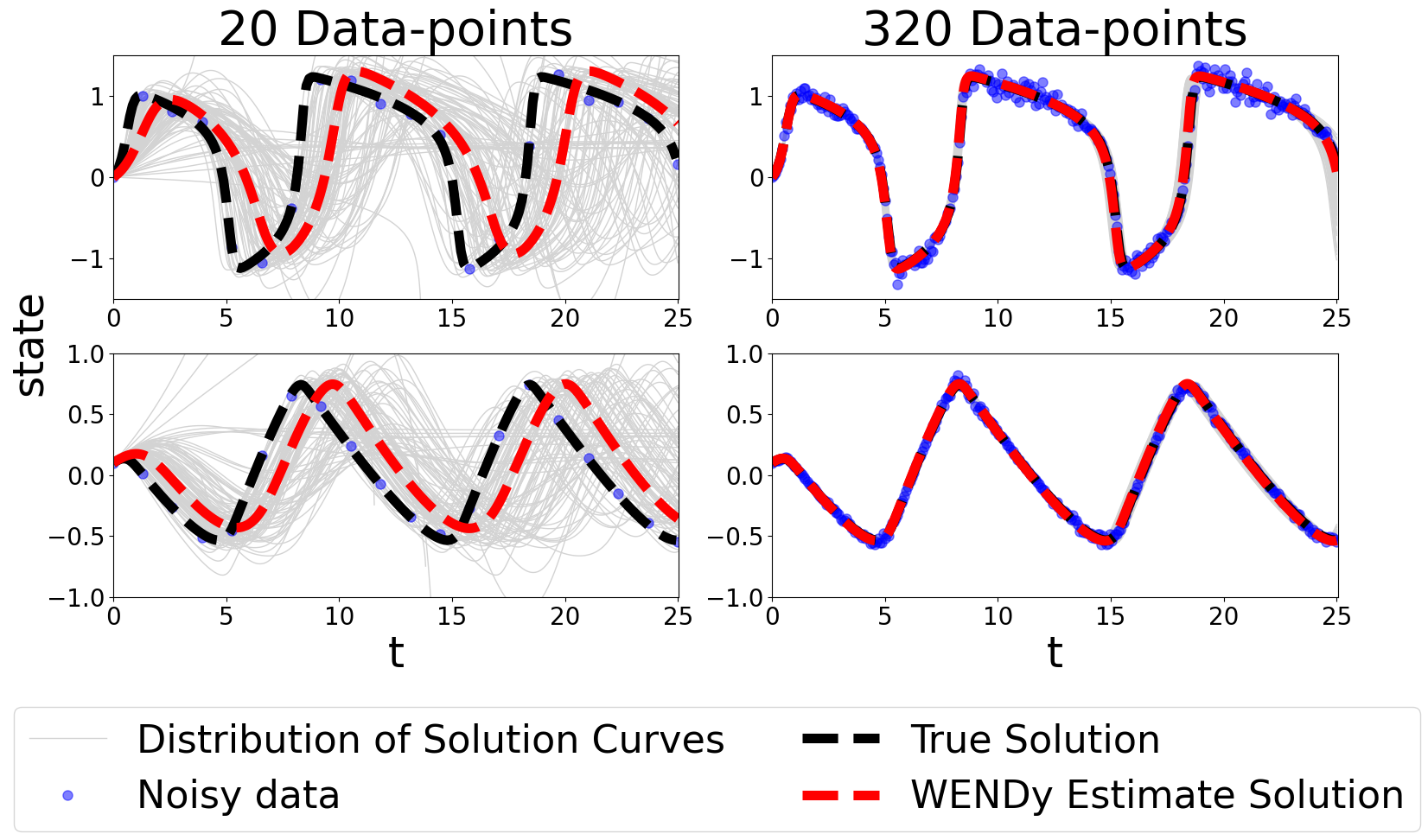} 
    \\
    \text{(a)}
    \\
    \includegraphics[width=1\linewidth]{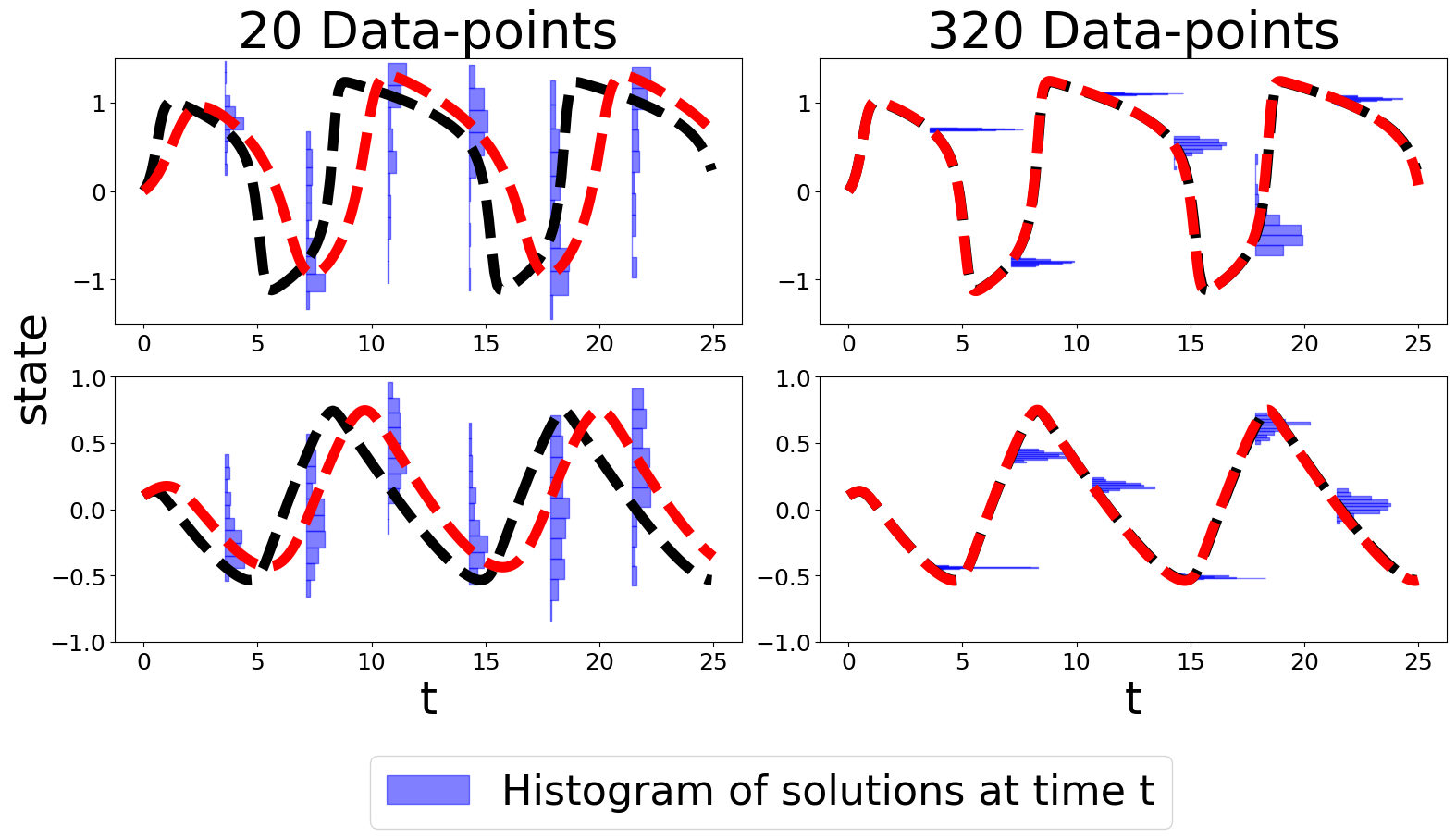}
    \\
    \text{(b)}
    \end{tabular}
     \caption{Top row (a): FitzHugh-Nagumo model parameter estimation example and uncertainty quantification on two datasets: one dataset with low resolution (left) and one dataset with resolution (right). The light gray curves are used to illustrate the uncertainty around the WENDy solutions; they are obtained via parametric bootstrap, as a sample of WENDy solutions corresponding to a random sample of 1000 parameters from their estimated asymptotic estimator distribution.  Bottom row (b): WENDy solution and histograms of state distributions across specific points in time for the datasets in (a).}
    \label{fig:SamplePlotResFHNLN}
\end{figure}

As shown in Figure~\ref{fig:SamplePlotResFHNLN}(a) and (b), the distribution of solution curves at lower resolution levels was wider than at higher resolutions and exhibited bimodality at certain time points. Compared to the normal noise case, the WENDy estimate fit captured the steep growth and decay in the FHN dynamics much better at 20 data points. This is likely due to the extremely low noise levels required to achieve reasonable parameter coverage.

\clearpage 
\subsection{Hindmarsh-Rose Model}
\subsubsection{Varying Noise Level}
\subsubsection*{Additive Normal Noise}
\begin{figure} 
    \centering
    \begin{tabular}{c}
{\includegraphics[width=0.8\linewidth]{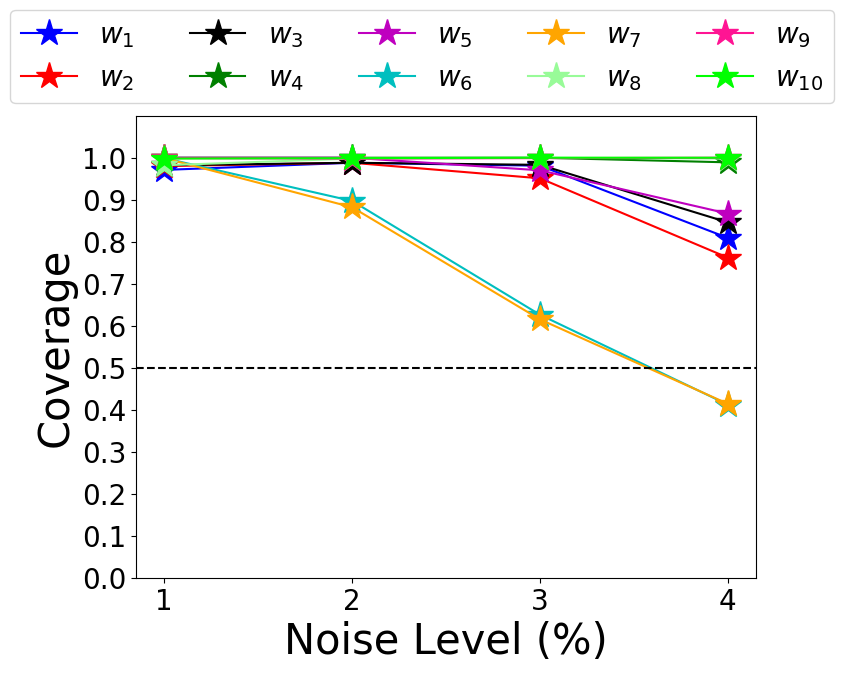}}
    \\
    \text{(a)}
    \\
    \includegraphics[width=1\linewidth]{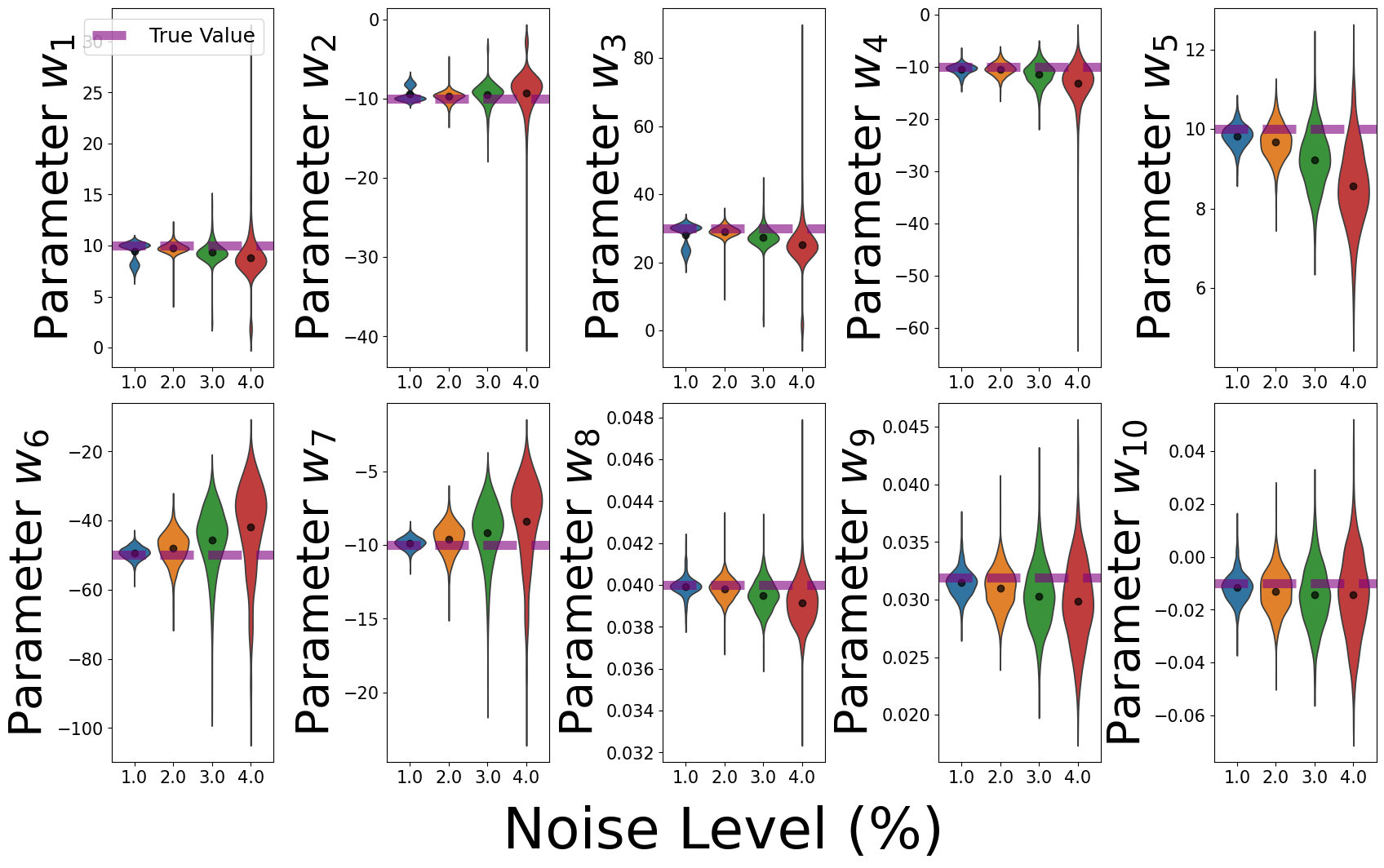}
    \\
      \text{(b)} 
    \end{tabular}
     \caption{Hindmarsh-Rose model parameter estimation performance with increasing ACN noise (1000 datasets per level, 205 data points each). (a) coverage across four noise levels. (b) violin plots of parameter estimates, with the dashed red line indicating the true parameter values.}
    \label{fig:CovBiasHMRTN}
\end{figure}

As shown in Figure~\ref{fig:CovBiasHMRTN}(a), the coverage of the 95\% confidence intervals for the $w_4$ and $w_{10}$ parameters remained slightly above the nominal 95\% level for all noise levels from 1\% to 4\%. The coverage for the $w_6$ and $w_7$ parameters decreased to below 50\% as noise increased to 4\%. Coverage for the remaining parameters fell below nominal but stayed above 50\% as noise increased. Unlike in the FHN model, despite the $u_1$ state containing a cubic term, the poorest coverage in the HMR model occurred for the $u_2$-state parameters.  

As shown in Figure~\ref{fig:CovBiasHMRTN}(b), the bias and variance for all parameters increased slightly with noise, as expected. For the $w_6$ and $w_7$ parameters, the violin plots indicate that their variance relative to other parameters did not increase as much, as shown by the shorter tails. This helps explain the poor coverage of these parameters at high noise levels.  

\begin{figure} 
    \centering
    \begin{tabular}{c}
    
    \includegraphics[width=1\linewidth]{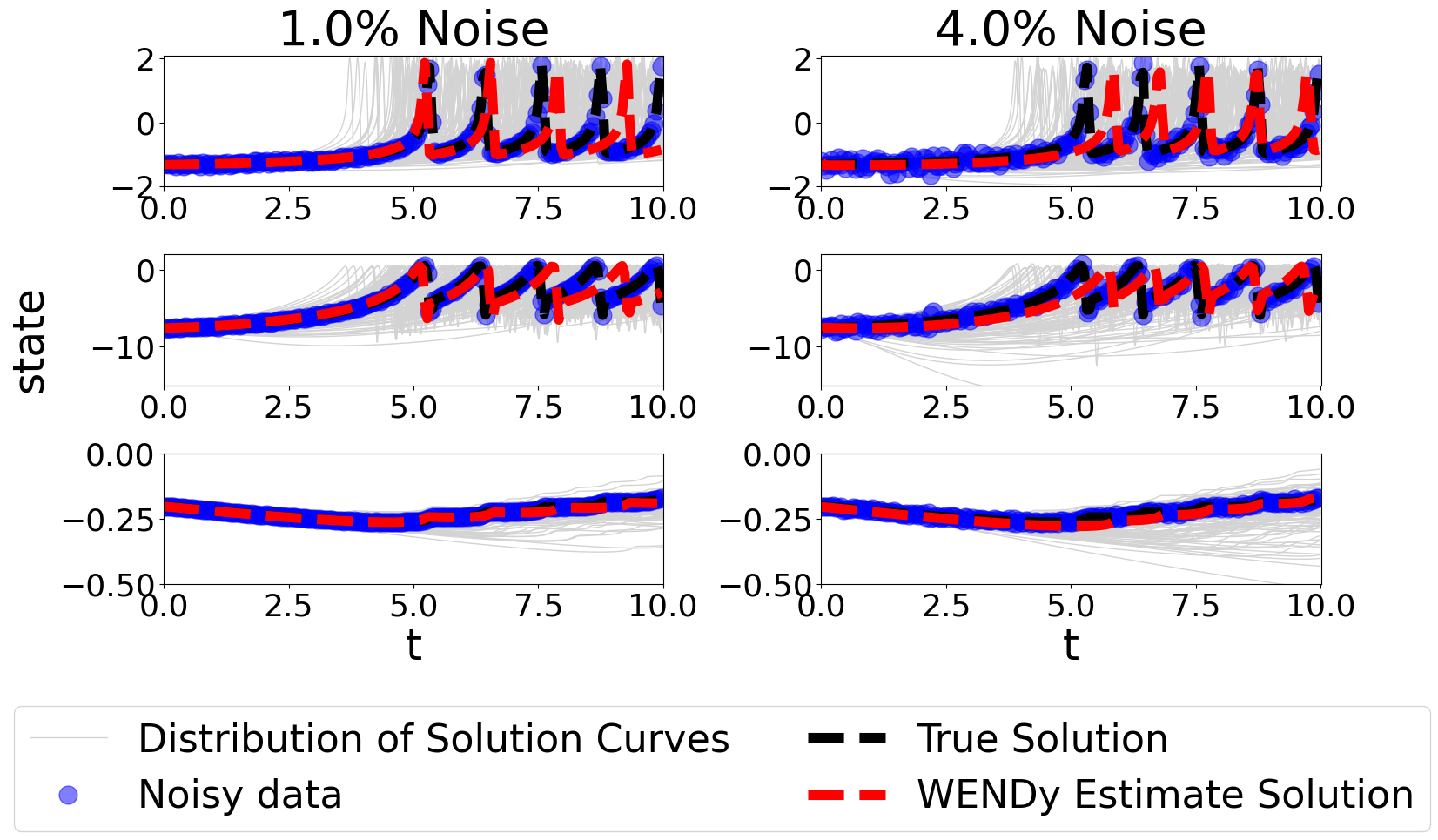} 
    \\
    \text{(a)}
    \\
    \includegraphics[width=1\linewidth]{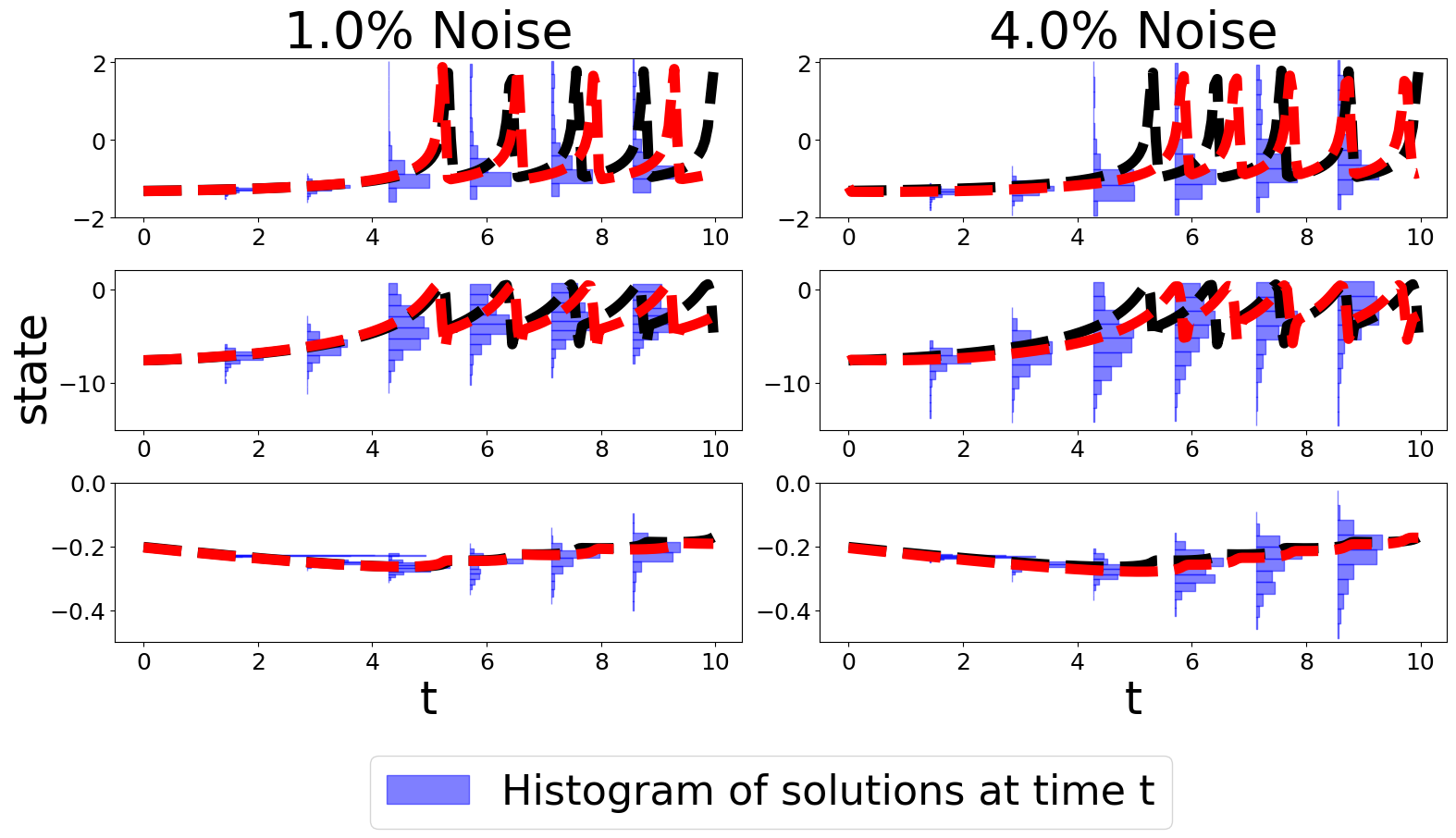}
    \\
    \text{(b)}
    \end{tabular}
     \caption{Top row (a): Hindmarsh-Rose model parameter estimation example and uncertainty quantification on two datasets: one dataset with low uncertainty and high coverage (left) and one dataset with high uncertainty and low coverage (right). The light gray curves are used to illustrate the uncertainty around the WENDy solutions; they are obtained via parametric bootstrap, as a sample of WENDy solutions corresponding to a random sample of 1000 parameters from their estimated asymptotic estimator distribution.  Bottom row (b): WENDy solution and histograms of state distributions across specific points in time for the datasets in (a).}
    \label{fig:SamplePlotHMRTN}
\end{figure}

As shown in Figure~\ref{fig:SamplePlotHMRTN}(a) and (b), the distribution of solution states (at selected time points) was wider at higher noise levels than at lower noise levels. The WENDy-estimated solutions at both noise levels also showed some phase shift from the true solution for the oscillatory states.

\clearpage
\subsubsection*{Multiplicative Log-normal Noise}
\begin{figure} 
    \centering
    \begin{tabular}{c}
{\includegraphics[width=0.8\linewidth]{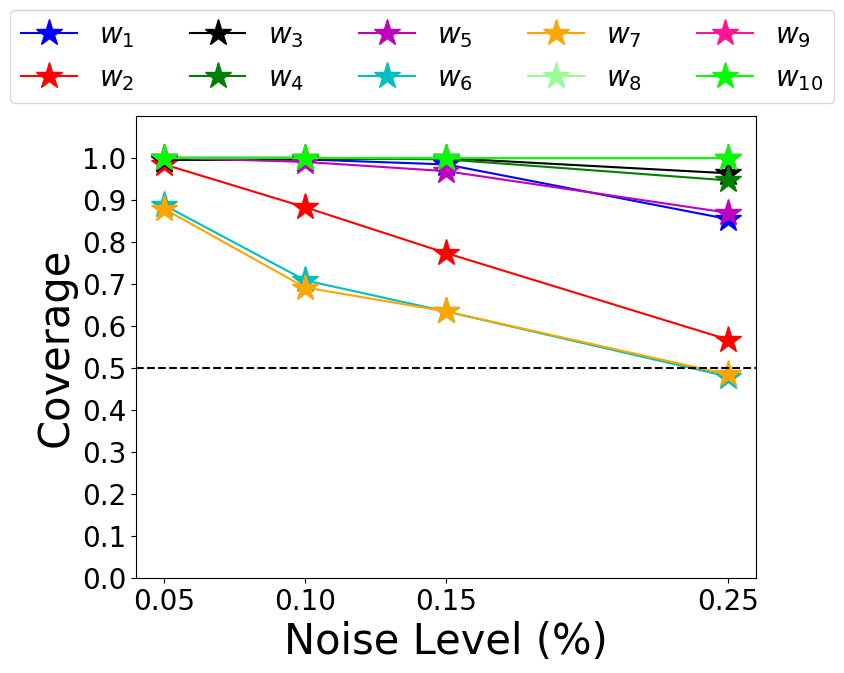}}
    \\
    \text{(a)}
    \\
    \includegraphics[width=1\linewidth]{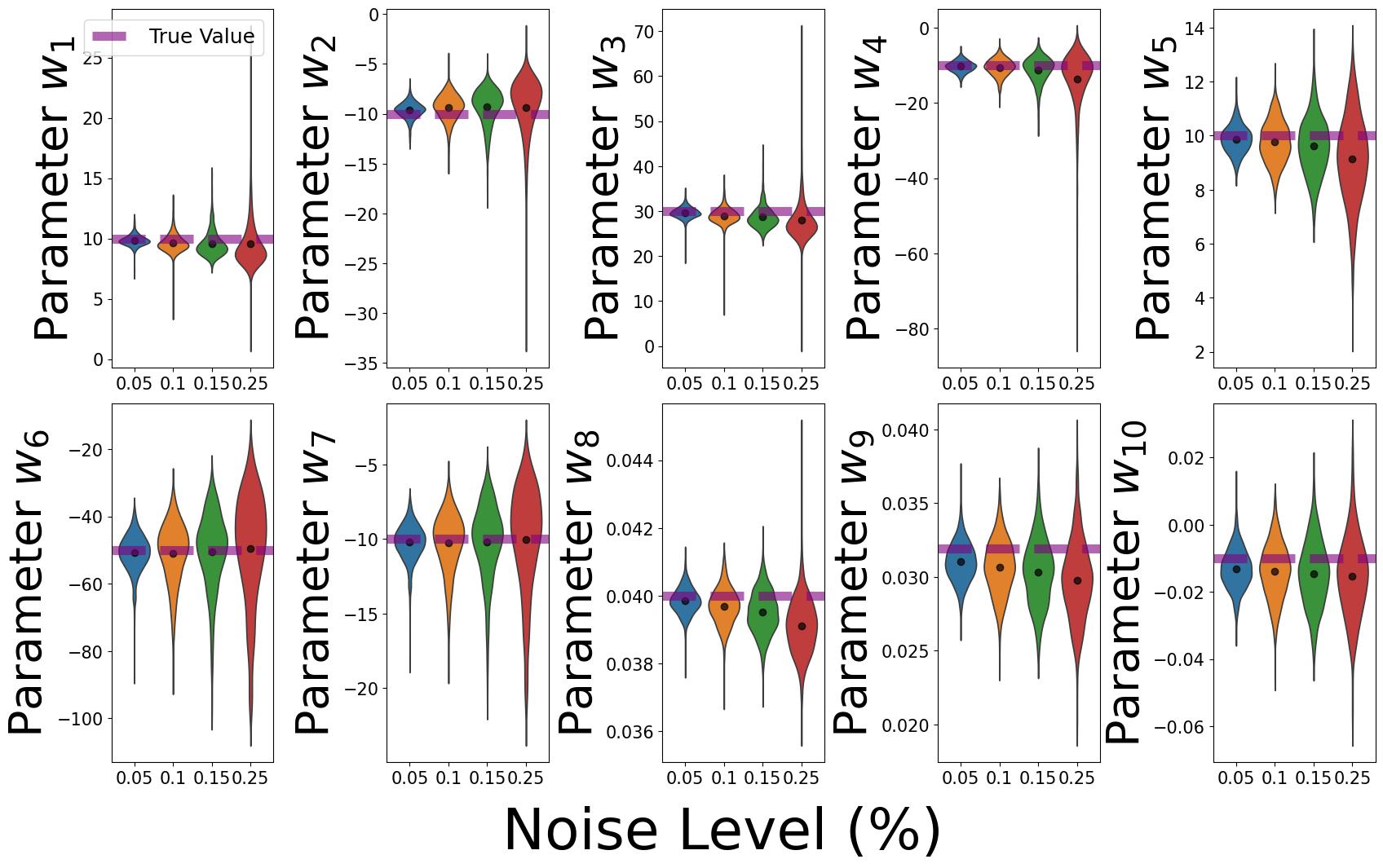}
    \\
      \text{(b)} 
    \end{tabular}
     \caption{Hindmarsh-Rose model parameter estimation performance with increasing  MLN noise (1000 datasets per level, 205 data points each). (a) coverage across four noise levels. (b) violin plots of parameter estimates, with the dashed red line indicating the true parameter values.}
    \label{fig:CovBiasHMRLN}
\end{figure}

As shown in Figure~\ref{fig:CovBiasHMRLN}(a), the coverage of the 95\% confidence intervals for the $w_3$, $w_4$, and $w_{10}$ parameters remained around the nominal 95\% level for all noise levels from 0.05\% to 0.25\%. The coverage for the $w_1$ and $w_5$ parameters decreased to just below nominal as noise increased to 0.25\%, while the coverage for the $w_2$ parameter dropped to just above 50\%. The $w_6$ and $w_7$ parameters again fell below 50\% coverage at 0.25\% noise.  

As shown in Figure~\ref{fig:CovBiasHMRLN}(b), the bias and variance for all parameters increased slightly with noise, as expected. A similar trend was observed for the $w_6$ and $w_7$ parameters as with additive normal noise, reflecting their poorer coverage compared to the other parameters. This suggests that these parameters are less robust to noise in the HMR model.  

\begin{figure} 
    \centering
    \begin{tabular}{c}
    
    \includegraphics[width=1\linewidth]{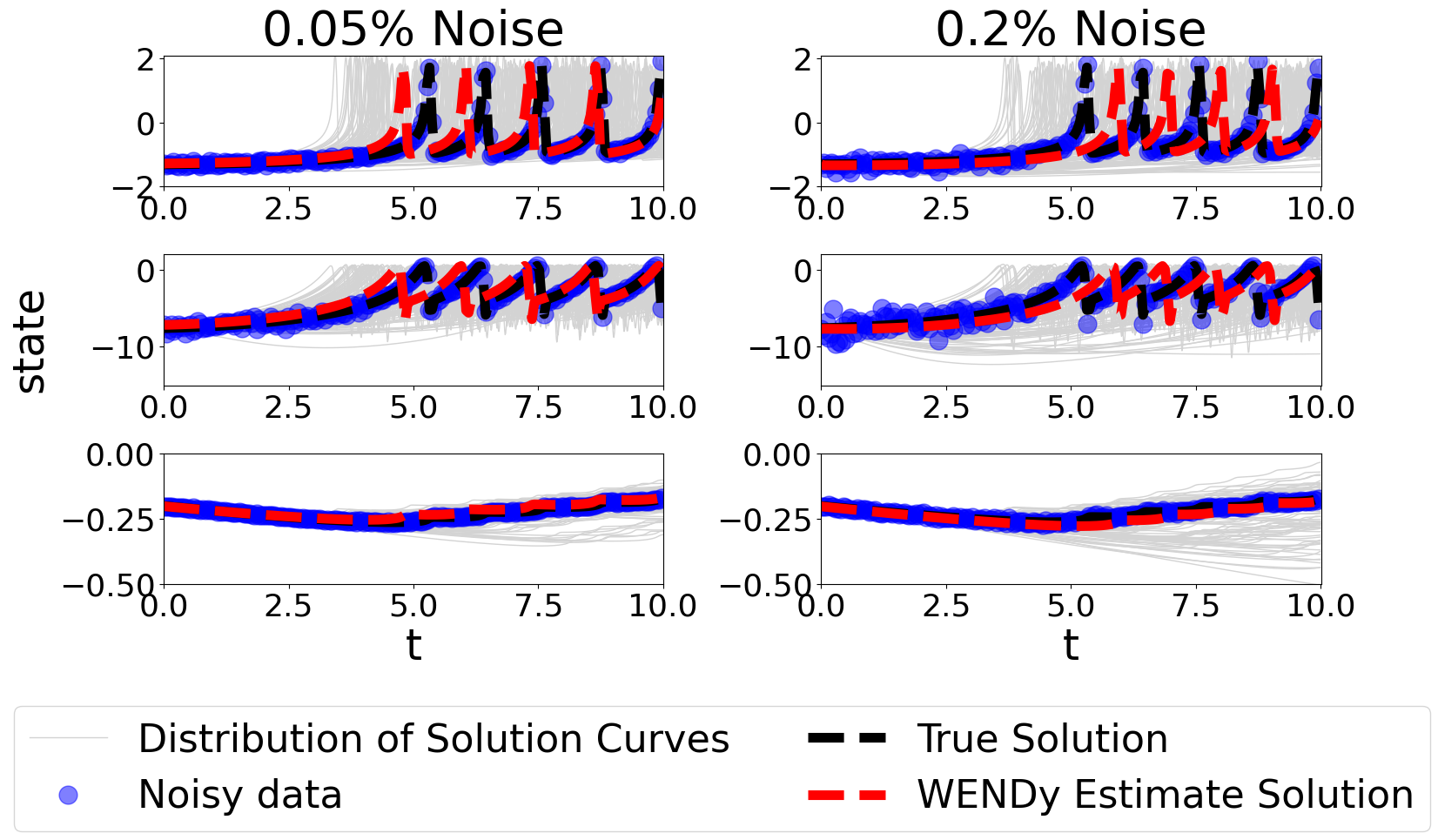} 
    \\
    \text{(a)}
    \\
    \includegraphics[width=1\linewidth]{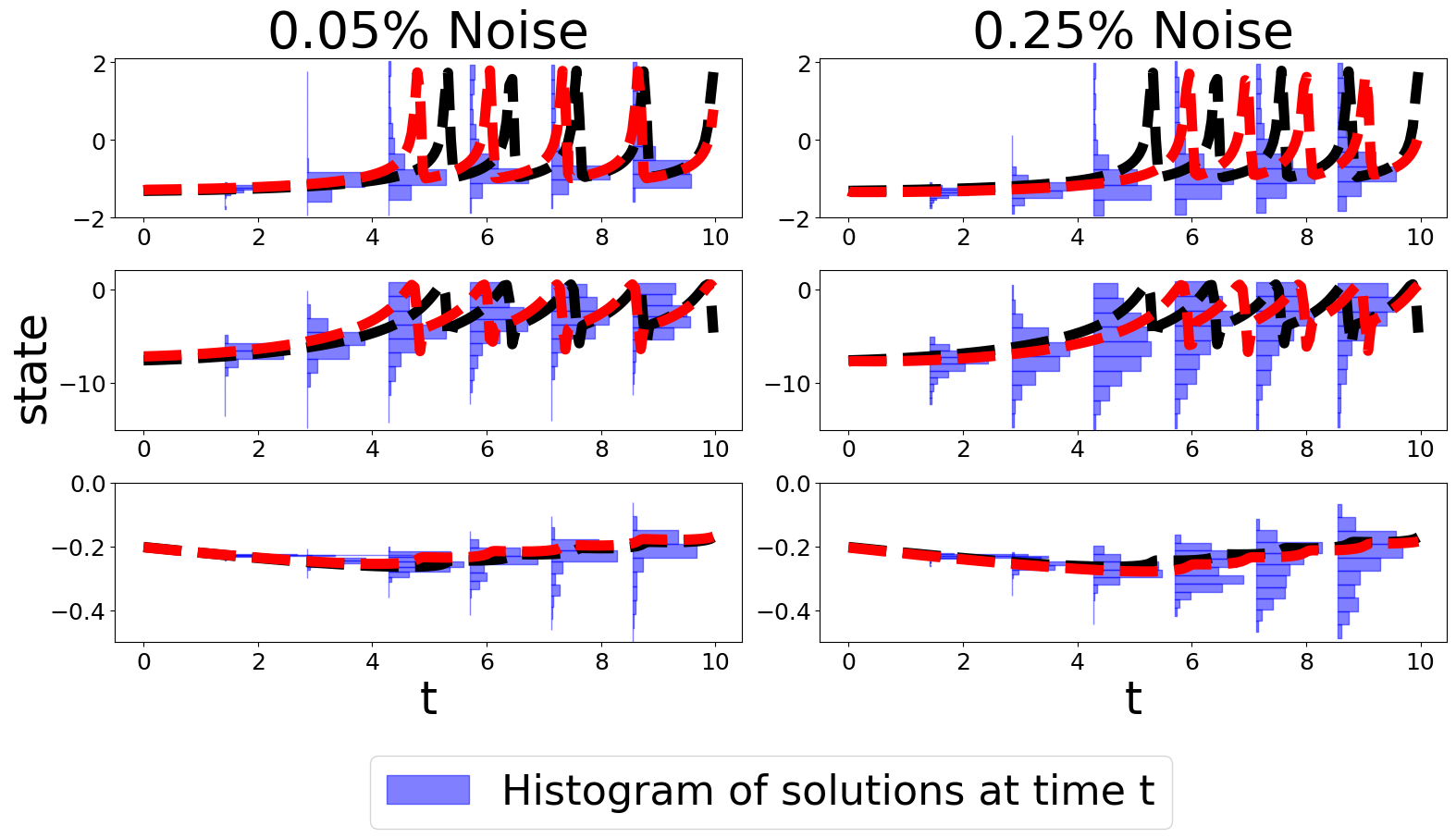}
    \\
    \text{(b)}
    \end{tabular}
     \caption{Top row (a): Hindmarsh-Rose model parameter estimation example and uncertainty quantification on two datasets: one dataset with low uncertainty and high coverage (left) and one dataset with high uncertainty and low coverage (right). The light gray curves are used to illustrate the uncertainty around the WENDy solutions; they are obtained via parametric bootstrap, as a sample of WENDy solutions corresponding to a random sample of 1000 parameters from their estimated asymptotic estimator distribution.  Bottom row (b): WENDy solution and histograms of state distributions across specific points in time for the datasets in (a).}
    \label{fig:SamplePlotHMRLN}
\end{figure}

As shown in Figure~\ref{fig:SamplePlotHMRLN}(a) and (b), the distribution of solution states (at selected time points) was wider at higher noise levels than at lower noise levels. The WENDy-estimated solutions for both noise levels also showed some phase shift from the true solution for the oscillatory states, similar to the additive normal noise case.

\clearpage
\begin{figure} 
    \centering
    \includegraphics[width=0.4\linewidth]{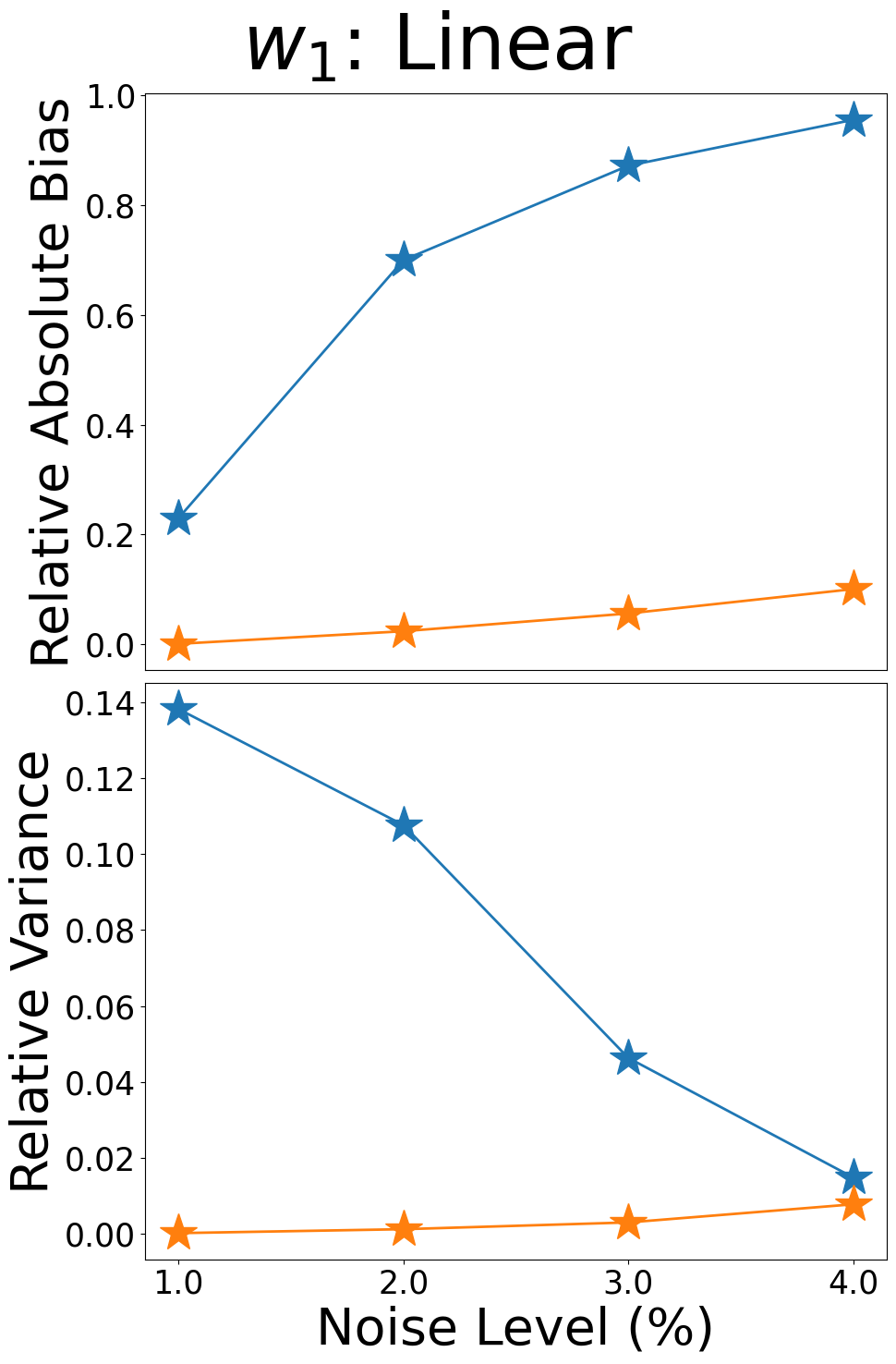}     \includegraphics[width=0.4\linewidth]{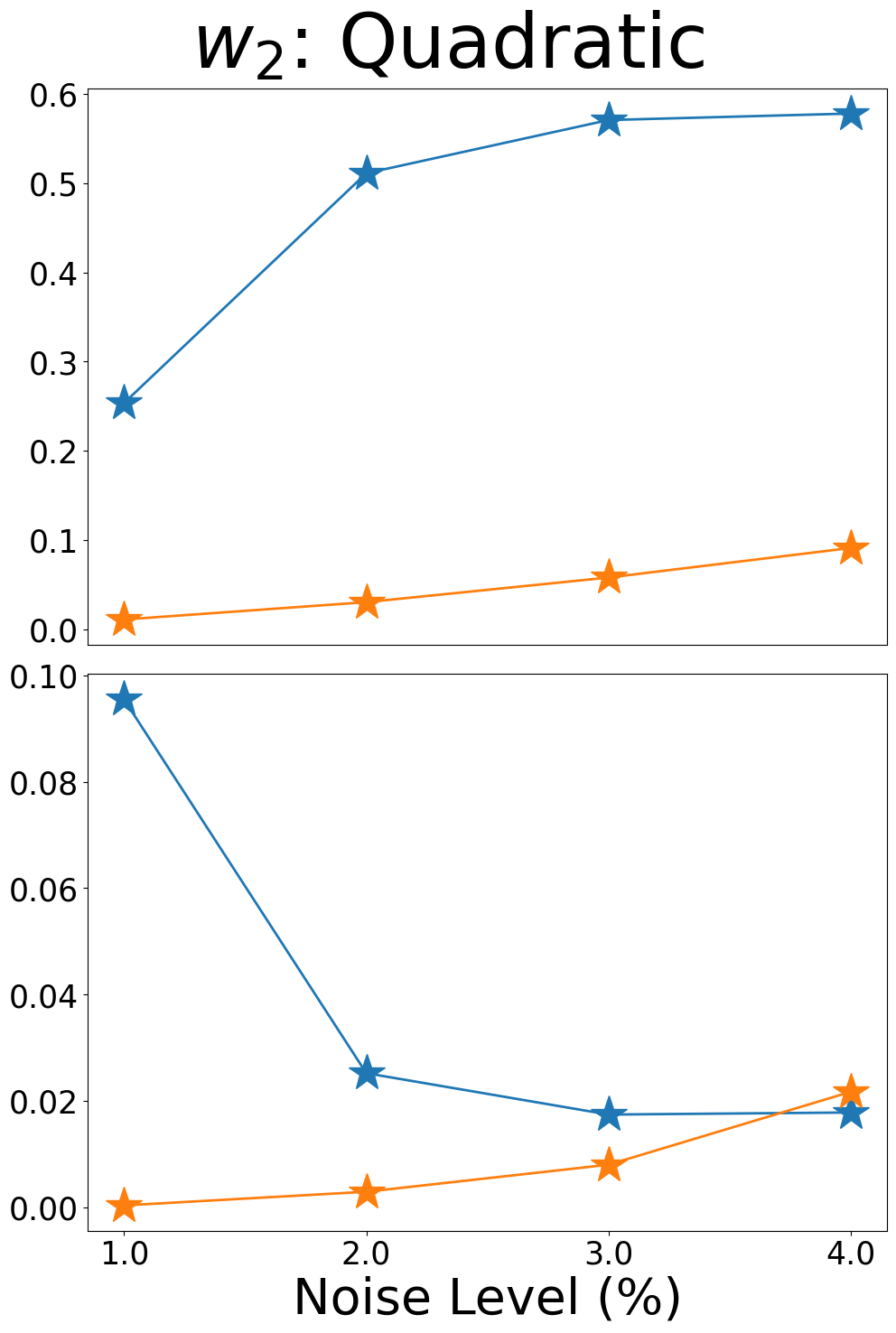}
    \includegraphics[width=0.4\linewidth]{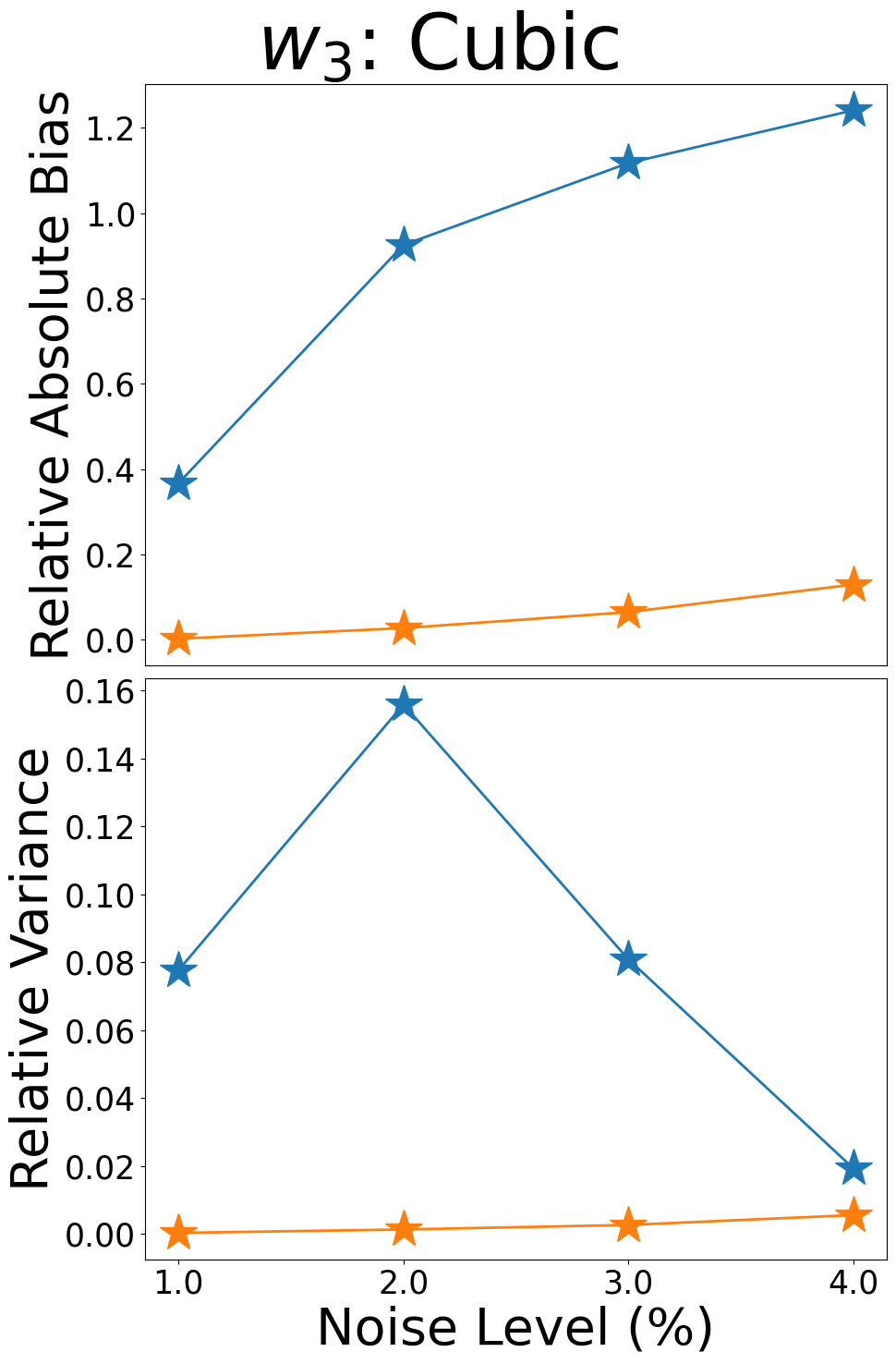}
    \includegraphics[width=0.4\linewidth]{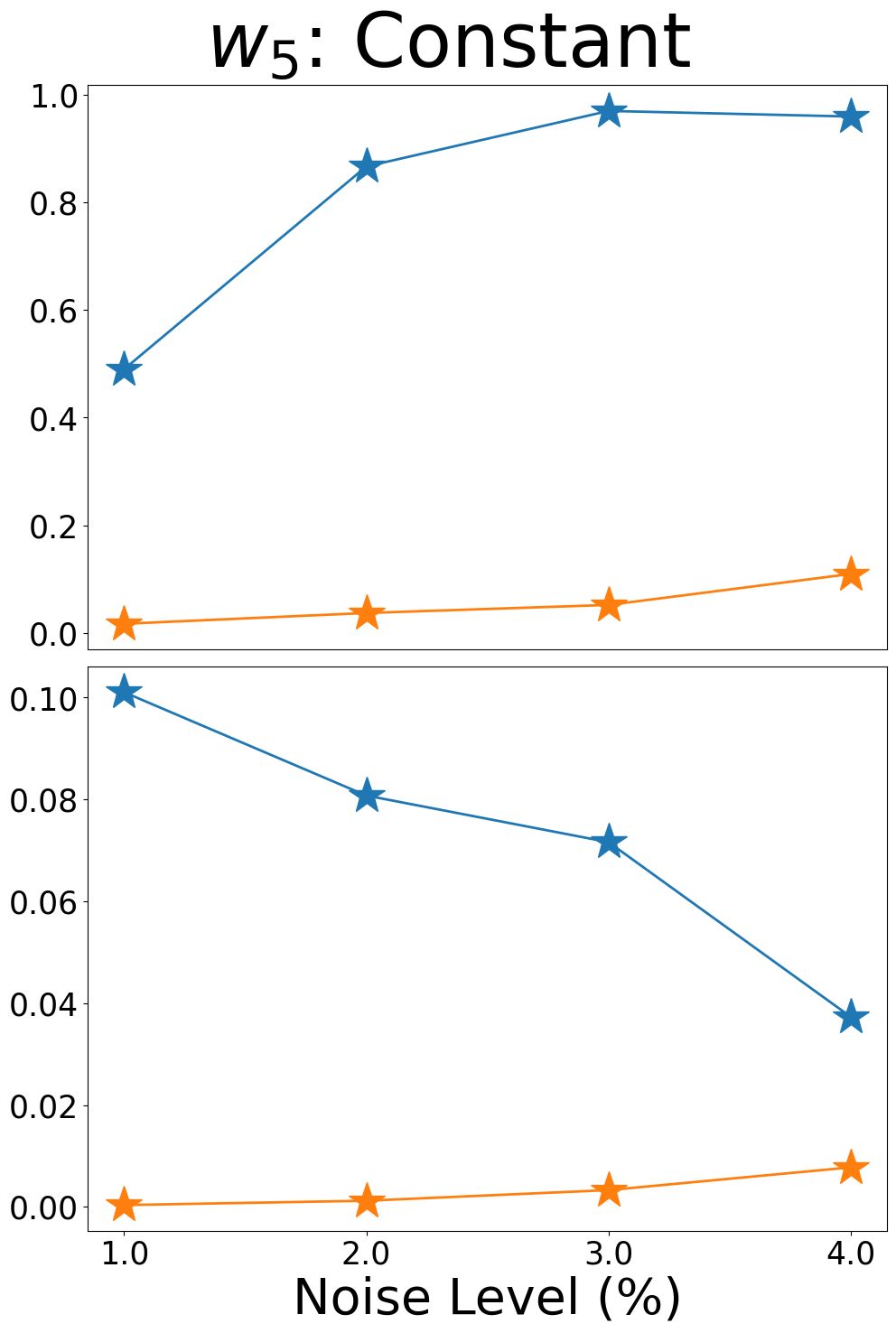}
    \\
    \includegraphics[width=0.3\linewidth]{NoiseDistrCompPlotsLogsitic/Legend2.png}
    \caption{Relative bias magnitude and variance for WENDy estimators of HMR model across 100 datasets with additive normal and MLN noise. Parameters modulating different order terms are selectively shown.}
    \label{fig:HMRBiasVarNoise}
\end{figure}
As seen in Figure~\ref{fig:HMRBiasVarNoise}, parameter estimators from datasets with MLN noise exhibited much higher relative bias and variance across all parameters compared to additive normal noise. In some cases, the relative bias was up to two orders of magnitude greater under MLN noise. This explains the poorer coverage observed for parameter estimators from MLN noise datasets. For the HMR model, applying a correction for the bias induced by MLN noise may therefore be worth considering.

\subsubsection{Varying Resolution Level}
\subsubsection*{Additive Normal Noise}
\begin{figure} 
    \centering
    \begin{tabular}{c}
{\includegraphics[width=0.8\linewidth]{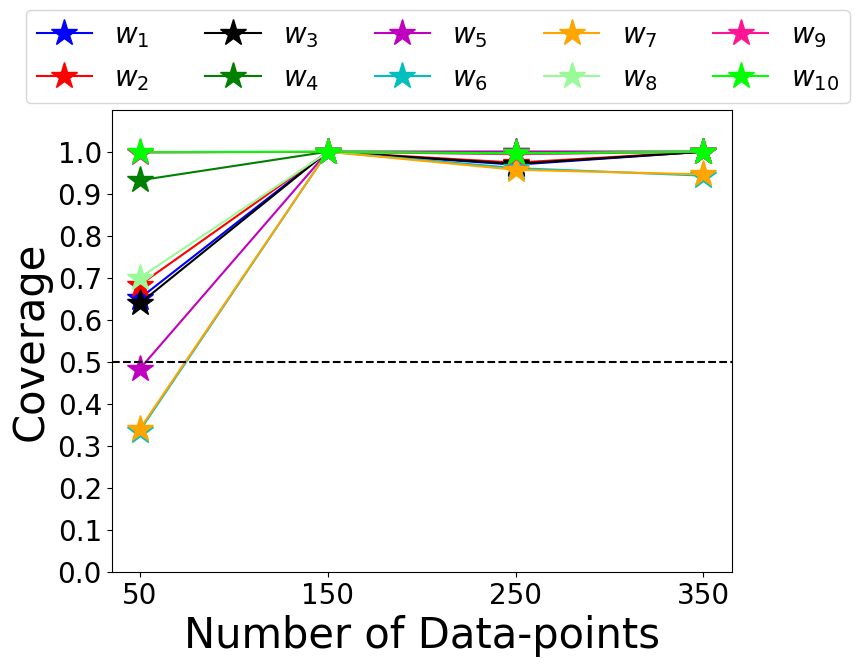}}
    \\
    \text{(a)}
    \\
\includegraphics[width=1\linewidth]{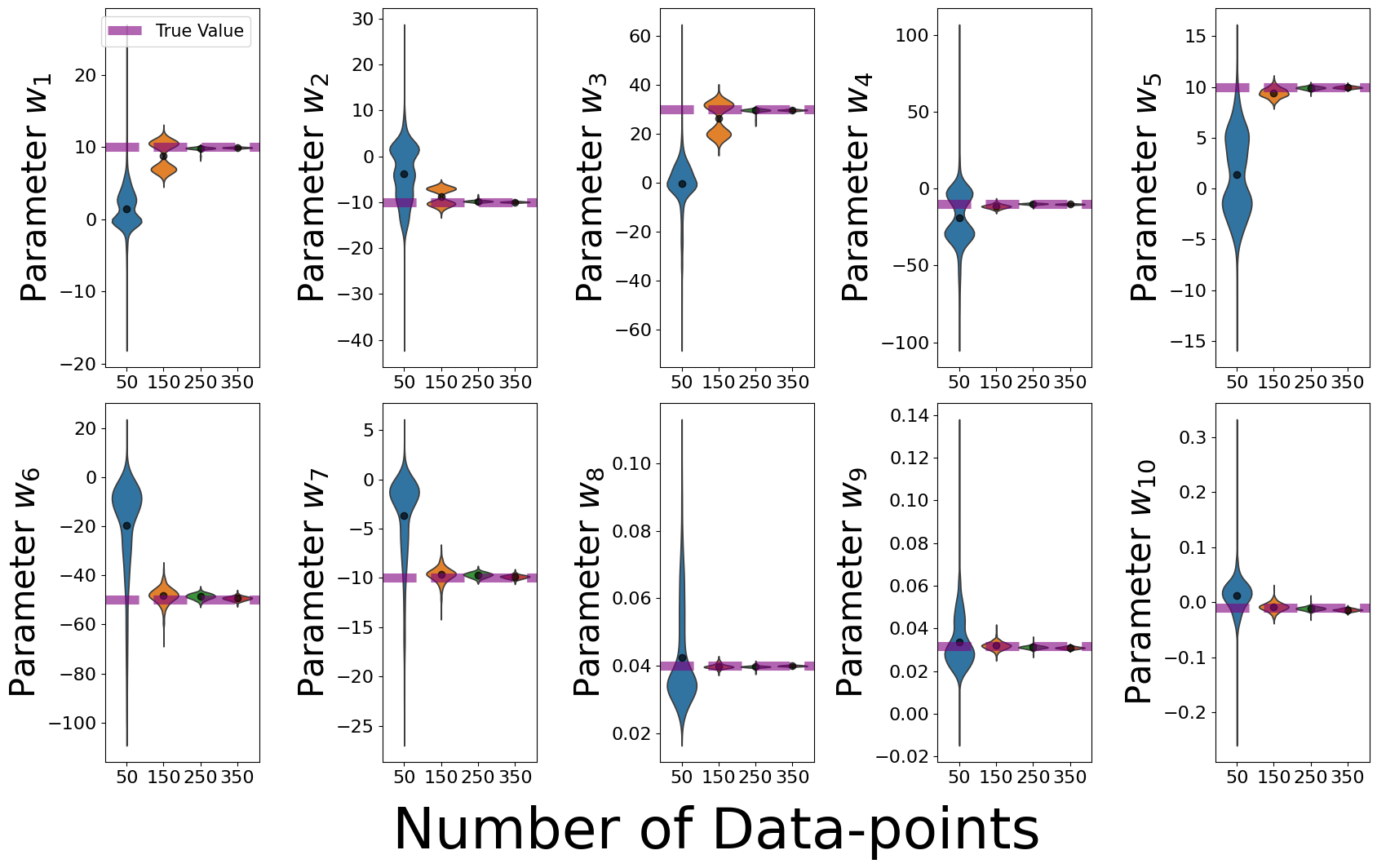}
    \\
      \text{(b)} 
    \end{tabular}
     \caption{Hindmarsh-Rose model parameter estimation performance with increasing data resolution (1000 datasets per level, 1\% normal noise). (a) coverage across four noise levels. (b) violin plots of parameter estimates, with the dashed red line indicating the true parameter values.}
    \label{fig:CovBiasResHMRTN}
\end{figure}

As shown in Figure~\ref{fig:CovBiasResHMRTN}(a), the coverage of the 95\% confidence intervals for the $w_4$, $w_9$, and $w_{10}$ parameters remained around the nominal 95\% level for all resolution levels from 50 to 350 data points. The coverage for the $w_6$, $w_7$, and $w_5$ parameters started below 50\% at 50 data points but increased to above nominal at 150 data points and higher. The coverage of the $w_1$, $w_2$, $w_3$, and $w_8$ parameters started slightly above 50\% and rose to nominal from 150 data points onward. As shown in Figure~\ref{fig:CovBiasResHMRTN}(b), the bias and variance for all parameters decreased slightly as resolution increased. The violin plots also indicated some bimodality for certain parameters at lower resolution levels.  

\begin{figure} 
    \centering
    \begin{tabular}{c}
    
    \includegraphics[width=1\linewidth]{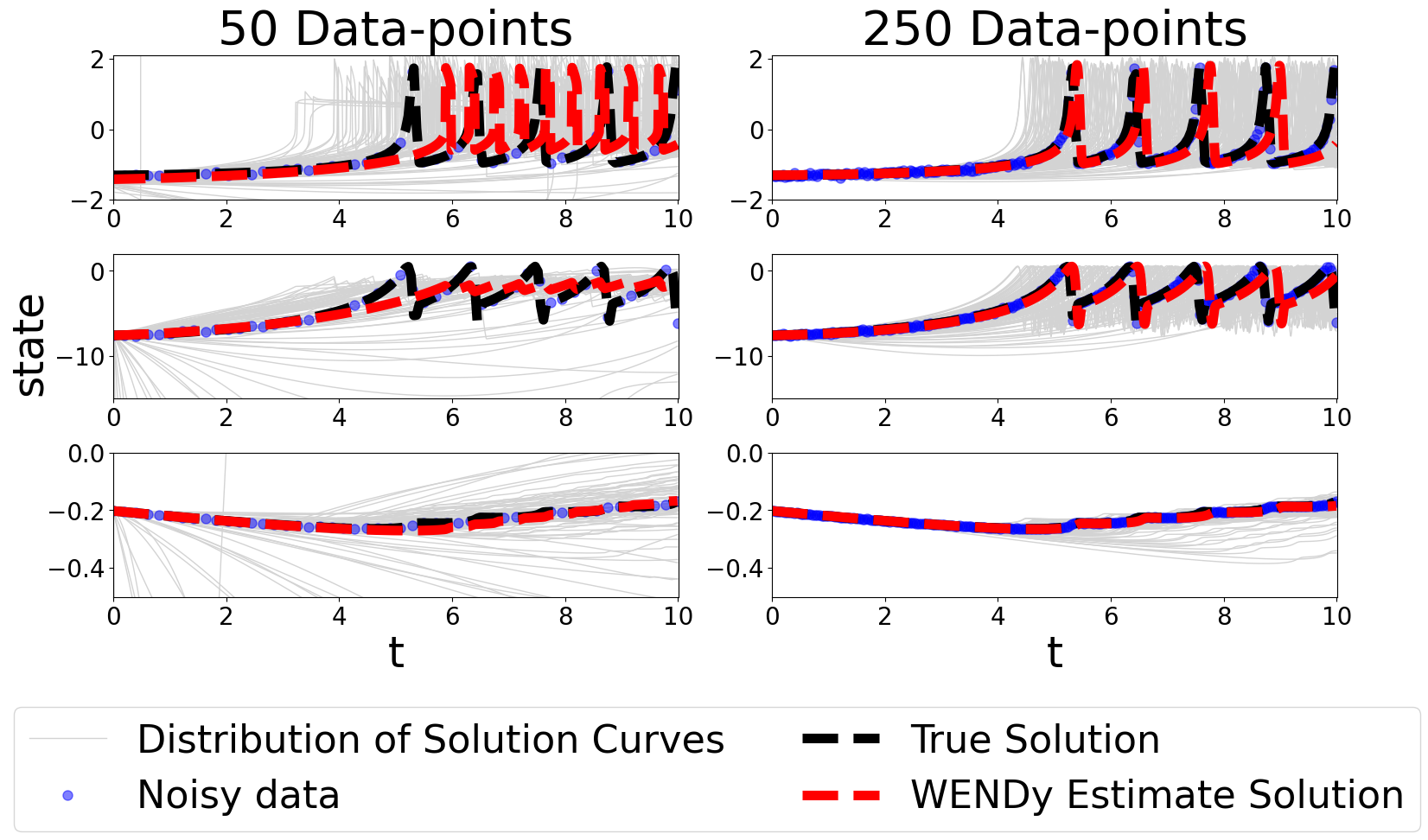} 
    \\
    \text{(a)}
    \\
    \includegraphics[width=1\linewidth]{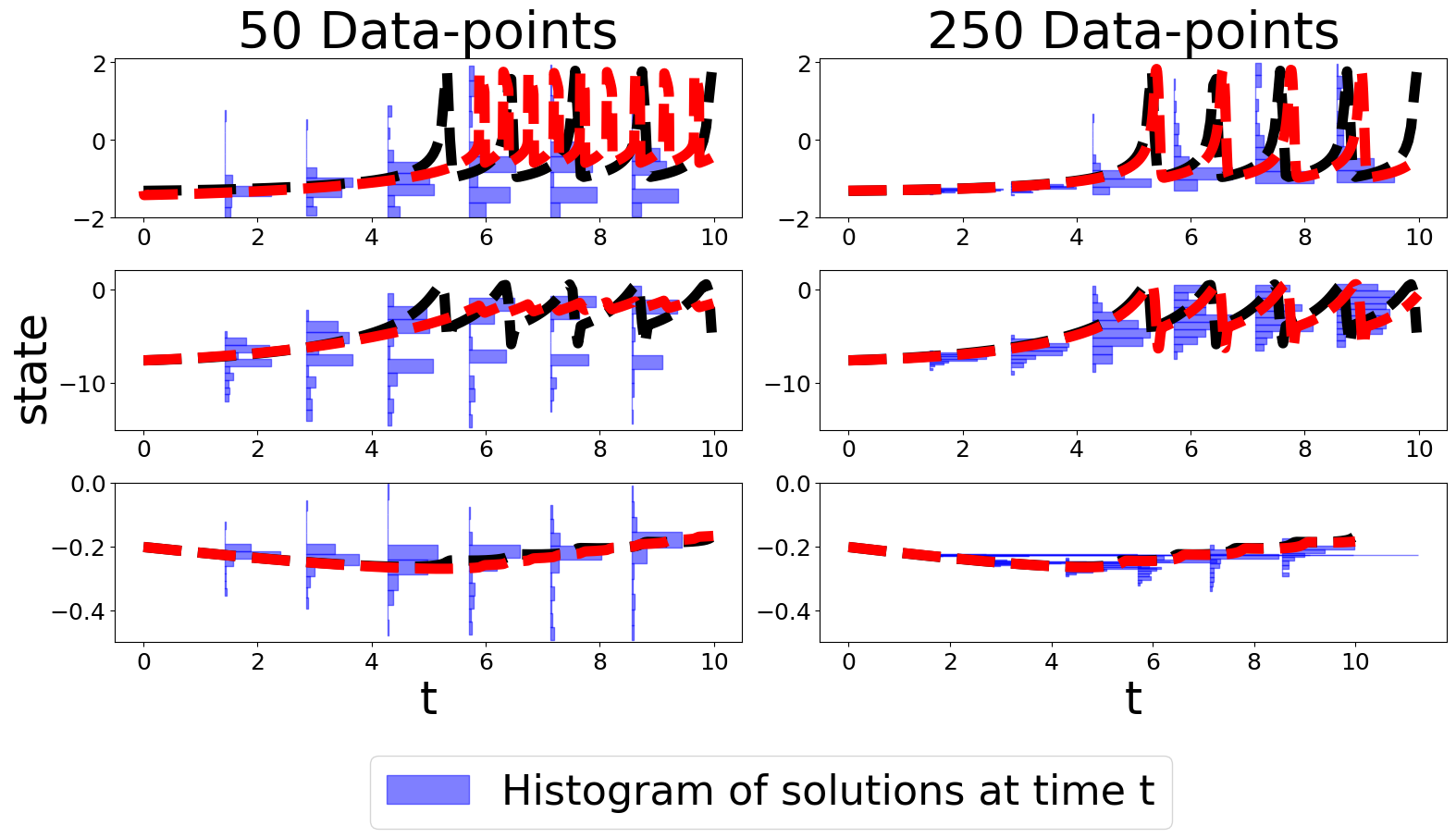}
    \\
    \text{(b)}
    \end{tabular}
     \caption{Top row (a): Hindmarsh-Rose model parameter estimation example and uncertainty quantification on two datasets: one dataset with low resolution (left) and one dataset with high resolution (right). The light gray curves are used to illustrate the uncertainty around the WENDy solutions; they are obtained via parametric bootstrap, as a sample of WENDy solutions corresponding to a random sample of 1000 parameters from their estimated asymptotic estimator distribution.  Bottom row (b): WENDy solution and histograms of state distributions across specific points in time for the datasets in (a).}
    \label{fig:SampleResHMRTN}
\end{figure}

As seen in Figure~\ref{fig:SampleResHMRTN}(a) and (b), the distribution of solution states (at selected time points) became more unimodal in the $u_1$ state as resolution increased. At low resolution, the WENDy estimators struggled to capture the frequency of the oscillatory states.

\clearpage 
\subsubsection*{Multiplicative Log-normal Noise}
\begin{figure} 
    \centering
    \begin{tabular}{c}
{\includegraphics[width=0.8\linewidth]{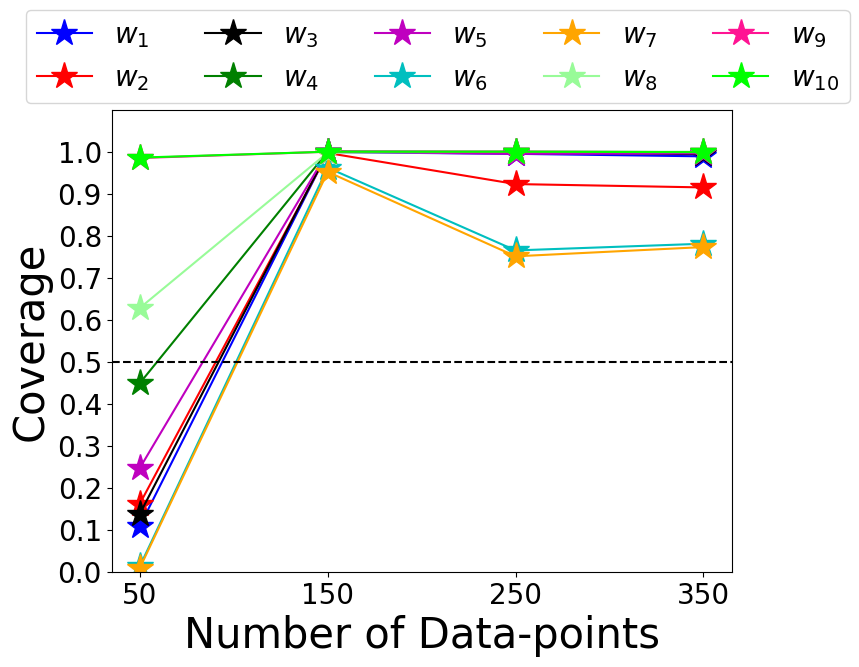}}
    \\
    \text{(a)}
    \\
    \includegraphics[width=1\linewidth]{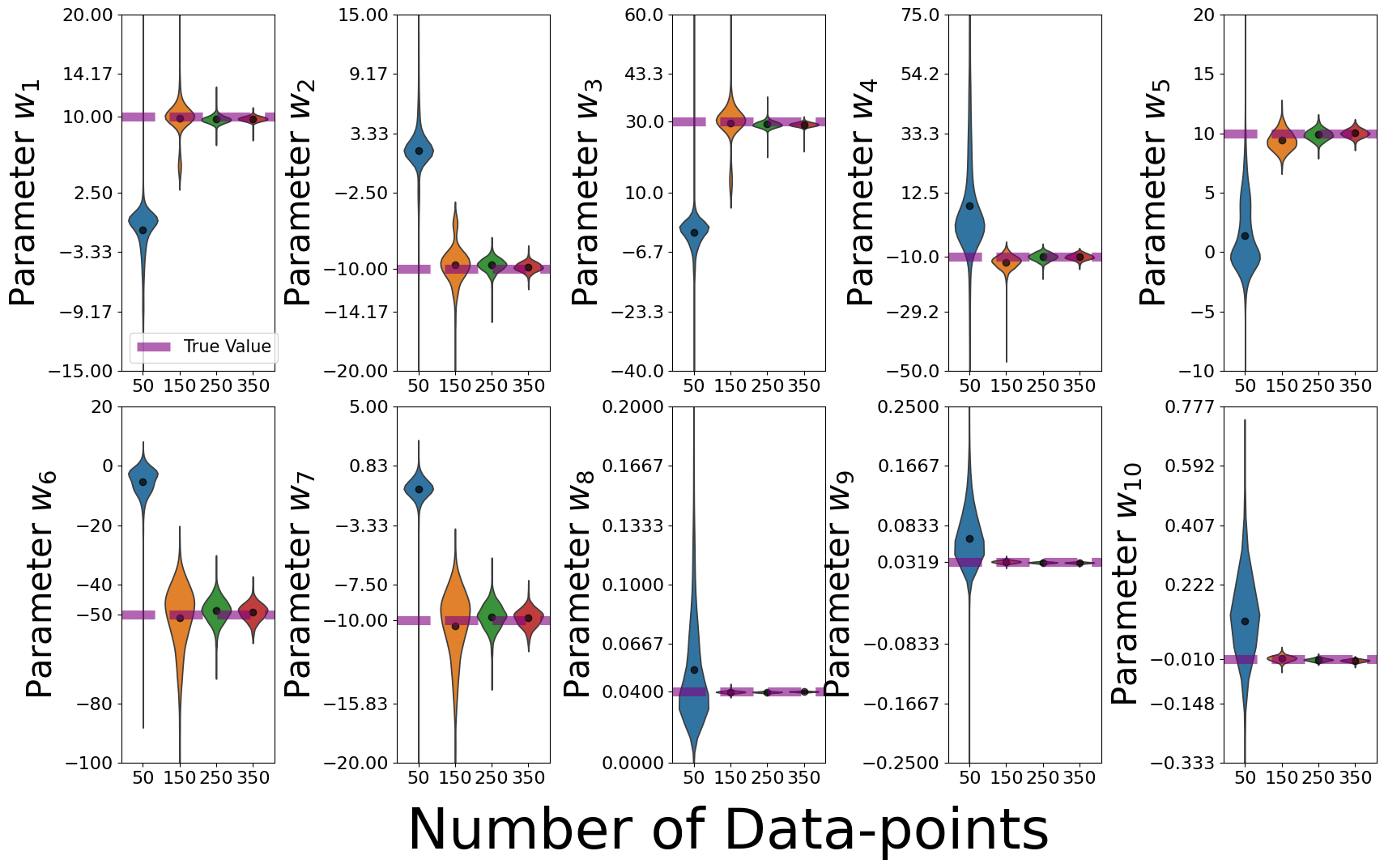}
    \\
      \text{(b)} 
    \end{tabular}
     \caption{Hindmarsh-Rose model parameter estimation performance with increasing data resolution (1000 datasets per level, 0.05\% normal noise). (a) coverage across four noise levels. (b) violin plots of parameter estimates, with the dashed red line indicating the true parameter values.}
    \label{fig:CovBiasResHMRLN}
\end{figure}

As shown in Figure~\ref{fig:CovBiasResHMRLN}(a), the coverage of the 95\% confidence intervals for the $w_9$ and $w_{10}$ parameters remained near the nominal 95\% level across all sample sizes from 50 to 350 data points. The coverage for the $w_8$ parameter began slightly above 50\% at 50 data points and increased to nominal by 150 data points. The remaining parameters all started below 50\% before rising to nominal at 150 data points. The coverage for the $w_2$, $w_6$, and $w_7$ parameters then decreased at 250 and 350 data points to around 75\%. As shown in Figure~\ref{fig:CovBiasResHMRLN}(b), the bias and variance for all parameters decreased slightly as resolution increased, as expected. Unlike under additive normal noise, no bimodality was observed at lower resolution levels.  

\begin{figure}
    \centering
    \begin{tabular}{c}
        \includegraphics[width=1\linewidth]{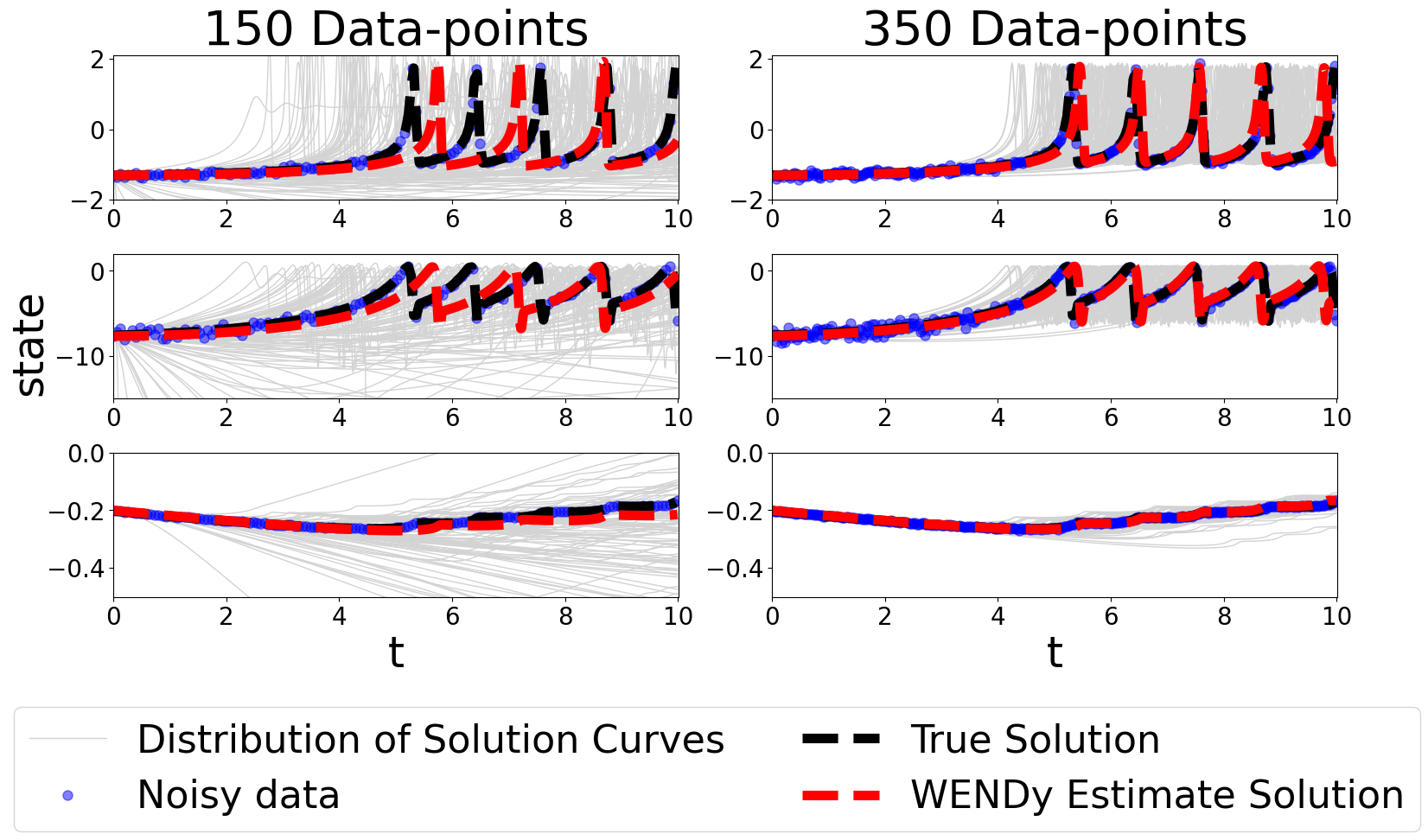} 
    \\
    \text{(a)}
    \\
    \includegraphics[width=1\linewidth]{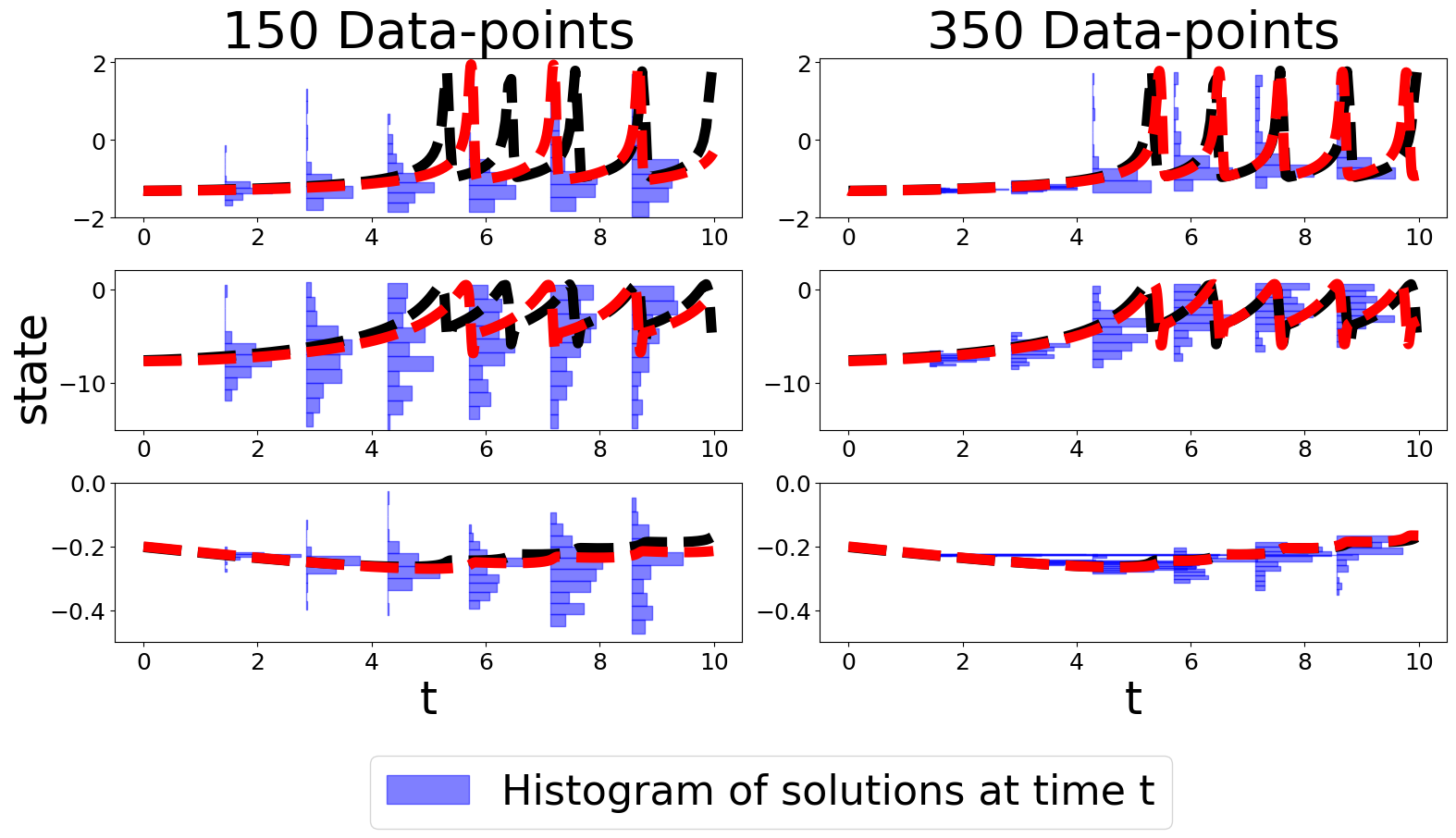}
    \\
    \text{(b)}
    \end{tabular}
     \caption{Top row (a): HMR model parameter estimation example and uncertainty quantification on two datasets: one dataset with low resolution (left) and one dataset with high resolution (right). The light gray curves are used to illustrate the uncertainty around the WENDy solutions; they are obtained via parametric bootstrap, as a sample of WENDy solutions corresponding to a random sample of 1000 parameters from their estimated asymptotic estimator distribution.  Bottom row (b): WENDy solution and histograms of state distributions across specific points in time for the datasets in (a).}
    \label{fig:SampleResHMRLN}
\end{figure}

As shown in Figure~\ref{fig:SampleResHMRLN}(a) and (b), the distribution of solution states (at selected time points) became more unimodal as resolution increased, but remained more skewed at lower resolution. The histograms were also wider at lower resolution, and the WENDy estimators, similar to the additive normal noise case, struggled to capture the frequency of the oscillatory states.

\clearpage
\subsection{Protein Transduction Benchmark Model}

\subsubsection*{Varying Noise Level}
\subsubsection{Additive Normal Noise}
\begin{figure} 
    \centering
    \begin{tabular}{c}
{\includegraphics[width=0.8\linewidth]{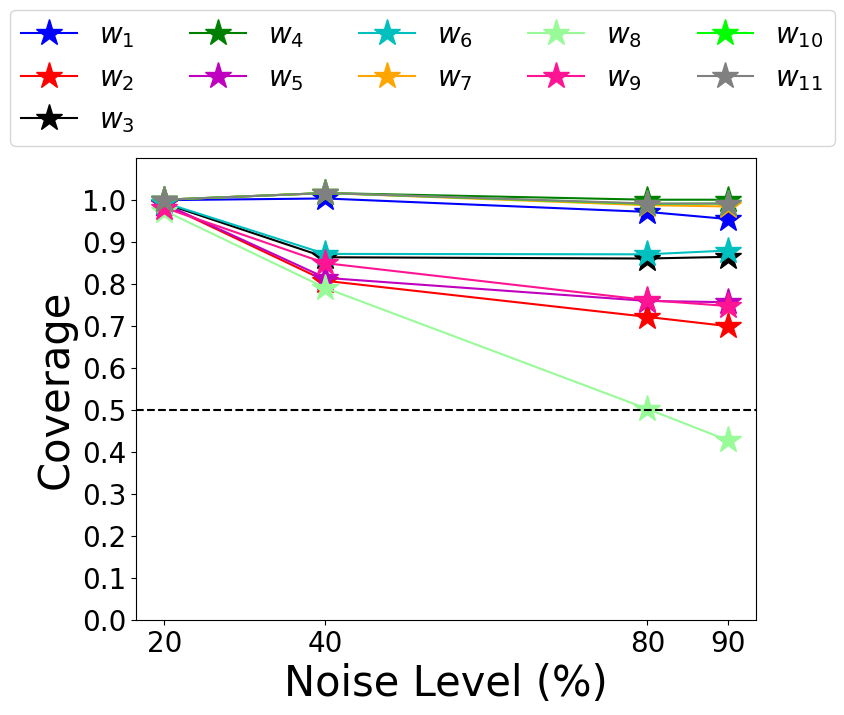}}
\\
\text{(a)}
\\
    \includegraphics[width=1\linewidth]{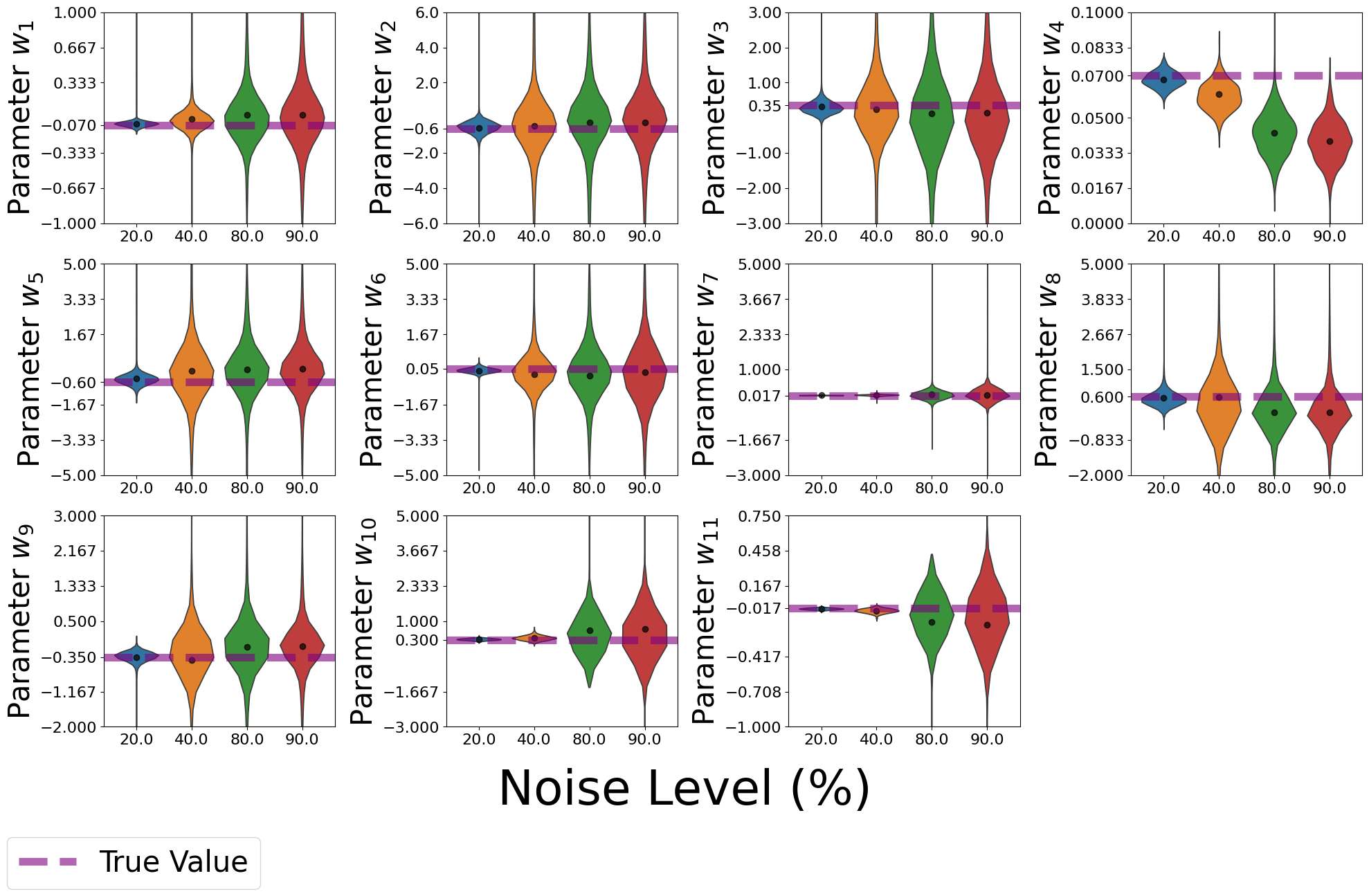}
    \\
      \text{(b)}
    \end{tabular}
     \caption{PTB model parameter estimation performance with increasing censored normal noise (1000 datasets per level, 205 data points each). (a) coverage across four noise levels. (b) violin plots of parameter estimates, with the dashed red line indicating the true parameter values.}
    \label{fig:CovBiasPTBN}
\end{figure}

As shown in Figure~\ref{fig:CovBiasPTBN}(a), the coverage of the $w_8$ parameter dropped below 50\% at around 90\% noise. The coverage for the $w_2$, $w_3$, $w_5$, $w_6$, and $w_9$ parameters also declined as noise increased, but did not fall below 50\%. The remaining parameters stayed near the nominal level across all noise levels. As seen in Figure~\ref{fig:CovBiasPTBN}(b), the bias and variance of all parameters increased as noise increased. Overall, the PTB parameter estimators obtained with WENDy demonstrated strong robustness to increasing noise, which is likely due to the compartmental structure of PTB and the relative absence of higher-order terms in the dynamics.  

\begin{figure} 
    \centering

    \includegraphics[width=1\linewidth]{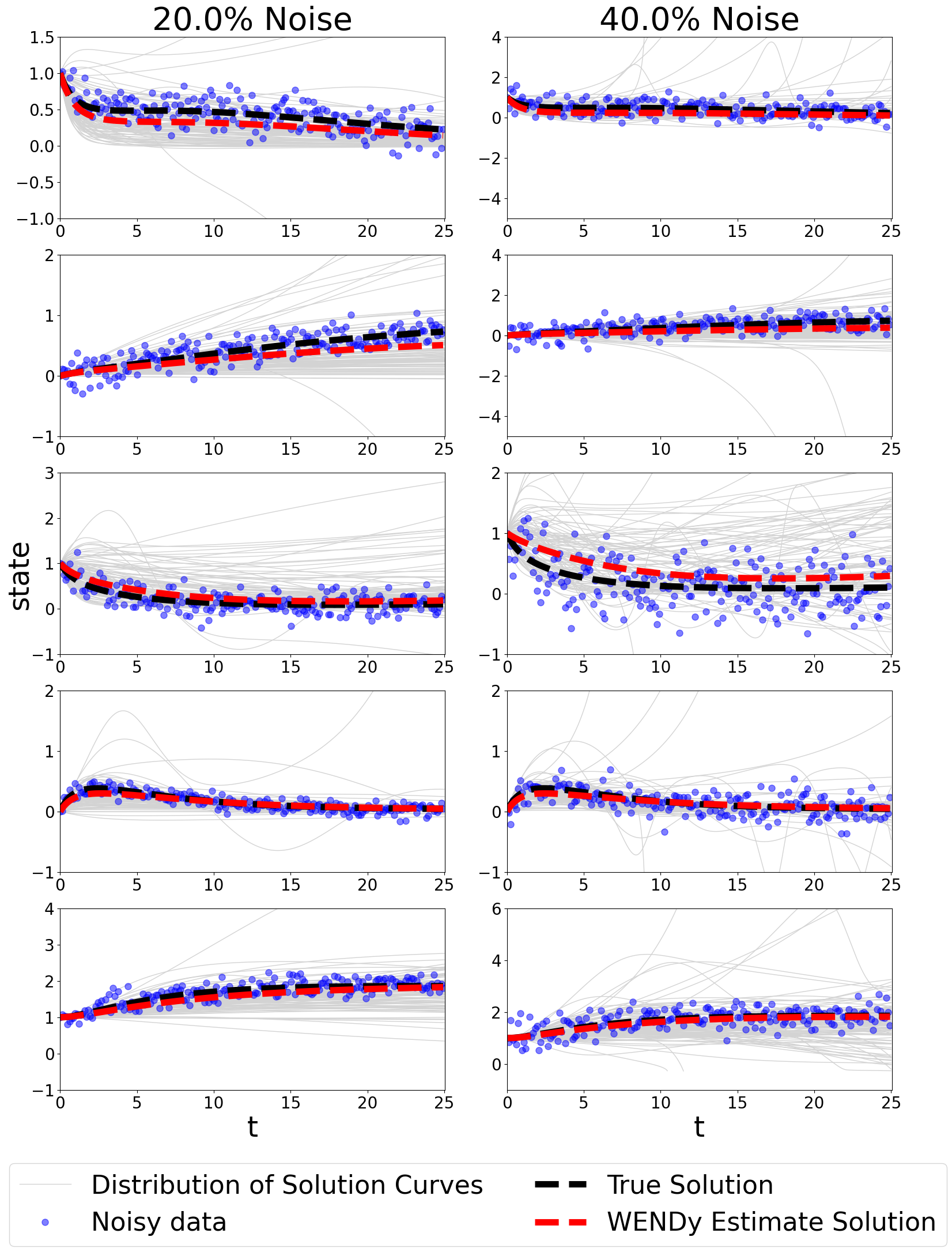} 
    
     \caption{PTB model parameter estimation example and uncertainty quantification on two datasets with additive normal noise: one dataset with low uncertainty and high coverage (left) and one dataset with high uncertainty and low coverage (right). The light gray curves are used to illustrate the uncertainty around the WENDy solutions; they are obtained via parametric bootstrap, as a sample of WENDy solutions corresponding to a random sample of 1000 parameters from their estimated asymptotic estimator distribution.}
    \label{fig:SamplePlotPTBN}
\end{figure}

\begin{figure} 
    \centering

    \includegraphics[width=1\linewidth]{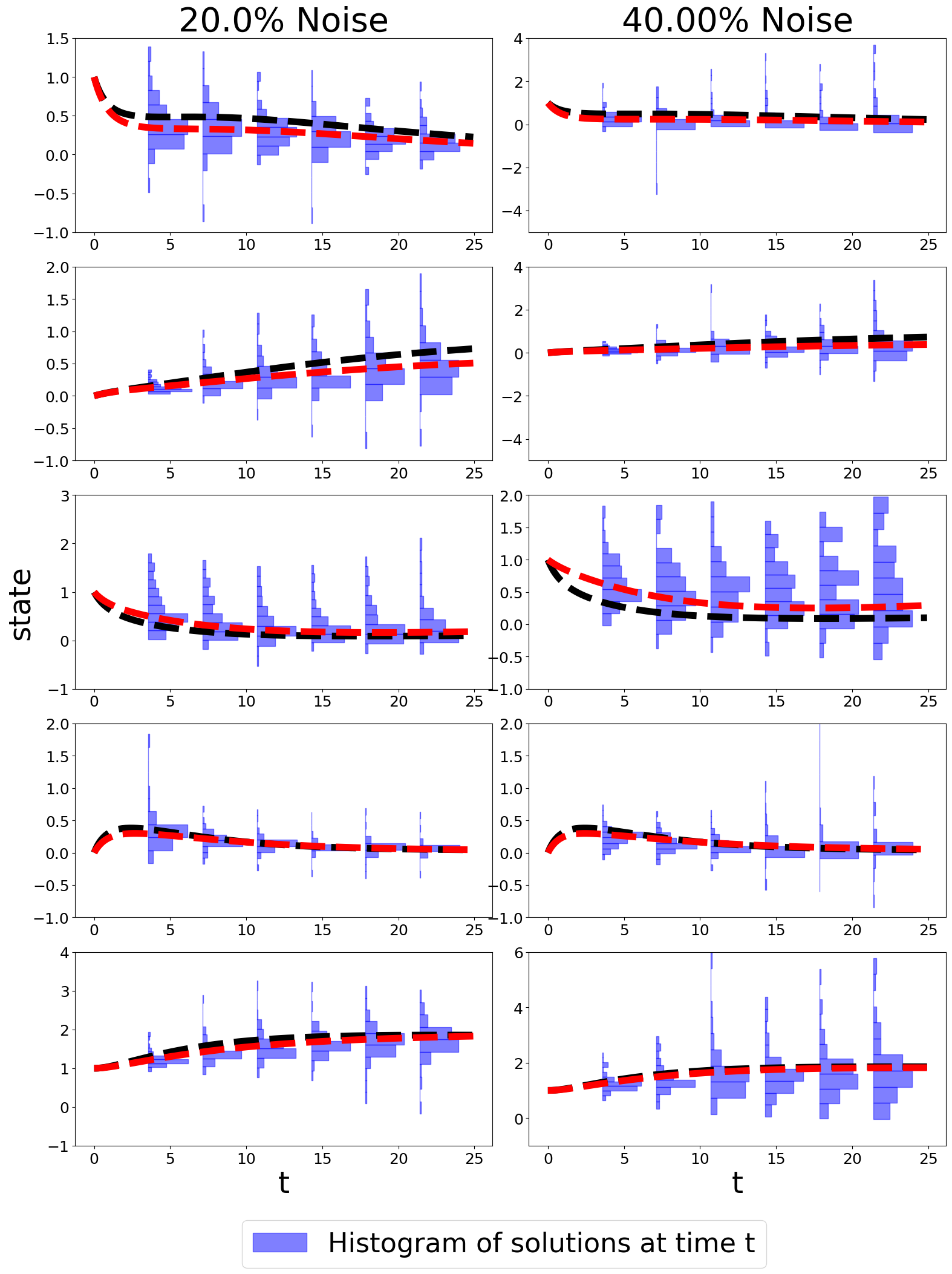}

     \caption{WENDy solution and histograms of state distributions across specific points in time for the datasets in Fig~\ref{fig:SamplePlotPTBN} with additive normal noise.}
    \label{fig:SampleHistPlotPTBN}
\end{figure}

As seen in Figure~\ref{fig:SamplePlotPTBN} and  ~\ref{fig:SampleHistPlotPTBN}, the distribution of solution states (at selected time points) became wider at higher noise levels. However, the histograms at those time points remained largely symmetric and unimodal across noise levels.
 
\subsubsection{Additive Censored Normal Noise}
\begin{figure} 
    \centering
    \begin{tabular}{c}
{\includegraphics[width=0.8\linewidth]{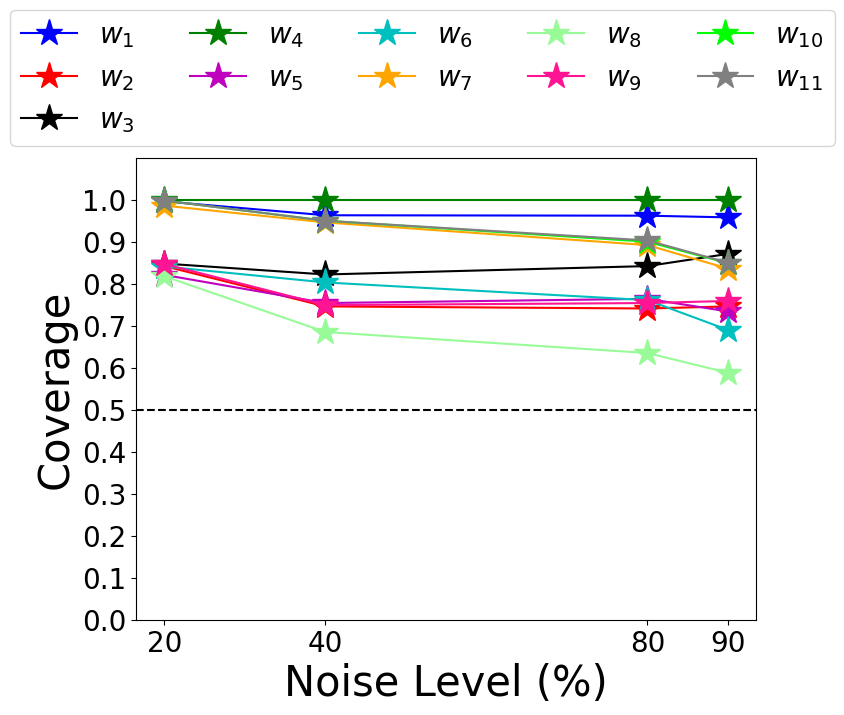}}
    \\
    \text{(a)}
    \\
    \includegraphics[width=1\linewidth]{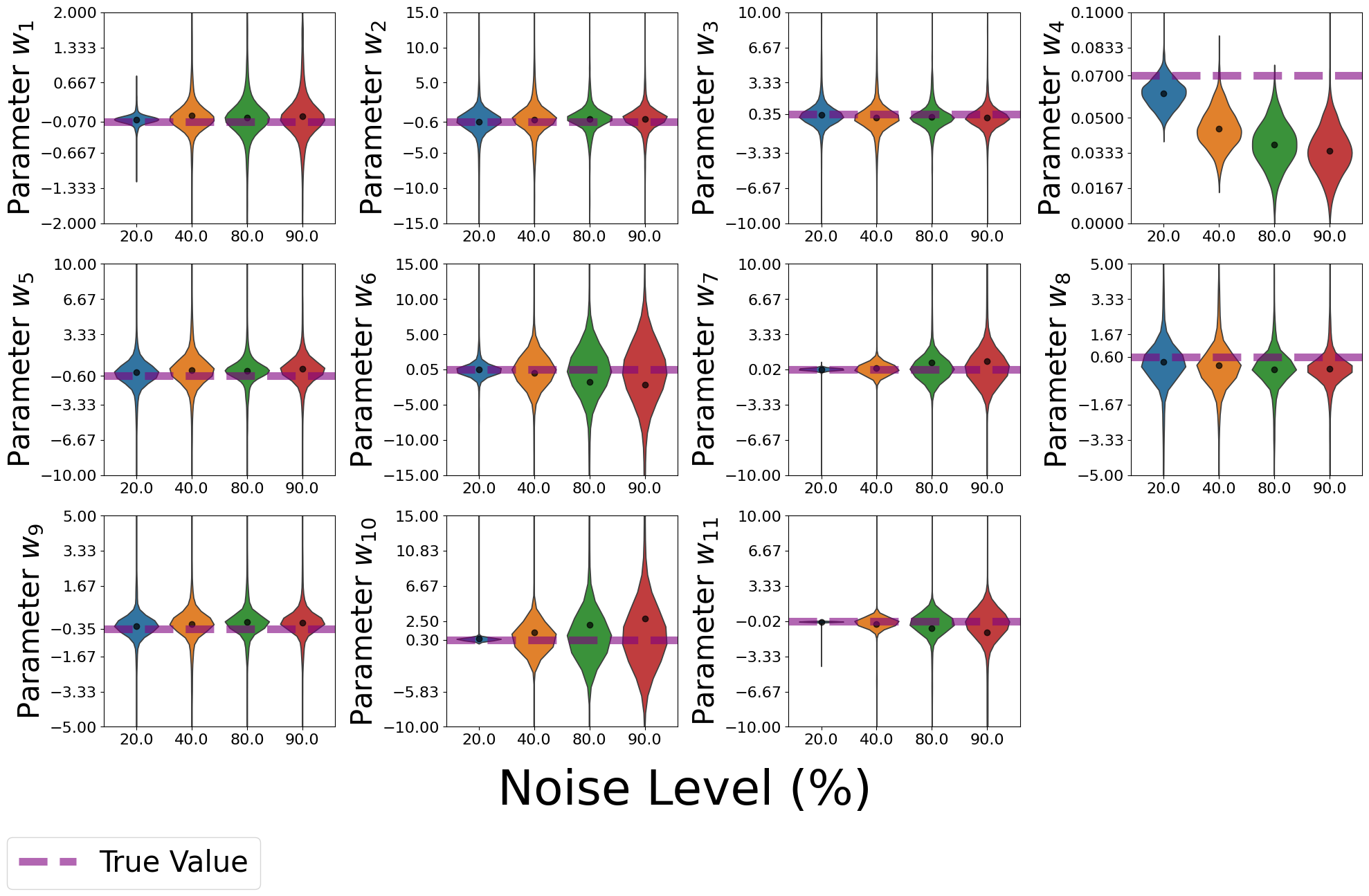}
    \\
     
      \text{(b)}
    \end{tabular}
     \caption{PTB model parameter estimation performance with increasing ACN noise (1000 datasets per level, 205 data points each). (a) coverage across four noise levels. (b) violin plots of parameter estimates, with the dashed red line indicating the true parameter values.}
    \label{fig:CovBiasPTBCN}
\end{figure}

As shown in Figure~\ref{fig:CovBiasPTBCN}(a), the coverage for all parameters remained above 50\% up to 90\% ACN noise. The coverage for the $w_1$ and $w_4$ parameters stayed near the nominal level across all noise levels, while the coverage for the $w_7$ and $w_{11}$ parameters dropped to below 90\%. The remaining parameters started slightly above 80\% coverage before declining as noise increased to 90\%. As seen in Figure~\ref{fig:CovBiasPTBCN}(b), the bias and variance for all parameters increased with noise, as expected.  

\begin{figure} 
    \centering

    \includegraphics[width=1\linewidth]{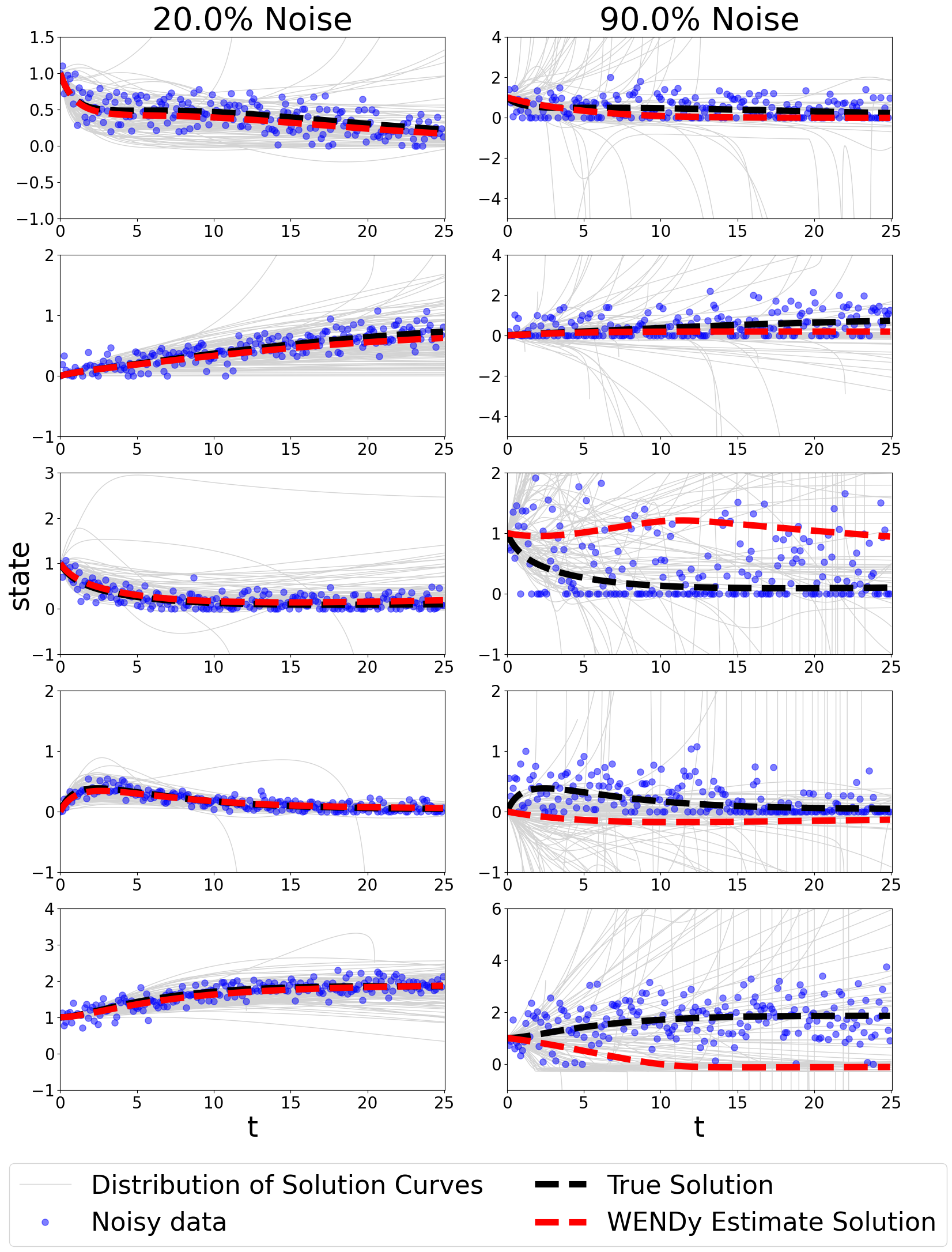} 
    
     \caption{PTB model parameter estimation example and uncertainty quantification on two datasets with ACN noise: one dataset with low uncertainty and high coverage (left) and one dataset with high uncertainty and low coverage (right). The light gray curves are used to illustrate the uncertainty around the WENDy solutions; they are obtained via parametric bootstrap, as a sample of WENDy solutions corresponding to a random sample of 1000 parameters from their estimated asymptotic estimator distribution.}
    \label{fig:SamplePlotPTBACN}
\end{figure}

\begin{figure} 
    \centering

    \includegraphics[width=1\linewidth]{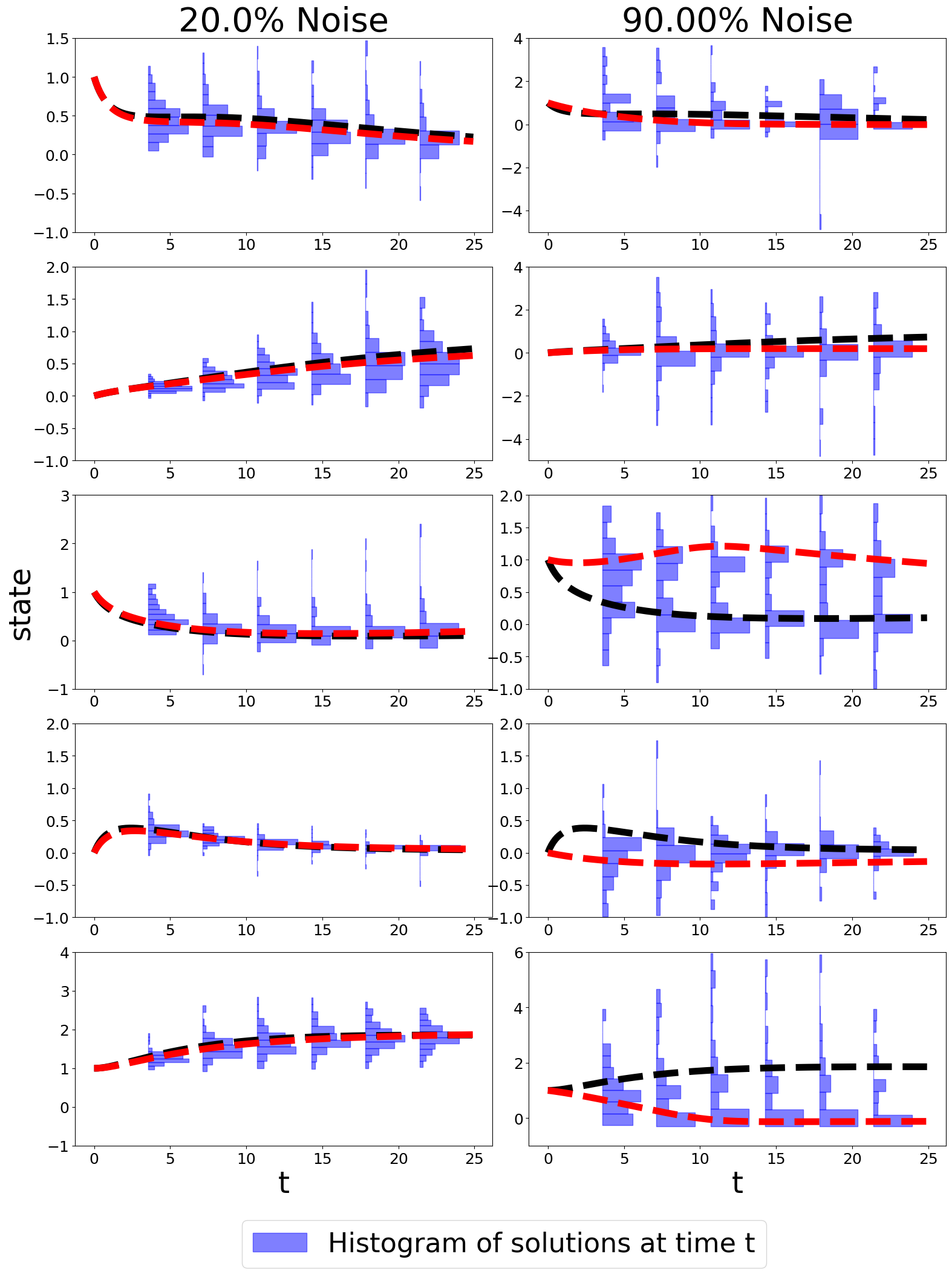}

     \caption{WENDy solution and histograms of state distributions across specific points in time for the datasets in Fig~\ref{fig:SamplePlotPTBACN} with ACN noise.}
    \label{fig:SampleHistPlotPTBACN}
\end{figure}

As shown in Figure~\ref{fig:SamplePlotPTBACN} and ~\ref{fig:SampleHistPlotPTBACN}, the distribution of solution states (at selected time points) became wider and more skewed at higher noise levels. Despite these high noise levels, WENDy estimates still produced a reasonable fit at 90\% ACN noise. This robustness may be attributed to the fact that ACN noise does not allow for negative states, unlike additive normal noise, thereby preserving more informative data points for WENDy estimation of the dynamics.

\subsubsection{Multiplicative Log-Normal Noise}
\begin{figure} 
    \centering
    \begin{tabular}{c}
{\includegraphics[width=0.8\linewidth]{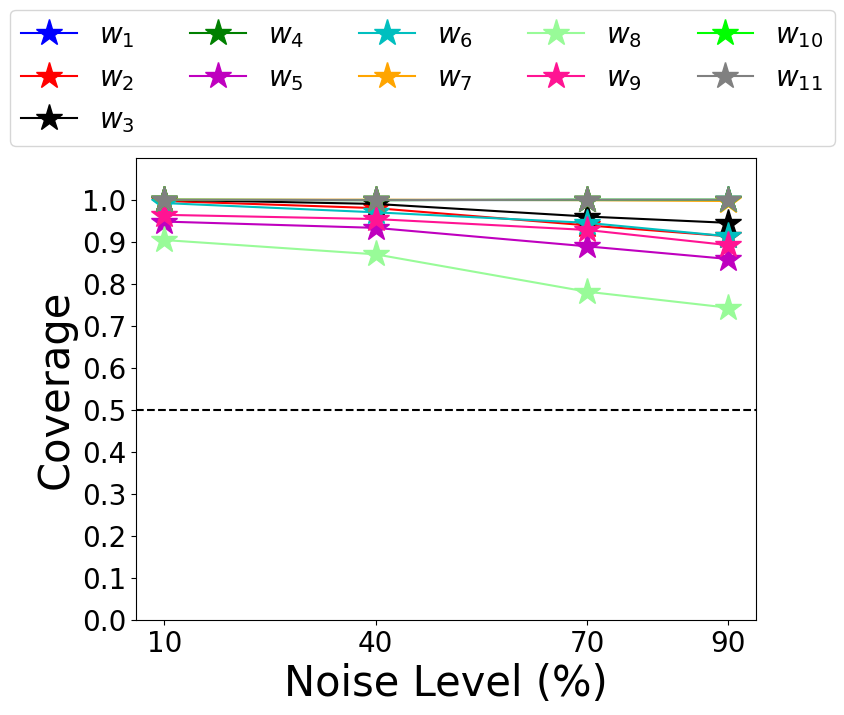}}
    \\
    \text{(a)}
    \\
    \includegraphics[width=0.8\linewidth]{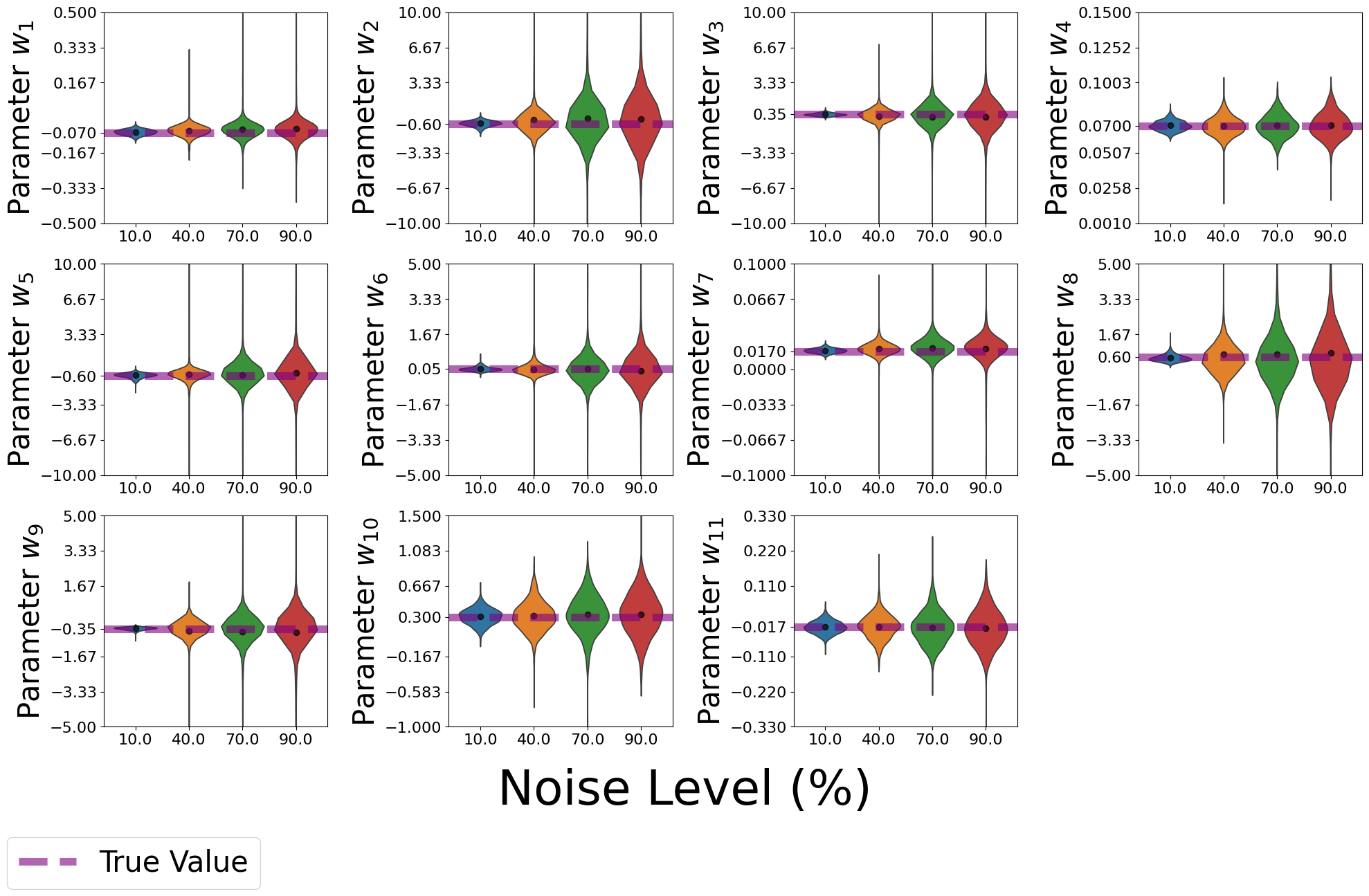}
    \\
      \text{(b)} 
    \end{tabular}
     \caption{PTB model parameter estimation performance with increasing MLN noise (1000 datasets per level, 205 data points each). (a) coverage across four noise levels. (b) violin plots of parameter estimates, with the dashed red line indicating the true parameter values.}
    \label{fig:CovBiasPTBLN}
\end{figure}

As shown in Figure~\ref{fig:CovBiasPTBLN}(a), the coverage of the parameters remained around 70\% for all noise levels up to 90\% MLN noise. The $w_8$ parameter showed the poorest coverage with increasing noise, decreasing to about 70\% at 90\% noise. As seen in Figure~\ref{fig:CovBiasPTBLN}(b), the bias and variance in all parameters increased with noise; however, some parameters, such as $w_{11}$, exhibited very little bias even at 90\% MLN noise.  

\begin{figure} 
    \centering

    \includegraphics[width=1\linewidth]{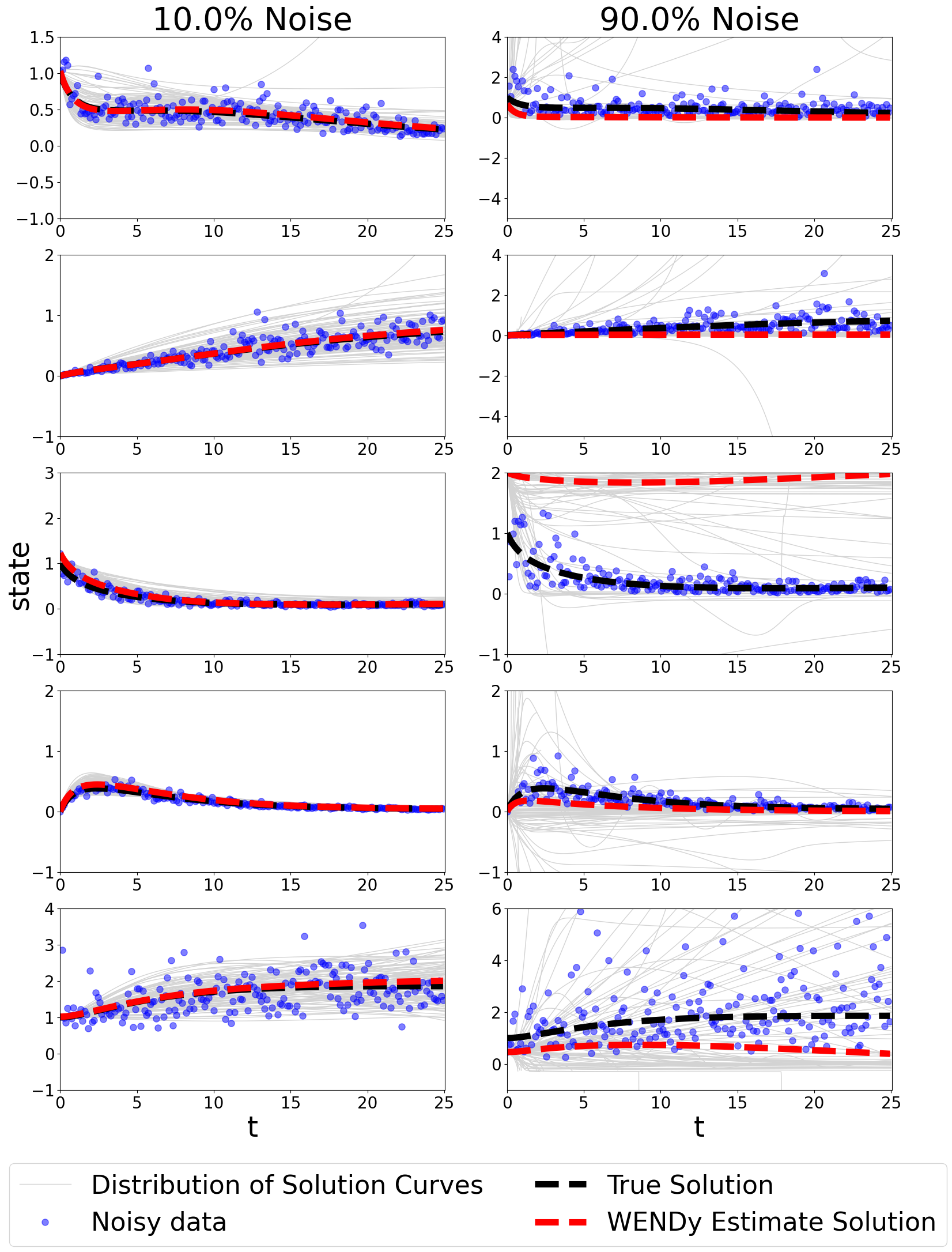} 
    
     \caption{PTB model parameter estimation example and uncertainty quantification on two datasets with MLN noise: one dataset with low uncertainty and high coverage (left) and one dataset with high uncertainty and low coverage (right). The light gray curves are used to illustrate the uncertainty around the WENDy solutions; they are obtained via parametric bootstrap, as a sample of WENDy solutions corresponding to a random sample of 1000 parameters from their estimated asymptotic estimator distribution.}
    \label{fig:SamplePlotPTBMLN}
\end{figure}

\begin{figure} 
    \centering

    \includegraphics[width=1\linewidth]{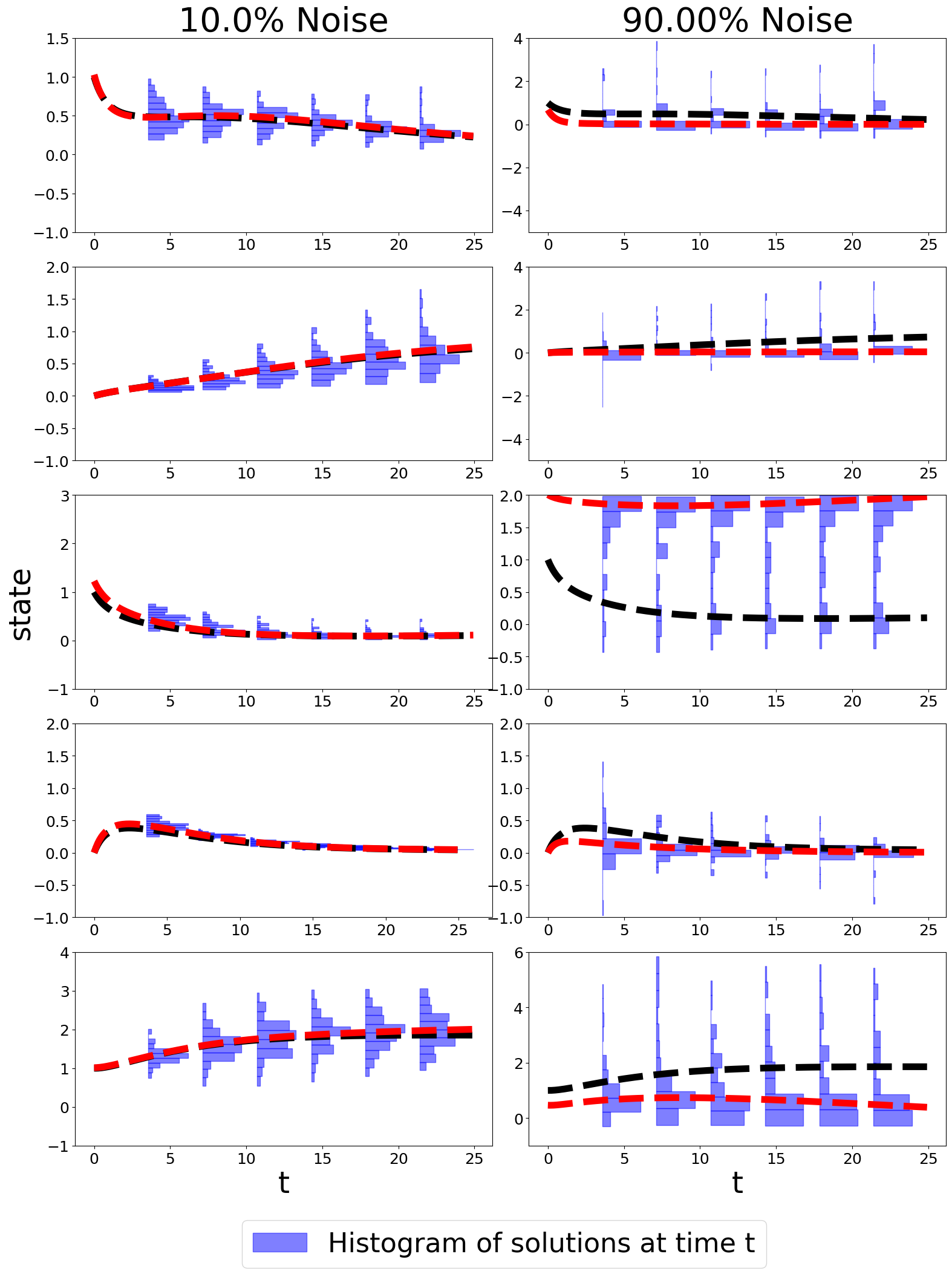}

     \caption{WENDy solution and histograms of state distributions across specific points in time for the datasets in Fig~\ref{fig:SamplePlotPTBMLN} with MLN noise.}
    \label{fig:SampleHistPlotPTBMLN}
\end{figure}

As shown in Figure~\ref{fig:SamplePlotPTBMLN} and \ref{fig:SampleHistPlotPTBMLN}, the distribution of solution states (at selected time points) became wider and more skewed at higher noise levels. Despite this, WENDy estimates still produced reasonable fits even at 90\% noise. This robustness is likely due to the low-magnitude state values in the chosen true dynamics. Since MLN noise is heteroskedastic, higher-magnitude states experience greater distortions, but in this case, the relatively small state magnitudes reduced the impact of noise, leading to more stable estimators.
  
\subsubsection{Additive Truncated Normal Noise}
\begin{figure} 
    \centering
    \begin{tabular}{c}
 {\includegraphics[width=0.8\linewidth]{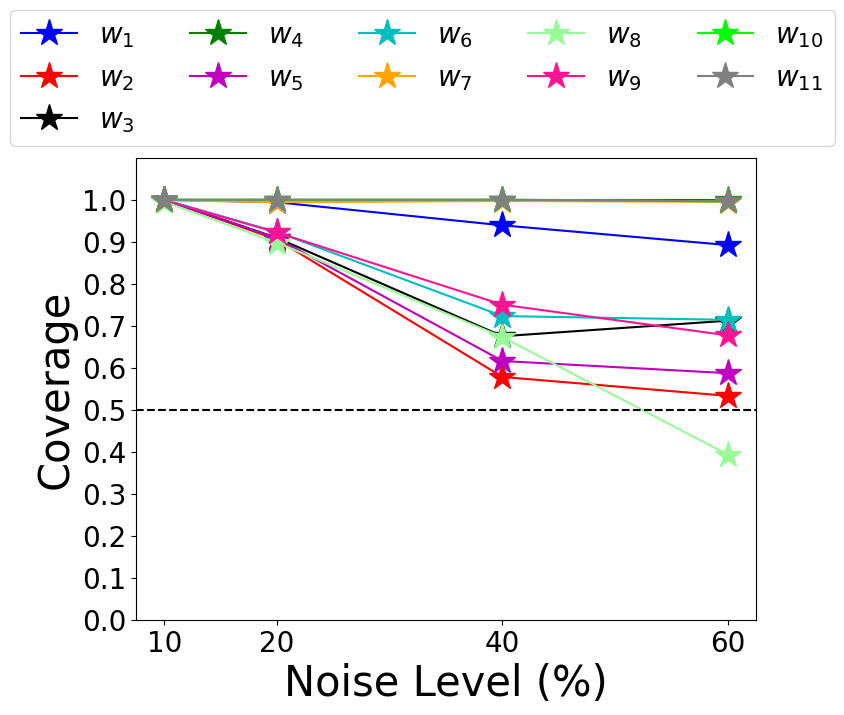}}
    \\
    \text{(a)} 
    \\
    \includegraphics[width=1\linewidth]{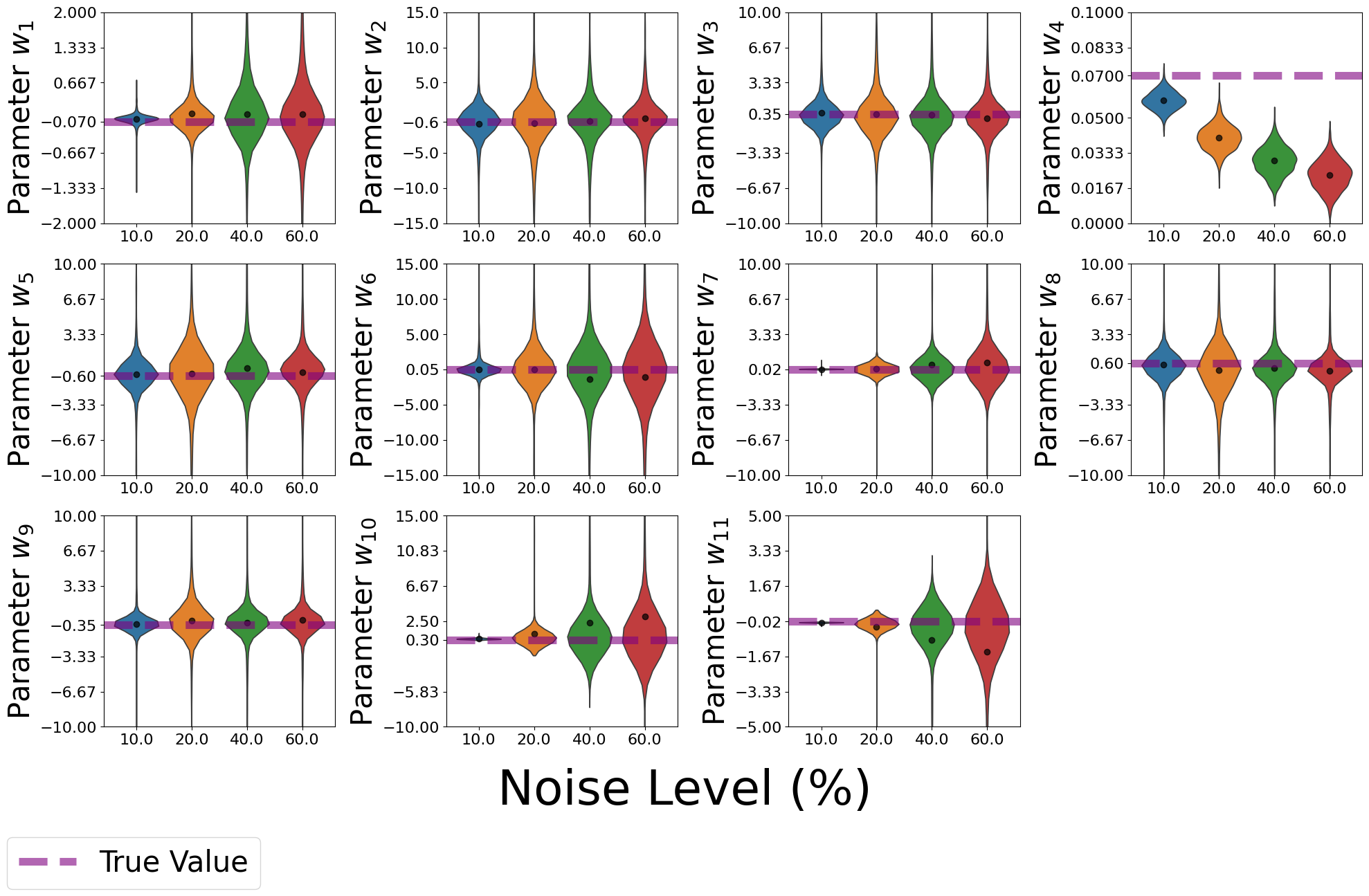}
    \\
    \text{(b)}
    \end{tabular}
     \caption{PTB model parameter estimation performance with increasing ATN noise (1000 datasets per level, 205 data points each). (a) coverage across four noise levels. (b) violin plots of parameter estimates, with the dashed red line indicating the true parameter values.}
    \label{fig:CovBiasPTBTN}
\end{figure}

As shown in Figure~\ref{fig:CovBiasPTBTN}(a), the coverage for the $w_7$ and $w_{11}$ parameters remained above nominal across all noise levels, while the coverage for the $w_8$ parameter dropped below 50\% at 60\% noise. The coverage of the remaining parameters decreased below nominal as noise increased, but never fell below 50\%. As seen in Figure~\ref{fig:CovBiasPTBTN}(b), the bias and variance for all parameters increased substantially with noise, with the exception of the $w_4$ parameter, whose variance did not increase sharply at higher noise levels.  

\begin{figure} 
    \centering

    \includegraphics[width=1\linewidth]{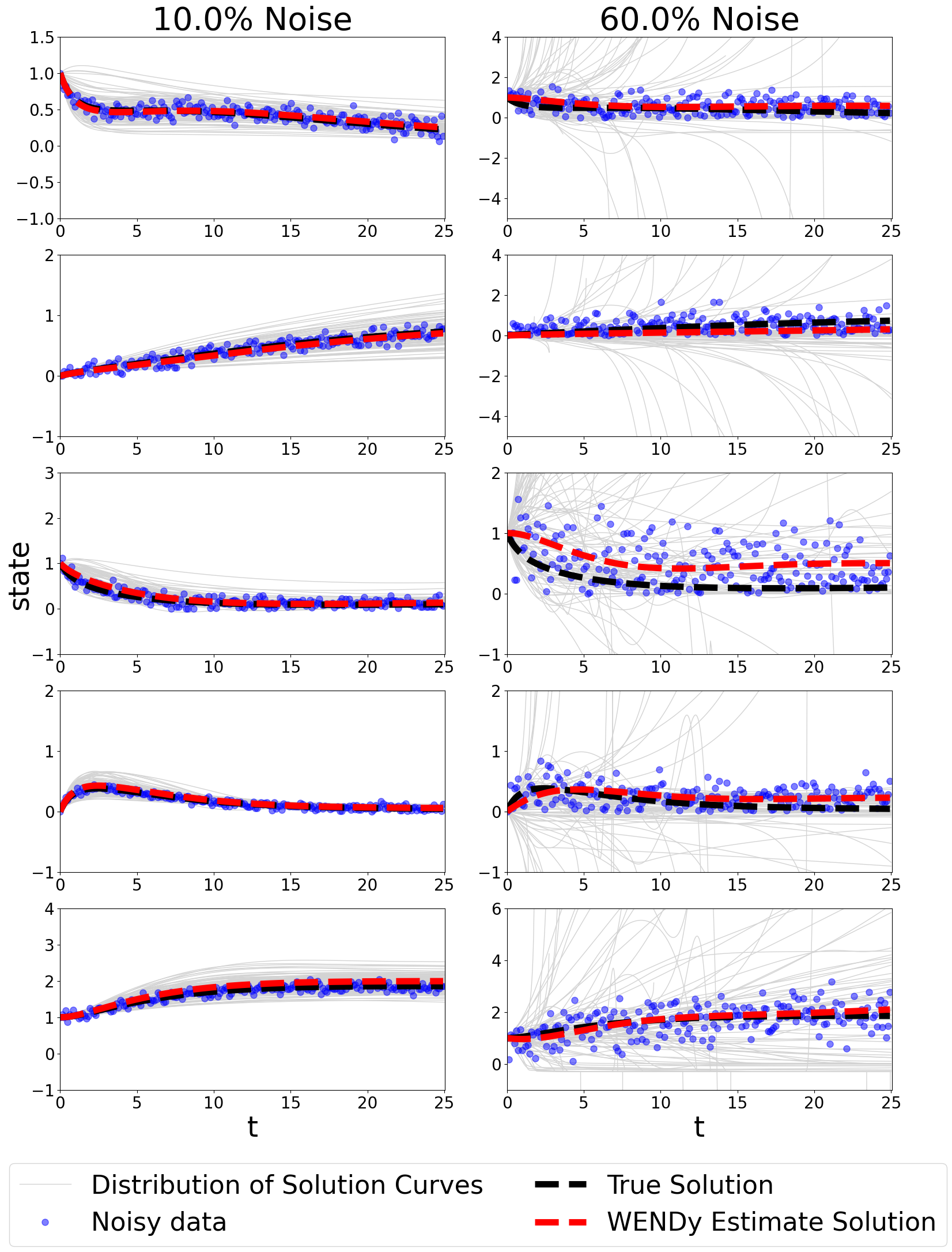} 
    
     \caption{PTB model parameter estimation example and uncertainty quantification on two datasets ATN noise: one dataset with low uncertainty and high coverage (left) and one dataset with high uncertainty and low coverage (right). The light gray curves are used to illustrate the uncertainty around the WENDy solutions; they are obtained via parametric bootstrap, as a sample of WENDy solutions corresponding to a random sample of 1000 parameters from their estimated asymptotic estimator distribution.}
    \label{fig:SamplePlotPTBTN}
\end{figure}

\begin{figure} 
    \centering

    \includegraphics[width=1\linewidth]{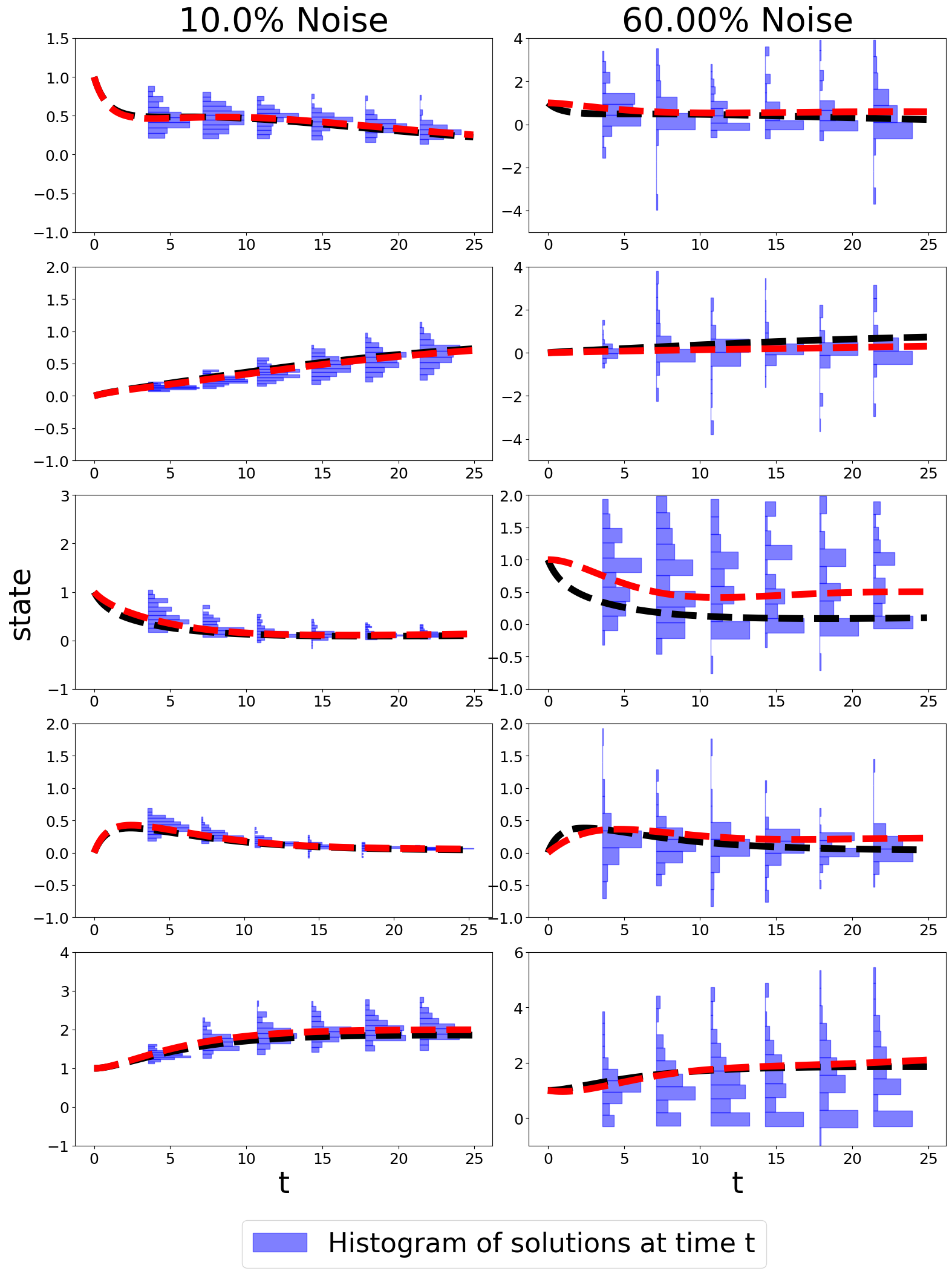}

     \caption{WENDy solution and histograms of state distributions across specific points in time for the datasets in Fig~\ref{fig:SamplePlotPTBN} ATN noise.}
    \label{fig:SampleHistPlotPTBTN}
\end{figure}

As shown in Figure~\ref{fig:SamplePlotPTBTN} and \ref{fig:SampleHistPlotPTBTN}, the distribution of solution states (at selected time points) became wider and more skewed as noise increased. At 60\% noise, some solutions deviated significantly from the true solution, with a subset diverging entirely.

\begin{figure} 
    \centering
    \includegraphics[width=0.4\linewidth]{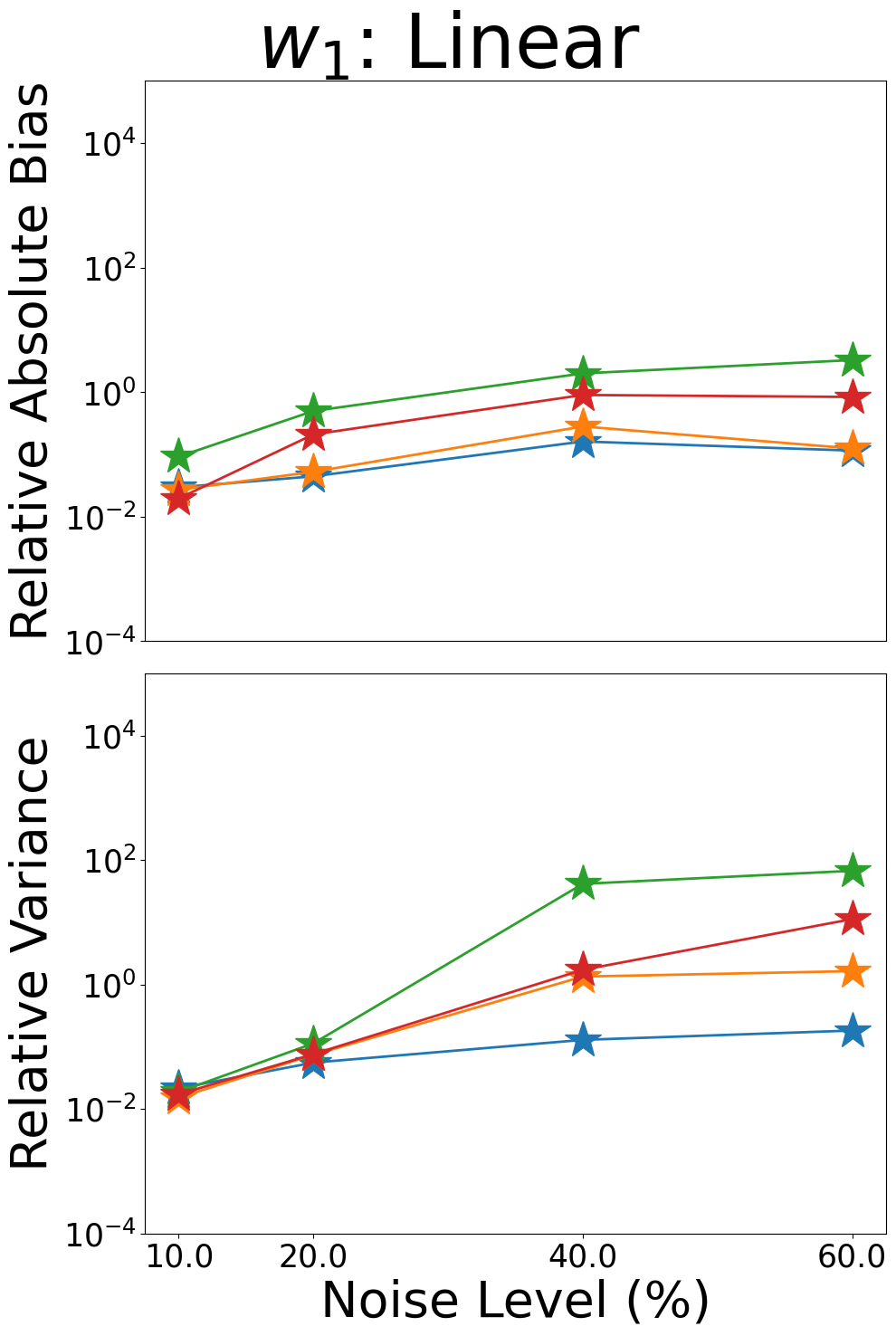}     \includegraphics[width=0.4\linewidth]{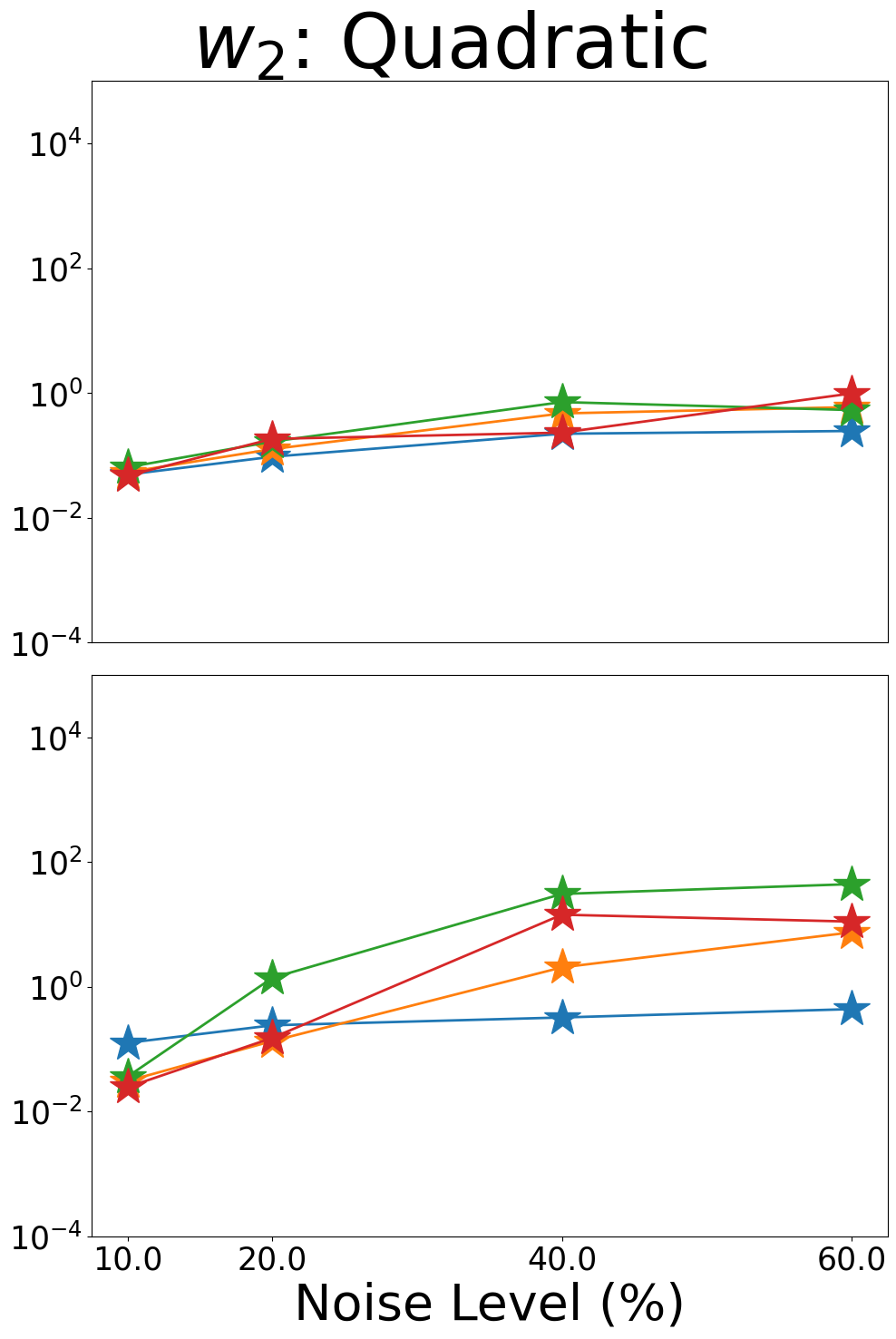}
    \includegraphics[width=0.4\linewidth]{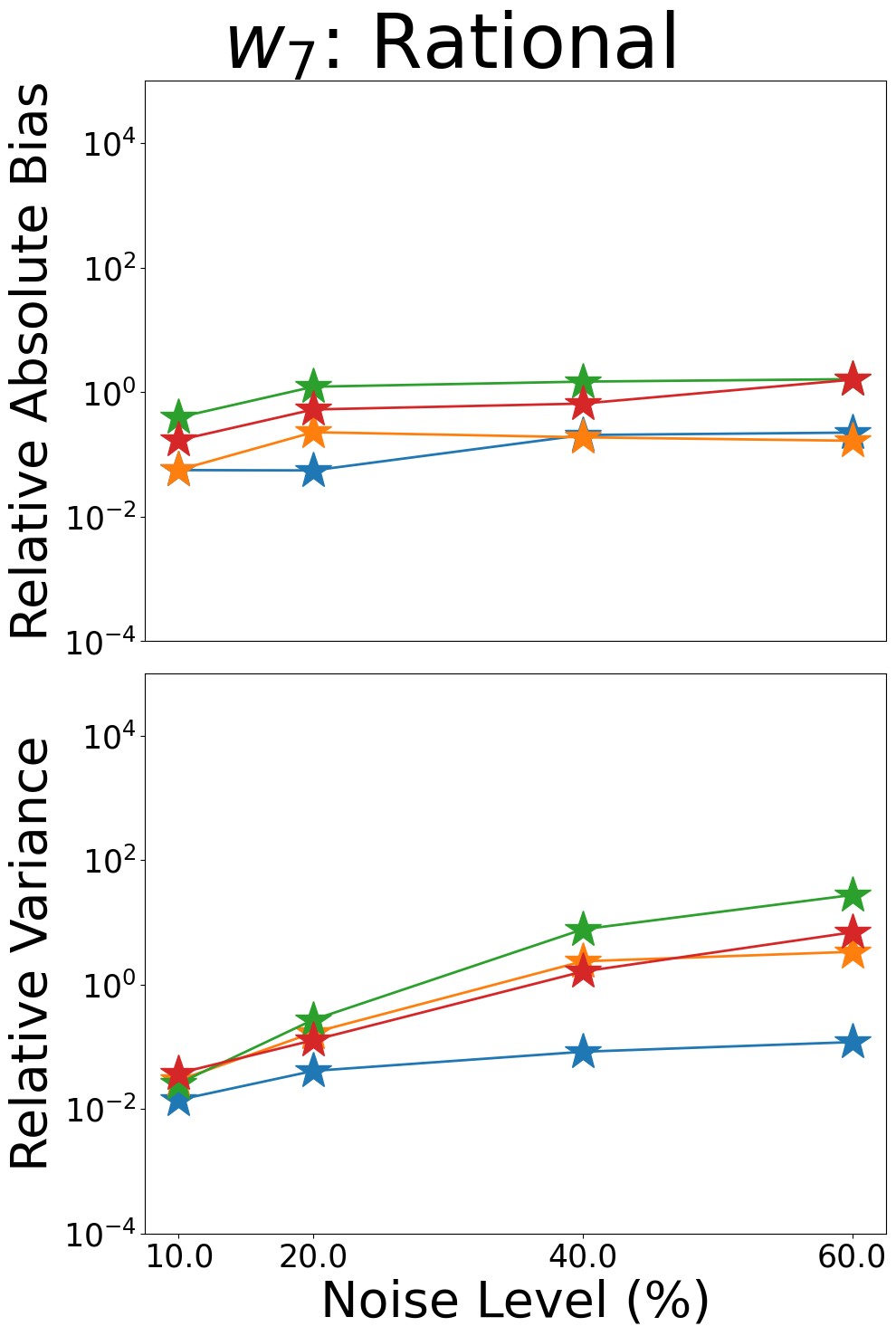}
    \\
    \includegraphics[width=0.3\linewidth]{NoiseDistrCompPlotsLogsitic/Legend4.png}
    \caption{Relative bias magnitude and variance for WENDy estimators of HMR model across 1000 datasets with additive normal, ACN, MLN, and ATN noise. Parameters modulating different terms are selectively shown.}
    \label{fig:PTBBiasVarNoise}
\end{figure}
As seen in Figure~\ref{fig:PTBBiasVarNoise}, when comparing the relative bias and variance across WENDy estimators from different noise distributions, ATN exhibited the highest relative bias and variance, except for the $w_2$ parameter. The lowest bias was consistently observed under MLN noise at high noise levels. Notably, the bias under MLN noise also varied less across noise levels compared to the other noise distributions for PTB parameter estimators. The high bias in ATN noise is further reflected in its coverage, as the $w_8$ parameter dropped below 50\% coverage at only 60\% ATN noise, whereas for the other noise distributions, coverage remained much higher up to 90\% noise.

\subsubsection*{Varying Resolution Level}
\subsubsection{Additive Normal Noise}
\begin{figure} 
    \centering
    \begin{tabular}{c}
{\includegraphics[width=0.8\linewidth]{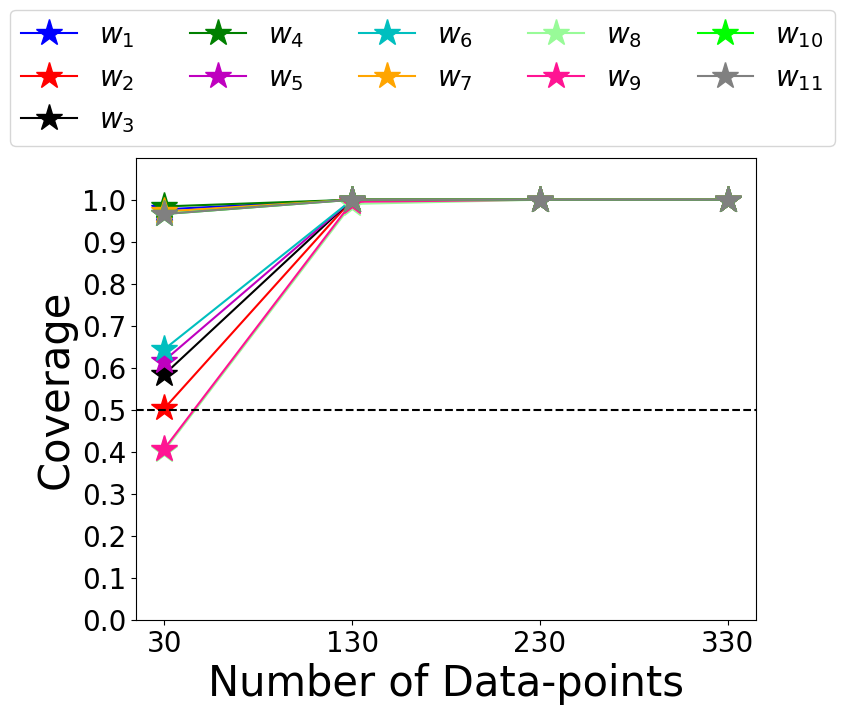}}
    \\
    \text{(a)}
    \\
    \includegraphics[width=1\linewidth]{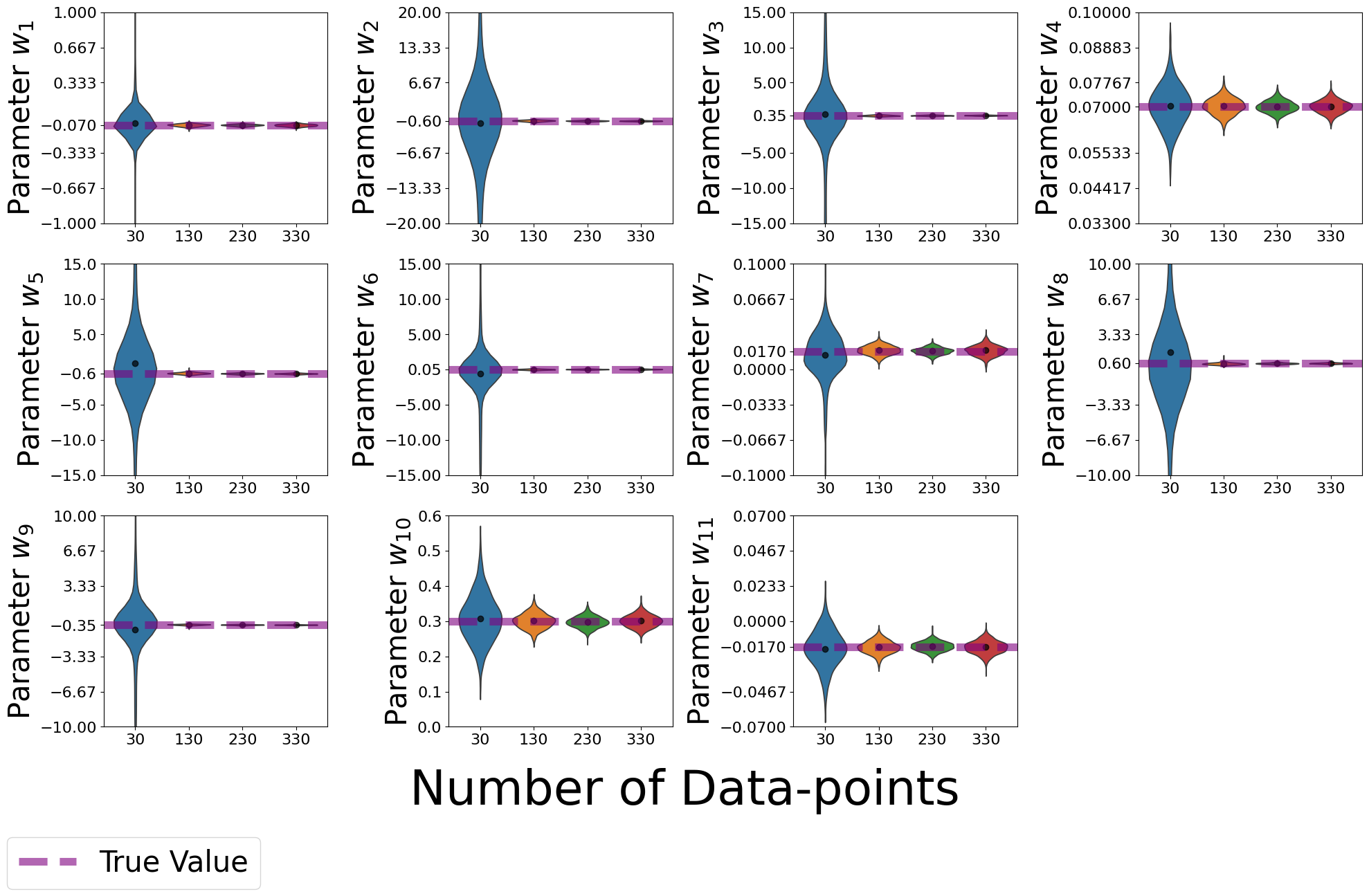}
    \\
      \text{(b)} 
    \end{tabular}
     \caption{PTB model parameter estimation performance with increasing resolution level (1000 datasets per level, 10\% additive normal noise). (a) coverage across four noise levels. (b) violin plots of parameter estimates, with the dashed red line indicating the true parameter values.}
    \label{fig:CovBiasResPTBN}
\end{figure}

As seen in Figure~\ref{fig:CovBiasResPTBN}(a), the coverage for the $w_2$, $w_3$, $w_5$, $w_6$, and $w_9$ parameters started below nominal at a resolution of 30 data points, with $w_9$ falling below 50\% coverage, before rising to nominal at 130 data points and higher. The remaining parameters started above nominal at 30 data points and remained above as resolution increased. As shown in Figure~\ref{fig:CovBiasResPTBN}(b)  bias and variance for all parameters decreased substantially as the resolution increased from 30 to 130 data points, as expected.  

\begin{figure} 
    \centering

    \includegraphics[width=1\linewidth]{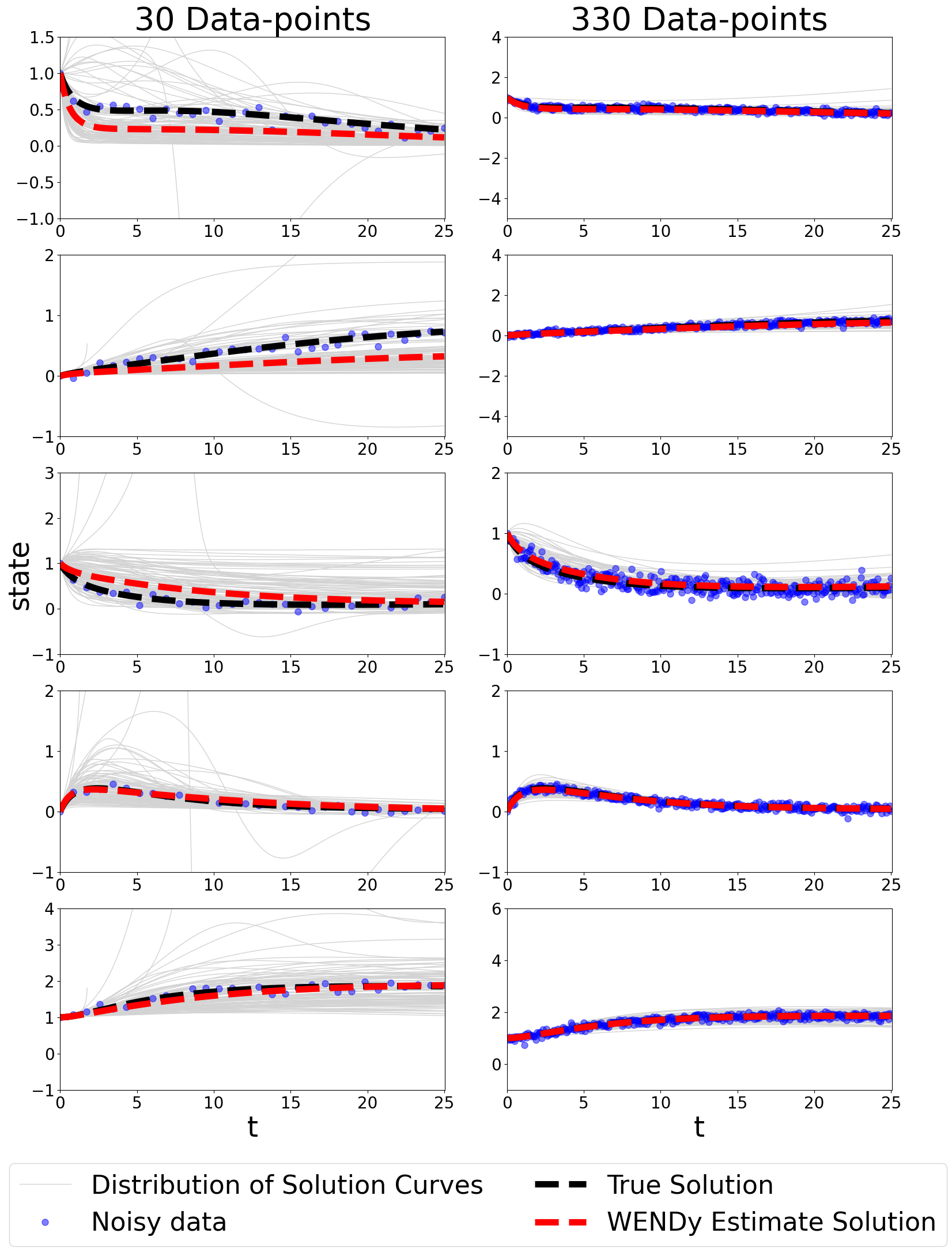} 
    
     \caption{PTB model parameter estimation example and uncertainty quantification on two datasets with additive normal noise: one dataset with low resolution (left) and one dataset with high resolution (right). The light gray curves are used to illustrate the uncertainty around the WENDy solutions; they are obtained via parametric bootstrap, as a sample of WENDy solutions corresponding to a random sample of 1000 parameters from their estimated asymptotic estimator distribution.}
    \label{fig:SamplePlotResPTBN}
\end{figure}

\begin{figure} 
    \centering
   
    \includegraphics[width=1\linewidth]{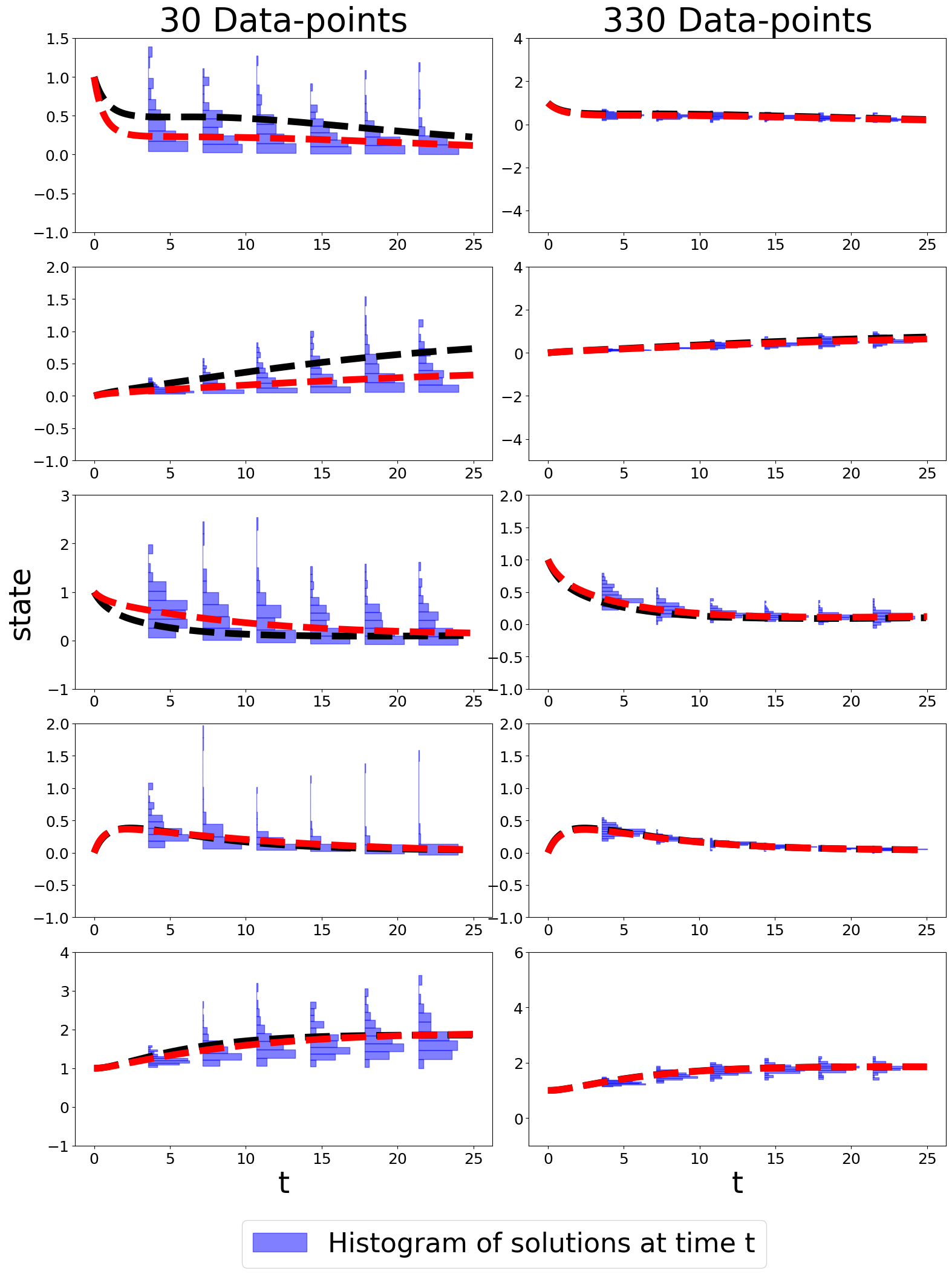}

     \caption{WENDy solution and histograms of state distributions across specific points in time for the datasets in ~\ref{fig:SamplePlotResPTBN} with additive normal noise.}
    \label{fig:SamplePlotResHistPTBN}
\end{figure}

Consistent with previous findings, the $w_9$ parameter was the most sensitive to resolution, whereas under noise variation the $w_8$ parameter was observed to be the most sensitive. As shown in Figure~\ref{fig:SamplePlotResPTBN} and ~\ref{fig:SamplePlotResHistPTBN}, the distribution of solution states (at selected time points) was wider at lower resolution, as expected. Nevertheless, even with only 30 data points, WENDy estimates produced a reasonable fit, likely due to the relatively simple dynamics of the PTB model, which lacks oscillatory states.

\clearpage

\subsubsection{Additive Censored Normal Noise}
\begin{figure} 
    \centering
    \begin{tabular}{c}
{\includegraphics[width=0.8\linewidth]{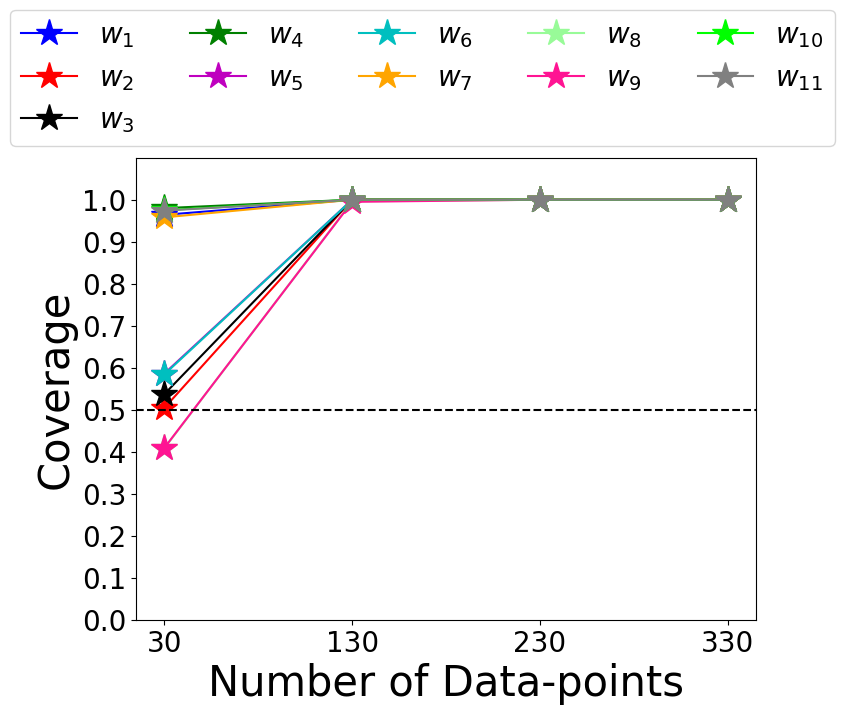}}
    \\
    \text{(a)}
    \\
    \includegraphics[width=1\linewidth]{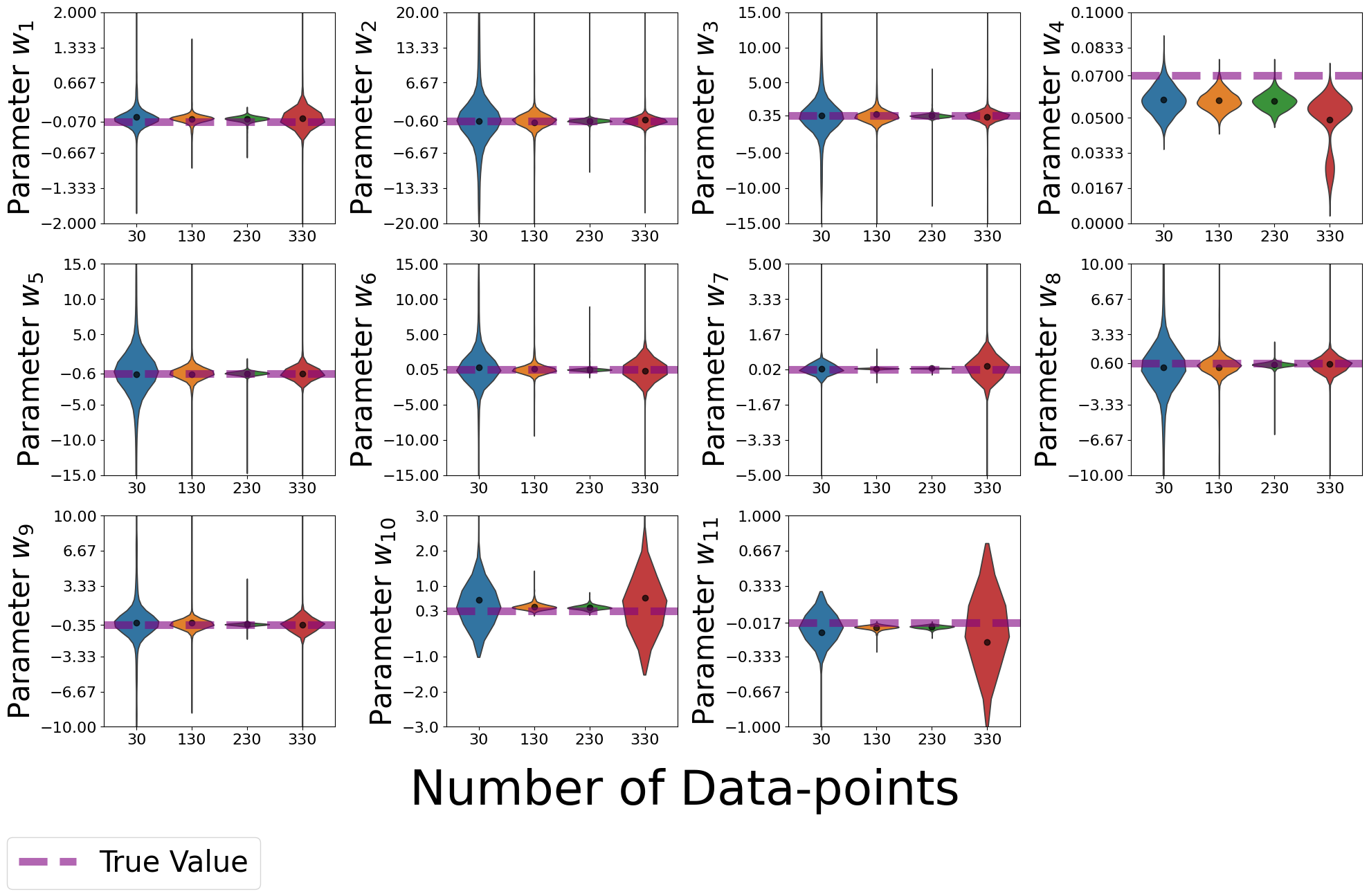}
    \\
      \text{(b)}
    \end{tabular}
     \caption{PTB model parameter estimation performance with increasing resolution level (1000 datasets per level, 10\% ACN). (a) coverage across four noise levels. (b) violin plots of parameter estimates, with the dashed red line indicating the true parameter values.}
    \label{fig:CovBiasResPTBCN}
\end{figure}

As seen in Figure~\ref{fig:CovBiasResPTBCN}(a), the coverage for parameters $w_2$, $w_3$, $w_6$, and $w_9$ started below nominal at 30 data points, with $w_9$ falling to less than 50\%. Coverage for these parameters rose to nominal levels as resolution increased to 130 data points and higher, while the remaining parameters began at nominal coverage and remained there across all resolution levels. As shown in Figure~\ref{fig:CovBiasResPTBCN}(b), both the bias and variance of all parameters decreased substantially with increasing resolution, as expected. Once again, the $w_9$ parameter emerged as the most sensitive to resolution, consistent with trends observed under additive normal noise.  

\begin{figure} 
    \centering

    \includegraphics[width=1\linewidth]{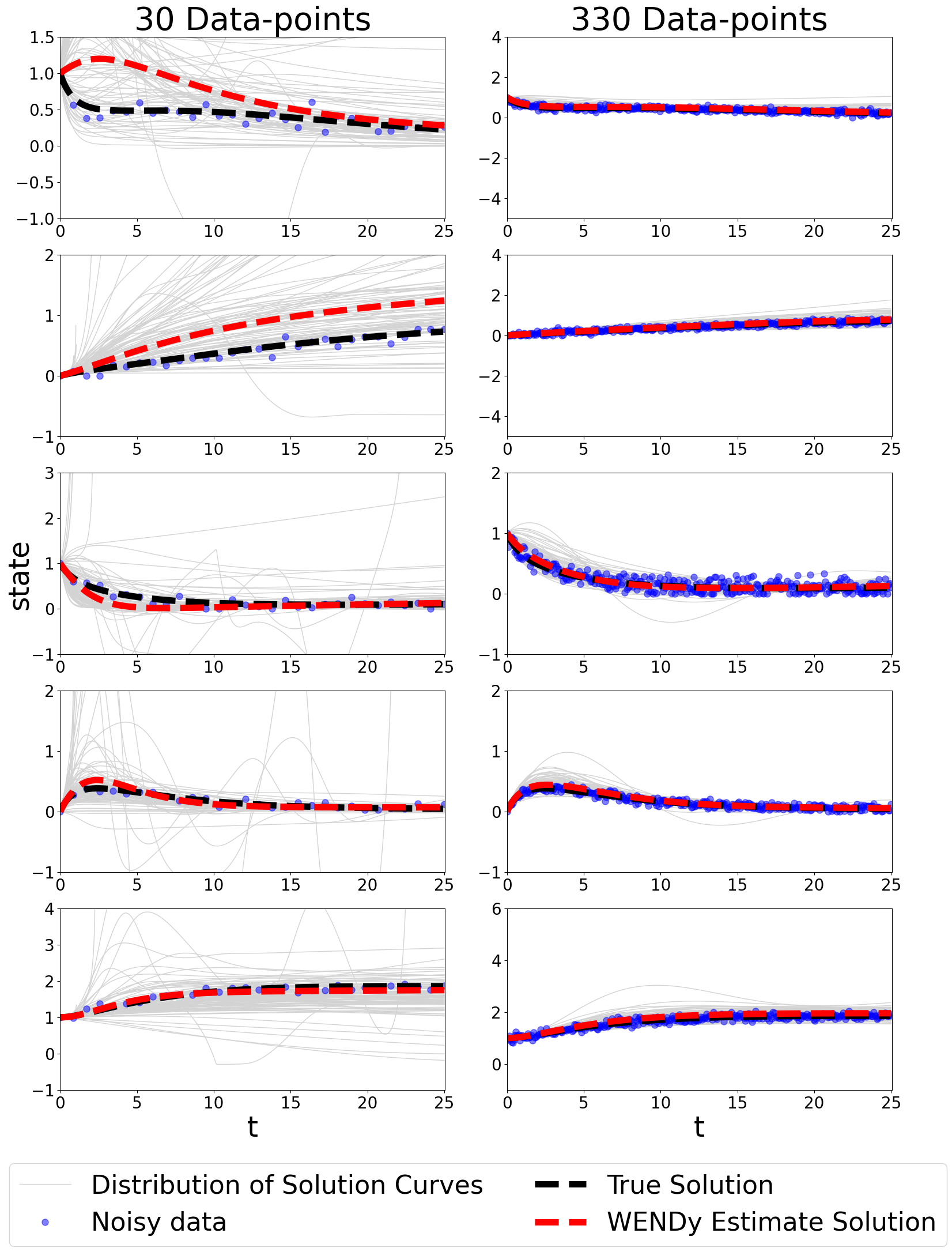} 
    
     \caption{PTB model parameter estimation example and uncertainty quantification on two datasets with ACN noise: one dataset with low resolution (left) and one dataset with high resolution (right). The light gray curves are used to illustrate the uncertainty around the WENDy solutions; they are obtained via parametric bootstrap, as a sample of WENDy solutions corresponding to a random sample of 1000 parameters from their estimated asymptotic estimator distribution.}
    \label{fig:SamplePlotResPTBACN}
\end{figure}

\begin{figure} 
    \centering
   
    \includegraphics[width=1\linewidth]{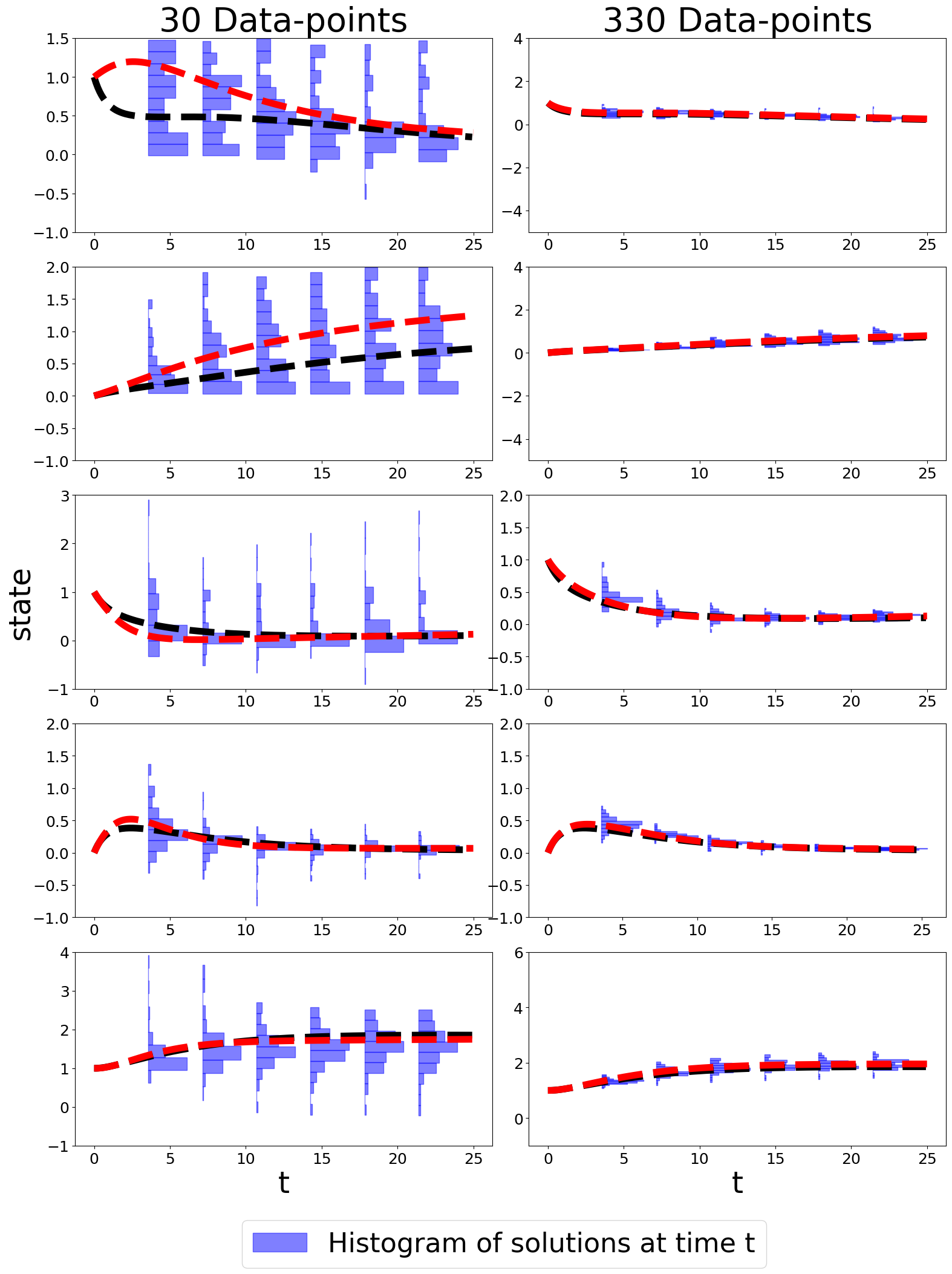}

     \caption{WENDy solution and histograms of state distributions across specific points in time for the datasets in ~\ref{fig:SamplePlotResPTBACN} with ACN noise.}
    \label{fig:SamplePlotResHistPTBACN}
\end{figure}

As illustrated in Figure~\ref{fig:SamplePlotResPTBACN} and ~\ref{fig:SamplePlotResHistPTBACN}, the distribution of solution states (at selected time points) was wider at low resolution, as expected. However, in contrast to the normal noise case, the WENDy-estimated solution produced a noticeably poorer fit to the true dynamics, with many trajectories skewed toward zero. This effect likely arises from the censoring of data points to zero under the ACN noise structure, which biases the solution in that direction.


\subsubsection{Multiplicative Log-normal Noise}
\begin{figure} 
    \centering
    \begin{tabular}{c}
{\includegraphics[width=0.8\linewidth]{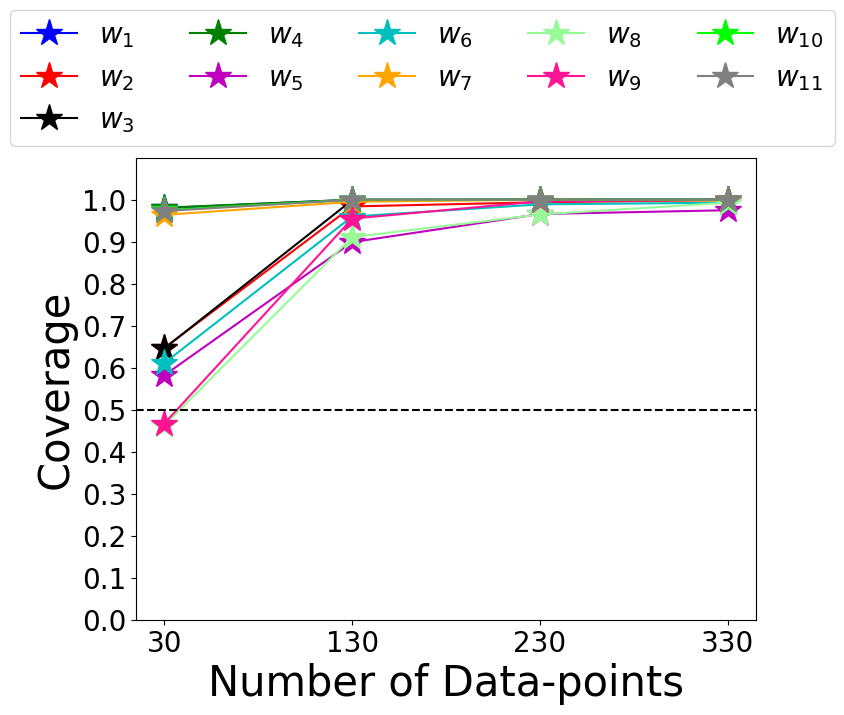}}
    \\
    \text{(a)}
    \\
    \includegraphics[width=1\linewidth]{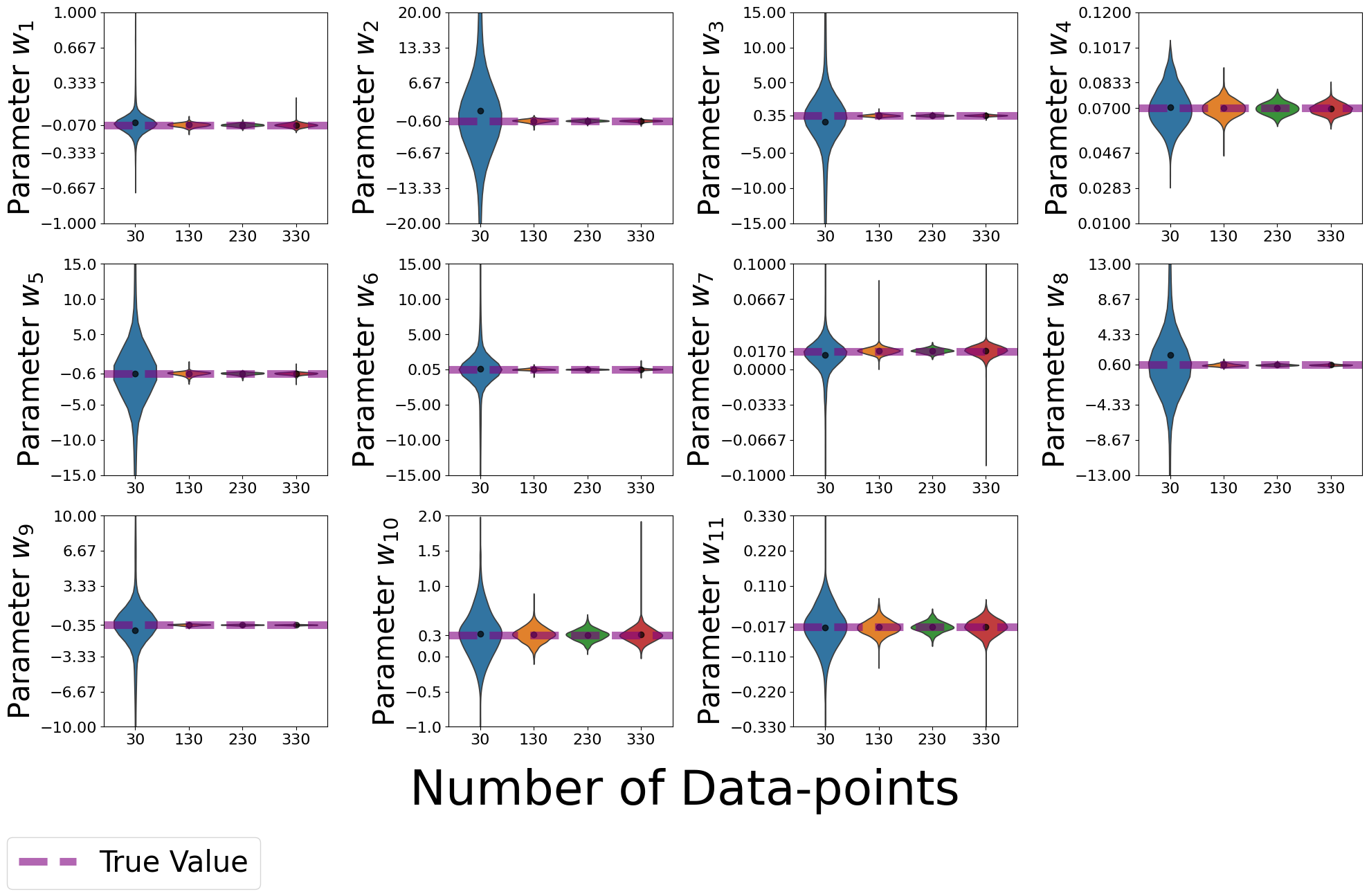}
    \\
      \text{(b)}
    \end{tabular}
     \caption{PTB model parameter estimation performance with increasing resolution level (1000 datasets per level, 10\% MLN). (a) coverage across four noise levels. (b) violin plots of parameter estimates, with the dashed red line indicating the true parameter values.}
    \label{fig:CovBiasResPTBLN}
\end{figure}

As seen in Figure~\ref{fig:CovBiasResPTBLN}(a), the coverage for parameters $w_2$, $w_3$, $w_5$, $w_6$, $w_8$, and $w_9$ started below nominal at 30 data points, with $w_8$ and $w_9$ falling below 50\%. Coverage for these parameters gradually increased, reaching just under nominal at 130 data points and achieving nominal levels only at 230 data points. The remaining parameters began at nominal coverage and remained stable as resolution increased. As shown in Figure~\ref{fig:CovBiasResPTBLN}(b), both bias and variance decreased substantially with increasing resolution, as expected. Notably, in this case, both $w_8$ and $w_9$ fell below 50\% coverage, and considerably more data points were required for all parameters to converge to nominal coverage compared to the other noise structures.  

\begin{figure} 
    \centering

    \includegraphics[width=1\linewidth]{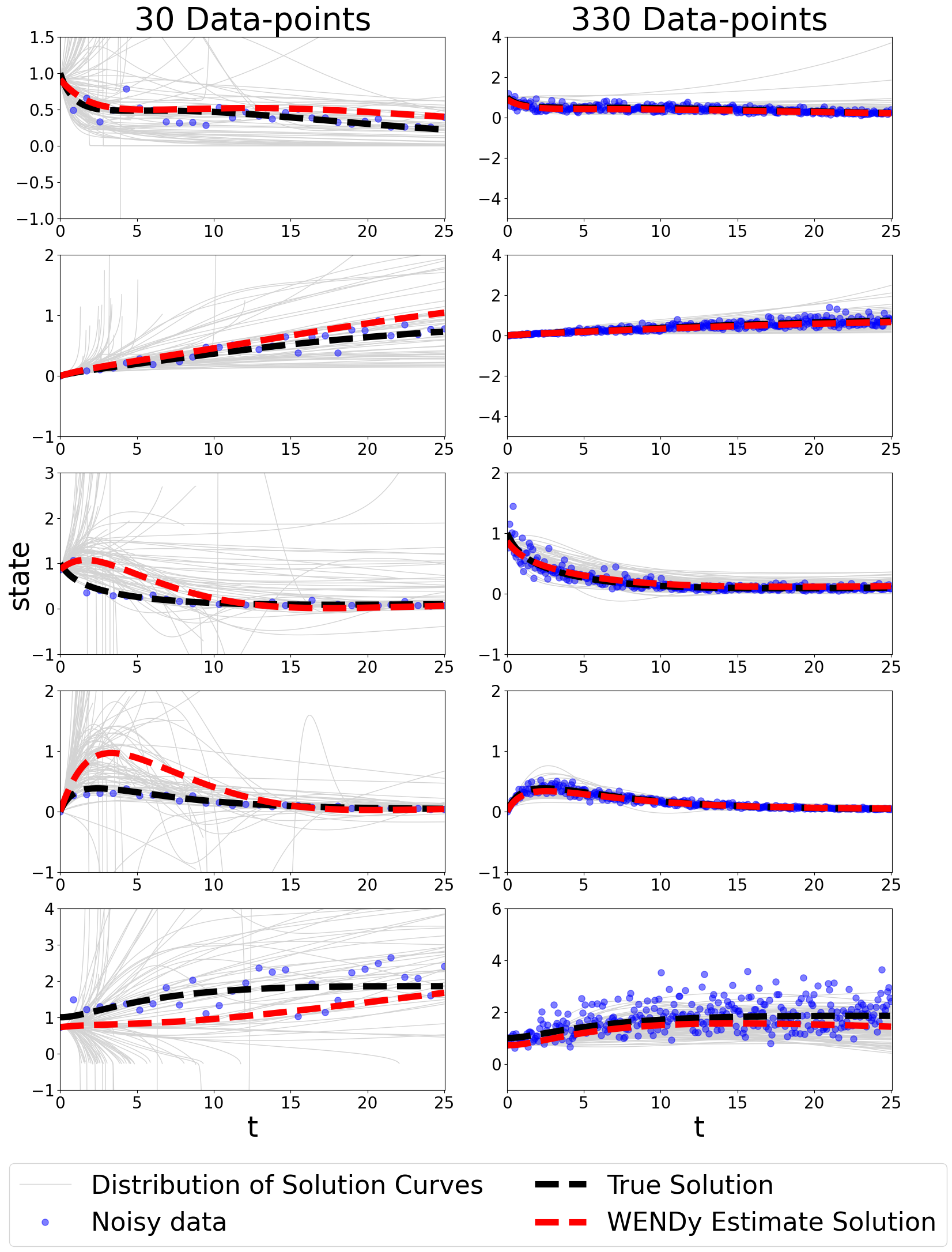} 
    
     \caption{PTB model parameter estimation example and uncertainty quantification on two datasets with MLN noise: one dataset with low resolution (left) and one dataset with high resolution (right). The light gray curves are used to illustrate the uncertainty around the WENDy solutions; they are obtained via parametric bootstrap, as a sample of WENDy solutions corresponding to a random sample of 1000 parameters from their estimated asymptotic estimator distribution.}
    \label{fig:SamplePlotResPTBMLN}
\end{figure}

\begin{figure} 
    \centering
   
    \includegraphics[width=1\linewidth]{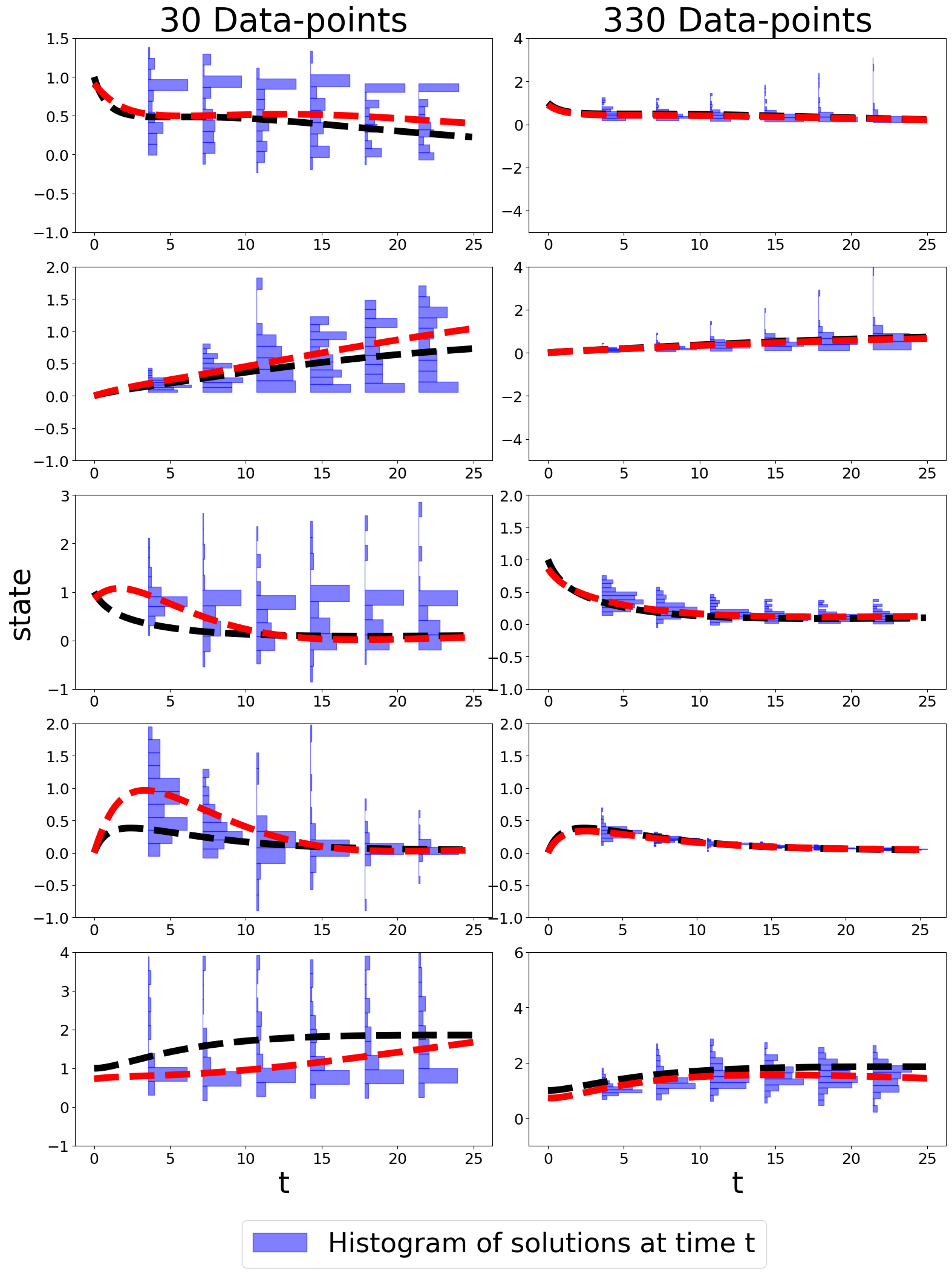}

     \caption{WENDy solution and histograms of state distributions across specific points in time for the datasets in ~\ref{fig:SamplePlotResPTBMLN} with ACN noise.}
    \label{fig:SamplePlotResHistPTBMLN}
\end{figure}

As illustrated in Figure~\ref{fig:SamplePlotResPTBMLN} and ~\ref{fig:SamplePlotResHistPTBMLN}, the distribution of solution states (at selected time points) was wider at low resolution, as expected. However, the quality of the fits at low resolution was noticeably worse than under the other noise structures. Combined with the slower recovery of coverage, this suggests that WENDy estimators are less robust to data resolution under MLN noise than under either ACN or additive normal noise. This reduced robustness likely stems from the heteroskedastic nature of MLN noise, which amplifies variability when data are sparse.  

\subsubsection{Additive Truncated Normal Noise}
\begin{figure} 
    \centering
    \begin{tabular}{c}
{\includegraphics[width=0.8\linewidth]{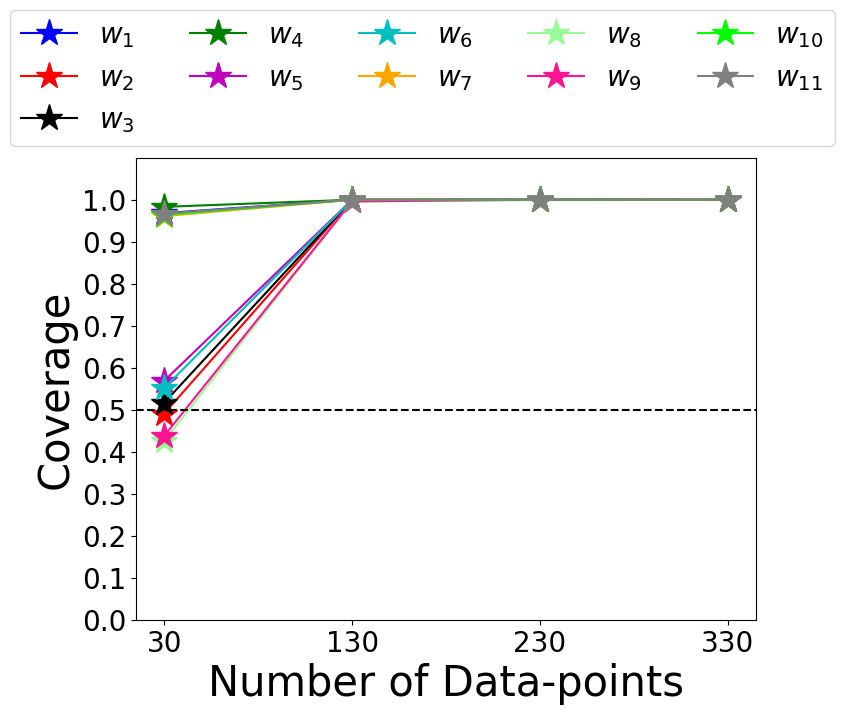}}
    \\
    \text{(a)}
    \includegraphics[width=0.8\linewidth]{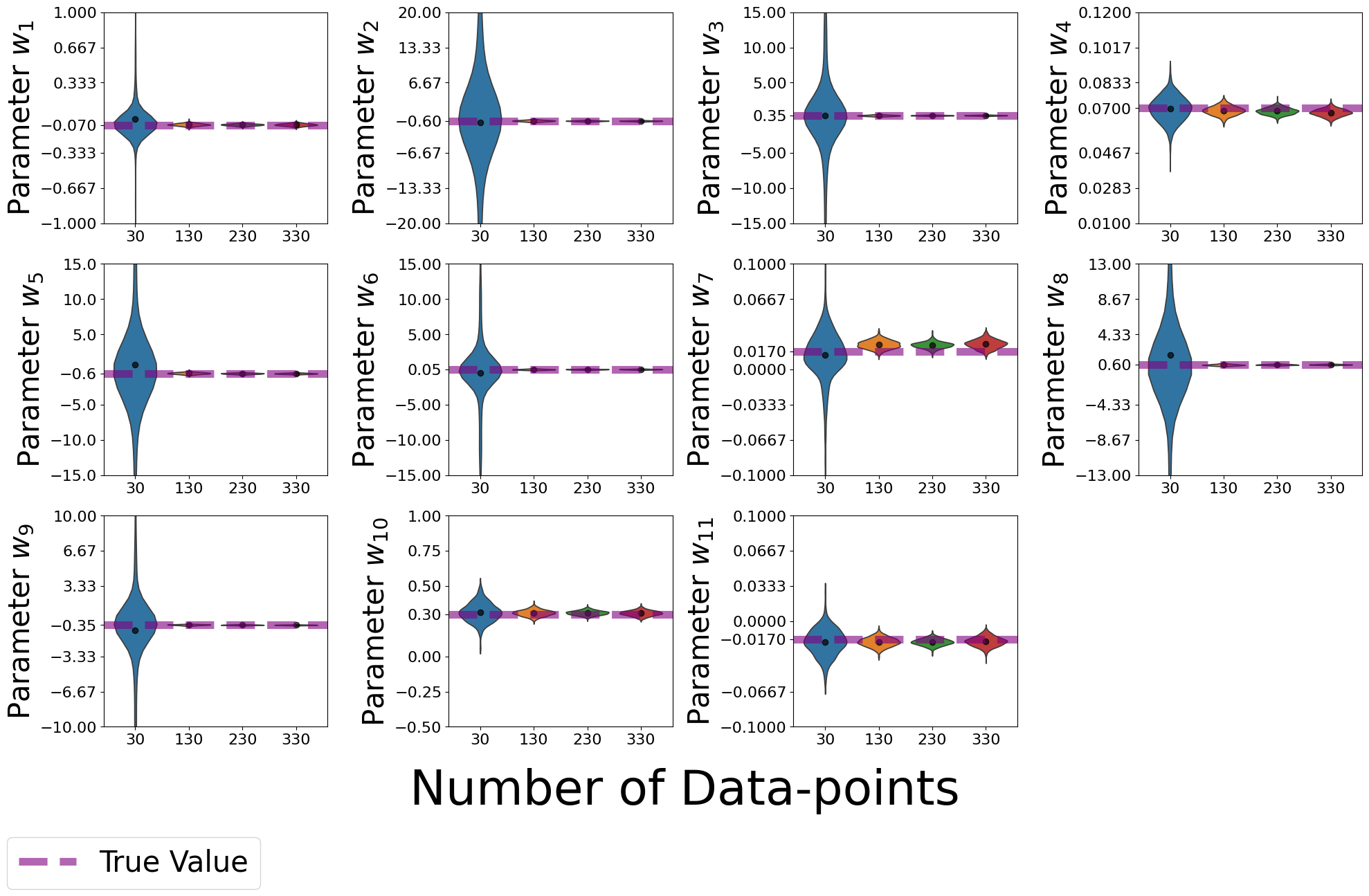}
    \\
      \text{(b)} 
    \end{tabular}
     \caption{PTB model parameter estimation performance with increasing resolution level (1000 datasets per level, 10\% ATN). (a) coverage across four noise levels. (b) violin plots of parameter estimates, with the dashed red line indicating the true parameter values.}
    \label{fig:CovBiasResPTBTN}
\end{figure}

As seen in Figure~\ref{fig:CovBiasResPTBTN}(a), the coverage for parameters $w_2$, $w_3$, $w_5$, $w_6$, $w_8$, and $w_9$ started below nominal at 30 data points, with $w_2$, $w_8$, and $w_9$ falling below 50\%. Coverage for these parameters rose to nominal levels as resolution increased to 130 data points and higher. The remaining parameters began at nominal coverage and remained stable as resolution increased. As shown in Figure~\ref{fig:CovBiasResPTBTN}(b), both bias and variance decreased with increasing resolution, as expected. At 30 data points, a larger number of parameters started below nominal compared to the other noise structures, with $w_2$, $w_8$, and $w_9$ showing the lowest coverage.  

\begin{figure} 
    \centering

    \includegraphics[width=1\linewidth]{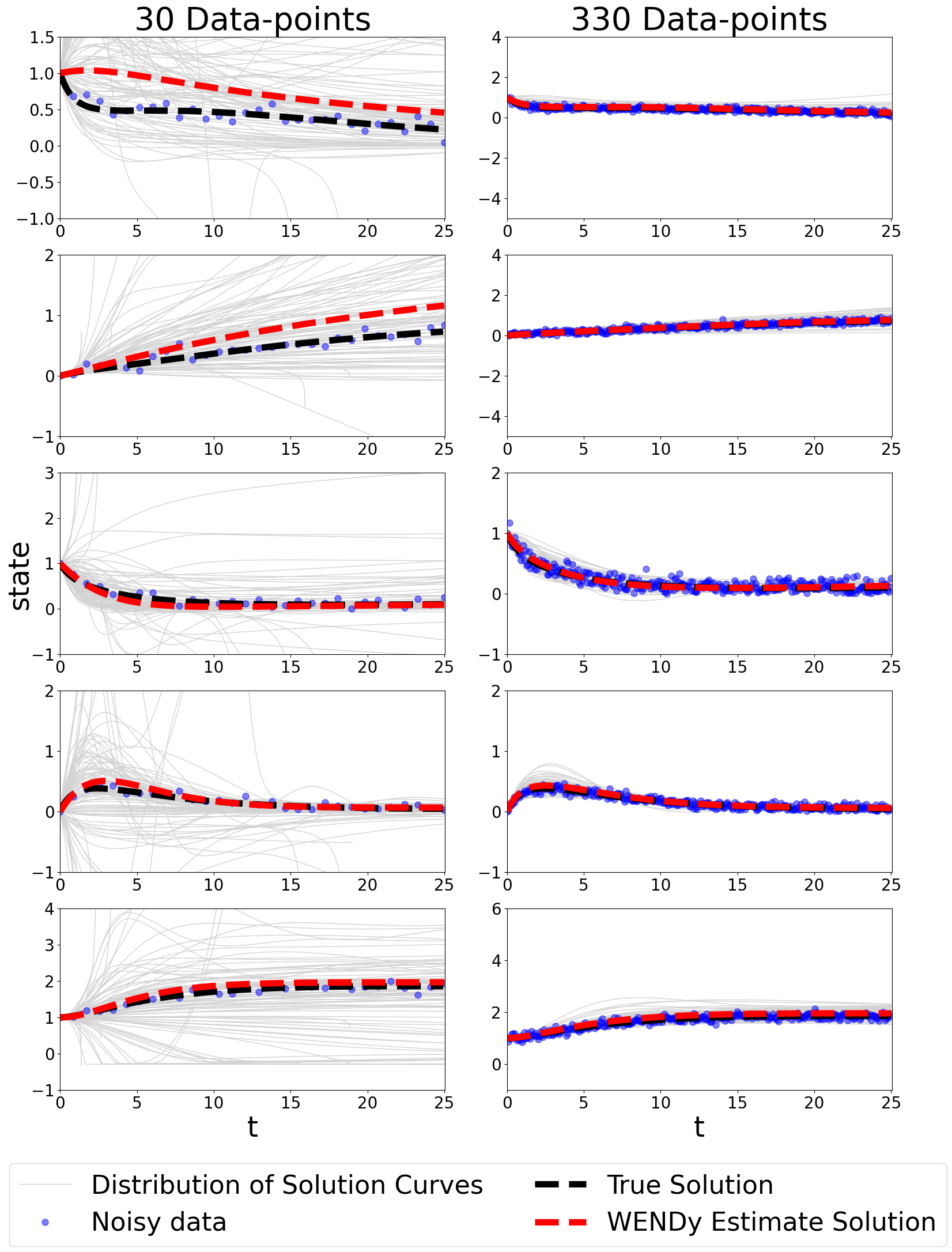} 
    
     \caption{PTB model parameter estimation example and uncertainty quantification on two datasets with ATN noise: one dataset with low resolution (left) and one dataset with high resolution (right). The light gray curves are used to illustrate the uncertainty around the WENDy solutions; they are obtained via parametric bootstrap, as a sample of WENDy solutions corresponding to a random sample of 1000 parameters from their estimated asymptotic estimator distribution.}
    \label{fig:SamplePlotResPTBATN}
\end{figure}

\begin{figure} 
    \centering
   
    \includegraphics[width=1\linewidth]{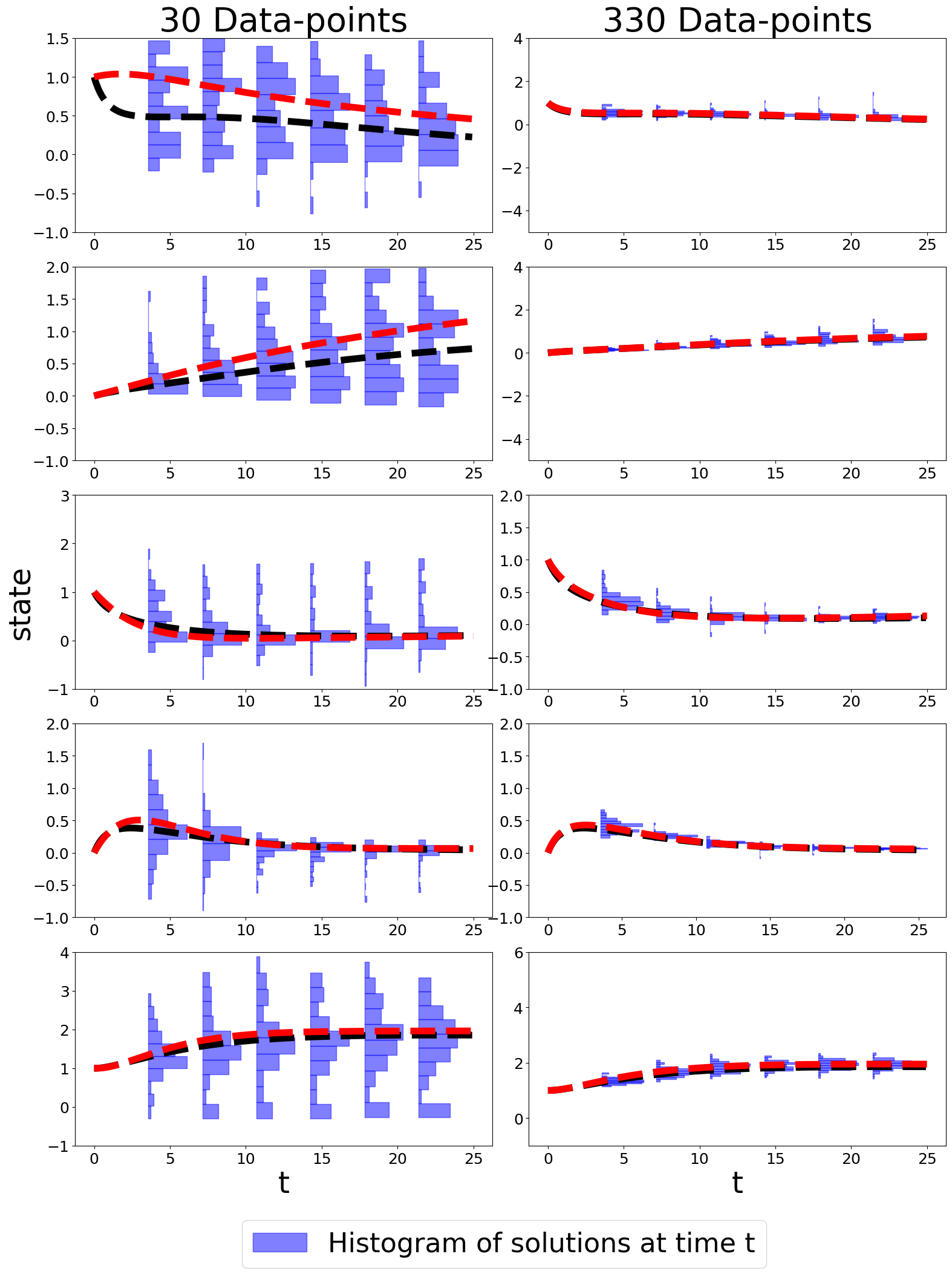}

     \caption{WENDy solution and histograms of state distributions across specific points in time for the datasets in ~\ref{fig:SamplePlotResPTBACN} with ACN noise.}
    \label{fig:SamplePlotResHistPTBATN}
\end{figure}

As illustrated in Figures~\ref{fig:SamplePlotResPTBATN} and ~\ref{fig:SamplePlotResHistPTBATN}, the distribution of solution states (at selected time points) was wider at low resolution, as expected. The WENDy-estimated fit was also somewhat worse than under additive normal noise, consistent with the lower coverage observed.  


\section{Discussion}

This is the first systematic study to evaluate the coverage and bias properties of state and parameter estimators from the original WENDy algorithm \citep{BortzMessengerDukic2023BullMathBiol} across five benchmark models under 4 different distributional forms and varying levels of noise, as well as different data resolutions. 
While previous work \citep{MessengerDwyerDukic2024JRSocInterface} has explored the performance of WENDy estimators for states and their functionals (namely, the amplitude and period) using parametric bootstrap, that was done for a specific ecological hybrid system, and not the 5 benchmark problems considered here.
Furthermore, the work here presents a study of the WENDy estimator performance across a wide range of noise levels, reaching much higher levels than what other algorithms have explored. For instance, while the sparse regression-based equation learning algorithm SINDy \citep{BruntonProctorKutz2016ProcNatlAcadSci} can also be used to estimate parameters, SINDy's use of approximate derivatives results in a high sensitivity to noise; the algorithm can break at noise levels as low as 0.5\% \citep{RudyBruntonProctorEtAl2017SciAdv}.  In contrast, the noise levels used in this paper can reach noise-to-signal ratio of 80\% in some examples. Note that the upper noise level in each example in this paper was chosen simply when the coverage of the 95\% confidence interval for at least one parameter fell below 50\%, and not because of the algorithm's breaking or failing to produce estimates. On the other hand, while the weak form extension (WSINDy \citep{MessengerBortz2021JComputPhys,MessengerBortz2021MultiscaleModelSimul}) is highly robust to noise, the parameter estimates exhibit bias in general \citep{MessengerBortz2024IMAJNumerAnal}.\footnote{Note that using WSINDy to estimate parameters is asymptotically unbiased in the limit of continuum data, but only for a restricted class of differential equations. See \citep{MessengerBortz2024IMAJNumerAnal} for more details.} 

In general, WENDy estimators exhibited higher coverage, lower bias, and lower uncertainty when noise levels were lower and data resolution was high. They showed remarkable robustness to high levels of noise and low levels of data resolution for some models  and noise distributions. For the Logistic, LV, and PTB models, which were tested with additive normal, ACN, MLN, and ATN noise, WENDy estimators had nominal coverage for high levels of noise. For each type of noise structure, each model consistently had the same few parameter estimators that would drop below 50\% coverage, showcasing WENDy's consistency in parameter coverage regardless of noise structure. 

WENDy's coverage performance declined more significantly for the FHN and HMR models. Coverage for at least one parameter dropped below 50\% at noise levels above 7\% for FHN and 4\% for HMR. Moreover, the HMR model required substantially higher data resolution -- over 125 data points per state -- to achieve parameter coverage exceeding 50\%. This difficulty is likely due to the presence of cubic nonlinearities in these models, which create complex dynamics that are harder to capture, especially under high noise or low resolution. Based on these findings, it is recommended to use at least 60 data points per state and at least 125 per state for FHN and HMR, respectively, for low-noise datasets. Significantly more data points may be needed at higher noise levels, and this is simply a lower bound.

The PTB model allowed the highest noise, with parameter coverage dipping below 50\% at approximately 80\% normal noise for the first time. This may be due to the fact that PTB is a compartmental model with mostly linear terms and simple dynamics, lacking any periodicity. However, the spread in estimated solutions at the highest noise level was quite large; thus, although WENDy can technically function under such noise, the resulting uncertainty may render the estimates impractical. In such cases with high noise, it may be worth switching to non-linear WENDy to get better estimators. Note, however, that the noise level of 80\% would be considered quite large by any standards, and would not normally be encountered in experimental and lab data.

Also tested was the bias correction for datasets with log-normal noise, showing promising preliminary results on the Logistic and LV models. The future plans are to expand upon this technique as a way to supplement WENDy estimated solution curves in the presence of non-normal noise structures. 

In conclusion, WENDy algorithm exhibits remarkable stability, with low variance and bias, under various noise levels and distributions, showing robustness to noise levels that are orders of magnitude larger than what would be considered reasonable in most scientific lab data \citep{BortzMessengerDukic2023BullMathBiol}. While extensions to WENDy exist \citep{RummelMessengerBeckerEtAl2025arXiv250208881}, WENDy remains an effective and fast algorithm of choice for parameter and state inference in dynamical systems.

\section{Acknowledgments}
This research was supported in part by the NIFA Biological Sciences Grant 2019-67014-29919 (to VD), in part by the NSF Division Of Environmental Biology Grant 2109774 (to VD), and in part by the NIH-NIGMS Division of Biophysics, Biomedical Technology and Computational Biosciences grant R35GM149335 (to DMB).

\bibliographystyle{spbasic}  
\bibliography{WENDyCovBias}

\begin{thebibliography}{24}
\providecommand{\natexlab}[1]{#1}
\providecommand{\url}[1]{{#1}}
\providecommand{\urlprefix}{URL }
\expandafter\ifx\csname urlstyle\endcsname\relax
  \providecommand{\doi}[1]{DOI~\discretionary{}{}{}#1}\else
  \providecommand{\doi}{DOI~\discretionary{}{}{}\begingroup
  \urlstyle{rm}\Url}\fi
\providecommand{\eprint}[2][]{\url{#2}}

\bibitem[{Bortz et~al(2023)Bortz, Messenger, and
  Dukic}]{BortzMessengerDukic2023BullMathBiol}
Bortz DM, Messenger DA, Dukic V (2023) Direct {{Estimation}} of {{Parameters}}
  in {{ODE Models Using WENDy}}: {{Weak-form Estimation}} of {{Nonlinear
  Dynamics}}. Bull Math Biol 85(110), \doi{10.1007/S11538-023-01208-6}

\bibitem[{Brunton et~al(2016)Brunton, Proctor, and
  Kutz}]{BruntonProctorKutz2016ProcNatlAcadSci}
Brunton SL, Proctor JL, Kutz JN (2016) Discovering governing equations from
  data by sparse identification of nonlinear dynamical systems. Proc Natl Acad
  Sci 113(15):3932--3937, \doi{10.1073/pnas.1517384113}

\bibitem[{Dukic et~al(2012)Dukic, Lopes, and
  Polson}]{DukicLopesPolson2012JAmStatAssoc}
Dukic V, Lopes HF, Polson NG (2012) Tracking {{Epidemics With Google Flu Trends
  Data}} and a {{State-Space SEIR Model}}. J Am Stat Assoc 107(500):1410--1426,
  \doi{10.1080/01621459.2012.713876}

\bibitem[{Elderd et~al(2006)Elderd, Dukic, and
  Dwyer}]{ElderdDukicDwyer2006ProcNatlAcadSciUSA}
Elderd BD, Dukic VM, Dwyer G (2006) Uncertainty in predictions of disease
  spread and public health responses to bioterrorism and emerging diseases.
  Proc Natl Acad Sci USA 103(42):15,693--15,697, \doi{10.1073/pnas.0600816103}

\bibitem[{FitzHugh(1961)}]{FitzHugh1961BiophysicalJournal}
FitzHugh R (1961) Impulses and {{Physiological States}} in {{Theoretical
  Models}} of {{Nerve Membrane}}. Biophysical Journal 1(6):445--466,
  \doi{10.1016/S0006-3495(61)86902-6}

\bibitem[{Greenberg(1951)}]{Greenberg1951NACATN2340}
Greenberg H (1951) A survey of methods for determining stability parameters of
  an airplane from dynamic flight measurements. Tech. Rep. NACA TN 2340, Ames
  Aeronautical Laboratory, Moffett Field, CA

\bibitem[{Hindmarsh and Rose(1984)}]{HindmarshRose1984ProcRSocLondBBiolSci}
Hindmarsh JL, Rose RM (1984) A model of neuronal bursting using three coupled
  first order differential equations. Proc R Soc Lond B Biol Sci
  221(1222):87--102, \doi{10.1098/rspb.1984.0024}

\bibitem[{Loeb and Cahen(1963)}]{LoebCahen1963Automatisme}
Loeb JM, Cahen GM (1963) Extraction a partir des enregistrements de mesures,
  des parametres dynamiques d'un systeme. Automatisme 8:479--486

\bibitem[{Loeb and Cahen(1965)}]{LoebCahen1965IEEETransAutomControl}
Loeb JM, Cahen GM (1965) More about process identification. IEEE Trans Autom
  Control 10(3):359--361, \doi{10.1109/TAC.1965.1098172}

\bibitem[{Lotka(1978)}]{Lotka1978TheGoldenAgeofTheoreticalEcology1923-1940}
Lotka AJ (1978) The Growth of Mixed Populations: {{Two}} Species Competing for
  a Common Food Supply, vol~22, Springer Berlin Heidelberg, Berlin, Heidelberg,
  pp 274--286. \doi{10.1007/978-3-642-50151-7_12}

\bibitem[{McGoff et~al(2015)McGoff, Mukherjee, and
  Pillai}]{McGoffMukherjeePillai2015StatistSurv}
McGoff K, Mukherjee S, Pillai N (2015) Statistical inference for dynamical
  systems: {{A}} review. Statist Surv 9(none), \doi{10.1214/15-SS111}

\bibitem[{Messenger and
  Bortz(2021{\natexlab{a}})}]{MessengerBortz2021JComputPhys}
Messenger DA, Bortz DM (2021{\natexlab{a}}) Weak {{SINDy For Partial
  Differential Equations}}. J Comput Phys 443:110,525,
  \doi{10.1016/j.jcp.2021.110525}

\bibitem[{Messenger and
  Bortz(2021{\natexlab{b}})}]{MessengerBortz2021MultiscaleModelSimul}
Messenger DA, Bortz DM (2021{\natexlab{b}}) Weak {{SINDy}}: {{Galerkin-Based
  Data-Driven Model Selection}}. Multiscale Model Simul 19(3):1474--1497,
  \doi{10.1137/20M1343166}

\bibitem[{Messenger and Bortz(2024)}]{MessengerBortz2024IMAJNumerAnal}
Messenger DA, Bortz DM (2024) Asymptotic consistency of the {{WSINDy}}
  algorithm in the limit of continuum data. IMA J Numer Anal p drae086,
  \doi{10.1093/imanum/drae086}

\bibitem[{Messenger et~al(2024{\natexlab{a}})Messenger, Burby, and
  Bortz}]{MessengerBurbyBortz2024SciRep}
Messenger DA, Burby JW, Bortz DM (2024{\natexlab{a}}) Coarse-{{Graining
  Hamiltonian Systems Using WSINDy}}. Sci Rep 14(14457):1--24,
  \doi{10.1038/s41598-024-64730-0}

\bibitem[{Messenger et~al(2024{\natexlab{b}})Messenger, Dwyer, and
  Dukic}]{MessengerDwyerDukic2024JRSocInterface}
Messenger DA, Dwyer G, Dukic V (2024{\natexlab{b}}) Weak-form inference for
  hybrid dynamical systems in ecology. J R Soc Interface 21(221):20240,376,
  \doi{10.1098/rsif.2024.0376}

\bibitem[{Minor et~al(2025)Minor, Messenger, Dukic, and
  Bortz}]{MinorMessengerDukicEtAl2025JGeophysResMachLearnComput}
Minor S, Messenger DA, Dukic V, Bortz DM (2025) Learning {{Physically
  Interpretable Atmospheric Models}} from {{Data}} with {{WSINDy}}. J Geophys
  Res Mach Learn Comput 2(3):e2025JH000,602, \doi{10.1029/2025JH000602}

\bibitem[{Nardini and Bortz(2019)}]{NardiniBortz2019InverseProbl}
Nardini JT, Bortz DM (2019) The influence of numerical error on parameter
  estimation and uncertainty quantification for advective {{PDE}} models.
  Inverse Probl 35(6):065,003, \doi{10.1088/1361-6420/ab10bb}

\bibitem[{Rudy et~al(2017)Rudy, Brunton, Proctor, and
  Kutz}]{RudyBruntonProctorEtAl2017SciAdv}
Rudy SH, Brunton SL, Proctor JL, Kutz JN (2017) Data-driven discovery of
  partial differential equations. Sci Adv 3(4):e1602,614,
  \doi{10.1126/sciadv.1602614}

\bibitem[{Rummel et~al(2025)Rummel, Messenger, Becker, Dukic, and
  Bortz}]{RummelMessengerBeckerEtAl2025arXiv250208881}
Rummel N, Messenger DA, Becker S, Dukic V, Bortz DM (2025) {{WENDy}} for
  {{Nonlinear-in-Parameter ODEs}}. arXiv:250208881 \eprint{2502.08881}

\bibitem[{Schoeberl et~al(2002)Schoeberl, {Eichler-Jonsson}, Gilles, and
  M{\"u}ller}]{SchoeberlEichler-JonssonGillesEtAl2002NatBiotechnol}
Schoeberl B, {Eichler-Jonsson} C, Gilles ED, M{\"u}ller G (2002) Computational
  modeling of the dynamics of the {{MAP}} kinase cascade activated by surface
  and internalized {{EGF}} receptors. Nat Biotechnol 20(4):370--375,
  \doi{10.1038/nbt0402-370}

\bibitem[{Shinbrot(1954)}]{Shinbrot1954NACATN3288}
Shinbrot M (1954) On the analysis of linear and nonlinear dynamical systems for
  transient-response data. Tech. Rep. NACA TN 3288, Ames Aeronautical
  Laboratory, Moffett Field, CA

\bibitem[{Shinbrot(1957)}]{Shinbrot1957TransAmSocMechEng}
Shinbrot M (1957) On the {{Analysis}} of {{Linear}} and {{Nonlinear Systems}}.
  Trans Am Soc Mech Eng 79(3):547--551, \doi{10.1115/1.4013092}

\bibitem[{Tran et~al(2024)Tran, He, Messenger, Choi, and
  Bortz}]{TranHeMessengerEtAl2024ComputMethodsApplMechEng}
Tran A, He X, Messenger DA, Choi Y, Bortz DM (2024) Weak-{{Form Latent Space
  Dynamics Identification}}. Comput Methods Appl Mech Eng 427:116,998,
  \doi{10.1016/j.cma.2024.116998}

\end{thebibliography}



\end{document}